\documentclass[12pt]{article}
\usepackage{amssymb}
\usepackage{amsmath}
\usepackage{graphicx,color,xcolor}
\usepackage{amsfonts}
\usepackage[margin=1in]{geometry}
\usepackage{url}
\usepackage[sectionbib]{natbib}\usepackage[colorlinks,citecolor=blue!60!black,linkcolor=blue!60!black,urlcolor=blue!60!black]{hyperref}
\usepackage{subcaption}
\usepackage[ruled,linesnumbered]{algorithm2e}
\usepackage{xcolor}
\graphicspath{ {figures/} }
\usepackage{comment}
\usepackage{enumitem}

\usepackage{amsthm}

\usepackage[normalem]{ulem}
\newcommand{\stkout}[1]{\ifmmode\text{\sout{\ensuremath{#1}}}\else\sout{#1}\fi}
\usepackage{setspace}
\usepackage{afterpage}

\usepackage{tikz}
\usetikzlibrary{patterns,patterns.meta}

\usepackage{amsthm}
\usepackage{setspace}
\usepackage{afterpage}

\usepackage{chngcntr}
\usepackage{apptools}
\usepackage{makecell}

\usepackage{xr}
\makeatletter
\newcommand*{\addFileDependency}[1]{
  \typeout{(#1)}
  \@addtofilelist{#1}
  \IfFileExists{#1}{}{\typeout{No file #1.}}
}
\makeatother

\DeclareMathOperator*{\argmin}{arg\,min}

\newcommand{\blue}[1]{\textcolor{blue}{#1}}

\newcommand{\RNum}[1]{\uppercase\expandafter{\romannumeral #1\relax}}

\newtheorem{theorem}{Theorem}

\newtheorem{condition}{Condition}

\newtheorem{definition}{Definition}

\newtheorem{lemma}{Lemma}

\newtheorem{proposition}{Proposition}
\newtheorem{remark}{Remark}

\newcommand{\EXPT}{\mathbb{E}}

\newcommand{\VAR}{\mathrm{var}}
\newcommand{\mytrans}{\top}

\newcommand{\DZt}{\mathcal{D}_{\Theta Z}}
\newcommand{\DZ}{\mathcal{D}_{Z\Theta}}
\newcommand{\UZ}{\mathcal{U}_Z}

\newcommand{\Dmut}{\mathcal{D}_{\mu_t}}
\newcommand{\Dwt}{\mathcal{D}_{W_t \Theta}}

\newcommand{\Dwtt}{\mathcal{D}_{\Theta W_t}}

\newcommand{\ksum}{k_{\operatorname{sum}}}
\newcommand{\dmax}{d_{\operatorname{max}}}
\newcommand{\DnT}{\mathcal{D}_{nT}}

\title{\vspace{-2.5em}Efficient Analysis of Latent Spaces in Heterogeneous Networks  }
\date{} 
 { 
 \author{Yuang Tian$^{1}$, Jiajin Sun$^{2}$, Yinqiu He$^{3,}$\thanks{Corresponding Author}}
 \date{\vspace{-1em}
     $^1${\small {Department of Mathematics, Hong Kong University of Science and Technology, \hyperref[yatian@ust.hk
 ]{yatian@ust.hk
 }}}\\[-9pt]
     $^2${\small {Department of Statistics, Florida State University, \hyperref[jsun5@fsu.edu
 ]{jsun5@fsu.edu
 } }}\\[-5pt]
     $^3${\small {Department of Statistics, University of Wisconsin-Madison, \hyperref[yinqiu.he@wisc.edu]{yinqiu.he@wisc.edu}}}
 }
 }
\begin{document}

\maketitle

 \vspace{-2em}
\begin{abstract}
 This work proposes a unified framework for efficient estimation under  latent space modeling of  heterogeneous networks. We consider a class of latent space models that decompose latent vectors into shared and network-specific  components across   networks. 
 We develop a novel procedure that first identifies the shared latent vectors and further refines estimates through efficient score equations to achieve statistical efficiency. 
Oracle error rates for estimating the shared and heterogeneous latent vectors are established simultaneously. 
 The analysis framework offers remarkable flexibility, accommodating various types of edge weights under {general distributions}. 
\end{abstract}
\vspace{-0.8em}

\vspace{-0.7em}
\section{Introduction} 
In recent years, 
analyses of multiview or multiplex networks \citep{kivela2014multilayer} open up new opportunities for understanding complex systems across diverse
scientific endeavors 
\citep{salter2017latent}, including social science \citep{lazega2001collegial,banerjee2013diffusion},  microbiome  \citep{gould2018microbiome}, and neuroscience \citep{wen2022characterizing}. 
In these applications, it is common to observe multiple networks sharing a common set of nodes, with each edge representing specific types of 
relationships between node pairs.  
Studying multiple networks 
enables a unified understanding of multifaceted and  interconnected systems.

Latent space modeling has emerged as an important approach  in network analysis  
 \citep{hoff2002latent,matias2014modeling}. In this framework, each node is mapped to a  vector in a latent space, and the connectivity strength  between two nodes is parametrized through a function between two corresponding latent vectors. 
While latent spaces can be discrete, they are 
more commonly referred to as stochastic block models and exhibit unique properties  from discreteness 
\citep{holland1983stochastic}.  
Throughout this paper, we focus on continuous latent vectors and consider latent spaces as  Euclidean spaces  following the convention in the literature \citep{hoff2002latent}. 

For multiple or multilayer networks, various  developments under latent space modeling have been proposed, including Bayesian approaches \citep{hoff2011hierarchical,salter2017latent}, and frequentist approaches \citep{arroyo2021inference,jones2020multilayer,nielsen2018multiple,zheng2022limit,zhang2020flexible,macdonald2022latent}. Our work aligns with the latter,  focusing on  estimating fixed latent vectors. 



Existing studies have consistently  shown that
multiple networks   often share  underlying structures  while simultaneously exhibiting significant network-specific heterogeneity 
\citep{wang2019common,zhang2020flexible,arroyo2021inference,chen2022global,he2023semiparametric,macdonald2022latent}. 
When analyzing a collection of heterogeneous networks, 
 leveraging their shared information is crucial for enhancing statistical  efficiency,
  whereas accurately characterizing unique features is equally important. 
 The  coexistence of shared and unique patterns calls for analysis capable of disentangling the intertwined  structural features and still fully exploiting statistical efficiency.  
Recently, 
\cite{macdonald2022latent}  introduced  
a general class of  latent space models  for  multiple networks that decompose  latent vectors into shared and network-specific components.  
A critical statistical question 
for jointly analyzing  multiple networks
is: Can pooling heterogeneous networks  improve the efficiency of estimating latent vectors, 
and if so, how can such gains be effectively achieved? 
\cite{macdonald2022latent} developed pioneering theoretical analysis showing efficiency improvement under Gaussian assumptions.  
However, it remains an open question whether and how similar efficiency gains can be established  under more general distributional scenarios.



To address this, 
we develop a unified framework of efficient analysis  under  the latent space modeling of multiple heterogeneous networks.  
We propose a novel strategy that first hunts for the shared latent space across networks and then applies a  refinement to achieve oracle efficiency. 
In this way, challenges from identifying shared space and achieving efficiency can be tackled separately, enhancing 
analytical flexibility.   
To theoretically quantify the efficiency gain, 
we develop novel techniques that  
disentangle the distinct oracle estimation error rates of shared and network-specific spaces through analyzing complex efficient score equations.  
Our proposed framework is versatile and encompasses  a broad range of edge weight distributions. 




The rest of the paper is organized as follows. Section \ref{sec:model} introduces the model and discusses statistical challenges  of efficient estimation.
Section \ref{sec:estall} presents our proposed procedures, and 
Section \ref{sec:theory} establishes the corresponding estimation error rates.  
Sections \ref{sec:simulation} and \ref{sec:data} present simulations and data analysis, respectively. 
Section \ref{sec:discussion} discusses several extensions of the proposed framework. 
Additional details and proofs are deferred to the Supplementary Material.

We summarize notations used below. 
Given two sequences of real numbers $\{g_n\}$ and $\{h_n\}$, 
$g_n = O(h_n) $ and $g_n\lesssim h_n $
represent $|g_n|\leqslant c_1|h_n|$ for a constant $c_1$, {$g_n\ll h_n$} and $g_n=o(h_n)$ mean $\lim_{n\to \infty} g_n/h_n=0$, and {$g_n\gg h_n$ means $\lim_{n\to \infty}h_n/g_n=0$}. 
For a sequence of random variables $\{X_n\}$, 
write $X_n=O_p(g_n)$ if for any $\epsilon>0$, there exists finite $M>0$ such that $\sup_n \Pr(|X_n/g_n|>M)<\epsilon$;  
{write $X_n=o_p(g_n)$ if for any $\epsilon>0$, $\lim_{n\to \infty}\Pr(|X_n/g_n|>\epsilon) =0$.}  
For $x = (x_i)_{i=1}^n$ and $y=(y_i)_{i=1}^n \in \mathbb R^n$, 
 {$\langle x, y\rangle = \sum_{i=1}^n x_iy_i$},  $\|x\|_2 = \sqrt{\langle x, x\rangle}$, and $\|x\|_{\infty} = \max_{1\leqslant i \leqslant n}|x_i|$. \label{page:notation}
For $X = (x_{ij})_{1\leqslant i\leqslant n, 1\leqslant j\leqslant m} \in \mathbb R^{n \times m}$, 
 $\|X\|_{\mathrm{F}} = \sqrt{\sum_{i = 1}^n \sum_{j = 1}^m x_{ij}^2}$, $\|X\|_{\operatorname{op}} = \sup_{\|v\|_2 = 1} \|Xv\|_2$,  and $\|X\|_{2 \to \infty} = \sup_{\|v\|_2 = 1} \|Xv\|_\infty$.
{Let $\mathrm{I}_k$ denote $k\times k$ identity matrix, and $\mathcal{O}(k)=\left\{Q \in \mathbb{R}^{k \times k}: Q Q^{\top}=\mathrm{I}_k\right\}$ represent the set of $k \times k$ orthogonal matrices.}

\section{Model for heterogeneous networks }\label{sec:model}

\subsection{Latent space model with shared factors}

This paper focuses on  multiple networks where each network is observed on a common set of $n$ nodes without inter-layer connections. 
Suppose we observe $T$ undirected networks on a common set of $n$ nodes, represented by $n\times n$ adjacency  matrices $\{\mathbf{A}_t = (A_{t,ij})_{1 \leqslant i,j \leqslant n} : t=1,\ldots, T\}$. 
For weighted networks, $A_{t,ij}$ is the weight value, which could be continuous or count-valued. For unweighted networks, $A_{t,ij}$ is binary indicating whether the connection between nodes $i$ and $j$ exists or not.  

To model interactions between nodes, 
we adopt the popular latent space modeling framework \citep{hoff2002latent}. Assume there exist latent vectors $y_{t,i} \in \mathbb{R}^{d_t}$ for $i=1,\ldots, n$ and $t=1,\ldots, T$ such that conditioning on $y_{t,i}$'s, network edges $A_{t,ij}$ are independently generated.
To flexibly model different types of weight values, 
we allow $A_{t,ij}$ to be generated from a
 {generic parametric distribution $p(\, \cdot \mid \theta \, )$ characterized by a  parameter $\theta$, i.e., } 
\begin{align} \label{eq:model}
    A_{t,ij} = A_{t,ji}\ \sim  \ p(\, \cdot \mid \Theta_{t,ij} \,), \quad \quad 1 \leqslant i \leqslant j \leqslant n,\ 1 \leqslant t \leqslant T,
\end{align} 
independently, 
where 
$\Theta_{t,ij}=\langle y_{t,i},y_{t,j} \rangle $ models the interaction effect between nodes $i$ and $j$ similarly to the popular low-rank  models in the literature \citep{hoff2002latent,ma2020universal}. 
{Examples of $p(\,\cdot \mid \theta\,)$ include Bernoulli, Gaussian,  and Poisson distributions, commonly used to model edge weights that are binary, continuous, and counts, respectively.} Model \eqref{eq:model} posits a common $p(\,\cdot \mid \theta\,)$ across networks for notational  simplicity. 
Our proposed analysis framework can be readily generalized when networks follow different types of distributions.   

For each node $i$, we assume that the latent vectors $\{y_{t,i}:t=1,\ldots,T\}$ may 
include shared and heterogeneous components over $T$ networks following  \cite{macdonald2022latent}. Specifically, 
$y_{t,i}=(z_i^{\top}, w_{t,i}^{\top})^{\top} \in \mathbb{R}^{k+k_t}$, 
where
$k+k_t=d_t$, 
$z_i\in \mathbb{R}^k$  represents a shared component over $T$ networks, 
and $w_{t,i}\in \mathbb{R}^{k_t}$ represents a heterogeneous component that may differ across $1\leqslant t\leqslant T$. Then $  \Theta_{t,ij}=\langle y_{t,i},y_{t,j} \rangle = \langle z_i, z_j\rangle + \langle w_{t,i}, w_{t,j}\rangle.  $

Let $\Theta_t=(\Theta_{t,ij})_{1\leqslant i,j\leqslant n} \in \mathbb{R}^{n\times n}$ denote the parameter matrix for the $t$-th network.  We define $ Y_t=[Z,\, W_t] \in \mathbb{R}^{n\times d_t}, $ $ Z=\left[z_1, \ldots, z_n\right]^{\mytrans} \in \mathbb{R}^{n \times k},$ and $ W_t=\left[w_{t,1}, \ldots, w_{t,n}\right]^{\mytrans} \in \mathbb{R}^{n \times k_t}. $ 
Then our model has an equivalent low-rank matrix representation after  link transformation
\begin{align}\label{eq:model2}
 \mu^{-1}\{\EXPT( \mathbf{A}_t ) \} =
 \Theta_t = Y_t Y_t^{\top} = ZZ^{\top}+W_tW_t^{\top}, \hspace{1.5em} 1\leqslant t\leqslant T, 
\end{align}
{where $\mu(\cdot)$ 
denotes the unique invertible function linking $\theta$ and variable expectation under $p(\,\cdot \mid \theta\,)$,}
and it is applied entrywise to its matrix argument in \eqref{eq:model2}. 

\begin{remark}
{
The core model \eqref{eq:model} 
is flexible and can be readily adapted to accommodate diverse network properties. 
For example, to allow flexible node degrees,  an overall degree factor varying with respect to $(n,T)$ can be added. 
Also, \eqref{eq:model} can be  augmented with an extra probability component through mixture modeling, enabling features such as link missingness or disconnections in weighted networks. 
Our developments can be similarly established under these generalized models, exemplifying  the broader applicability of our framework.
Detailed results are provided in Sections \ref{sec:sparse_extend}--\ref{sec:extend_full} of the Supplementary Material.}
\end{remark}

\subsection{Identifiability}  \label{sec:ident}
By \eqref{eq:model2}, the parameters $Z$ and $W_t$ can only be identified up to orthogonal transformation, i.e.,  
$ZZ^{\top}=ZQQ^{\top} Z^{\top}$ and $W_t W_t^{\top}=W_tQ_tQ_t^{\top}  W_t^{\top}$ for any $Q\in \mathcal{O}(k)$ and $Q_t \in \mathcal{O}(k_t)$.
We say that parameters $Z$ and $W_t$ in the model \eqref{eq:model2} are \textit{identifiable up to orthogonal group transformation} if, for any two sets of parameters $\{Z, W_1,\ldots, W_T\}$ and $\{Z^\prime, W_1^\prime, \ldots, W_T^\prime\}$ yielding the same model \eqref{eq:model2}, 
there exist $Q\in \mathcal{O}(k)$ and $Q_t\in \mathcal{O}(k_t)$ 
such that $Z=Z' Q$ and $W_t=W'_t Q_t$ for $1\leqslant t \leqslant T$. 
We next establish a sufficient condition for the identifiability of the model \eqref{eq:model2}. 

\begin{proposition} \label{prop:identify}
The model \eqref{eq:model2} is identifiable up to orthogonal group transformation if 
\begin{itemize}\setlength{\itemsep}{0pt}
    \item[(i)] for any $1 \leqslant t  \leqslant T$, the columns of $Y_t=[Z, W_{t}]$
    are linearly independent;
    \item[(ii)] there exist $1 \leqslant t < s \leqslant T$ such that the columns of  $[ Z,  W_{t}, W_{s}]$ are linearly independent. 
\end{itemize}
\end{proposition} 





\begin{remark}
Proposition \ref{prop:identify} is adapted from \cite{macdonald2022latent} 
but 
relaxes the connectivity assumption of the inter-layer graphs.
Given (i) and (ii) in Proposition \ref{prop:identify}, the column span of $Z$ equals the intersection of all  column spans of $Y_t$ across $1\leqslant t\leqslant T$. 
\end{remark}



\subsection{Efficient analysis of latent spaces and challenges} \label{sec:challenges}

Estimating latent spaces can reveal intrinsic connectivity patterns underlying noisy networks. 
Analyzing estimation efficiency provides crucial insights into the quality of obtained representations and is vital for downstream tasks. 
While \cite{macdonald2022latent} introduced the general model, their theoretical analysis was  constrained to Gaussian distribution with identity link. There still lacks comprehensive  understanding into efficiency gain from pooling heterogeneous networks under general distributions.   

For longitudinal networks with  baseline  node degrees varying over time, \cite{he2023semiparametric}   established semiparametric efficient estimation for shared latent spaces. 
Specifically, \cite{he2023semiparametric} 
showed that the aggregated squared estimation error of  shared latent factors can reach the oracle rate that is inverse proportional to the number of networks $T$, whereas the estimation errors of baseline nuisance parameters do not decrease with respect to $T$. 

For estimating the shared factor $Z$ under \eqref{eq:model2}, we expect a similar semiparametric oracle error rate. 
As an illustration, 
consider an oracle scenario  when heterogeneous (nuisance) parameters $w_{t,i}$'s are known. 
In this case, 
each shared (target) parameter $z_i$ is measured by $nT$ independent edges $\{A_{t,ij}: j=1,\ldots, n; t=1,\ldots, T\}$.    
The oracle squared estimation error rate of each latent vector $z_i$ is  expected to be of the order of $O_p(k/(nT))$, the ratio between the parameter number and the effective sample size  \citep{portnoy1988asymptotic}. Thus the aggregated oracle estimation error of $n$ latent vectors $Z$ is expected to be $O_p(n\times  k/(nT))=O_p(1/T)$ under fixed $k$. 
{Theoretically, achieving this oracle error rate requires approximating Cram\'er-Rao lower bound for high-dimensional $Z$; more discussions on the notion of efficiency are given in Remark \ref{rm:l2normvhat} of the Supplementary Material.} 
Nevertheless, to establish the result, technical challenges similar to those in \cite{he2023semiparametric}  persist. 
\begin{enumerate}[leftmargin=17pt]
\setlength{\itemsep}{0pt}
    \item \textit{Oracle error rates for estimating target  and nuisance parameters are different, but their errors are entangled in the analysis.} 
   Intuitively, each nuisance $w_{t,i}$ is measured by $n$ independent edges: $\left\{A_{t, i j}: j=1, \ldots, n\right\}$, so   
 the oracle squared error rate of each $w_{t,i}$ is expected to be $O_p(1/n)$, larger than $O_p(1/(nT))$ of each $z_i$. 
In theoretical analysis,
it is difficult to separate the error of $z_i$ from that of $w_{t,i}$ due to several reasons. 
One is that the individual components $w_{t,i}$'s 
encode both node-specific heterogeneity over $i$ and  network-specific heterogeneity over $t$. 
As a result, classical partial likelihood approach \citep{andersen1982cox} cannot directly remove the  heterogeneous factors.  
Moreover, when the link function in \eqref{eq:model2} is non-linear, such as the logistic or exponential function, 
semiparametric oracle rates for $Z$ cannot be easily obtained by  
spectral-based analyses \citep{arroyo2021inference} or analysis under Gaussian distribution \citep{macdonald2022latent}. 
\item \textit{Target parameters are  unidentifiable.} 
As  argued in Section  \ref{sec:ident}, the shared (target) latent space $Z$ is unidentifiable in the Euclidean space. 
Actually, the intrinsic space of  $Z$ is a quotient set with the equivalence relation up  to the orthogonal group transformation. To evaluate the discrepancy between a target  matrix $Z^{\star}$ and an estimate $\hat{Z}$,  a natural distance is $\mathrm{dist}(\hat{Z}, Z^{\star}) = \min_{Q\in \mathcal{O}(k) }\|\hat{Z}-Z^{\star}Q\|_{\mathrm{F}}$, instead of the Euclidean distance.
\item \textit{Both target and nuisance parameters are high-dimensional.} Specifically, under \eqref{eq:model2}, the dimension of $Z$ increases with respect to the network size $n$, and the total number of parameters in $[W_1,\ldots, W_T]$ grows with  respect to both $n$ and $T$. 
\end{enumerate}
\noindent On the other hand, our model \eqref{eq:model2} posits  \textit{unique}  challenges and properties, 
making existing analyses 
inapplicable and necessitating new developments of methodology and theory. We summarize  critical distinctions into four key aspects. \vspace{-5pt}
\begin{enumerate}[leftmargin=17pt]
\setlength{\itemsep}{0pt} 
\item \textit{Cross-network identification.} Identifying the shared latent space  under \eqref{eq:model2} requires joint information across multiple  networks, as shown in Proposition \ref{prop:identify}. This fundamentally distinguishes from \cite{he2023semiparametric} that only needs a single network to identify the target latent space. 
{Under \eqref{eq:model2},  \cite{macdonald2022latent} proposed to  identify shared latent space through nuclear norm penalties, which may not directly adapt to  the heterogeneity level of factors, as will be further illustrated in Section \ref{sec:simulation}.} 
Therefore, new procedures are required to effectively extract the shared latent space, and  theoretical analysis must properly integrate multiple networks together. 
\item \textit{Intricate factor interrelationships.} The relationship between the shared (target) space $Z$ and individual (nuisance) spaces $W_t$'s exhibits unprecedented complexity. 
In  \cite{he2023semiparametric}, 
the target and nuisance parameters are 
 latent factors and baseline degrees, respectively,
which  characterize intrinsically orthogonal  information that can enhance theoretical analysis. 
In contrast, factor interrelationships in \eqref{eq:model2} defy simple characterizations. {To derive theory under \eqref{eq:model2}, \cite{macdonald2022latent} imposed restrictions on factor interrelationships, which will be further elaborated in Remark \ref{rmk:compare_peter}.} In this work, we aim to consider general interrelationship that would greatly amplify the challenge of separating  estimation errors across different $1+T$ factor groups.  
\item \textit{Unidentifiable nuisance parameters.} Similar to the target $Z$, the network-specific nuisance parameters $W_t$'s are unidentifiable in  Euclidean space. 
This contrasts sharply with \cite{he2023semiparametric}, which worked with identifiable nuisance degree parameters.  
Consequently, we develop entirely novel proof techniques to tackle the unidentifiability of nuisance parameters and establish  uniform error control across them. 
\item \textit{Distributional generalization.} 
The model  \eqref{eq:model} accommodates a broad family of distributions, significantly extending beyond the Poisson edge  distributions  in \cite{he2023semiparametric} {or Gaussian edge distributions required in the theory of \cite{macdonald2022latent}}. 
This distributional flexibility  leads to additional analytical complexity. 
\end{enumerate}

\section{Estimation of the latent spaces} \label{sec:estall}

In the following, we assume the observed data  follow the model \eqref{eq:model2} with parameters $(Z,W_t)=(Z^{\star},W_t^{\star})$ over $1\leqslant t \leqslant T$. 
We  aim to estimate the underlying true parameters $Z^{\star}$ and $ W^{\star}_t$ from the observed data. 
To address all the challenges discussed in Section \ref{sec:challenges},  we propose a three-stage estimation procedure:

\begin{itemize}
\setlength{\itemsep}{0pt} 
    \item[(a)] for  $1\leqslant t\leqslant T$,  estimate concatenated latent factors  $Y_t^{\star}=[Z^{\star}, W_t^{\star}]$ up to an orthogonal group transformation in each $t$-th network  individually;  
    \item[(b)] separate  shared factors $Z^{\star}$ and heterogeneous factors ${W}_t^{\star}$  based on joint information across the $T$ estimates in Step (a) above;   
    \item[(c)] refine the estimators with likelihood information to achieve oracle efficiency.  
\end{itemize} 
{As Step (a) analyzes each network individually,   $Y_t^{\star}$ can only be identified  up to an orthogonal group  transformation $\mathcal{O}(d_t)$ under \eqref{eq:model2}. Consequently, column spaces of $Z^{\star}$ and $W_t^{\star}$ cannot be separated, necessitating a joint analysis across $T$ networks in Steps (b) and (c). 
%
As Section \ref{sec:ident} suggests, $Z^{\star}$ and  $W_t^{\star}$ can only be estimated up to  factor-specific orthogonal group transformations $\mathcal{O}(k)$ and $\mathcal{O}(k_t)$, respectively. 
For simplicity, we refer to estimating  $Z^{\star}$ and  $W_t^{\star}$ without  explicitly emphasizing these group  transformations when there is no ambiguity.} 
In the following, 
we assume the latent dimensions $k$ and $k_t$'s are fixed and given, and discuss how to consistently estimate them in Section \ref{sec:estk} of the Supplementary Material.



\subsection{Individual estimation}\label{sec:estimateyt}


We first estimate each $Y_t^{\star}$ up to an  orthogonal group transformation, which  can be achieved by examining each single network individually. 
For a single network, 
there have been extensive studies to estimate latent spaces for both random dot product models and latent space models with non-linear link function \citep{chatterjee2015matrix,ma2020universal,zhang2020flexible}. 
For the subsequent analysis in Sections \ref{sec:spectmethod} and \ref{sec:likeestimator}, it suffices to obtain an estimator $\mathring{Y}_t$ satisfying $\mathrm{dist}^2(\mathring{Y}_t, Y_t^{\star}) $
being of the order of $ O_p(1)$ up to logarithmic factors. 

The desired error rate is consistent with existing results in the literature and may be achieved by  various methods.  
In this work, we adopt a strategy that maximizes the likelihood function of $Y_t^{\star}$ via projected gradient descent for its simplicity and computational efficiency. 
The method generalizes that in  \cite{ma2020universal} for a binary network to models under a general edge distribution.  
As we consider a general link function in the model \eqref{eq:model2},
different adjustments for links  are needed are in both methodology and theory. We defer the details to Section \ref{sec:estyt} in  the Supplementary Material. 
Notably, 
any other methods that can estimate $Y_t^{\star}$ with the desired error rate can also be used in our subsequent analysis.  

\subsection{Shared space hunting}\label{sec:spectmethod}

Although estimating $Y_t^{\star}=[Z^\star, W_t^\star]$ from each network  is well-established, 
$Y_t^{\star}$ is only identifiable up to an unknown orthogonal transformation under a single-network model. As a result, $Z^{\star}$ cannot be directly obtained. 
To address this issue, we leverage joint information of  $Y_t^{\star}$'s and  develop a novel and efficient  spectral method below. 
To motivate our proposed construction, 
we first introduce an algebraic result in an oracle scenario below. 

\begin{proposition}\label{prop:yqtoz} Consider an oracle scenario where 
\begin{itemize}\setlength{\itemsep}{0pt}
    \item[(i)] we obtain estimators ${Y}_t \in \mathbb{R}^{n\times d_t}$ satisfying   ${Y}_t{Y}_t^{\top}=Y_t^{\star}Y_t^{\star\top}$ for all $1\leqslant t \leqslant T$; 
    \item[(ii)] there exist  $1\leqslant t< s \leqslant T$ such that  columns of $[Z^{\star}, W_t^{\star},  W_s^{\star}]$ are linearly independent. 
\end{itemize} 
Let $\mathcal{T} = \{(t,s): 1\leqslant t <s\leqslant T \text{ and Assumption (ii)  holds for } (t,s)  \}$. 
For $(t,s)\in \mathcal{T}$, 
\begin{align}\label{eq:yqretrievez_lm}
   Z^{\star}  Z^{\star\top} = Y_{t,-s}\, V_{t,s} \, V_{t,s}^{\top}\, Y_{t,-s}^{\top}/2 , 
\end{align}
where 
we let $Y_{t,-s}=[Y_t,\, - Y_s] \in \mathbb{R}^{n\times (d_t+d_s)}$, and 
$V_{t,s}\in \mathbb{R}^{(d_t+d_s) \times k}$ is a matrix whose columns can be any set of basis of the null space of $Y_{t,s} = [Y_t, Y_s] \in \mathbb{R}^{n\times (d_t+d_s)}$. 


\end{proposition}

Assumption (i) in Proposition \ref{prop:yqtoz} posits an oracle situation where we can accurately estimate each $Y_t^{\star}$ individually up to the orthogonal group  transformation $\mathcal{O}(d_t)$.
Assumption (ii) in Proposition \ref{prop:yqtoz} is the same as Assumption (ii) in Proposition \ref{prop:identify}.
Eq. \eqref{eq:yqretrievez_lm} provides the theoretical foundation for identifying  $Z^{\star}$,
which is unachievable when only investigating each $Y_t$ individually. 
However,  \eqref{eq:yqretrievez_lm} may not be directly applicable to estimating $Z^{\star}$ in practice because an  estimator $\mathring{Y}_t$ obtained in Step (a) may not exactly satisfy $\mathring{Y}_t\mathring{Y}_t^{\top} =  Y_t^{\star}Y_t^{\star\top}$, and $\mathcal{T}$ relies on unknown true parameters. 


To address these issues, we develop practical procedures based on spectral information of the estimates $\mathring{Y}_t$'s in Step (a).
We first estimate  $\mathcal{T}$.  
Let $\sigma_i(B)$ denote the $i$-th largest singular value of a matrix $B$, and 
define 
$\mathcal{R}_{i,j}(B)=\sigma_i(B)/\sigma_j(B) $. 
Then we have
\begin{align*}
 \sigma_{k+k_t+k_s}\big(\, [Y_t^{\star}, Y_s^{\star}] \, \big)   \  \begin{cases}
  \ =     0,  & \ (t,s)\not \in \mathcal{T};\\
 \   \neq 0, & \     (t,s) \in \mathcal{T}.
    \end{cases}
\end{align*}
When the estimates $\mathring{Y}_t$'s  are close to true $Y_t^{\star}$'s, 
we expect that for $(t,s) \not\in \mathcal{T}$, $\sigma_{k+k_t+k_s}( \mathring{Y}_{t,s})$ with $\mathring{Y}_{t,s}=[\mathring{Y}_t, \mathring{Y}_s]$ is close to 0,  
and then $ \mathcal{R}_{1, k+k_t+k_s}(\mathring{Y}_{t,s})$ would diverge to infinity quickly. 
Based on this idea,  
we construct an estimated set $\hat{\mathcal{T}}$ by letting  $(t,s)\in \hat{\mathcal{T}}$ if  the corresponding  ratio $ \mathcal{R}_{1, k+k_t+k_s}(\mathring{Y}_{t,s})$ is below an appropriate threshold.  

Moreover,  
as \eqref{eq:yqretrievez_lm} 
cannot hold exactly due to the estimation error in $\mathring{Y}_t$, 
we stabilize the result by  investigating the averaged terms over index pairs in $\hat{\mathcal{T}}$, i.e., we define 
\begin{align}\label{eq:gringdef}
    \mathring{F} =\frac{1}{2|\hat{\mathcal{T}} |} \sum_{(t,s)\in \hat{\mathcal{T}}}  \mathring Y_{t,-s} \, \mathring V_{t,s} \, \mathring  V_{t,s}^\mytrans \, \mathring Y_{t,-s}^{\mytrans}, 
\end{align}
where $\mathring{V}_{t,s}\in \mathbb{R}^{(d_t+d_s) \times k}$ is a matrix  consisting of the right null singular vectors of $\mathring Y_{t,s} $, and $\mathring Y_{t,-s} = [\mathring Y_{t}, -\mathring Y_{s}].$
By \eqref{eq:yqretrievez_lm},  we estimate $Z^{\star}$ by squared root of the top $k$ eigen-decomposition of $\mathring{F}$, represented as $\mathcal{S}_k(\mathring{F})$, which is formally defined in  \eqref{eq:skmap} of the Supplementary Material. 

\subsection{Likelihood-based refinement} \label{sec:likeestimator}

The spectral method in Section \ref{sec:spectmethod} is computationally efficient and could be used under    weak assumptions on data distribution. 
However, when the data distribution is available, 
the spectral method may not be optimal for estimating the shared space  $Z^{\star}$ in the sense of statistical efficiency discussed in Section \ref{sec:challenges}, as the joint likelihood information is not fully exploited. 
We next propose a likelihood-based refinement procedure 
and theoretically demonstrate its optimality in Section \ref{sec:theory}.


Under the model \eqref{eq:model}, the log-likelihood function with respect to $Z$ and $W_t$'s satisfies
\begin{align}\label{eq:zwlikelihood}
    \ell(Z, W) =  \sum_{t=1}^T\sum_{1 \leqslant i \leqslant j \leqslant n}l\big( \langle z_i, z_j\rangle + \langle w_{t,i}, w_{t,j}\rangle; A_{t,ij} \big), 
\end{align}
where $l(\theta;x) = \log p(\,x\mid \theta\,)$ and  $W = [W_1, \ldots, W_T]$. 
Our refinement procedure  first conducts the projected gradient descent \citep{chen2015fast}, primarily using first-order derivatives of the likelihood function that can be efficiently computed. Specifically,  the algorithm descends  parameters along their gradient directions with pre-specified step sizes 
 and then projects  updated estimates to  pre-specified constraint sets for parameters. 
  The detailed  steps are summarized in Algorithm  \ref{algor:refine} with notations explained in Section \ref{sec:algodetail}.   
Let $\check{Z}$ and $\check{W}_t$'s denote the estimates 
after sufficient iterations of the projected gradient descent.


Then the refinement procedure   further constructs a  second-order update, which  utilizes  efficient influence function
and provides convenience for   theoretically establishing desired  statistical error rate. 
{More discussions on its theoretical usefulness are provided in Remark \ref{rm:initialrelax} of the Supplementary Material.}   
For the ease of introducing formula below, we denote a vectorization of all the entrywise parameters of $[Z, W_1,\ldots ,W_T]$  as   
\begin{align}\label{eq:vectorization}
    v= \left[z_1^{\top},\ldots, z_n^{\top},w_{1,1}^{\top},\ldots, w_{1,n}^{\top},\ldots, w_{T,1}^{\top},\ldots, w_{T,n}^{\top}\right]^{\top}  \in \mathbb{R}^{ n(k+\ksum )\times 1},  
\end{align}
where $\ksum=k_1+\cdots+k_T$, 
and similarly define $\check{v}$ as the vectorization of $[\check{Z},\check{W}_1,\ldots, \check{W}_T]$.  
Then $\ell(Z,W)=\ell(v)$, and let  $\dot{\ell}(v)$ denote the partial derivative of $\ell(v)$ with respect to $v$.
We construct a second-order update as 
\begin{align} \label{eq:newtononev}
    \hat{v} = \check{v} + I(\check{v})^{+}\dot{\ell}(\check{v}), 
\end{align}
where $B^{+}$ represents the pseudo-Moore-Penrose inverse of a matrix $B$, which is uniquely defined, and 
$ I(v)=\EXPT\{ \dot{\ell}(v) \dot{\ell}(v)^{\top}\} $  is the Fisher information matrix of $v$ with   expectation taken under the model \eqref{eq:model} with parameters $v$. An analytical formula for $I(v)$  is derived for practical use in Section \ref{sec:newtonform} of the Supplementary Material. 
The final estimators $\hat{Z}$ and $\hat{W}_t$'s are obtained through realigning $\hat{v}$ into matrices. 

\begin{remark}\label{rm:onestepiv}
Our proof shows $I(\check{v})$ in \eqref{eq:newtononev}  is singular, necessitating the use of pseudo inverse in \eqref{eq:newtononev}. Intuitively, this  singularity arises from the unidentifiability of latent factors $Z$ and $W_t$ \citep{little2010parameter}. 
It makes the theoretical analysis challenging and conclusions for classical one-step estimator \citep{van2000asymptotic} or a single network \cite{xie2023efficient}  inapplicable; {more details are discussed in Remark \ref{rm:strongeff} of the Supplementary Material}. 
Moreover, 
\eqref{eq:newtononev} 
is simultaneously  constructed for $1+T$ groups of 
  latent factors $[Z,W_1,\ldots, W_T]$.  
 This markedly differs from \cite{he2023semiparametric} that constructs one-step estimator for shared latent factors only and  requires network-specific parameters to be identifiable.  
As a result, our theoretical analysis pioneers a new approach that jointly analyzes 
all the parameters in \eqref{eq:vectorization}  and  tackles inherent singularity of the information matrix from multiple groups of latent factors. More details are given in  Section \ref{sec:proplemmaiv} of the Supplementary Material. 
\end{remark} 

\subsection{Algorithm details} 
\label{sec:algodetail}

Algorithms  \ref{algor:initial} and \ref{algor:refine} summarize the construction of the space-hunting estimator in Section  \ref{sec:spectmethod}  and the likelihood-refinement estimator in Section \ref{sec:likeestimator}, respectively.  
We next introduce detailed notations and discuss the choice of  hyperparameters in the  algorithms. 

{ 

\begin{algorithm}
\caption{Spectral-based shared space hunting.}
\label{algor:initial}
\setstretch{1.5}
\KwIn{Individual estimates: $\mathring Y_t$ for $1 \leqslant t\leqslant T$. Parameter: {$\tau_{1}$} (threshold).}
\KwOut{$\mathring{Z}$ and $\mathring{W}_t$ for $1\leqslant t\leqslant T$.}

Let $\hat{\mathcal{T}} $ be an empty set. 

\For{$1\leqslant t < s \leqslant T$}{
Let $\mathring Y_{t,s} = [\mathring Y_{t},\, \mathring Y_{s}]$. 

\If{$\mathcal{R}_{1, \, k + k_{t} + k_{s}} (\mathring{Y}_{t,s}) \leqslant \tau_1$}{
    Let $\hat{\mathcal{T}} =\hat{\mathcal{T}} \cup \{(t, s)\}$.

    Construct $\mathring{V}_{t,s}$ as the matrix whose columns consist of the {right singular vectors} of $\mathring Y_{t,s}$ corresponding to its $k$ smallest singular values. 

     
    }
}


Let $\mathring Z = \mathcal{S}_k( \mathring{F})$ for $\mathring{F}$ in \eqref{eq:gringdef} and $\mathcal{S}_k(\cdot)$ defined in \eqref{eq:skmap}.

Let $\mathring{W}_t =  \mathcal{S}_{k_t}( \mathring{Y}_t\mathring{Y}_t^{\top}-\mathring{F})$ for $1\leqslant t\leqslant T$, where $\mathcal{S}_{k_t}(\cdot)$ is defined similarly to  \eqref{eq:skmap}. 


\end{algorithm}
\begin{algorithm}
	\caption{Likelihood-based refinement.}
	\label{algor:refine}
    \setstretch{1.5}
	\KwIn{Data: $\mathbf A_1, \ldots, \mathbf A_T \in \mathbb{R}^{n\times n}$. Initial estimates: $\mathring Z$ and $ \mathring W_t$ for $1\leqslant t\leqslant T$.\\
    \hspace{3.3em} Parameters: $\eta_Z$,  $\eta_W$ (step sizes), $R$ (number of iterations), \\
    \hspace{9.3em}$\mathcal{C}_Z,  \mathcal{C}_{W_t}$ for $1\leqslant t\leqslant T$ (constraint sets for projection).}

\KwOut{$\hat{Z}$ and $\hat{W}_t$ for $1\leqslant t\leqslant T$.}


Let $Z^0 = \mathring{Z}$ and $W_t^0 = \mathring{W}_t$ for $1\leqslant t\leqslant T$.
	
\For{$r = 0, \ldots, R - 1$}{

Let $\tilde{Z}^{r+1}=Z^{r} + {\eta_Z}\, \dot{\ell}_{Z}(Z^r,W^r) $ and   $Z^{r+1}={\mathcal{P}_{\mathcal C_Z}}(\tilde{Z}^{r+1})$.  

   Let $\tilde{W}_t^{r+1}=W_t^{r} + \eta_W\,  \dot{\ell}_{W_t}(Z^r,W^r)$ and $W_t^{r+1}=\mathcal{P}_{\mathcal C_{W_t}}(\tilde{W}_t^{r+1})$ for $1\leqslant t\leqslant T$.

}

{Construct $\hat{v}$ by \eqref{eq:newtononev} with $(\check{Z},\check{W})=(Z^R, W^R)$, and let $[\hat{Z},\hat{W}]$ be its matrix version.} 

\end{algorithm}

}



	







  \emph{In Algorithm \ref{algor:initial}:} The threshold $\tau_1$ is chosen to estimate $\mathcal{T}$. 
As Section \ref{sec:spectmethod} suggests,  the choice of $\tau_1$ depends on statistical properties of the input initial estimates $\mathring{Y}_t$. 
 In this paper, we construct $\mathring{Y}_t$ by the method in Section \ref{sec:estimateyt}, whose estimation error is established in Section \ref{sec:estyt_theory} of the Supplementary Material. To achieve theoretical guarantee, it suffices to set the threshold to satisfy $1 \ll \tau_1 \lesssim \sqrt{\log n}$. 
 Our numerical implementation sets $\tau_1 = \sqrt{2 \log n}$.



 \emph{In Algorithm \ref{algor:refine}:} 
During the gradient descent, we let $\dot{\ell}_{Z}(Z,W) \in \mathbb{R}^{n\times k}$ and $\dot{\ell}_{W_t}(Z,W)\in \mathbb{R}^{n\times k_t}$ represent the partial derivatives of $\ell(Z,W)$ with respect to matrices $Z $ and $ W_t$, respectively.  Let  $\mathcal{P}_{\mathcal{C}}(\cdot)$ denote a projection operator given a pre-specified  constraint set $\mathcal{C}$, and  we consider two constraint sets 
\begin{align}\label{eq:constraintsets}
    \mathcal{C}_Z = \big\{Z \in \mathbb R^{n\times k }: \|Z\|_{2 \to \infty} \leqslant M_1\big\} \quad\text{and}\quad \mathcal{C}_{W_t} = \big\{W_t \in \mathbb R^{n\times k_t }: \|W_t\|_{2 \to \infty} \leqslant M_1\big\},
\end{align}
for $Z$ and $W_t$, respectively. 
Eq.\ \eqref{eq:constraintsets} constraints the two-to-infinity norms of $Z$ and $W_t$ to be  bounded by $M_1$, which also  
  corresponds to the constraints on the true parameters $(Z^{\star}, W_t^{\star})$  in Condition \ref{cond:truevalue}  below. 
The projection steps and second-order update in Algorithm \ref{algor:refine} are  required for the convenience of the proof, and skipping them can yield equally good numerical performance. 
{More details are discussed in Section \ref{sec:pseudolik}  of the Supplementary Material and Remark \ref{rmk:pseudolik}.} 
In our numerical implementation, we use Barzilai-Borwein step sizes \citep{barzilai1988two} and set $R = 1000$. 
To establish theoretical results in Section \ref{sec:theory}, Condition \ref{cond:tuning12} below summarizes the  requirements on the hyperparameters in Algorithms   \ref{algor:initial} and \ref{algor:refine}.  

\begin{condition}\label{cond:tuning12}
Assume the hyperparameters in Algorithms   \ref{algor:initial} and \ref{algor:refine}  satisfy:  
(i) $1 \ll \tau_1 \lesssim \sqrt{\log n}$.   (ii) Step sizes  $ \eta_Z = \eta/(nT)$ and $\eta_W = \eta/n$ for a sufficiently small constant $\eta > 0$. (iii) Constraint sets $\mathcal{C}_Z$ and $\mathcal{C}_{W_t}$ are chosen as in \eqref{eq:constraintsets}. (iv) Number of iterations $R \gg \log(nT)$. 
\end{condition}




\subsection{Related works}\label{sec:relatedwork}
The proposed estimation scheme in Section \ref{sec:estall}  shares connections with, yet notably differs from, existing methods.  
The shared space hunting Algorithm \ref{algor:initial} is developed based on spectral properties of latent factors. While spectral-based analysis is common in  network studies  
\citep{rohe2011spectral,chatterjee2015matrix,arroyo2021inference}, 
our procedure utilizes unique properties of multiple networks to separate the shared and distinct  latent spaces, and thus cannot be directly implied by existing methods.
The likelihood refinement Algorithm \ref{algor:refine} utilizes the projected gradient descent, 
which is known to be computationally efficient for maximizing  the non-convex likelihood functions 
for both single network \citep{ma2020universal} and multiple networks \citep{zhang2020flexible}.
Nevertheless, the oracle estimation error rates under our model cannot be directly attained through  existing  technical developments of those methods. 
To address the theoretical challenges, Algorithm  \ref{algor:refine} 
further constructs a novel second-order update estimator  \eqref{eq:newtononev} for multiple heterogeneous networks. 
The construction is motivated from the classical one-step estimator, typically obtained  through a linear approximation of score equation \citep{van2000asymptotic}. 
Similar idea has also been recently used to construct efficient estimators for latent factors in network analysis, including the random dot product model for a single network \citep{xie2023efficient}  and a semiparametric longitudinal network model \citep{he2023semiparametric}. 
However, our model \eqref{eq:model} substantially differs from those models in that it embraces a general edge distribution and also allows distinct latent factors over multiple networks. 
The extended flexibility posits unique challenges in identifying shared spaces and obtaining efficient  estimation errors, necessitating   novel developments of efficient score updating formula and theory. 
The model \eqref{eq:model}  is motivated from 
\cite{macdonald2022latent}, which 
similarly considers \eqref{eq:model} and assumes  that part of the latent vectors are shared across the networks, i.e., $y_{t,i}=(z_i^{\top},w_{t,i}^{\top})^{\top}$.  
The difference is that in their work $\Theta_{t,ij}=\kappa(y_{t,i},y_{t,j})$ 
with 
$\kappa(x, y) = x^{\top}\mathrm{I}_{p,q}y$, where $\mathrm{I}_{p,q}$ represents the block-diagonal matrix with $\mathrm{I}_p$ and $-\mathrm{I}_{q}$ on the upper left and lower right blocks, respectively.  
Although a more general similarity function is proposed,
their theoretical analysis is restricted to Gaussian distribution with identity link, which  can be  inappropriate for network data with binary edges or count edges. 
Alternatively, we will next establish theoretical results that encompass various exponential family distributions. 
Please also see Remark \ref{rmk:compare_peter} for more discussions on the comparison of technical conditions and results.  

\begin{table}
\caption{Comparison of  different models and results: link functions and the corresponding  estimation error rates of shared  and individual parameters (ER-Shared and ER-Individual). 
}   
\label{table::summary_multilayer} 
\renewcommand{\arraystretch}{2}
\setlength{\tabcolsep}{6pt} 
\footnotesize    
\vspace{-15pt}
\begin{center}
\begin{tabular}{l|c|l|l|l}
\hline
 Model  & \makecell[c]{Link $\mu(\theta)$}  &   \makecell[c]{ER-Shared}  & \makecell[c]{ER-Individual} & \makecell[c]{Constraints on $(n,T)$} \\[3pt] \hline
\cite{arroyo2021inference} 
& $\theta$  &  $\displaystyle O_p\biggr(\frac{1}{n}+\frac{1}{T} \biggr)$  & $-$   & $-$\\[4pt] 
\cite{macdonald2022latent} 
& $ \theta$ &  $\displaystyle  O_p\biggr(\frac{1}{T}\biggr)
$ &  $\displaystyle  O_p(1)
$  &   $T=o(n^{{1}/{2}})$
\\[4pt]  
\cite{zhang2020flexible}
& 
$ \operatorname{expit}(\theta)$  
& $\displaystyle O_p\biggr(1+\frac{1}{T}\biggr)$  & $-$ & $T=O(n)$    \\[4pt]  
\cite{he2023semiparametric}
& $\exp (\theta)$  & $\displaystyle O_p\biggr(\frac{1}{n}+\frac{1}{T}\biggr)$ & $\displaystyle O_p(1)$   &  {$\mathrm{polylog}(T) = O(n)$} \\[4pt]  \textbf{This work}  & \textbf{General} &   $\displaystyle O_p\biggr(\frac{1}{n}+\frac{1}{T}\biggr)$ & $\displaystyle O_p(1)$ & {$\mathrm{polylog}(T) = O(n)$}  \\[4pt]  \hline
\end{tabular} 
\end{center}
\vspace{-8pt}
{\footnotesize(Error rates above are simplified to emphasize their orders with respect to $n$ and $T$ up to logarithmic factors and also rescaled to be comparable. ``$-$'' represents inapplicable  information.  See more details in Section \ref{sec:detailtbcompare} of the Supplementary Material.})
\end{table}

Besides the studied model  \eqref{eq:model}, there are various other models extracting node-wise latent embeddings from multiple networks. 
One research line extends the random dot product graphs \citep{young2007random,athreya2018survey} to multilayer networks \citep{arroyo2021inference,zheng2022limit}. 
These models typically corresponds to an identity link function in \eqref{eq:model2}, and thus developments  cannot directly address the challenges under models with general non-linear link functions. 
Another class of models adopt non-linear canonical link functions   
\citep{hoff2002latent,ma2020universal} and develop extensions for multiple networks \citep{zhang2020flexible,macdonald2022latent,he2023semiparametric}. 
Our developments under 
  \eqref{eq:model} 
advance the  literature by  establishing a unified theoretical framework that can encompass a variety of general link functions and allow heterogeneous latent factors with complex relationships.  
We summarize properties and results under the related models in  Table \ref{table::summary_multilayer},
where our theory will be formally established in Section \ref{sec:theory}. 
Notably, oracle error rate  of the shared parameters in this paper is consistent with earlier work in \cite{arroyo2021inference}, \cite{macdonald2022latent}, and  \cite{he2023semiparametric}.

Targeting at distinct properties and challenges, researchers have also proposed other models for multiple networks. 
Examples include 
multilayer stochastic block models  for community detection \citep{han2015consistent, paul2016consistent, lei2023bias}, and tensor-decomposition-based models for effective dimension reduction \citep{lyu2023latent,zhangefficient}, etc. 
Although these models may also incorporate shared and heterogeneous structures across multiple networks, 
their analytical objectives and challenges differ from our focus on estimating continuous node-wise latent embeddings and thus are not directly compared in Table \ref{table::summary_multilayer}; {see more discussions in Section \ref{sec:detailtbcompare} of the Supplementary Material}.

\section{Theory for estimating latent spaces} \label{sec:theory} 
In this section, we establish estimation error rates of  Algorithms \ref{algor:initial} and \ref{algor:refine} in Theorems \ref{thm:initial} and \ref{thm:onestep}, respectively.  
For technical developments, 
we impose regularity 
assumptions on the true parameters in Condition \ref{cond:truevalue} and edge-wise distribution  in Condition \ref{cond:parfunction} below. 


\begin{condition} \label{cond:truevalue}
For a matrix $X \in \mathbb R^{n \times m}$, let $\mathcal{G}(X) = X^\mytrans X / n \in \mathbb R^{m \times m}$ denote the Gram matrix of $X$ standardized by its row number, and let $\sigma
_{\operatorname{min}}(X)$ denote the minimum singular
value of $X$. 
Assume that there exist positive constants $M_1$, $M_2$, and $M_3$ such that
\begin{itemize}\setlength{\itemsep}{0pt}
\item[(i)]  $\|Z^{\star}\|_{2 \to \infty} \leqslant  M_{1}$ and for any $1\leqslant t \leqslant T$, $  \|W_{t}^{\star}\|_{2 \to \infty}
\leqslant  M_{1}$;

\item[(ii)] for any $1 \leqslant t \leqslant T$, 
$\sigma_{\operatorname{min}}(G_{t}^{\star})\geqslant M_2 $ for $G_t^{\star}=\mathcal{G}([Z^\star, W_t^\star]) $; 

\item[(iii)] there exist $1 \leqslant t < s \leqslant T$ such that $\sigma_{\operatorname{min}}(G_{t,s}^{\star} ) \geqslant M_3 $ for $G_{t,s}^{\star}=\mathcal{G}([Z^\star, W_{t}^\star, W_{s}^\star])$.
\end{itemize}
\end{condition}

{We describe Condition \ref{cond:truevalue} on one set of true parameters $(Z^{\star}, W_t^{\star})$ for convenience and clarity.
It is equivalent to impose Condition \ref{cond:truevalue}  on the equivalence classes of $Z^{\star}$ and $ W_t^{\star}$ that are identifiable up to orthogonal transformations. 
In particular,  Condition \ref{cond:truevalue} is satisfied for $Z^{\star}$ and $W_t^{\star}$ if and only if it is satisfied for $Z^{\star}Q$ and $W_t^{\star}Q_t$ with $Q\in \mathcal{O}(k)$ and $Q_t\in \mathcal{O}(k_t)$ over $1\leqslant t\leqslant T$, since the matrix norms remain the same.} 

 Condition \ref{cond:truevalue} (i) implies that  $\ell_2$-norms of the latent vector $z_i^\star$ and $w_{t,i}^\star$  are bounded over all $i$ uniformly.
 It constrains all the parameters to be in a bounded space, 
 which is a prevalent mild assumption in high-dimensional network analysis \citep{ ma2020universal, zheng2022limit}. 
 Condition \ref{cond:truevalue} (ii) and (iii) impose  constraints on the scaled  Gram matrices of the latent factors.
They imply that the columns of $[Z^{\star}, W_t^{\star}]$ (for any $t$) and $[Z^{\star}, W_t^{\star}, W_s^{\star}]$ (for one pair of $\{t,s\}$) are linearly independent, ensuring that the true model parameters are identifiable by Proposition \ref{prop:identify}. 
For Condition \ref{cond:truevalue} (ii),  note that 
\begin{align*}
    \mathcal{G}([Z^\star, W_t^\star]) = \frac{1}{n} \begin{pmatrix}
        Z^{\star \top} Z^\star & Z^{\star \top} W_t^\star\\
        W_t^{\star \top} Z^\star &  W_t^{\star \top} W_t^\star
    \end{pmatrix} 
\end{align*}
is a two-by-two block matrix. Condition \ref{cond:truevalue} (ii) can be satisfied by directly imposing conditions on each block separately, and similarly for   Condition \ref{cond:truevalue} (iii).
In particular, we establish Proposition  \ref{prop:condtion1} below providing sufficient conditions on the sub-matrices for Condition   \ref{cond:truevalue}. 

\begin{proposition} \label{prop:condtion1}
If there exists a  constant $M_4>0$ such that, 
\begin{itemize}\setlength{\itemsep}{0pt}
\item[(a)] for any $1 \leqslant t \leqslant T$, 
$\sigma_{\operatorname{min}}(Z^{\star\top} Z^{\star}/n) \geqslant M_4$ and $\sigma_{\operatorname{min}}(W_t^{\star\top} W_t^{\star}/n) \geqslant M_4$, 

\item[(b)] for any $1 \leqslant t \leqslant T$, $\|Z^{\star\mytrans} W_t^{\star}/n\|_{\operatorname{op}} \leqslant  M_4/4$,
\item[(c)] and there exist $1 \leqslant t < s \leqslant T$ such that $\|W_t^{\star\mytrans} W_s^\star/n\|_{\operatorname{op}} \leqslant M_4/4$, 
\end{itemize}
then Condition \ref{cond:truevalue} (ii) and (iii) are satisfied with $M_2$ and $M_3$ that are fixed functions of $M_4$. 
\end{proposition}


\begin{remark} \label{rmk:compare_peter}
\cite{macdonald2022latent} studies a similar model with generalized inner product \citep{rubin2022statistical}. Their theoretical analysis focuses on edge-wise normal distribution with $\mu(\theta)=\theta$, and  
their imposed  conditions imply  that as $n\to \infty$, 
\begin{align}\label{eq:condpeter2}
    \|Z^{\star\mytrans} W_t^{\star}/n\|_{\operatorname{op}}\to 0 , \quad \quad \|W_t^{\star \mytrans} W_s^{\star}/n\|_{\operatorname{op}}\to 0.   
\end{align} 
Proposition \ref{prop:condtion1} suggests  that  Condition \ref{cond:truevalue} (ii) and (iii) can relax \eqref{eq:condpeter2}. 
As $\|Z^{\star\mytrans} W_t^{\star}/n\|_{\operatorname{op}} = 0$
means that the columns of $Z^{\star}$ and $W_t^{\star}$ are orthogonal, we refer to the decaying rate in \eqref{eq:condpeter2} as ``nearly orthogonal'' conditions. 
Such restrictions on the model parameters could limit the expressive power of the models for practical use. 
Methodologically, their proposed penalty  performs convex  relaxation of the rank constraints for $Z$ and all $W_t$'s individually. This strategy, intuitively, may not effectively take the relationship between latent vectors into account.  
As a result, method in \cite{macdonald2022latent} could lead to different interpretations of estimated latent vectors. 
See more numerical illustrations in Sections \ref{sec:simulation} and \ref{sec:data}.  
\end{remark}

\begin{condition} \label{cond:parfunction}
Let $\mathcal{X} = \{x \in \mathbb R : p(\,x\mid \theta\,) > 0\}$ denote the support of $p(\,\cdot \mid \theta\,)$. Assume $l(\theta;x)$ in \eqref{eq:zwlikelihood} belongs to the natural exponential family and satisfies the following conditions. 
\begin{itemize}\setlength{\itemsep}{0pt}

\item[(i)] For any fixed $x \in \mathcal{X}$, $l(\theta;x)$ is three times differentiable with respect to $\theta$, with its first to   third derivatives with respect to $\theta$ denoted by   $l^{\prime}(\theta;x)$, $l^{\prime \prime}(\theta;x)$, and $l^{\prime \prime \prime}(\theta;x)$, respectively. Moreover, there exist positive constants $\kappa_1$, $\kappa_2$, and $\kappa_3$ such that $\kappa_1 \leqslant -l^{\prime\prime}(\theta;x) \leqslant \kappa_2$ and $|l^{\prime\prime\prime}(\theta;x)| \leqslant \kappa_3$ for any $x \in \mathcal{X}$ and $|\theta| \leqslant 2M_1^2$, where $M_1$ is given in Condition \ref{cond:truevalue}. 

\item[(ii)] There exists a constant $L>0$ such that $\EXPT |l^{\prime}(\Theta_{t,ij}^\star;A_{t,ij})|^{m} \leqslant \VAR\{l^{\prime}(\Theta_{t,ij}^\star;A_{t,ij})\} L^{m -2} m ! /2$ for any  $1 \leqslant i \leqslant j \leqslant n$, $1 \leqslant t \leqslant T$, and any integer $m \geqslant 2$, where $m!=m(m-1)\cdots 1$ represents the factorial of $m$. 
\end{itemize}
\end{condition}

{ Condition \ref{cond:parfunction} can be naturally satisfied by most distributions under natural exponential families. For instance, 
$l(\theta;x)\propto \theta x - \log(1+\exp(\theta)) $ for Bernoulli distribution, and $l(\theta;x)\propto \theta x -\exp(\theta)$ for Poisson distribution \citep{efron2022exponential}. The assumption of natural exponential family is adopted here primarily for the ease of presentation and understanding.  We show that generalizations under more general distributions can be established following similar arguments in  Section \ref{sec:expfam} of the Supplementary Material.} 




Given the conditions, we first establish the estimation error rates for the shared space hunting algorithm in Section \ref{sec:spectmethod}. 

\begin{theorem} \label{thm:initial}
    Assume Conditions \ref{cond:tuning12}
    --\ref{cond:parfunction}. Let $(\mathring Z,\mathring W)$ be the estimators obtained through Algorithm \ref{algor:initial} with  $\mathring Y_t$ in Theorem \ref{thm:estY} as initialization. 
    For any constant $\varepsilon >0$, there exist positive constants $c_\varepsilon$ and $C_\varepsilon$ such that when $\log^{\dmax+2}(nT)/n \leqslant c_\varepsilon $ with $\dmax = \max_{1\leqslant t\leqslant T} d_t$, 
    $$
    \Pr\left[ \left\{\operatorname{dist}^2(\mathring Z, Z^\star) +  \max_{1\leqslant t \leqslant T} \operatorname{dist}^2(\mathring W_t, W_t^\star) \right\} > C_\varepsilon \log^2(nT)\right] = O((nT)^{-\varepsilon}).
    $$
\end{theorem}

Theorem \ref{thm:initial} suggests that with high probability, the overall estimation error rate of $\operatorname{dist}^2(\mathring Z, Z^\star)$ and $\operatorname{dist}^2(\mathring W_t, W_t^\star) $ are of the order of $O(1)$ up to logarithmic factors.  
This aligns with the expected oracle rate for $\Theta_t^{\star}=Z^{\star}Z^{\star\top}+W_t^{\star}W_t^{\star\top}$ as discussed in Section \ref{sec:challenges}.  

\begin{remark}
The shared space hunting Algorithm \ref{algor:initial} requires initial individual estimates $\mathring{Y}_t$ for  ${Y}_t^{\star}$ over $1\leqslant t\leqslant T$. As discussed in Section \ref{sec:estimateyt}, our analysis  focuses on  $\mathring{Y}_t$ obtained by Algorithm  \ref{algor:estY} for the ease of discussion. The estimation error rates of the proposed  $\mathring{Y}_t$'s are established in Theorem \ref{thm:estY}. Nevertheless,  Theorem \ref{thm:initial} can  be similarly established using other initial estimates $\mathring{Y}_t$, as long as they can achieve a similar rate to that in Theorem \ref{thm:estY}. 
\end{remark}



Theorem \ref{thm:initial} only provides an  upper bound on the estimation error rates  for shared and heterogeneous parameters simultaneously. 
Establishing the desired efficiency 
for each group of parameters separately is challenging, as discussed in Section \ref{sec:challenges}. 
To facilitate the theoretical analysis, 
we make an additional assumption on 
the heterogeneous parameters. 


\begin{condition} \label{cond:truevalue2} 
Assume $\|\mathcal{G} (W^\star) /T \|_{\operatorname{op}} \leqslant M_5$, where $M_5=M_2^2/(8M_1^2)$ and the constants $M_1$ and $M_2$ are given in Condition \ref{cond:truevalue}. 
\end{condition} 


For $W^{\star}=[W_1^{\star},\ldots,W_T^{\star}]$, $\mathcal{G}(W^{\star})$ consists of submatrices $W_t^{\star \top}W_s^\star/n$ for $1\leqslant t\leqslant s\leqslant T$. 
Similarly to Condition \ref{cond:truevalue},  
Proposition \ref{prop:condtion3} gives a sufficient condition for interpreting Condition \ref{cond:truevalue2} based on the submatrices of $\mathcal{G}(W^{\star})$. 

\begin{proposition} \label{prop:condtion3}
Condition \ref{cond:truevalue2} is satisfied if with $M_5=M_2^2/(8M_1^2)$,  

\begin{itemize}\setlength{\itemsep}{0pt}
\item[(a)] for any $1 \leqslant t \leqslant T$, $\|W_t^{\star\top}W_t^{\star}/n\|_{\operatorname{op}} \leqslant M_5T/2$; 

\item[(b)] for any $1 \leqslant t < s \leqslant T$, $\|W_t^{\star\mytrans} W_s^{\star}/n\|_{\operatorname{op}} \leqslant M_5/2$.

\end{itemize}
\end{proposition}

We compare Proposition \ref{prop:condtion3} with Proposition \ref{prop:condtion1} to gain insights into Condition \ref{cond:truevalue2}. 
First, (a) in Proposition \ref{prop:condtion3} specifies an upper bound on the singular values of $W_t^{\star\top}W_t^{\star}/n$, which is compatible with the lower bound (a) in Proposition \ref{prop:condtion1} when $T$ is sufficiently large. 
Second, (b) in Proposition \ref{prop:condtion3} specifies an upper bound on the singular values of $W_t^{\star\mytrans} W_s^{\star}/n$, which differs from (b) in Proposition  \ref{prop:condtion1}, as $M_5$ depends on  constants $(M_1,M_2)$ in Condition \ref{cond:truevalue}. 
Condition \ref{cond:truevalue2} is primarily a technical condition needed  to facilitate the control of  certain high-order error terms in the analysis. 
{Intuitively, it ensures that the column spaces of  $W_t^{\star}$ and $W_s^{\star}$  for most pairs $t\neq  s$ are sufficiently different}, making it easier to disentangle their respective estimation errors. 
{We point out that the constraint on $M_5$ may not limit the practical use of our method.  Numerical studies in Section \ref{sec:simulation} demonstrate its robust  performance even in the extreme case where  $W_t^{\star}$'s are identical. 
This empirical evidence suggests that the constraint on $M_5$ could potentially be relaxed under certain scenarios, which, however, is not pursued here to preserve the clarity and coherence of our presentation.} 





\begin{remark} \label{rmk:pseudolik}
The unidentifiability of parameters under  the joint log-likelihood function $ \ell(Z, W)$ makes theoretical analysis challenging. 
To overcome the issue, we replace the gradient descent objective $\ell(Z, W)$ {used in the lines 2--5 of Algorithm \ref{algor:refine}} with a pseudo log-likelihood  {$ p\ell(Z, W) = \sum_{t=1}^T \sum_{1 \leqslant i, j \leqslant n} l(\langle z_i, \mathring z_j \rangle + \langle w_{t,i}, \mathring w_{t,j}\rangle;A_{t,ij}),$} 
where $\mathring{z}_j$ and $\mathring{w}_{t,j}$ denote the estimates obtained in Section \ref{sec:spectmethod}. 
Compared to $\ell(Z,W)$ in \eqref{eq:zwlikelihood}, $ p\ell(Z, W) $ fixes half of $z_j$ and $w_{t,j}$ in the inner product to their corresponding estimates  $\mathring{z}_j$ and $\mathring{w}_{t,j}$.  
This is only a technical adjustment to facilitate the theory. 
With $\mathring{z}_j$ and $\mathring{w}_{t,j}$ being close enough to the true parameters, using $ p\ell(Z, W)$ and $ \ell(Z, W)$ in 
{the lines 2--5} of Algorithm \ref{algor:refine} can perform similarly. 
{Please also find details in Section \ref{sec:pseudolik} of the Supplementary Material.} 

\end{remark}

We next show that the likelihood-based refinement procedure can further improve the estimation error rate in Theorem \ref{thm:initial}.

\begin{theorem} \label{thm:onestep} 
Assume Conditions \ref{cond:tuning12}--\ref{cond:truevalue2}.
Let $(\hat Z, \hat{W})$ be the refined estimators from Algorithm \ref{algor:refine} with $(\mathring{Z},\mathring{W})$ in Theorem \ref{thm:initial} as initialization and the adjustment in Remark \ref{rmk:pseudolik}. 
For any constant $\varepsilon >0$, there exist positive constants $c_\varepsilon$ and $C_\varepsilon$ such that when $\log^{\varsigma}(nT)/n \leqslant c_\varepsilon $ with $\varsigma = \max\{\dmax + 2,8\}$, 
\begin{align}
\Pr\left[
\operatorname{dist}^2(\hat Z, Z^\star) >
   C_\varepsilon \max\left\{\frac{1}{T} ,\ \frac{1}{n}\right\} \log^{8}(nT)
\right] = &~O\big((nT)^{-\varepsilon}\big),\label{eq:zhaterr}\\
\Pr\left[ \max_{1\leqslant t \leqslant T} 
\operatorname{dist}^2(\hat W_t, W_t^\star) > 
   C_\varepsilon  \log(nT) 
\right] = &~O\big((nT)^{-\varepsilon}\big).\label{eq:wthaterr}
\end{align}

\end{theorem}


Theorem \ref{thm:onestep} suggests that with a high probability, up to logarithmic factors, 
the estimation error for $W_t^{\star}$ is of the order of $O(1)$, whereas 
the estimation error for $Z^{\star}$ is of the order of
\begin{align} \label{eq:onestepratesimplify}
  O\left[  \max \left\{\frac{1}{T},\, \frac{1}{n}\right\} \right]. 
\end{align}
When $1<T\lesssim n$, $\eqref{eq:onestepratesimplify} = O(1/T)$, showing that the semiparametric oracle rate for the shared factors $Z^{\star}$ discussed in Section \ref{sec:challenges} can be achieved. 
In this case,  $\hat{Z}$ and $\hat{W}_t$ can achieve the oracle estimation error rate for estimating $Z^{\star}$ and $W_t^{\star}$ up to logarithmic factors. 
On the other hand, when $T\gg n$, $\eqref{eq:onestepratesimplify} = O(1/n)$, which is referred to as a sub-optimal rate. 
Intuitively, when $T$ is too large, the degree of heterogeneity becomes larger, and the estimation of shared factor $Z^{\star}$ is more challenging. 
{The constraint of $T$ aligns with squared error requirement in classical semiparametric analysis \citep{murphy2000profile}, which is further explained in Remark \ref{rm:interprtTn} of the Supplementary Material. } 

 \begin{remark}\label{rm:theorynovel}
 As Table \ref{table::summary_multilayer} suggests, 
 Theorem \ref{thm:onestep} not only aligns with the best rates reported in the compared works but also advances the literature by accommodating the most flexible range of link functions. 
 While \cite{he2023semiparametric} reported similar error rates for both shared and individual parameters, 
our work distinguishes itself by 
addressing unique and previously insurmountable challenges  highlighted in Section \ref{sec:challenges}. 
We develop a comprehensive set of  analytical innovations  
including effective shared space identification validated by  Theorem \ref{thm:initial}, and 
a novel joint one-step construction framework addressing the multiplex  
 singularity issue of the efficient  information matrix  in Remark \ref{rm:onestepiv}. 
 These breakthroughs collectively establish a solid foundation for comprehensive analysis of heterogeneous networks, deepening our  understanding of efficiency gains beyond existing statistical paradigms. 
 \end{remark}

\section{Simulation studies} \label{sec:simulation}
This section demonstrates the performance of the proposed procedures through simulations. 
We first introduce the settings of  parameter generation and then present results. 


In Section \ref{sec:theory}, we highlight 
that  the proposed estimators do not require the columns of $Z^{\star}$ and $W_t^{\star}$ to be nearly orthogonal, characterized through $\mathcal{G}([Z^{\star}, W^{\star}])$.
To demonstrate that, 
we next generate true parameters satisfying 
$\mathcal{G}([Z^{\star}, W^{\star}])= \Omega \otimes \mathrm{I}_k/(2\sqrt{k}) $, 
where $\Omega = (\omega_{i,l})\in \mathbb R^{(1+T) \times (1+T)}$ with $\omega_{i,i}=1$, and 
$1/(2\sqrt{k})$ is a normalization scalar.
This Gram structure implies that columns of $Z^{\star}$ are orthogonal, columns of $W_t^{\star}$ are orthogonal, and 
$\cos(Z_j^{\star}, W_{t,j}^{\star} )=\omega_{1,1+t}$ and $\cos( W_{t,j}^{\star}, W_{s,j}^{\star} )= \omega_{1+t,1+s}$, where $\cos(a,b)=\langle a, b\rangle/(\|a\|_2\|b\|_2)$. 
When $\omega_{1,1+t}$ is larger, 
the two vectors $Z_j^{\star}$ and $W_{t,j}^{\star}$ are less orthogonal; 
similar interpretation applies to $\omega_{1+t,1+s}$. 
We next consider three cases (A)--(C) with $\Omega$ set as follows: 
\begin{table*}[!htbp]
\renewcommand{\arraystretch}{1}
\setlength{\tabcolsep}{12pt} 
    \centering
    \begin{tabular}{c|c c   c}
    &   (A)  &  (B) & (C) \\
$\Omega$ & $ \begin{pmatrix}
        1 & \mathbf{0}\\
        \mathbf{0} & \mathrm{I}_T
    \end{pmatrix}$ 
& $\begin{pmatrix}
       1 & \phi \mathbf{1}_{T}^{\top}\\
       \phi \mathbf{1}_{T} &  \Sigma_{T}(\rho)
    \end{pmatrix}$
    & $\begin{pmatrix}
        \mathrm{I}_{1+T_o} & \mathbf{0}\\
        \mathbf{0} & \Sigma_{T - T_o }(1)
    \end{pmatrix}$, 
    \end{tabular}
\end{table*}

\noindent where  $\mathbf{0}$ represents all-zero matrices whose dimensions are set such that $\Omega$ is of size $(1+T)\times (1+T)$, $\mathbf{1}_T$ represents a $T$-dimensional all-one column vector,  $\Sigma_T(\rho)$ denotes a $T \times T$  matrix with diagonal elements equal to $1$ and off-diagonal elements equal to $\rho$, i.e., with a compound symmetry structure. 
Under Case (A), all columns of $[Z^\star,W^\star]$  are orthogonal. 
Under Case (B),
$\cos( Z_j^{\star}, W_{t,j}^{\star}) =\phi $ and $\cos(W_{t,j}^{\star}, W_{s,j}^{\star}) = \rho$.  
In simulations, 
we take $\phi = 0.1$ and $\rho = 0.3$, implying 
an intermediate level of  non-orthogonality. 
Under Case (C),
the first $1+T_o$ columns of $[Z^{\star},W^{\star}]$ are orthogonal with the other columns, whereas 
$W_{t,j}^{\star}=W_{s,j}^{\star}$ for $T_o+1 \leqslant t,s\leqslant T$. In simulations, we take $T_o=4$, corresponding to an extreme scenario allowing part of the heterogeneous factors to overlap and thus are highly non-orthogonal.  
To randomly generate true parameters with a specified Gram matrix $\mathcal{G}^{\star}$, 
we first generate $\tilde {Z}$ and $\tilde {W}_t$ with rows independently sampled from a $k$-dimensional standard normal distribution restricted on the set $\{x \in \mathbb R^k: \|x\|_2^2 \leqslant k\}$, and  
 then set $[Z^\star, W^\star] = [\tilde Z, \tilde W] \tilde{\mathcal{G}}^{-1/2} (\mathcal{G}^\star)^{1/2}$, where $\tilde{\mathcal{G}} = \mathcal{G}([\tilde Z,\tilde W])$. 
Given true parameters in each case, we simulate data under three distributions: Bernoulli, Gaussian, and Poisson, all  with their canonical links. 

In each setting, we evaluate the performance of the proposed Algorithms \ref{algor:initial} and \ref{algor:refine}, referred to as SS-Hunting and SS-Refinement, respectively.
The two methods MultiNeSS and   MultiNeSS+ in \cite{macdonald2022latent} are compared  as our models are similar. 
Their errors are not presented under Poisson distribution as codes are unavailable. 
We next focus on illustrating 
the error rate of estimating $Z^{\star}$ with respect to $T$,
highlighting the improvement over using a single network. 
The error rates for estimating $W_t^{\star}$'s are not improved over $T$, which is consistent with our expected oracle rates in Section \ref{sec:challenges}, and detailed results are deferred to Section \ref{sec:suppsim} of the Supplementary Material.
For the ease of visualization, we vary $T \in \{5, 10, 20, 40, 80\}$ while fixing $n = 400$ and  $k_t = k = 2$ for $t=1,\ldots, T$. 
For a fair comparison with \cite{macdonald2022latent}, we present error  $\|\widehat{ZZ^\mytrans} - Z^\star Z^{\star \mytrans}\|_{\mathrm{F}}^2/n$, equivalent to $\operatorname{dist}^2(\hat Z, Z^\star)$ up to constants multiplied \citep{tu2016low}.

\begin{figure}
\centering
\begin{subfigure}{0.29\textwidth}
	\centering
	\includegraphics[width=1\linewidth]{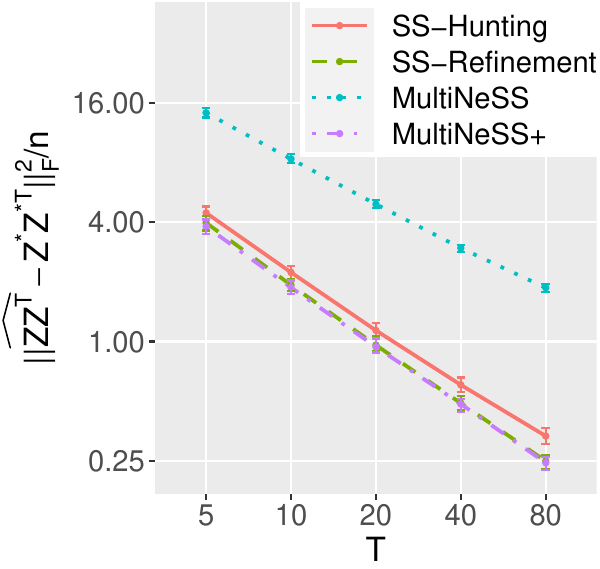}
    \caption{Bernoulli distribution}
\end{subfigure}
\begin{subfigure}{0.29\textwidth}
	\centering
	\includegraphics[width=1\linewidth]{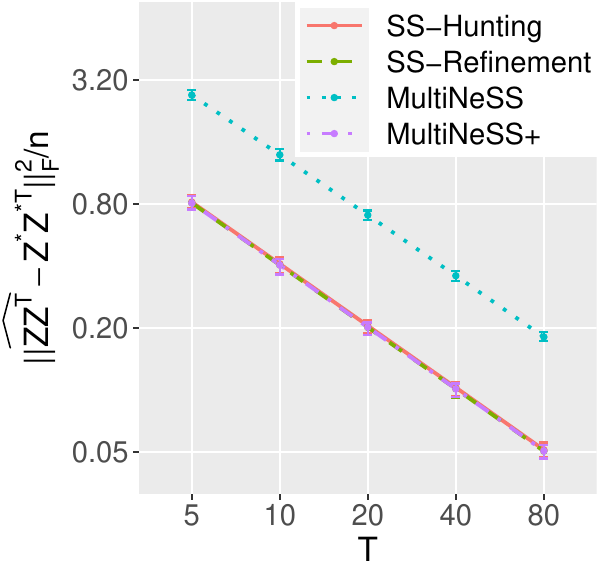}
    \caption{Gaussian distribution}
\end{subfigure}
\begin{subfigure}{0.29\textwidth}
	\centering
	\includegraphics[width=1\linewidth]{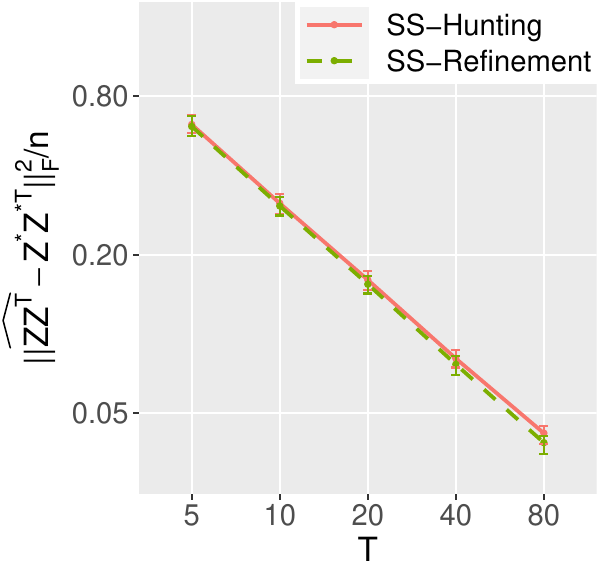}
    \caption{Poisson distribution}
\end{subfigure}
\caption{Empirical estimation errors versus $T$ under Case (A). Lines connect median estimation errors over 100 repetitions, and error bars are obtained from the 0.05 and 0.95 quantiles of those repetitions. Axes are in the log scale.}
\label{fig:simuA}
\end{figure}

\begin{figure}
\centering
\begin{subfigure}{0.29\textwidth}
	\centering
	\includegraphics[width=1\linewidth]{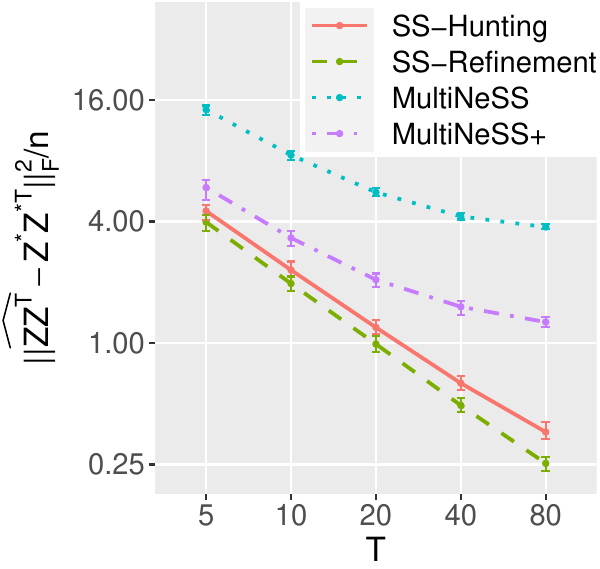}
    \caption{Bernoulli distribution}
\end{subfigure}
\begin{subfigure}{0.29\textwidth}
	\centering
	\includegraphics[width=1\linewidth]{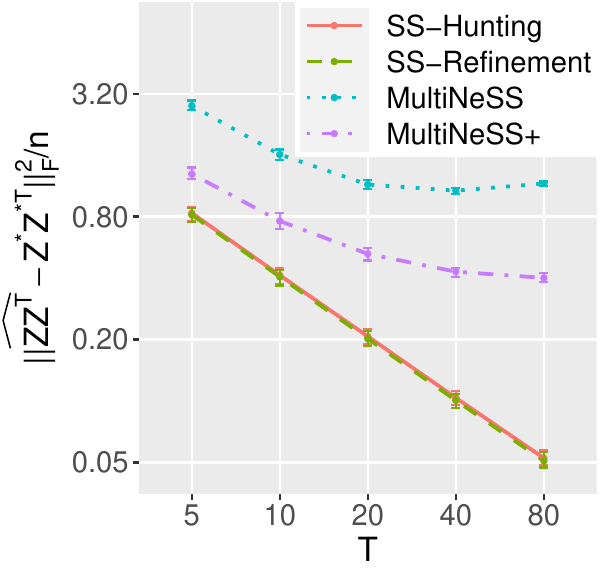}
    \caption{Gaussian distribution}
\end{subfigure}
\begin{subfigure}{0.29\textwidth}
	\centering
	\includegraphics[width=1\linewidth]{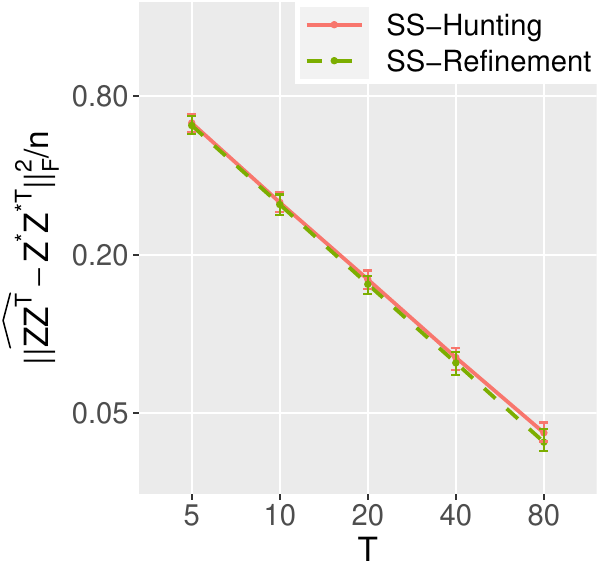}
    \caption{Poisson distribution}
\end{subfigure}
\caption{Empirical estimation errors  versus $T$ under Case (B). (Similar to Figure \ref{fig:simuA}.)}
\label{fig:simuB}
\end{figure}

\begin{figure}
\centering
\begin{subfigure}{0.29\textwidth}
	\centering
	\includegraphics[width=1\linewidth]{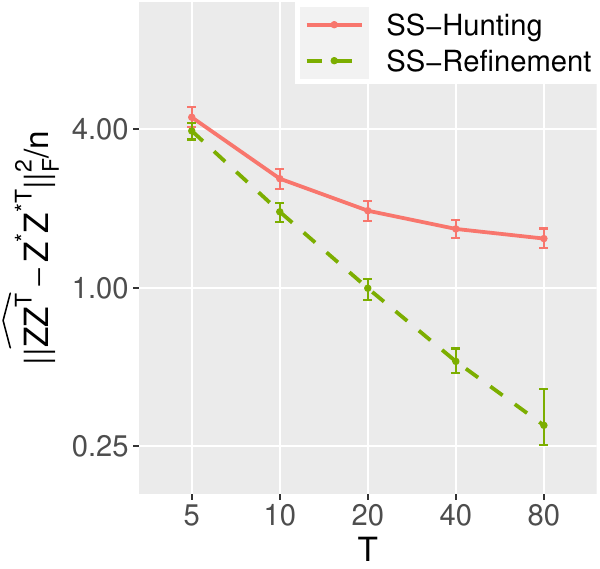}
    \caption{Bernoulli distribution}
\end{subfigure}
\begin{subfigure}{0.29\textwidth}
	\centering
	\includegraphics[width=1\linewidth]{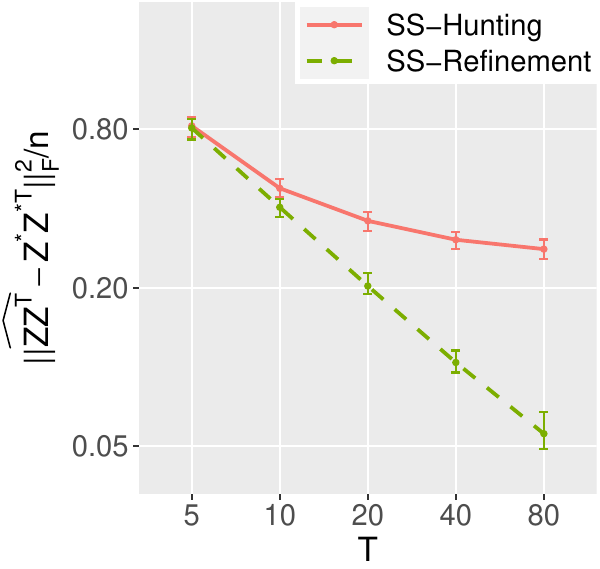}
    \caption{Gaussian distribution}
\end{subfigure}
\begin{subfigure}{0.29\textwidth}
	\centering
	\includegraphics[width=1\linewidth]{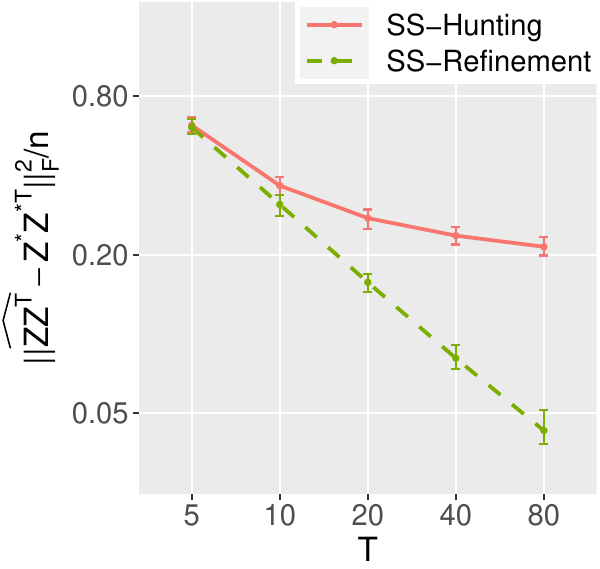}
    \caption{Poisson distribution}
\end{subfigure}
\caption{Empirical estimation errors versus $T$ under Case (C). (Similar to Figure \ref{fig:simuA}.)}
\label{fig:simuC}
\end{figure}

Figures \ref{fig:simuA}--\ref{fig:simuC} present  empirical estimation errors of the four estimators under Cases (A)--(C), respectively. 
In Figure \ref{fig:simuA}, all the errors are inverse proportional to $T$, achieving the oracle error rate discussed in Section \ref{sec:challenges}, and   SS-Refinement and MultiNeSS+ achieve the smallest errors. 
In Figure \ref{fig:simuB}, SS-Refinement achieves the smallest error across three distributions, 
while SS-Hunting performs similarly under Gaussian and Poisson distributions but not Bernoulli distribution. 
Different from Figure \ref{fig:simuA},
errors of MultiNeSS and MultiNeSS+ are no longer inverse proportional to $T$. 
Under Case (C), errors of MultiNeSS and MultiNeSS+ explode and are not presented in Figure \ref{fig:simuC} for visual clarity. {More details of their results are given in Section \ref{sec:suppsim_multiness} of the Supplementary Material.} 
Unlike in Figures \ref{fig:simuA} and \ref{fig:simuB},  
errors of SS-Hunting are not inverse proportional to $T$, 
whereas SS-Refinement still achieves that, demonstrating efficiency gain from the likelihood refinement. 

Intuitively, Cases (A)--(C) correspond to more and more challenging scenarios where the relationship between latent spaces become non-orthogonal and more complex. 
The non-ideal performances of MultiNeSS and MultiNeSS+ under Cases (B) and (C)
 align with Remark \ref{rmk:compare_peter} showing that our methods would require  less restriction on the latent spaces. 
For our two proposed estimators, 
SS-Refinement achieves the desired inverse proportional to $T$ error rate across all cases and distribution, whereas  SS-Hunting fails to achieve that in  challenging Case (C). 
This is consistent with our theoretical results in Theorems \ref{thm:initial} and \ref{thm:onestep} providing general error rates under weak assumptions about latent spaces. 
The results demonstrate the effectiveness of the proposed efficient estimation strategy. 

\section{Data analysis} \label{sec:data}
We analyze a multiple-network dataset of  lawyers \citep{lazega2001collegial},
which contains 
three types of connection relationships:  coworker, friendship, and advice,   between seventy-one lawyers in a 
US corporate law firm. 
Within each type of network, we consider binary and undirected edges between lawyers indicating whether 
connections of each type exist between them. 
Since there is no ground truth of latent spaces in practice, 
we compare the estimated latent embeddings with observed node-wise features for interpretation. 


As network edges are binary, 
we adopt Bernoulli distribution with logistic link 
and estimate latent dimensions by Algorithm \ref{algor:estk}, which gives  $\hat{k}_1 = 6$, $\hat{k}_2 = 4$, $\hat{k}_3 = 1$, and $\hat{k}=2$. 
Then we estimate  latent embedding  vectors under estimated latent dimensions by SS-Refinement.  
As  a comparison, we also apply   MultiNeSS+ in  \cite{macdonald2022latent} and  MASE in \cite{arroyo2021inference}  with the dimension of the shared space set to be two.  

\begin{figure}
\centering
\begin{subfigure}{0.28\textwidth}
	\centering
	\includegraphics[width=1\linewidth]{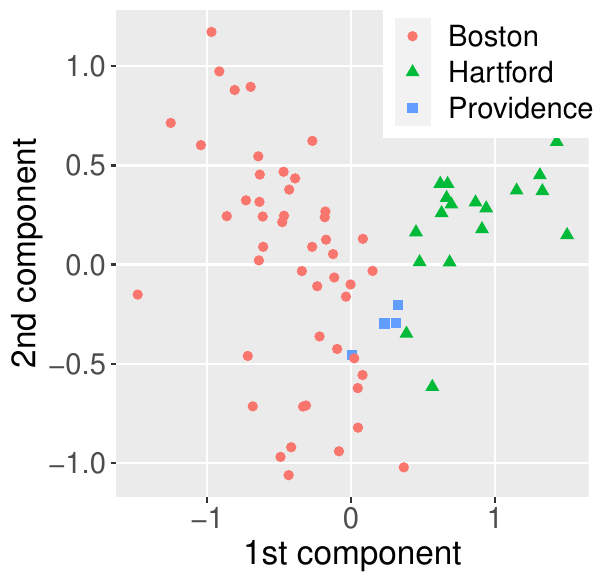}
    \caption{SS-Refinement}
\end{subfigure}\quad 
\begin{subfigure}{0.28\textwidth}
	\centering
	\includegraphics[width=1\linewidth]{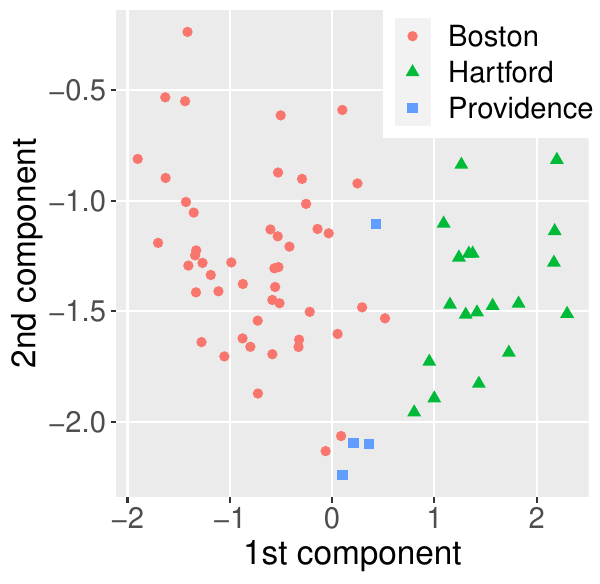}
    \caption{MultiNeSS+}
\end{subfigure}\quad 
\begin{subfigure}{0.28\textwidth}
	\centering
	\includegraphics[width=1\linewidth]{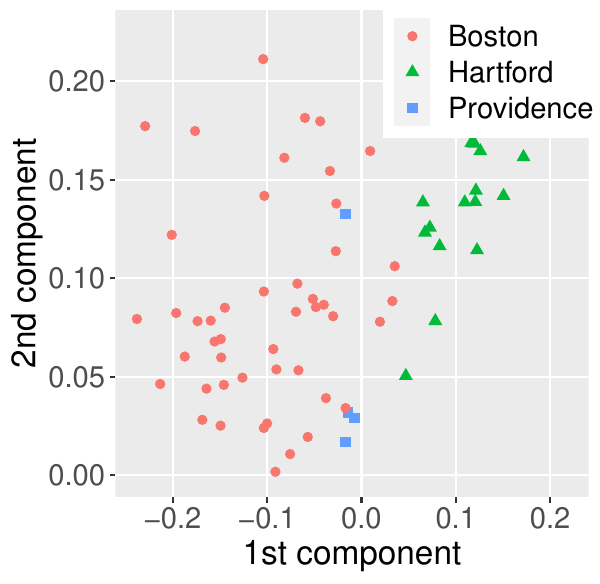}
    \caption{MASE}
\end{subfigure}
\caption{Estimated shared latent vectors $z_i$'s  using three methods.} 
\label{fig:Z}
\end{figure}

 We first illustrate the 
 shared latent embeddings estimated by the three methods. 
Figure \ref{fig:Z} (a) and (b) present  the leading two principal component scores of the shared latent component $Z$ estimated 
by SS-Refinement and MultiNeSS+, respectively. 
 Figure \ref{fig:Z} (c) displays latent embeddings estimated by MASE that are rotated to approximate (a) and (b) for the ease of comparison. 
In Figure \ref{fig:Z}, each point is colored according to the office location of the corresponding  lawyer.  
The results of all three methods show that the shared latent embeddings are correlated with the office locations of lawyers. 
This  suggests that office locations could play a shared role in forming multiple types of connections between lawyers.

\begin{figure}
\centering
\begin{subfigure}{0.28\textwidth}
	\centering
	\includegraphics[width=1\linewidth]{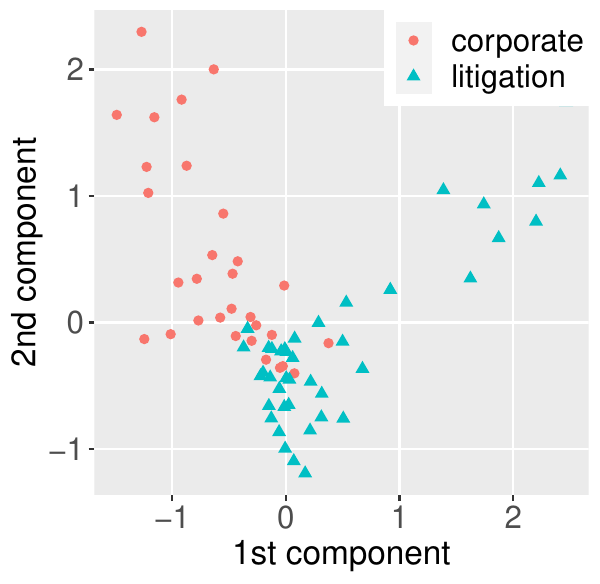}
    \caption{SS-Refinement}
\end{subfigure}\quad 
\begin{subfigure}{0.28\textwidth}
	\centering
	\includegraphics[width=1\linewidth]{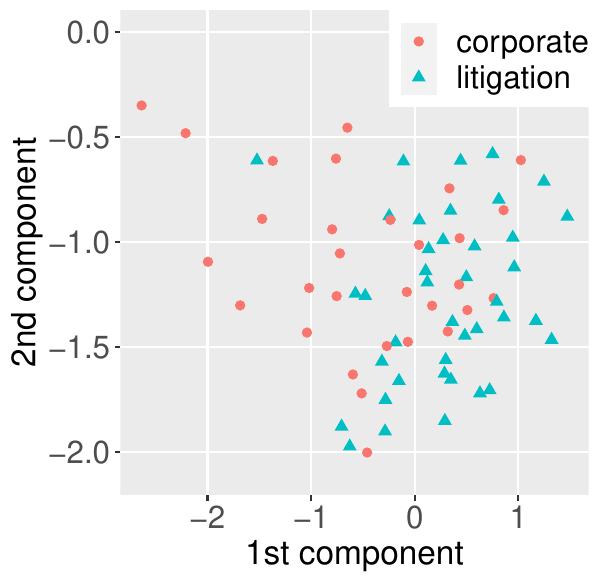}
    \caption{MultiNeSS+}
\end{subfigure}
\caption{Estimated individual latent vectors $w_{1,i}$'s  using two methods.}
\label{fig:estW1}
\end{figure}

\begin{figure}[htbp!]
\centering
\begin{subfigure}{0.28\textwidth}
	\centering
	\includegraphics[width=1\linewidth]{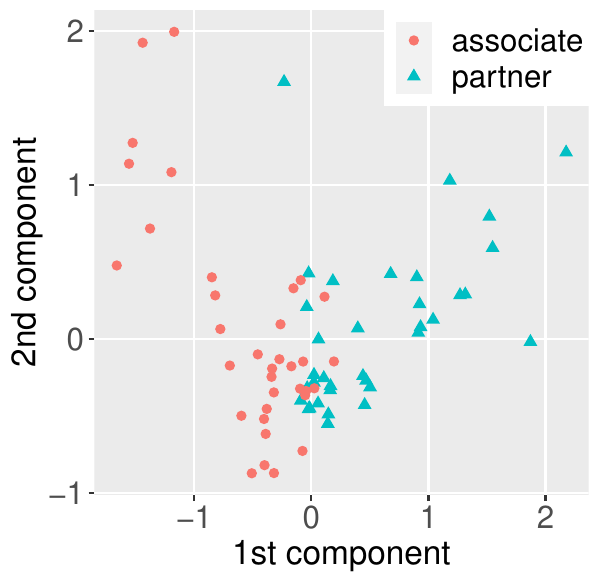}
    \caption{SS-Refinement}
\end{subfigure}\quad 
\begin{subfigure}{0.28\textwidth}
	\centering
	\includegraphics[width=1\linewidth]{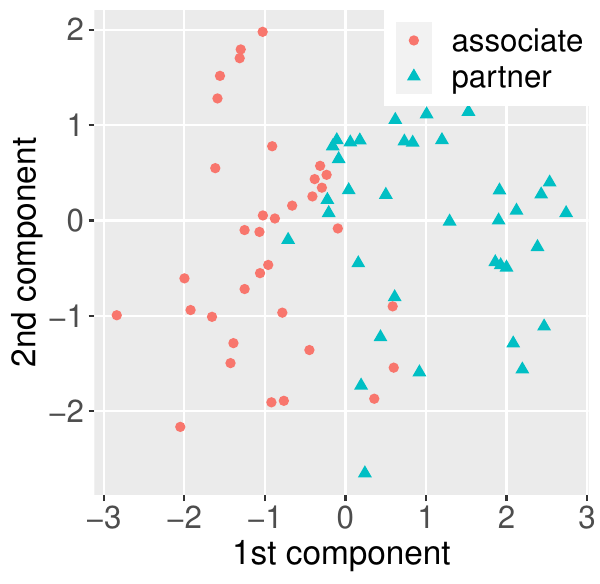}
    \caption{MultiNeSS+}
\end{subfigure}
\caption{Estimated individual latent vectors $w_{2,i}$'s  using two methods.}
\label{fig:estW2}
\end{figure}

We next compare the individual latent embeddings $W_t$'s estimated by SS-Refinement and MultiNeSS+, where we point out that MASE does not directly give  comparable estimates and thus is not included.     
Figures \ref{fig:estW1} and \ref{fig:estW2}  present the estimated individual latent component  after projected to its leading two principal components for the coworker and friendship networks, respectively;  
see more details in Section \ref{sec:embeddingdetails} of the Supplementary Material. 
In Figure \ref{fig:estW1}, points are colored according to lawyers’ practice. 
The results show that latent vectors estimated by both two methods  demonstrate association with lawyars' practice, whereas  the separation between litigation and corporate lawyers appears to be more  obvious in estimates given by SS-Refinement. 
This discrepancy might be attributed to the fact that MultiNeSS+ tends to encourage the column spaces of $Z$ and $W_t$'s to be orthogonal, which is not required in our approach and thus leads to distinct estimates in practice.   
In  Figure \ref{fig:estW2},  points are colored based on the lawyers’ status. 
Similarly to Figure \ref{fig:estW1}, 
latent vectors estimated by both two methods exhibit correlation with a common covariate, lawyers’ status,  even though the  patterns of points are not exactly the same.  

In summary, our analysis shows that investigating shared and individual latent embeddings  reveals key underlying factors driving the formation of heterogeneous multiplex networks. 
The proposed analysis paradigm could offer us a deeper understanding of multifaceted relationships and unveil  inherent structures of interlinked systems across various domains.  

\section{Discussions} \label{sec:discussion}

This work establishes   
a general analysis framework for  efficient estimation of node-wise latent embeddings in multiple heterogeneous networks. 
It shows that aggregating multiplex networks can improve the efficiency of estimating shared latent spaces. 
In particular, we develop 
  a  novel two-stage approach: the first stage 
 hunts for shared latent embedding spaces,
and the second stage improves the statistical efficiency with likelihood information. 
We establish estimation error bounds for the estimators of the two stages. The final estimator has been shown to achieve the oracle rates of component for both the shared and individual components according to their dimensions.  
  The new developments would provide statistical insights into how to efficiently  aggregate heterogeneous datasets and unravel fundamental structures in complex interconnected systems. 

The new developments pave the way for a broad range of future studies. 
First, accurate estimation of latent embedding spaces plays a crucial role in  subsequent analytical tasks including, but not limited to, understanding latent confounders across multifaceted interconnected systems,  adjusting dependence of observations for causal inference  \citep{mcfowland2023estimating,nath2022identifying,hayes2022estimating}, and  prediction with or on noisy network data \citep{ma2020universal,le2022linear,lunde2023conformal}.

Second, another interesting future research direction is to  extend the  analysis to various other network models. 
For instance, to analyze networks exhibiting  significant degree heterogeneity, we could add network-specific and node-specific degree heterogeneity parameters in current model, similarly to   \cite{zhang2020flexible} and \cite{he2023semiparametric}. 
Moreover, other forms of interactions could be considered, such as using  a general kernel function beyond inner product \citep{rubin2022statistical,macdonald2022latent},
or modeling directed network edges with asymmetric latent embeddings  \citep{perry2013point,yan2019statistical}.  
In addition, when there are observed covariates \citep{yan2021covariate,huang2018pairwise}, it would be worthwhile to devise appropriate procedures for  covariate adjustments. 
The  methods and theoretical tools developed in this work provide a versatile toolbox that could be useful under  diverse scenarios of network analysis.  

Finally, our current  analysis of efficient estimation  requires the likelihood to be correctly specified.
However, in many applications, 
distributions of noisy networks  may deviate from their prespecified forms. This misspecification might alter the interpretation of latent embeddings and impede efficient estimations.   Understanding the effect of distribution misspecification would be an important future research direction. Such studies could  provide us better understanding towards underlying data structures and lead to  reliable and robust conclusions \citep{rubin2020manifold}. 
\section*{Supplementary Material}
Due to space limitation, additional details, including individual estimation in Section \ref{sec:estimateyt}, estimating latent dimensions, and proofs are deferred to the Supplementary Material.

\section*{Acknowledgment} We are grateful to   the editors and two anonymous  reviewers for their valuable comments. The authors report there are no competing interests to declare. 

\bibliographystyle{apalike}
\begin{spacing}{1.5}
    \bibliography{Main.bbl}

\begin{thebibliography}{}

\bibitem[Abramowitz and Stegun, 1965]{abramowitz1965handbook}
Abramowitz, M. and Stegun, I.~A. (1965).
\newblock {\em Handbook of mathematical functions: with formulas, graphs, and mathematical tables}, volume~55.
\newblock Courier Corporation.

\bibitem[Agresti, 2015]{agresti2015foundations}
Agresti, A. (2015).
\newblock {\em Foundations of linear and generalized linear models}.
\newblock John Wiley \& Sons.

\bibitem[Andersen and Gill, 1982]{andersen1982cox}
Andersen, P.~K. and Gill, R.~D. (1982).
\newblock Cox's regression model for counting processes: a large sample study.
\newblock {\em Ann. Stat.}, 10(4):1100--1120.

\bibitem[Arroyo et~al., 2021]{arroyo2021inference}
Arroyo, J., Athreya, A., Cape, J., Chen, G., Priebe, C.~E., and Vogelstein, J.~T. (2021).
\newblock Inference for multiple heterogeneous networks with a common invariant subspace.
\newblock {\em J. Mach. Learn. Res.}, 22(142):1--49.

\bibitem[Athreya et~al., 2018]{athreya2018survey}
Athreya, A., Fishkind, D.~E., Tang, M., Priebe, C.~E., Park, Y., Vogelstein, J.~T., Levin, K., Lyzinski, V., Qin, Y., and Sussman, D.~L. (2018).
\newblock Statistical inference on random dot product graphs: a survey.
\newblock {\em J. Mach. Learn. Res.}, 18(226):1--92.

\bibitem[Athreya et~al., 2016]{athreya2016limit}
Athreya, A., Priebe, C.~E., Tang, M., Lyzinski, V., Marchette, D.~J., and Sussman, D.~L. (2016).
\newblock A limit theorem for scaled eigenvectors of random dot product graphs.
\newblock {\em Sankhya A}, 78:1--18.

\bibitem[Banerjee et~al., 2013]{banerjee2013diffusion}
Banerjee, A., Chandrasekhar, A.~G., Duflo, E., and Jackson, M.~O. (2013).
\newblock The diffusion of microfinance.
\newblock {\em Science}, 341(6144):1236498.

\bibitem[Barzilai and Borwein, 1988]{barzilai1988two}
Barzilai, J. and Borwein, J.~M. (1988).
\newblock Two-point step size gradient methods.
\newblock {\em IMA J. Numer. Anal.}, 8(1):141--148.

\bibitem[Bickel et~al., 1993]{bickel1993efficient}
Bickel, P.~J., Klaassen, C.~A., Ritov, Y., and Wellner, J.~A. (1993).
\newblock {\em Efficient and Adaptive Estimation for Semiparametric Models}.
\newblock Springer.

\bibitem[Bradic et~al., 2019]{bradic2019minimax}
Bradic, J., Chernozhukov, V., Newey, W.~K., and Zhu, Y. (2019).
\newblock Minimax semiparametric learning with approximate sparsity.
\newblock {\em arXiv preprint arXiv:1912.12213}.

\bibitem[Cai and Zhang, 2018]{cai2018rate}
Cai, T.~T. and Zhang, A. (2018).
\newblock Rate-optimal perturbation bounds for singular subspaces with applications to high-dimensional statistics.
\newblock {\em Ann. Stat.}, 46(1):60--89.

\bibitem[Chatterjee, 2015]{chatterjee2015matrix}
Chatterjee, S. (2015).
\newblock Matrix estimation by universal singular value thresholding.
\newblock {\em Ann. Stat.}, 43(1):177--214.

\bibitem[Chen et~al., 2022]{chen2022global}
Chen, S., Liu, S., and Ma, Z. (2022).
\newblock Global and individualized community detection in inhomogeneous multilayer networks.
\newblock {\em Ann. Stat.}, 50(5):2664--2693.

\bibitem[Chen and Wainwright, 2015]{chen2015fast}
Chen, Y. and Wainwright, M.~J. (2015).
\newblock Fast low-rank estimation by projected gradient descent: General statistical and algorithmic guarantees.
\newblock {\em arXiv preprint arXiv:1509.03025}.

\bibitem[Chernozhukov et~al., 2018]{chernozhukiv.ectj.12097}
Chernozhukov, V., Chetverikov, D., Demirer, M., Duflo, E., Hansen, C., Newey, W., and Robins, J. (2018).
\newblock {Double/debiased machine learning for treatment and structural parameters}.
\newblock {\em The Econometrics Journal}, 21(1):C1--C68.

\bibitem[Efron, 2022]{efron2022exponential}
Efron, B. (2022).
\newblock {\em Exponential Families in Theory and Practice}.
\newblock Cambridge University Press.

\bibitem[Gould et~al., 2018]{gould2018microbiome}
Gould, A.~L., Zhang, V., Lamberti, L., Jones, E.~W., Obadia, B., Korasidis, N., et~al. (2018).
\newblock Microbiome interactions shape host fitness.
\newblock {\em Proc. Natl. Acad. Sci.}, 115(51):E11951--E11960.

\bibitem[Han et~al., 2015]{han2015consistent}
Han, Q., Xu, K., and Airoldi, E. (2015).
\newblock Consistent estimation of dynamic and multi-layer block models.
\newblock In {\em Proc. ICML}, volume~37, pages 1511--1520. PMLR.

\bibitem[Hayes et~al., 2024]{hayes2022estimating}
Hayes, A., Fredrickson, M.~M., and Levin, K. (2024).
\newblock Estimating network-mediated causal effects via spectral embeddings.
\newblock {\em arXiv preprint arXiv:2212.12041}.

\bibitem[He et~al., 2025]{he2023semiparametric}
He, Y., Sun, J., Tian, Y., Ying, Z., and Feng, Y. (2025).
\newblock Semiparametric modeling and analysis for longitudinal network data.
\newblock {\em arXiv preprint arXiv:2308.12227}.

\bibitem[Hoff, 2011]{hoff2011hierarchical}
Hoff, P.~D. (2011).
\newblock Hierarchical multilinear models for multiway data.
\newblock {\em Comput. Stat. Data Anal.}, 55(1):530--543.

\bibitem[Hoff et~al., 2002]{hoff2002latent}
Hoff, P.~D., Raftery, A.~E., and Handcock, M.~S. (2002).
\newblock Latent space approaches to social network analysis.
\newblock {\em J. Am. Stat. Assoc.}, 97(460):1090--1098.

\bibitem[Holland et~al., 1983]{holland1983stochastic}
Holland, P.~W., Laskey, K.~B., and Leinhardt, S. (1983).
\newblock Stochastic blockmodels: First steps.
\newblock {\em Social networks}, 5(2):109--137.

\bibitem[Huang et~al., 2024]{huang2018pairwise}
Huang, S., Sun, J., and Feng, Y. (2024).
\newblock {PCABM}: Pairwise covariates-adjusted block model for community detection.
\newblock {\em J. Am. Stat. Assoc.}, 119(547):2092--2104.

\bibitem[Jones and Rubin-Delanchy, 2020]{jones2020multilayer}
Jones, A. and Rubin-Delanchy, P. (2020).
\newblock The multilayer random dot product graph.
\newblock {\em arXiv preprint arXiv:2007.10455}.

\bibitem[Kivel{\"a} et~al., 2014]{kivela2014multilayer}
Kivel{\"a}, M., Arenas, A., Barthelemy, M., Gleeson, J.~P., Moreno, Y., and Porter, M.~A. (2014).
\newblock Multilayer networks.
\newblock {\em J. Complex Netw.}, 2(3):203--271.

\bibitem[Lazega, 2001]{lazega2001collegial}
Lazega, E. (2001).
\newblock {\em The collegial phenomenon: The social mechanisms of cooperation among peers in a corporate law partnership}.
\newblock Oxford University Press, USA.

\bibitem[Le and Li, 2022]{le2022linear}
Le, C.~M. and Li, T. (2022).
\newblock Linear regression and its inference on noisy network-linked data.
\newblock {\em J. R. Stat. Soc. (Series B)}, 84(5):1851--1885.

\bibitem[Lei and Lin, 2023]{lei2023bias}
Lei, J. and Lin, K.~Z. (2023).
\newblock Bias-adjusted spectral clustering in multi-layer stochastic block models.
\newblock {\em Journal of the American Statistical Association}, 118(544):2433--2445.

\bibitem[Li et~al., 2020]{li2020network}
Li, T., Levina, E., and Zhu, J. (2020).
\newblock Network cross-validation by edge sampling.
\newblock {\em Biometrika}, 107(2):257--276.

\bibitem[Little et~al., 2010]{little2010parameter}
Little, M.~P., Heidenreich, W.~F., and Li, G. (2010).
\newblock Parameter identifiability and redundancy: theoretical considerations.
\newblock {\em PloS one}, 5(1):e8915.

\bibitem[Lunde et~al., 2023]{lunde2023conformal}
Lunde, R., Levina, E., and Zhu, J. (2023).
\newblock Conformal prediction for network-assisted regression.
\newblock {\em arXiv preprint arXiv:2302.10095}.

\bibitem[Lyu et~al., 2023]{lyu2023latent}
Lyu, Z., Xia, D., and Zhang, Y. (2023).
\newblock Latent space model for higher-order networks and generalized tensor decomposition.
\newblock {\em J. Comput. Graph. Stat.}, 32(4):1320--1336.

\bibitem[Ma et~al., 2020]{ma2020universal}
Ma, Z., Ma, Z., and Yuan, H. (2020).
\newblock Universal latent space model fitting for large networks with edge covariates.
\newblock {\em J. Mach. Learn. Res.}, 21(1):86--152.

\bibitem[MacDonald et~al., 2022]{macdonald2022latent}
MacDonald, P.~W., Levina, E., and Zhu, J. (2022).
\newblock Latent space models for multiplex networks with shared structure.
\newblock {\em Biometrika}, 109(3):683--706.

\bibitem[Matias and Robin, 2014]{matias2014modeling}
Matias, C. and Robin, S. (2014).
\newblock Modeling heterogeneity in random graphs through latent space models: a selective review.
\newblock {\em ESAIM Proc. Surv.}, 47:55--74.

\bibitem[McFowland~III and Shalizi, 2023]{mcfowland2023estimating}
McFowland~III, E. and Shalizi, C.~R. (2023).
\newblock Estimating causal peer influence in homophilous social networks by inferring latent locations.
\newblock {\em J. Am. Stat. Assoc.}, 118(541):707--718.

\bibitem[Murphy and Van~der Vaart, 2000]{murphy2000profile}
Murphy, S.~A. and Van~der Vaart, A.~W. (2000).
\newblock On profile likelihood.
\newblock {\em Journal of the American Statistical Association}, 95(450):449--465.

\bibitem[Nath et~al., 2024]{nath2022identifying}
Nath, S., Warren, K., and Paul, S. (2024).
\newblock Identifying peer influence in therapeutic communities adjusting for latent homophily.
\newblock {\em arXiv preprint arXiv:2203.14223}.

\bibitem[Nesterov, 2004]{nesterov2003introductory}
Nesterov, Y. (2004).
\newblock {\em Introductory lectures on convex optimization: A basic course}, volume~87.
\newblock Springer Science \& Business Media.

\bibitem[Nielsen and Witten, 2018]{nielsen2018multiple}
Nielsen, A.~M. and Witten, D. (2018).
\newblock The multiple random dot product graph model.
\newblock {\em arXiv preprint arXiv:1811.12172}.

\bibitem[Paul and Chen, 2016]{paul2016consistent}
Paul, S. and Chen, Y. (2016).
\newblock {Consistent community detection in multi-relational data through restricted multi-layer stochastic blockmodel}.
\newblock {\em Electron. J. Stat.}, 10(2):3807 -- 3870.

\bibitem[Perry and Wolfe, 2013]{perry2013point}
Perry, P.~O. and Wolfe, P.~J. (2013).
\newblock Point process modelling for directed interaction networks.
\newblock {\em J. R. Stat. Soc. (Series B)}, 75(5):821--849.

\bibitem[Portnoy, 1988]{portnoy1988asymptotic}
Portnoy, S. (1988).
\newblock Asymptotic behavior of likelihood methods for exponential families when the number of parameters tends to infinity.
\newblock {\em Ann. Stat.}, 16(1):356--366.

\bibitem[Rohe et~al., 2011]{rohe2011spectral}
Rohe, K., Chatterjee, S., and Yu, B. (2011).
\newblock Spectral clustering and the high-dimensional stochastic blockmodel.
\newblock {\em Ann. Stat.}, 39(4):1878--1915.

\bibitem[Rubin-Delanchy, 2020]{rubin2020manifold}
Rubin-Delanchy, P. (2020).
\newblock Manifold structure in graph embeddings.
\newblock {\em NeurIPS}, 33:11687--11699.

\bibitem[Rubin-Delanchy et~al., 2022]{rubin2022statistical}
Rubin-Delanchy, P., Cape, J., Tang, M., and Priebe, C.~E. (2022).
\newblock A statistical interpretation of spectral embedding: the generalised random dot product graph.
\newblock {\em J. R. Stat. Soc. (Series B)}, 84(4):1446--1473.

\bibitem[Salter-Townshend and McCormick, 2017]{salter2017latent}
Salter-Townshend, M. and McCormick, T.~H. (2017).
\newblock Latent space models for multiview network data.
\newblock {\em Ann. Appl. Stat.}, 11(3):1217--1244.

\bibitem[Tang and Priebe, 2018]{tang2018limit}
Tang, M. and Priebe, C.~E. (2018).
\newblock Limit theorems for eigenvectors of the normalized laplacian for random graphs.
\newblock {\em The Annals of Statistics}, 46(5):2360 -- 2415.

\bibitem[Tropp, 2012]{tropp2012user}
Tropp, J.~A. (2012).
\newblock User-friendly tail bounds for sums of random matrices.
\newblock {\em Foundations of computational mathematics}, 12(4):389--434.

\bibitem[Tu et~al., 2016]{tu2016low}
Tu, S., Boczar, R., Simchowitz, M., Soltanolkotabi, M., and Recht, B. (2016).
\newblock Low-rank solutions of linear matrix equations via procrustes flow.
\newblock In {\em Proc. ICML}, volume~48, pages 964--973. PMLR.

\bibitem[Van~der Vaart, 2000]{van2000asymptotic}
Van~der Vaart, A.~W. (2000).
\newblock {\em Asymptotic Statistics}, volume~3.
\newblock Cambridge university press.

\bibitem[Wang et~al., 2019]{wang2019common}
Wang, L., Zhang, Z., and Dunson, D. (2019).
\newblock Common and individual structure of brain networks.
\newblock {\em Ann. Appl. Stat.}, 13(1):85--112.

\bibitem[Wedin, 1973]{wedin1973perturbation}
Wedin, P.-{\AA}. (1973).
\newblock Perturbation theory for pseudo-inverses.
\newblock {\em BIT Numerical Mathematics}, 13:217--232.

\bibitem[Wellner, 2005]{wellner2005empirical}
Wellner, J.~A. (2005).
\newblock {\em Empirical processes: Theory and applications}.
\newblock Notes for a course given at Delft University of Technology.

\bibitem[Wen et~al., 2022]{wen2022characterizing}
Wen, J., Fu, C.~H., Tosun, D., Veturi, Y., Yang, Z., Abdulkadir, A., et~al. (2022).
\newblock Characterizing heterogeneity in neuroimaging, cognition, clinical symptoms, and genetics among patients with late-life depression.
\newblock {\em JAMA Psychiatry}, 79(5):464--474.

\bibitem[Xie and Xu, 2023]{xie2023efficient}
Xie, F. and Xu, Y. (2023).
\newblock Efficient estimation for random dot product graphs via a one-step procedure.
\newblock {\em J. Am. Stat. Assoc.}, 118(541):651--664.

\bibitem[Yan and Sarkar, 2021]{yan2021covariate}
Yan, B. and Sarkar, P. (2021).
\newblock Covariate regularized community detection in sparse graphs.
\newblock {\em J. Am. Stat. Assoc.}, 116(534):734--745.

\bibitem[Yan et~al., 2019]{yan2019statistical}
Yan, T., Jiang, B., Fienberg, S.~E., and Leng, C. (2019).
\newblock Statistical inference in a directed network model with covariates.
\newblock {\em J. Am. Stat. Assoc.}, 114(526):857--868.

\bibitem[Young and Scheinerman, 2007]{young2007random}
Young, S.~J. and Scheinerman, E.~R. (2007).
\newblock Random dot product graph models for social networks.
\newblock In {\em Proc. Int. Workshop Alg. Models Web-Graph}, pages 138--149. Springer.

\bibitem[Zhang and Wang, 2023]{zhangefficient}
Zhang, H. and Wang, J. (2023).
\newblock Efficient estimation for longitudinal network via adaptive merging.
\newblock {\em arXiv:2211.07866}.

\bibitem[Zhang et~al., 2020]{zhang2020flexible}
Zhang, X., Xue, S., and Zhu, J. (2020).
\newblock A flexible latent space model for multilayer networks.
\newblock In {\em Proc. ICML}, volume 119, pages 11288--11297. PMLR.

\bibitem[Zhang et~al., 2017]{zhang2017estimating}
Zhang, Y., Levina, E., and Zhu, J. (2017).
\newblock Estimating network edge probabilities by neighbourhood smoothing.
\newblock {\em Biometrika}, 104(4):771--783.

\bibitem[Zheng and Tang, 2022]{zheng2022limit}
Zheng, R. and Tang, M. (2022).
\newblock Limit results for distributed estimation of invariant subspaces in multiple networks inference and {PCA}.
\newblock {\em arXiv preprint arXiv:2206.04306}.

\end{thebibliography}
\end{spacing}

\appendix

\begin{center}
\Large\bfseries Supplementary Material of ``Efficient Analysis of Latent Spaces in Heterogeneous Networks''
\end{center}

\renewcommand{\thetable}{S\arabic{table}}
\renewcommand{\thefigure}{S\arabic{figure}}
\numberwithin{equation}{section}

\paragraph{Notation} We summarize notations that will be used throughout the Supplementary Material but have not been explained in the main text. Given two sequences of real numbers $\{g_n\}$ and $\{h_n\}$, 
we write $g_n\asymp h_n$ if there exist constants $c_1,c_2>0$ such that $c_2h_n\leqslant  g_n\leqslant c_1  h_n $. 
Given a matrix $X\in \mathbb{R}^{n\times m}$, let $\|X\|_*$ denote its nuclear norm. For two symmetric matrices $X,Y \in \mathbb R^{n \times n}$, we write $X \preceq Y$ if $Y-X$ is positive semidefinite.
For two matrices $X,Y \in  \mathbb R^{n \times m}$, we denote by $X\odot Y $ their Hadamard product, i.e., $(X\odot Y)_{ij} = X_{ij} \times Y_{ij} $ for $1\leqslant i \leqslant n, 1\leqslant j \leqslant m $. Let $\mathbb I(\cdot)$ denote the indicator function, which takes the value $1$ if its argument is true and $0$ otherwise.
Moreover, in the following proofs, we use $C_{\varepsilon}$ and $c_{\varepsilon}$ to represent   generic constants that  depend    on $\varepsilon$ but are independent with  $(n,T)$. With a slight abuse of notation, the specific values of $C_{\varepsilon}$ and $c_{\varepsilon}$  may vary across different contexts.


\setcounter{algocf}{0}
\renewcommand{\thealgocf}{A.\arabic{algocf}}

\section{Results for Estimating Individual Networks}
\label{sec:estyt}
In this section, we present detailed methods and results for estimating latent spaces of each network individually in Section \ref{sec:estimateyt}. 
Our strategy generalizes the analysis of single  binary network in \cite{ma2020universal} to networks with general edge distributions. In particular, we use universal singular value thresholding (Algorithm  \ref{algor:estTheta1}) followed by  projected gradient descent (Algorithm \ref{algor:estY}). 
More generally, the procedure of obtaining $\mathring{Y}_t$ can be replaced by any other estimators that achieve the same error rates as in Theorem \ref{thm:estY}. 
The theoretical results for Steps (b) and (c) in Section  \ref{sec:estall} follow similarly. 

\subsection{Algorithms} \label{sec:ytestappendx}
\begin{algorithm}[!htbp]
\caption{Estimation of $\Theta_t^{\star}$.(Universal singular value thresholding) }
\label{algor:estTheta1}
\setstretch{1.5}
\KwIn{Data: $\mathbf A_t\in \mathbb{R}^{n\times n}$. Parameters: $\tau_4$ (threshold),  $\mathcal{C}_{E} $ (constraint set).}
\KwOut{$\breve{\Theta}_t.$}


    Let $\sum_{i=1}^{n} d_{t,i} u_{t,i}v_{t,i}^{\top}$ denote the singular value decomposition of $\mathbf A_{t}$. 
     
    Let $\tilde{E}_{t}=\sum_{\{i:\, d_{t,i} > \tau_{4}\}} d_{t,i} u_{t,i} v_{t,i}^{\top}$ and  $\breve{E}_{t} = \mathcal{P}_{\mathcal{C}_E}(\tilde{E}_{t})$. 
 
    Let $\tilde{\Theta}_{t} = \mu^{-1}(\breve{E}_{t})$ and $\breve{\Theta}_t = \mathcal{P}_{\mathbb S^n_+}(\tilde{\Theta}_t)$, where $\mathbb S^n_+$ represents the class of $n \times n$ positive semidefinite matrices.
\end{algorithm} 
        

\begin{algorithm}[!htbp]
\caption{Estimation of $Y_t^{\star}$. (Projected gradient descent)}
\label{algor:estY}
\setstretch{1.5}
\KwIn{Data: $\mathbf A_t \in \mathbb{R}^{n\times n}$. \quad Initial estimate: $\breve{\Theta}_t \in \mathbb{R}^{n\times n}$.\\
\hspace{3.3em} Parameters: $\eta_{Y}$ (step size), $R$ (number of iterations), $\mathcal{C}_{Y_t}$ (constraint set).}
\KwOut{$\mathring Y_t.$}
    Let $Y_t^0 = \mathcal{S}_{d_t}(\breve{\Theta}_t)$, where $\mathcal{S}_{d_t}(\cdot)$ is defined as in  \eqref{eq:skmap}.

    \For{$r = 0, \ldots, R - 1$}{
			$\tilde{Y}_t^{r+1}= Y_t^{r} + \eta_{Y}\,  l^{\prime}(Y_t^r Y_t^{r \mytrans}){Y_t^r} $, where $l^{\prime}(Y_t^r Y_t^{r \mytrans}) = \big[l^{\prime}( \langle y_{t,i}^r, y_{t,j}^r \rangle; A_{t,ij})\big]_{1 \leqslant i,j \leqslant n}$.

            $Y_t^{r+1}=\mathcal{P}_{\mathcal{C}_{Y_t}}(\tilde{Y}_t^{r+1})$.
	}
 Let $\mathring Y_t = Y_t^R$.
\end{algorithm}

We specify the formal definition of $\mathcal{S}_{d_t}(\cdot)$ used in Algorithm \ref{algor:estY} below. 
\begin{definition}
    Consider a symmetric matrix $B$ with eigen-decomposition $B=UDU^{\top} \in \mathbb{R}^{n\times n}$. For an integer $k\leqslant n$, let $D_k$ denote a diagonal matrix with the largest $k$ eigenvalues of  $B$, and let $U_k$ represent the corresponding eigenvectors. Define $\mathcal{S}_k: \mathbb{R}^{n\times n} \to \mathbb{R}^{n\times k}$ as 
\begin{align}\label{eq:skmap}
    \mathcal{S}_k(B) = U_kD_{k,+}^{1/2}, 
\end{align}
where $D_{+}$ is a diagonal matrix obtained by taking the maximum over entries in $D$ and 0 elementwisely, and $D_+^{1/2}$ represents taking the matrix square root of $D_+$.  
\end{definition}

In  Algorithms \ref{algor:estTheta1} and \ref{algor:estY}, the hyperparameters are required to satisfy Condition   \ref{cond:parestTheta1} below to establish theoretical results. 
In our numerical implementation, for Algorithm \ref{algor:estTheta1}, we set $\tau_{4} = \sqrt{\sum_{1\leqslant i,j \leqslant n}A_{t,ij}/n}$, and $\mathcal{C}_E = \{E \in \mathbb R^{n \times n}:  E_{ij} \in \mu([-5,5]) \text{ for } 1\leqslant i,j \leqslant n\}$, where $\mu([a,b]) = \{\mu(\theta): \theta \in [a,b]\}$ denotes the image of $[a,b]$ under the function $\mu$. For Algorithm \ref{algor:estY}, we use Barzilai-Borwein step sizes \citep{barzilai1988two}, set $R = 1000$, and skip the projection step in line 4.

\begin{condition} \label{cond:parestTheta1}\label{cond:parestY}
In Algorithm \ref{algor:estTheta1}, 
(i) $\sqrt{n}\log^{1/2}(nT) \ll \tau_4 \lesssim  \sqrt{n}\log(nT) $
and (ii) $\mathcal{C}_E = \{E \in \mathbb R^{n \times n}: 
E_{ij} \in \mu([-2M_1^2,2M_1^2])
\text{ for } 1\leqslant i,j \leqslant n\}$, where $M_1$ is given in Condition \ref{cond:truevalue}.
In Algorithm \ref{algor:estY},   (i) step size $\eta_Y = \eta/n$  for a sufficiently small constant $\eta > 0$, (ii)  number of iterations $R \gg \log(nT)$, and (iii) constraint set $\mathcal{C}_{Y_t} = \{Y_t \in \mathbb R^{n\times d_t }: \|Y_t\|_{2 \to \infty} \leqslant 2M_1\} $, where $M_1$ is given in Condition \ref{cond:truevalue}. 
\end{condition} 


\subsection{Theoretical Results} \label{sec:estyt_theory}

We next present theoretical results for Algorithms \ref{algor:estTheta1} and \ref{algor:estY} in Theorems \ref{thm:estTheta} and \ref{thm:estY}, respectively.  
 
        

\begin{theorem} \label{thm:estTheta}
 Assume Conditions \ref{cond:truevalue}, \ref{cond:parfunction}, and \ref{cond:parestTheta1}. Let $\breve \Theta_t$ be the individual estimator obtained through Algorithm \ref{algor:estTheta1}. For any constant $\varepsilon >0$, there exist positive constants $c_\varepsilon$ and $C_\varepsilon$ such that when $\log(nT)/n \leqslant c_\varepsilon $, 
    $$
    \Pr\left[ \bigcup_{1\leqslant t \leqslant T} \left\{\big\|\breve \Theta_t - \Theta_t^\star\big\|_{\mathrm{F}}/n > C_\varepsilon n^{-\frac{1}{2 d_t + 4}}\log^{\frac{1}{2}}(nT)\right\}\right] = O\big((nT)^{-\varepsilon}\big).
    $$
\end{theorem}

\begin{theorem} \label{thm:estY}
 Assume Conditions \ref{cond:truevalue}, \ref{cond:parfunction}, and \ref{cond:parestY}. Let $\mathring Y_t$ be the output of Algorithm \ref{algor:estY} when using $\breve \Theta_t$ in Theorem \ref{thm:estTheta} as initialization. For any constant $\varepsilon >0$, there exist positive constants $c_\varepsilon$ and $C_\varepsilon$ such that when $\log^{\dmax+2}(nT)/n \leqslant c_\varepsilon $, 
    $$
    \Pr\left[ \max_{1\leqslant t \leqslant T} \left\{\operatorname{dist}^2(\mathring Y_t, Y_t^\star) \right\} > C_\varepsilon \log(nT)\right] = O\big( (nT)^{-\varepsilon}\big).
    $$
\end{theorem}

Error rates in Theorems \ref{thm:estTheta} and \ref{thm:estY} are consistent with the  probabilistic errors established in \cite{ma2020universal} for a single network with binary edges. Here we need an extra $\log (T)$ term to achieve  uniform error control over $T$ networks. 

\newpage
\setcounter{algocf}{0}
\renewcommand{\thealgocf}{B.\arabic{algocf}}
\section{Results for Estimating Latent Dimensions}  \label{sec:estk}

\subsection{Algorithm}\label{sec:estk_algor}

 In the main text, 
 we have treated the latent dimensions $d_t, k,$ and $k_t$ as prior information. 
 %
But this information is often unknown in real-world applications. 
 We next propose a spectral-based approach that can consistently estimate the latent space dimensions. 
The proposed approach consists of two steps: the first step estimates $d_t$ for $t=1,\ldots, T$, and the second  step estimates $k$. 
Then $k_t$ can be estimated as the differences between the estimates for $d_t$ and $k$.  We next describe the two steps in detail. 


\smallskip 
\emph{(a) Step 1: Estimate $d_t$.}\ 
Consider a sufficiently good estimator for $ \Theta_t^\star$, e.g., $\breve \Theta_t$ by Algorithm \ref{algor:estTheta1}. 
As  the rank of  $\Theta_t^{\star}=Y_t^{\star}Y_t^{\star\top}$ equals $d_t$,  we expect a significant gap between the $d_t$-th and the $(d_t+1)$-th  singular values of $\breve \Theta_t$.
Therefore, 
we can estimate $d_t$ as the smallest index such that the ratio of consecutive singular values drops below an appropriate threshold.  
In particular, we construct 
\begin{align}\label{eq:dthat}
    \hat{d}_t = \min\big\{1\leqslant m \leqslant n-1: \mathcal{R}_{m+1, \, m }(\breve{\Theta}_{t})\leqslant \tau_{2,m}\}  
\end{align}
with the choice of $\tau_{2,m}$ discussed in Condition \ref{cond:choicetaupar} below. 

\smallskip 
 \emph{(b) Step 2: Estimate $k$ and $k_t$.} \ 
 In Step 2, we estimate $k$ and $k_t$ based on the estimates in Step 1. 
To achieve this, we utilize the following fact. 
\begin{proposition} \label{prop:dimensionk}
Let $K_{t,s}^r$ denote the column rank of $[Y_{t}^{\star}, Y_{s}^{\star}]$,
and let $K_{t,s}=d_t+d_s-K_{t,s}^r$ denote its nullity. 
\begin{enumerate}
    \item[(i)] $k \leqslant  K_{t,s}$ for all $1\leqslant t < s\leqslant T$,    
    \item[(ii)]  $k = K_{t,s}$ holds for $(t,s)\in \mathcal{T}$, where $\mathcal{T}$ is defined in Proposition \ref{prop:yqtoz}. 
\end{enumerate}
\end{proposition}
Proposition \ref{prop:dimensionk} suggests that 
we can estimate $k$ as the minimum over all estimates of  $K_{t,s}$ for $1\leqslant t < s\leqslant T$.
To estimate $K_{t,s}$, since $K_{t,s}=d_t+d_s-K_{t,s}^r$ and $d_t$'s have been estimated in Step 1, 
it remains to estimate $K_{t,s}^r$, which can be achieved by examining the number of significant singular values of a good estimate of $[Y_t^\star, Y_s^\star]$. 
The details are provided in Algorithm \ref{algor:estk}, where $\mathring{Y}_t$ represents an estimate of $Y_t^\star$ and we use the output of Algorithm \ref{algor:estY}. 

\begin{algorithm}[!htbp]
\caption{Estimation of the latent dimensions $k$ and $k_t$'s.}
\label{algor:estk}
\setstretch{1.5}

\KwIn{Estimates: $\hat{d}_t$  in Step 1 and $\mathring{Y}_t$ for $1\leqslant t \leqslant T$. \ Parameter: {$\tau_{3}$} (threshold).} 
\KwOut{$\hat{k}$ and $\hat{k}_t$ for $1\leqslant t \leqslant T$.}


\For{$1\leqslant t < s \leqslant T$}{

    Let $\hat{K}_{t,s}^r = \min\big\{1\leqslant m \leqslant \hat{d}_{t} + \hat{d}_{s}-1: \mathcal{R}_{m+1,1}(\mathring{Y}_{t,s})  \leqslant \tau_3\big\}$ with $\mathring Y_{t,s} = [\mathring Y_{t}, \mathring Y_{s}]$.
    
    Let $\hat{K}_{t,s} = \hat{d}_{t} + \hat{d}_{s} - \hat{K}_{t,s}^r$.
}

Let $\hat{k} = \min \big\{ \hat{K}_{t,s} : 1 \leqslant t < s \leqslant T\big\}$ and 
 $\hat{k}_t = \hat{d}_t - \hat{k}$ for $1\leqslant t \leqslant T$. 
\end{algorithm}

Steps 1--2 above depend on hyperparameters $\tau_{2,m}$ and $\tau_3$, which are required to satisfy the rates in Condition \ref{cond:choicetaupar} below to establish theoretical guarantee. 
In numerical implementation, we set $\tau_{2,m} = n^{-1/(4m+8)}$ and $\tau_3 = n^{-1/8}$, giving reasonably good finite-sample performance. 

\begin{condition}\label{cond:choicetaupar} 
Assume the hyperparameters satisfy that 
 (i) $\tau_{2,m} \ll 1$ for $1 \leqslant m \leqslant d_t-1$ and $\tau_{2,d_t} \gg n^{-1/(2 d_t + 4)} \log(nT)$, 
and (ii)  $n^{-1/2} \log(nT)  \ll \tau_3 \ll 1$ as  $nT\to \infty$. 
\end{condition}


\subsection{Theoretical Results}

We next establish theoretical consistency of the proposed estimation procedure. 

\begin{theorem} \label{thm:estk}
Assume Conditions \ref{cond:truevalue}, \ref{cond:parfunction}, and \ref{cond:choicetaupar}. Let $\hat{k}$ and $\hat{k}_t$ be the estimated latent dimensions via the procedure proposed in Section \ref{sec:estk_algor}.  
For any constant $\varepsilon > 0$, there exists a constant $c_\varepsilon > 0$ such that when $\log^{\dmax+2}(nT)/n \leqslant c_\varepsilon$, 
$$
    \Pr\big(\hat{k} = k, \hat{k}_t = k_t \ \operatorname{for}\  1\leqslant t \leqslant T\big) = 1 - O((nT)^{-\varepsilon}) .
$$ 

\end{theorem}

The above procedure is developed utilizing  singular value gaps of matrices. The spectral-based estimators are considered due to its computational advantage and theoretical guarantee. More generally, we can replace $\breve \Theta_t$ and $\mathring{Y}_{t,s}$ by any other estimators $\hat{\Theta}_t$ and $\hat{Y}_{t,s}$ satisfying $\|\hat{\Theta}_t - \Theta_t^\star\|_{\operatorname{op}} = o_p(n)$ and $\|\hat{Y}_{t,s} - {Y}^\star_{t,s}\|_{\operatorname{op}} = o_p(n^{1/2})$. The true latent dimensions can be consistently estimated with appropriate thresholds. 

\subsection{Simulation Results}\label{sec:simudim}

We next investigate the numerical performance of  the  procedure for estimating latent dimensions. 
Similarly to Section \ref{sec:simulation}, we set  $n = 400$ and   $T\in \{5, 10, 20, 40,80\}$. We let $k = k_t \in \{2,3,4\}$ for $t=1,\ldots,T$ to vary the estimation targets.  
Table \ref{tab:estk} presents empirical accuracy, evaluated by the proportions of the event  $\{\hat k = k$ and $\hat k_t = k_t$ for $1 \leqslant t \leqslant T\}$ over 100 Monte Carlo repetitions. 

The proposed procedure accurately recovers $k$ and $k_t$'s in almost all scenarios except Case (B) of Bernoulli distribution. 
In that scenario, we observe Step 1 mistakenly estimates $d_t$.  This may be due to the fact that the non-orthogonality between columns of $Z^\star$ and $W_t^\star$ creates a large eigengap between the $(d_t-1)$-th and the $d_t$-th eigenvalues of $\Theta_t^{\star}$. Consequently, the empirical choice of hyperparameters mentioned in Section \ref{sec:estk_algor} no longer works well. 
The performance may be improved by developing tuning strategies that are adaptive to the relationship between latent spaces. 

\begin{table}[!htbp]
\centering
\begin{tabular}{ccccccccccccc}
\hline
 &  & \multicolumn{3}{c}{Bernoulli distribution} &  & \multicolumn{3}{c}{Gaussian distribution} &  & \multicolumn{3}{c}{Poisson distribution} \\ \cline{3-5} \cline{7-9} \cline{11-13} 
$T$ & Case & $k=2$ & $k=3$ & $k=4$ &  & $k=2$ & $k=3$ & $k=4$ &  & $k=2$ & $k=3$ & $k=4$ \\ \hline
5 & (A) & 1.00 & 1.00 & 1.00 &  & 1.00 & 1.00 & 1.00 &  & 1.00 & 1.00 & 1.00 \\
 & (B) & 1.00 & 1.00 & 0.99 &  & 1.00 & 1.00 & 1.00 &  & 1.00 & 1.00 & 1.00 \\
 & (C) & 1.00 & 1.00 & 1.00 &  & 1.00 & 1.00 & 1.00 &  & 1.00 & 1.00 & 1.00 \\
10 & (A) & 1.00 & 1.00 & 1.00 &  & 1.00 & 1.00 & 1.00 &  & 1.00 & 1.00 & 1.00 \\
 & (B) & 1.00 & 1.00 & 1.00 &  & 1.00 & 1.00 & 1.00 &  & 1.00 & 1.00 & 1.00 \\
 & (C) & 1.00 & 1.00 & 1.00 &  & 1.00 & 1.00 & 1.00 &  & 1.00 & 1.00 & 1.00 \\
20 & (A) & 1.00 & 1.00 & 1.00 &  & 1.00 & 1.00 & 1.00 &  & 1.00 & 1.00 & 1.00 \\
 & (B) & 1.00 & 1.00 & 0.99 &  & 1.00 & 1.00 & 1.00 &  & 1.00 & 1.00 & 1.00 \\
 & (C) & 1.00 & 1.00 & 1.00 &  & 1.00 & 1.00 & 1.00 &  & 1.00 & 1.00 & 1.00 \\
40 & (A) & 1.00 & 1.00 & 1.00 &  & 1.00 & 1.00 & 1.00 &  & 1.00 & 1.00 & 1.00 \\
\multicolumn{1}{l}{} & (B) & 1.00 & 1.00 & 0.96 &  & 1.00 & 1.00 & 1.00 &  & 1.00 & 1.00 & 1.00 \\
\multicolumn{1}{l}{} & (C) & 1.00 & 1.00 & 1.00 &  & 1.00 & 1.00 & 1.00 &  & 1.00 & 1.00 & 1.00 \\ 
80 & (A) & 1.00 & 1.00 & 1.00 &  & 1.00 & 1.00 & 1.00 &  & 1.00 & 1.00 & 1.00 \\
\multicolumn{1}{l}{} & (B) & 1.00 & 1.00 & 0.88 &  & 1.00 & 1.00 & 1.00 &  & 1.00 & 1.00 & 1.00 \\
\multicolumn{1}{l}{} & (C) & 1.00 & 1.00 & 1.00 &  & 1.00 & 1.00 & 1.00 &  & 1.00 & 1.00 & 1.00 \\\hline
\end{tabular}
\caption{Empirical proportions of simulations satisfying $\{\hat k = k$ and $\hat k_t = k_t$ for $1 \leqslant t \leqslant T\}$ over $100$ repetitions when $n=400$ and $T \in \{5, 10, 20, 40,80\}$.}
\label{tab:estk}
\end{table}

\newpage
\section{Proofs of Propositions}

\subsection{Proof of Proposition \ref{prop:identify}}
Suppose two sets of parameters $\{Z,W_1,\ldots, W_T\}$ and $\{Z^{\prime},W_1^{\prime},\ldots, W_T^{\prime}\}$ yield the same model \eqref{eq:model2}.
We next show that, if they satisfy the conditions of Proposition \ref{prop:identify}, there exist $Q_z \in \mathcal{O}(k)$ and $Q_{wt} \in \mathcal{O}(k_t)$ such that $Z = Z^\prime Q_z$ and $W_t = W_t^\prime Q_{wt}$ for $1 \leqslant t \leqslant T$.

First note that the two sets of parameters produce the same $\EXPT (\mathbf A_t)$ for $1 \leqslant t \leqslant T$. Then we have 
\begin{align} \label{eq:prop1-1}
    Y_t Y_t^\mytrans = Y_t^\prime Y_t^{\prime \mytrans} \quad\text{ for }\quad 1\leqslant t \leqslant T.
\end{align}
Combining \eqref{eq:prop1-1} with condition (i) and Lemma \ref{adlem:tu2016}, there exists $Q_t \in \mathcal{O}(d_t)$ such that 
\begin{align} \label{eq:prop1-2}
    Y_t = Y_t^\prime  Q_t \quad\text{ for }\quad 1 \leqslant t \leqslant T.  
\end{align}
Assume $Q_t$ has the block structure
\begin{align*}
     Q_t = \left[\begin{array}{cc}
        Q_{t,11} &  Q_{t,12} \\
        Q_{t,21} &  Q_{t,22}
    \end{array}\right] \in \mathbb R^{(k+k_t) \times (k + k_t)}.
\end{align*}
Then for any $(t,s)$ pair satisfying condition (ii) of Proposition \ref{prop:identify}, we have
\begin{align} \label{eq:prop1-3}
    Z^\prime \big(Q_{t,11} - Q_{s,11}\big) + W_t^\prime  Q_{t,21} - W_s^\prime Q_{s,21} =  \mathbf 0,
\end{align}
by \eqref{eq:prop1-2} and the left half of $Y_t$ and $Y_s$ are equal. Since the columns of $[Z^\prime, W_t^\prime, W_s^\prime]$ are linearly independent, \eqref{eq:prop1-3} indicates $ Q_{t,11} =  Q_{s,11}$ and $ Q_{t,21} =  Q_{s,21} =  \mathbf 0$. Since $ Q_t \in \mathcal{O}(d_t)$, $ Q_{t,21} = \mathbf 0$ implies $Q_{t,12} = \mathbf 0$, $ Q_{t,11} \in \mathcal{O}(k)$, and $ Q_{t,22} \in \mathcal{O}(k_t)$. Combining these facts with  \eqref{eq:prop1-2} we obtain $Z = Z^\prime Q_{t,11}$, and consequently $ZZ^\mytrans = Z^\prime Z^{\prime \mytrans}$. Finally, combining with \eqref{eq:prop1-1} we obtain
\begin{align*}
    W_tW_t^\mytrans = W_t^\prime W_t^{\prime \mytrans} \quad\text{ for }\quad 1\leqslant t \leqslant T,
\end{align*}
which indicates that there exists $Q_{wt} \in \mathcal{O}(k_t)$ such that $W_t = W_t^\prime Q_{wt}$.

\subsection{Proof of Proposition \ref{prop:yqtoz}}
By Lemma \ref{adlem:tu2016}, assumption (i) suggests that 
there exists an unknown orthogonal matrix $Q_t \in \mathcal{O}(d_t)$  such that 
$ Y_t Q_t = Y_t^\star = [Z^\star, W_t^\star]$.
It follows that 
\begin{align} \label{eq:prop2-1}
    Z^\star = &~ Y_t Q_{t} L_{t,z}  \hspace{5.7em}\text{and}\hspace{1.7em} W_{t}^\star = Y_t Q_{t} L_{t,w},  \\[3pt] \text{ with }\quad  \ L_{t,z}=&~\begin{bmatrix}
        \mathrm{I}_k\\
        \mathbf{0}
    \end{bmatrix}\in \mathbb{R}^{d_t\times k}
    \hspace{3.4em} \text{and} \hspace{1.5em} L_{t,w}=\begin{bmatrix}
        \mathbf{0}\\
        \mathrm{I}_{k_t}
    \end{bmatrix}\in \mathbb{R}^{d_t \times k_t}.\notag
\end{align}
Then for any $(t, s) \in \mathcal{T}$, define 
\begin{align} \label{eq:prop2-2}
    V_{t,s}= \frac{1}{\sqrt{2}}\begin{bmatrix}
    Q_t & \mathbf{0}\\
    \mathbf{0} & Q_s
\end{bmatrix} \begin{bmatrix}
     L_{t,z}\\
     -L_{s,z}
 \end{bmatrix}  \quad\text{ and }\quad V_{t,s}^{\perp} =  \begin{bmatrix}
    Q_t & \mathbf{0}\\
    \mathbf{0} & Q_s
\end{bmatrix} \begin{bmatrix}
      L_{t,z}/\sqrt{2} &  L_{t,w} &  \mathbf{0}\\
      L_{s,z}/\sqrt{2}  &  \mathbf{0} &  L_{s,w}
    \end{bmatrix}.
\end{align}
It is straightforward to verify that the columns of $[V_{t,s}, V_{t,s}^{\perp}]$ form an orthonormal basis of $\mathbb R^{d_t + d_s}$.
Combining \eqref{eq:prop2-1} and \eqref{eq:prop2-2} shows 
\begin{align} \label{eq:prop2-3}
    {Y}_{t,s}\, V_{t,s}  = \frac{1}{\sqrt{2}}\, Y_t Q_t L_{t,z} - \frac{1}{\sqrt{2}}\, Y_s Q_s L_{s,z} = \mathbf{0} \quad\text{ and }\quad 
{Y}_{t,s}\, V_{t,s}^{\perp} = \big[\sqrt{2}Z^\star, W_t^\star, W_s^\star\big].
\end{align}
Since the columns of $[Z^\star, W_t^\star, W_s^\star]$ are linearly independent, \eqref{eq:prop2-3} indicates that the column space of $V_{t,s}$ equals the right null space of ${Y}_{t,s}$. For any other matrix $\tilde{V}_{t,s} \in \mathbb R^{(d_t + d_s) \times k}$ that has the same column space as $V_{t,s}$, we can express it as 
\begin{align*}
    \tilde{V}_{t,s} = V_{t,s} Q_v  \quad\text{ with }\quad Q_v \in \mathcal{O}(k).
\end{align*}
Similarly by \eqref{eq:prop2-1} and \eqref{eq:prop2-2}, we have 
\begin{align*}
Z^\star= \frac{1}{\sqrt{2}}\, {Y}_{t,-s} \, V_{t,s} = \frac{1}{\sqrt{2}}\, {Y}_{t,-s} \, \tilde{V}_{t,s} Q_v^{-1}, 
\end{align*}
which implies
\begin{align*}
    Z^\star Z^{\star \mytrans} = Y_{t,-s} \tilde{V}_{t,s} \tilde{V}_{t,s}^\mytrans Y_{t,-s}^\mytrans /2.
\end{align*}

\subsection{Proof of Proposition \ref{prop:condtion1}}
We first show Condition \ref{cond:truevalue} (ii) is satisfied under the conditions of Proposition \ref{prop:condtion1}. For any $v = (v_z, v_w) \in \mathbb R^{k + k_t}$, 
\begin{align*}
    v^\mytrans G_t^\star v &= v_z^\mytrans \big(Z^{\star \mytrans} Z^\star/n\big) v_z  + v_w^\mytrans \big(W_t^{\star \mytrans} W_t^\star/n\big) v_w  + 2 v_z^\mytrans \big(Z^{\star \mytrans} W_t^\star/n\big) v_w \\ 
    &\geqslant M_4 \|v_z\|_2^2 + M_4 \|v_w\|_2^2 - (M_4/2) \|v_z\|_2 \|v_w\|_2 \\
    &\geqslant M_4 \|v_z\|_2^2 + M_4 \|v_w\|_2^2 - (M_4/4) \|v_z\|_2^2 - (M_4/4) \|v_w\|_2^2 \\
    &= (3M_4/4) \|v\|_2^2,
\end{align*}
which indicates that $\sigma_{\operatorname{min}}(G_t^\star) \geqslant 3M_4/4$ for any $1 \leqslant t \leqslant T$.

\bigskip

Similarly, for any $(t,s)$ pair satisfying  condition (c) of Proposition \ref{prop:condtion1}, we have
\begin{align*}
    v^\mytrans G_{t,s}^\star v &= v_z^\mytrans \big(Z^{\star \mytrans} Z^\star/n\big) v_z  + v_t^\mytrans \big(W_t^{\star \mytrans} W_t^\star/n\big) v_t  + v_s^\mytrans \big(W_s^{\star \mytrans} W_s^\star/n\big) v_s \\
    &\quad +2 v_z^\mytrans \big(Z^{\star \mytrans} W_t^\star/n\big) v_t + 2 v_z^\mytrans \big(Z^{\star \mytrans} W_s^\star/n\big) v_s + 2 v_t^\mytrans \big(W_t^{\star \mytrans} W_s^\star/n\big) v_s\\
    &\geqslant M_4 \|v_z\|_2^2 + M_4 \|v_t\|_2^2 + M_4 \|v_s\|_2^2\\
    &\quad- (M_4/2) \|v_z\|_2 \|v_t\|_2 - (M_4/2) \|v_z\|_2 \|v_s\|_2 - (M_4/2) \|v_t\|_2 \|v_s\|_2\\
    &\geqslant M_4 \|v_z\|_2^2 + M_4 \|v_t\|_2^2 + M_4 \|v_s\|_2^2 - (M_4/2) \|v_z\|_2^2 - (M_4/2) \|v_t\|_2^2 - (M_4/2) \|v_s\|_2^2 \\
    &= (M_4/2) \|v\|_2^2
\end{align*}
for any $v = (v_z, v_t, v_s) \in \mathbb R^{k+k_t+k_s}$. Then we obtain $\sigma_{\operatorname{min}}(G_{t,s}^\star) \geqslant M_4/2$.

\subsection{Proof of Proposition \ref{prop:condtion3}}
For any $v = (v_1, \ldots, v_n) \in \mathbb R^{\ksum}$,
\begin{align*}
    v^\mytrans \mathcal{G}(W^\star) v &= \sum_{1 \leqslant t \leqslant T} v_t^\mytrans \big(W_t^{\star \mytrans} W_t^\star/n\big) v_t + 2\sum_{1 \leqslant t < s \leqslant T} v_t^\mytrans \big(W_t^{\star \mytrans} W_s^\star/n\big) v_s \\
    & \leqslant (M_5T/2) \sum_{1 \leqslant t \leqslant T} \|v_t\|_2^2 + M_5 \sum_{1 \leqslant t < s \leqslant T} \|v_t\|_2 \|v_s\|_2 \\
    &\leqslant  (M_5T/2) \sum_{1 \leqslant t \leqslant T} \|v_t\|_2^2 + (M_5/2) \sum_{1 \leqslant t  \leqslant T} T\|v_t\|_2^2 \\
    &\leqslant M_5T \|v\|_2^2,
\end{align*}
which implies $\|\mathcal{G}(W^\star)\|_{\operatorname{op}} \leqslant M_5T$.

\subsection{Proof of Proposition \ref{prop:dimensionk} }
Similar to \eqref{eq:prop2-2} and \eqref{eq:prop2-3}, we have
\begin{align*} 
&Y_{t,s}^\star V_{t,s} = \mathbf 0 \hspace{4.6em} \text{ and }\quad Y_{t,s}^\star V_{t,s}^\perp = \big[\sqrt{2}Z^\star, W_t^\star, W_s^\star\big] \\
  \text{ with }\quad   &V_{t,s}= \frac{1}{\sqrt{2}}\begin{bmatrix}
     L_{t,z}\\
     -L_{s,z}
 \end{bmatrix}  \quad\text{ and }\quad V_{t,s}^{\perp} = \begin{bmatrix}
      L_{t,z}/\sqrt{2} &  L_{t,w} &  \mathbf{0}\\
      L_{s,z}/\sqrt{2}  &  \mathbf{0} &  L_{s,w}
    \end{bmatrix}
\end{align*}
for any $1 \leqslant t < s \leqslant T$. Then \begin{enumerate}
    \item[(i)] $\operatorname{col}(V_{t,s}) \subset \operatorname{null}(Y_{t,s}^\star)$ for all $1\leqslant t < s\leqslant T$,    
    \item[(ii)]  $\operatorname{col}(V_{t,s}) = \operatorname{null}(Y_{t,s}^\star)$ holds for $(t,s)\in \mathcal{T}$, where $\mathcal{T}$ is defined in Proposition \ref{prop:yqtoz}, 
\end{enumerate}
which imply (i) and (ii) in Proposition \ref{prop:dimensionk} since $\operatorname{dim}(\operatorname{col}(V_{t,s})) = k$. 

\newpage

\section{Proofs of Theorems} \label{sec:proof}


\begin{remark}\label{rm:lprimenotation}
For the simplicity of notation, we  let
$l^{\prime}(\Theta_{t,ij})$  and $l^{\prime \prime}(\Theta_{t,ij})$ 
be shorthands for  $l^{\prime}(\Theta_{t,ij};A_{t,ij})$ and $l^{\prime \prime}(\Theta_{t,ij};A_{t,ij})$, respectively, throughout this section. Moreover, let $l^{\prime}(\Theta_{t}) = (l^{\prime}(\Theta_{t,ij}))_{1\leqslant i, j\leqslant n} \in \mathbb{R}^{n\times n}$. Notably, under Condition \ref{cond:parfunction} (i.e.,  $l(\theta;x)$ belongs to the natural exponential families),  $l''(\Theta_{t,ij})$ does not depend on $A_{t,ij}$ given fixed $\Theta_{t,ij}$ and $  \mathbb{E}\{l'(\Theta_{t,ij}) l'(\Theta_{t,ij})\} = -l''(\Theta_{t,ij})$.  
\end{remark}

\subsection{Preliminary Lemmas} 
In this section, we present two lemmas for matrix concentration inequalities that will be used in our proofs. 
\begin{lemma} \label{lem:concentration}
    Assume Conditions \ref{cond:truevalue}--\ref{cond:parfunction}. For any constant $\varepsilon >0$, there exist positive constants $c_\varepsilon$ and $C_\varepsilon$ such that when $\log(nT)/n \leqslant c_\varepsilon $, 
    \begin{align*}
        \Pr\left[ \max_{1 \leqslant t \leqslant T} \left\{\big\|l^{\prime}(\Theta_t^\star)\big\|_{\operatorname{op}}^2\right\} \geqslant C_\varepsilon n \log(nT)\right] \leqslant (nT)^{-\varepsilon}.
    \end{align*}
\end{lemma}

\begin{proof}
    For any $1 \leqslant t \leqslant T$, we can decompose $l^{\prime}(\Theta_t^\star)$ as 
    \begin{align*}
        l^{\prime}(\Theta_t^\star) = \sum_{1 \leqslant i < j \leqslant n} l^{\prime}(\Theta_{t,ij}^\star) (\boldsymbol\delta_{ij} + \boldsymbol\delta_{ji}) + \sum_{1 \leqslant i \leqslant n } l^{\prime}(\Theta_{t,ii}^\star) \,\boldsymbol\delta_{ii},
    \end{align*}
    where $\boldsymbol\delta_{ij} \in \mathbb R^{n \times n}$ denotes the indicator matrix with a $1$ at position $(i,j)$ and $0$ elsewhere. 
     By Condition \ref{cond:parfunction} (ii), $l^{\prime}(\Theta_{t,ij}^\star) (\boldsymbol\delta_{ij} + \boldsymbol\delta_{ji}), 1 \leqslant i < j \leqslant n$ and $l^{\prime}(\Theta_{t,ii}^\star)\,\boldsymbol{\delta}_{ii}, 1 \leqslant i \leqslant n$ are
 independent, zero mean, and symmetric random matrices  satisfying 
 \begin{align*}
     \mathbb{E}\left[l^{\prime}(\Theta_{t,ij}^\star)\left(\boldsymbol\delta_{i j}+\boldsymbol\delta_{j i}\right)\right]^{m}  &\preceq \frac{m!}{2} L^{m-2} \kappa_2\left(\boldsymbol\delta_{i j}+\boldsymbol\delta_{j i}\right)  \\[3pt]
     \text{ and }\quad\quad\quad\quad\quad \mathbb{E}\left[l^{\prime}(\Theta_{t,ii}^\star)\,\boldsymbol\delta_{i i}\right]^{m}  &\preceq \frac{m!}{2} L^{m-2} \kappa_2\,\boldsymbol\delta_{ii}
 \end{align*}
 for any integer $m \geqslant 2$. Applying Lemma \ref{adlem:tropp2012} by specifying $X_{ij}$ as the aforementioned random matrices indicates that for any $x \geqslant 0$,
 \begin{align*}
     \Pr\Big(\big\|l^{\prime}(\Theta_t^\star)\big\|_{\operatorname{op}} \geqslant x\Big) \leqslant 2n \exp\Big(\frac{-x^2/2}{\kappa_2 n  + Lx}\Big) \quad\text{ for }\quad 1 \leqslant t \leqslant T.
 \end{align*}
 As a result, 
 \begin{align*}
     \Pr\Big( \max_{1 \leqslant t \leqslant T} \big\|l^{\prime}(\Theta_t^\star)\big\|_{\operatorname{op}}\geqslant x \Big) \leqslant \sum_{t = 1 }^T \Pr\Big(\big\|l^{\prime}(\Theta_t^\star)\big\|_{\operatorname{op}} \geqslant x\Big) \leqslant 2nT\exp\Big(\frac{-x^2/2}{\kappa_2 n  + Lx}\Big).
 \end{align*}
 Set $x = 8(\varepsilon + 1) \max\{\sqrt{\kappa_2 n \log(nT)},L\log(nT)\}$ in the above inequality we obtain
\begin{align*}
        \Pr\left[ \max_{1 \leqslant t \leqslant T} \left\{\big\|l^{\prime}(\Theta_t^\star)\big\|_{\operatorname{op}}^2\right\} \geqslant C_\varepsilon n \log(nT)\right] \leqslant (nT)^{-\varepsilon}
    \end{align*}
as $\log(nT)/n \leqslant c_\varepsilon$ 
for a constant $c_\varepsilon $. 
\end{proof}

\begin{lemma} \label{lem:concentration2}
    Assume Conditions \ref{cond:truevalue}--\ref{cond:parfunction}. For any constant $\varepsilon >0$, there exists a constant $C_\varepsilon > 0$ such that 
    \begin{align*}
        \Pr\left[ \max_{1 \leqslant t \leqslant T} \left\{\big\|l^{\prime}(\Theta_t^\star)\big\|_{2 \to \infty}^2\right\} \geqslant C_\varepsilon n \log^2(nT)\right] &\leqslant (nT)^{-\varepsilon}, \\
        \Pr\left[ \max_{1 \leqslant t \leqslant T} \left\{\big\|l^{\prime}(\Theta_t^\star) Y_t^\star\big\|_{2 \to \infty}^2\right\} \geqslant C_\varepsilon n \log^2(nT)\right] &\leqslant  (nT)^{-\varepsilon}.
    \end{align*}
\end{lemma}
\begin{proof}
Lemma \ref{adlem:wellner2005} indicates that for any $1 \leqslant t \leqslant T$, $1 \leqslant i,j \leqslant n$, $x \geqslant 0$,
\begin{align*}
    \Pr\Big(\big| l^{\prime}(\Theta_{t,ij}^\star) \big| \geqslant x\Big) \leqslant 2\exp\Big(\frac{-x^2/2}{\kappa_2 + Lx}\Big).
\end{align*}
Consequently, 
\begin{align*}
    \Pr\Big(\max_{1\leqslant t \leqslant T, 1\leqslant i,j \leqslant n} \big| l^{\prime}(\Theta_{t,ij}^\star) \big| \geqslant x \Big) \leqslant \sum_{t =1 }^T \sum_{i,j =1}^n \Pr\Big(\big| l^{\prime}(\Theta_{t,ij}^\star) \big| \geqslant x\Big) 
    &\leqslant 2n^2T  \exp\Big(\frac{-x^2/2}{\kappa_2 + Lx}\Big).
\end{align*}
Set $x = 8(\varepsilon + 2) L\log(nT)$ in the above inequality gives 
\begin{align*}
    \Pr\Big(\max_{1\leqslant t \leqslant T, 1\leqslant i,j \leqslant n} \big| l^{\prime}(\Theta_{t,ij}^\star) \big|^2 \geqslant C_\varepsilon \log^2(nT) \Big) \leqslant (nT)^{-\varepsilon},
\end{align*}
and the first argument of Lemma \ref{lem:concentration2} follows from $\|l^{\prime}(\Theta_t)\|_{2 \to \infty}^2 \leqslant n \max_{1 \leqslant i,j \leqslant n}|l^{\prime}(\Theta_{t,ij})|^2$.

Similarly by Lemma \ref{adlem:wellner2005}, for any $1 \leqslant t \leqslant T$, $1 \leqslant i \leqslant n$, $1 \leqslant m \leqslant d_t$, $x \geqslant 0$,
\begin{align*}
    \Pr\Big(\Big|\sum_{j = 1}^n l^{\prime}(\Theta_{t,ij}^\star) Y_{t,jm}^\star\Big| \geqslant x \Big)  \leqslant 2 \exp\Big(\frac{-x^2/2}{M_1^2\kappa_2 n + M_1Lx}\Big), 
\end{align*}
then 
\begin{align*}
    \Pr\Big(\max_{1 \leqslant t \leqslant T, 1 \leqslant i \leqslant n, 1 \leqslant m \leqslant d_t}\Big|\sum_{j = 1}^n l^{\prime}(\Theta_{t,ij}^\star) Y_{t,jm}^\star\Big| \geqslant x \Big)  \leqslant 2 nT \dmax \exp\Big(\frac{-x^2/2}{M_1^2\kappa_2 n + M_1Lx}\Big).
\end{align*}
Set $x = 8(\varepsilon + 2) n^{1/2} \log(nT)$ we obtain
\begin{align*}
    \Pr\Big(\max_{1 \leqslant t \leqslant T, 1 \leqslant i \leqslant n, 1 \leqslant m \leqslant d_t}\Big|\sum_{j = 1}^n l^{\prime}(\Theta_{t,ij}^\star) Y_{t,jm}^\star\Big|^2 \geqslant C_\varepsilon n \log^2(nT) \Big) \leqslant  (nT)^{-\varepsilon}, 
\end{align*}
which guarantees a similar upper bound for $\max_{1 \leqslant t \leqslant T}\|l^{\prime}(\Theta_t^\star)Y_t^\star\|_{2 \to \infty}^2$. 
\end{proof}

\subsection{Proof of Theorem \ref{thm:estTheta}} \label{sec:pf_thmA1}

{To prove Theorem \ref{thm:estTheta}, by applying the union bound, 
it suffices to show 
\begin{align*}
    \max_{1\leqslant t\leqslant T}\Pr \left\{\big\|\breve \Theta_t - \Theta_t^\star\big\|_{\mathrm{F}}/n > C_\varepsilon n^{-\frac{1}{2 d_t + 4}}\log^{\frac{1}{2}}(nT)\right\}  = O\big((nT)^{-\varepsilon-1}\big).
\end{align*}
By the construction of $\breve \Theta_t$ in Algorithm \ref{algor:estTheta1}, we have
\begin{align*}
    \big\|\breve \Theta_t - \Theta_t^\star\big\|_{\mathrm{F}} \leqslant \big\|\tilde \Theta_t - \Theta_t^\star\big\|_{\mathrm{F}} \leqslant \kappa_1^{-1} \big\|\breve E_t - \mu(\Theta_t^\star)\big\|_{\mathrm{F}} \leqslant \kappa_1^{-1} \big\|\tilde E_t - \mu(\Theta_t^\star)\big\|_{\mathrm{F}}.
\end{align*} 
where the first and the last inequalities follow from the projection property, and the second inequality follows from the $\kappa_1^{-1}$-Lipschitz property of $\mu^{-1}(\cdot)$ by Condition \ref{cond:parfunction}. 
Therefore, it suffices to prove  
\begin{align} \label{eq:tildeet_bound}
    \max_{1\leqslant t\leqslant T}\Pr   \left\{\big\|\tilde E_t - \mu(\Theta_t^\star)\big\|_{\mathrm{F}}/n > C_\varepsilon n^{-\frac{1}{2 d_t + 4}}\log^{\frac{1}{2}}(nT)\right\} = O((nT)^{-\varepsilon-1}).
\end{align} 
}

{As $\tilde E_t$ is constructed from thresholding singular values, we bound $\|\tilde E_t - \mu(\Theta_t^\star)\|_{\mathrm{F}}$ by Lemma \ref{adlem:chatterjee2015}  with $A = \mathbf{A}_t$, $B = \mu(\Theta_t^\star)$, and $\delta = \tau_4/\|A - B\|_{\operatorname{op}}$.  Notably, when $n$ is sufficiently large,  the condition $\delta>2$ in Lemma  \ref{adlem:chatterjee2015}  is satisfied with probability $1 - (nT)^{-\varepsilon-1}$. This is  because we choose  $\tau_4 \gg n^{1/2} \log^{1/2}(nT)$ by Condition \ref{cond:parestTheta1},  $\mathbf A_t - \mu(\Theta_t^\star) = l^{\prime}(\Theta_t^\star)$ by Condition \ref{cond:parfunction}, and with  $\log(nT)/n \leqslant c_\varepsilon$ for a  constant $c_{\varepsilon}>0$,
Lemma \ref{lem:concentration} gives 
\begin{align} \label{eq:at_bernstein_a1}
        \Pr\left[ \max_{1 \leqslant t \leqslant T} \left\{\big\|\mathbf A_t - \mu(\Theta_t^\star)\big\|_{\operatorname{op}}\right\} \geqslant C_\varepsilon n^{\frac{1}{2}} \log^{\frac{1}{2}}(nT)\right] \leqslant (nT)^{-\varepsilon-1}.
\end{align}
Consequently,  with probability  $1 - (nT)^{-\varepsilon-1}$, 
 Lemma \ref{adlem:chatterjee2015} can be applied to give
\begin{align} \label{eq:thm4-1}
    \big\|\tilde E_t - \mu(\Theta_t^\star)\big\|_{\mathrm{F}}^2  &\lesssim \delta \,\big\|\mathbf A_t - \mu(\Theta_t^\star)\big\|_{\operatorname{op}} \big\|\mu(\Theta_t^\star)\big\|_* = \tau_4 \, \big\|\mu(\Theta_t^\star)\big\|_* \notag \\
    &\lesssim n^{\frac{1}{2}} \log(nT) \big\|\mu(\Theta_t^\star)\big\|_*.
\end{align}
where the last inequality follows as  $\tau_4\lesssim n^{1/2} \log(nT)$ by Condition \ref{cond:parestTheta1}.}

To prove \eqref{eq:tildeet_bound}, we next {show 
\begin{align} \label{eq:thm4-5}
    \big\|\mu(\Theta_t^\star)\big\|_* \leqslant C\, n^{\frac{3}{2} - \frac{1}{d_t + 2}}
\end{align}
for a universal constant $C > 0$ through a one-step discretization.}  To do this, we define $\mathcal{N}_\delta$ as the smallest $\delta$-net of the set $\{y \in \mathbb R^{d_t}: \|y\|_2^2 \leqslant 2M_1^2\}$ and select $y_{t,i}^\dagger$ as the closest element to $y_{t,i}^\star$ in $\mathcal{N}_\delta$. Let $Y_t^\dagger = (y_{t,1}^\dagger, \ldots, y_{t,n}^\dagger)^\mytrans$ and $\Theta_t^\dagger = Y_t^\dagger Y_t^{\dagger \mytrans}$. Note that  when $y_{t,i}^\dagger$ and $y_{t,j}^\dagger$ belong to the same element of $\mathcal{N}_\delta$, the $i$-th row and $j$-th row of $\mu(\Theta_t^\dagger)$  are identical. Then by applying the standard covering number upper bound for a $2$-norm ball, we obtain
\begin{align} \label{eq:thm4-2}
\operatorname{rank}\big(\mu(\Theta_t^\dagger)\big) \leqslant \big|\mathcal{N}_\delta\big| \leqslant \Big(\frac{3M_1}{\delta}\Big)^{d_t}
\end{align}
when $\delta$ is sufficiently small. Also, $\mu(\Theta_t^\dagger)$ is closed to $\mu(\Theta_t^\star)$ since
\begin{align} \label{eq:thm4-3}
    \big|\mu(\Theta_{t,ij}^\dagger) - \mu(\Theta_{t,ij}^\star)\big| \leqslant \kappa_2 \big|\Theta_{t,ij}^\dagger - \Theta_{t,ij}^\star\big| \leqslant 2\sqrt{2} M_1 \kappa_2 \delta \quad\text{ for }\quad 1\leqslant i,j\leqslant n,
\end{align}
where the first inequality follows from the Lipschitz property of $\mu$, as $|\mu^\prime(\theta)| = |l^{\prime \prime}(\theta)| \leqslant \kappa_2$ by Condition \ref{cond:parfunction}, and the second inequality follows from  $\|y_{t,i}^\dagger - y_{t,i}^\star\|_2 \leqslant \delta$. 
Therefore,
\begin{align} \label{eq:thm4-4}
    \big\|\mu(\Theta_t^\star)\big\|_* \leqslant  \big\|\mu(\Theta_t^\star) - \mu(\Theta_t^\dagger)\big\|_* + \big\|\mu(\Theta_t^\dagger)\big\|_* 
    \leqslant 2\sqrt{2} M_1 \kappa_2 \delta n^{\frac{3}{2}} + \mu(2M_1^2)\, n \Big(\frac{3M_1}{\delta}\Big)^{\frac{d_t}{2}},
\end{align}
where the last inequality follows from \eqref{eq:thm4-2}, \eqref{eq:thm4-3}, and $\|X\|_* \leqslant \|X\|_{\mathrm{F}} \sqrt{\operatorname{rank}(X)}$. Specifying $\delta \asymp n^{-1/(d_t + 2)}$ in \eqref{eq:thm4-4} shows \eqref{eq:thm4-5}. 

{In summary, \eqref{eq:tildeet_bound} is obtained since  \eqref{eq:thm4-1} and \eqref{eq:thm4-5} hold   uniformly across $1 \leqslant t \leqslant T$. 
}


\subsection{Proof of Theorem \ref{thm:estY}} \label{sec:pf_thm5}

We first outline the key steps of the proof, followed by the technical details.

\paragraph{Outline.} Recall that $\mathring{Y}_t = Y_t^R$ is the output from a projected gradient descent procedure. We will analyze the estimation errors with respect to iterations in the procedure. For each iteration $r = 0, \ldots, R$, we define the error term 
\begin{align*}
    \epsilon_t^r = \big\|Y_t^r - Y_t^\star Q_t^r\big\|_{\mathrm{F}}^2 \quad\text{ with }\quad Q_t^r = \arg \min_{Q \in \mathcal{O}(d_t)} \big\|Y_t^r - Y_t^\star Q\big\|_{\mathrm{F}}^2.
\end{align*}
By Lemma \ref{adlem:tu2016}, across all $1 \leqslant t \leqslant T$,
\begin{align} \label{eq:pfb2-1}
    \epsilon_t^0  &\leqslant \frac{2}{M_2 n} \big\|Y_t^0 Y_t^{0 \mytrans} -  Y_t^\star Y_t^{\star \mytrans}\big\|_{\mathrm{F}}^2  
     \leqslant \frac{{8}}{M_2 n} \big\|\breve \Theta_t  -  \Theta_t^\star\big\|_{\mathrm{F}}^2 \leqslant C_\varepsilon n^{1-\frac{1}{d_t + 2}} \log(nT)
\end{align}
{with probability $1- O((nT)^{-\varepsilon})$,  where the second inequality is obtained through 
\begin{align}\label{eq:bound_EYMtheorem}
    \big\|Y_t^0 Y_t^{0 \mytrans} -  Y_t^\star Y_t^{\star \mytrans}\big\|_{\mathrm{F}} \leqslant \big\|Y_t^0 Y_t^{0 \mytrans} -  \breve \Theta_t \big\|_{\mathrm{F}} + \big\|\breve \Theta_t -  \Theta_t^\star \big\|_{\mathrm{F}} \leqslant 2 \big\|\breve \Theta_t -  \Theta_t^\star \big\|_{\mathrm{F}},
\end{align}
obtained by  the triangle inequality, 
Eckart–Young–Mirsky theorem, and $Y_t^0 = \mathcal{S}_{d_t}(\breve \Theta_t)$, and the third inequality follows by  Theorem \ref{thm:estTheta}.} 

{Consider a fixed constant $c_0$; for clarity, we set $c_0 = \kappa_1M_2/(32\kappa_2)$ as an example below, though the proof holds for any sufficiently small $c_0$.} 
To prove Theorem \ref{thm:estY}, given any constant $\varepsilon > 0$, it suffices to consider  $(n,T)$  such that  $\log^{\dmax + 2}(nT)/n \leqslant c_\varepsilon$ for a constant $c_{\varepsilon}$. Specifically, we choose  $c_\varepsilon = \{c_0/(2 C_ \varepsilon)\}^{\dmax+2}$ with $C_{\varepsilon}$ representing a constant that satisfies \eqref{eq:pfb2-1}, so that \eqref{eq:pfb2-1} implies 
\begin{align} \label{eq:pfb2-1.0}
    \Pr\Big(\max_{1 \leqslant t \leqslant T} \epsilon_t^0 > c_0 n/2\Big) = O((nT)^{-\varepsilon}).
\end{align}
We next prove Theorem \ref{thm:estY} through induction argument.  
{At each step $0 \leqslant r_0\leqslant R-1$, assume
\begin{align} \label{eq:pfb2-3}
    \Pr\Big(\max_{0\leqslant r\leqslant r_0,\   1\leqslant t\leqslant T} \, \epsilon_t^{r}  > c_0 n\Big) = O((nT)^{-\varepsilon}).
\end{align}
In Section \ref{sec:pfb2-2},  we will prove 
there exist constants $c \in (0,1)$ and $C>0$ such that 
\begin{align}
       & \epsilon_t^{r+1} \leqslant (1 - c)\, \epsilon_t^r + \frac{C}{n}\, \big\|l^\prime(\Theta_t^\star)\big\|_{\operatorname{op}}^2 \hspace{2em} \text{for all } 0\leqslant r\leqslant r_0, \label{eq:pfb2-2} 
\end{align}  
with probability $1-O((nT)^{-\varepsilon})$. By iteratively using \eqref{eq:pfb2-2}, we have for all  $0 \leqslant r\leqslant r_0$, 
\begin{align*} 
    \epsilon_t^{r+1} \leqslant (1-c)^{r+1} \epsilon_t^0 + \sum_{i = 1}^{r+1}  \frac{C}{n} \big\|l^\prime(\Theta_t^\star)\big\|_{\operatorname{op}}^2 (1-c)^{i-1} \leqslant  \epsilon_t^0 + \frac{C}{cn} \big\|l^\prime(\Theta_t^\star)\big\|_{\operatorname{op}}^2
\end{align*} 
with probability $1 - O((nT)^{-\varepsilon})$. Therefore, when $n$ is sufficiently large,
\begin{align*}
    \max_{0\leqslant r\leqslant r_0+1,\   1\leqslant t\leqslant T} \, \epsilon_t^{r} \leqslant    \max_{1\leqslant t\leqslant T}\epsilon_t^0 + \frac{C}{cn} \max_{1\leqslant t\leqslant T} \big\|l^\prime(\Theta_t^\star)\big\|_{\operatorname{op}}^2 \leqslant c_0n
\end{align*}
with probability $1 - O((nT)^{-\varepsilon})$ by \eqref{eq:pfb2-1.0} and Lemma \ref{lem:concentration}. Then we can obtain \eqref{eq:pfb2-2} for $0\leqslant r \leqslant r_0+1$ following the same arguments. Keep iterating the arguments, and when $r_0=R-1$, we obtain that 
\begin{align*} 
     \epsilon_t^R \leqslant (1-c)^R \epsilon_t^0 + \sum_{i = 1}^R  \frac{C}{n} \big\|l^\prime(\Theta_t^\star)\big\|_{\operatorname{op}}^2 (1-c)^{i-1} \leqslant (1-c)^R \epsilon_t^0 + \frac{C}{cn} \big\|l^\prime(\Theta_t^\star)\big\|_{\operatorname{op}}^2
\end{align*}
with probability $1 - O((nT)^{-\varepsilon})$.
Combining the above inequality with \eqref{eq:pfb2-1}, $R \gg \log(nT)$, and Lemma \ref{lem:concentration} gives the conclusion of Theorem \ref{thm:estY}.
 }

\subsubsection{Proof of \eqref{eq:pfb2-2}.} \label{sec:pfb2-2}
Denote $\tilde Q_t^{r+1} = \arg \min_{Q \in \mathcal{O}(d_t)} \|\tilde Y_t^{r+1} - Y_t^\star Q\|_{\mathrm{F}}^2$. Then, by the definitions of $Q_t^{r+1}$ and $\tilde Q_t^{r+1}$,
\begin{align} \label{eq:thm5-1}
    \epsilon_t^{r+1} &= \big\|Y_t^{r+1} - Y_t^\star Q_t^{r+1}\big\|_{\mathrm{F}}^2 \leqslant \big\|Y_t^{r+1} - Y_t^\star \tilde Q_t^{r+1}\big\|_{\mathrm{F}}^2 \notag\\
    &\leqslant \big\|\tilde Y_t^{r+1} - Y_t^\star \tilde Q_t^{r+1}\big\|_{\mathrm{F}}^2 \leqslant \big\|\tilde Y_t^{r+1} - Y_t^\star Q_t^{r}\big\|_{\mathrm{F}}^2.
\end{align}
Plugging $\tilde Y_t^{r+1} = Y_t^r + \eta_Y\,l^\prime(\Theta_t^r)Y_t^r$ with $\Theta_t^r = Y_t^r Y_t^{r \mytrans}$ into \eqref{eq:thm5-1} shows
\begin{align} \label{eq:thm5-2}
    \epsilon_t^{r+1} \leqslant \epsilon_t^r + 2 \eta_Y  \big\langle l^\prime(\Theta_t^r)Y_t^r, Y_t^r - Y_t^\star Q_t^r\big\rangle + \eta_Y^2 \big\|l^\prime(\Theta_t^r) Y_t^r\big\|_{\mathrm{F}}^2
\end{align}
Further note that 
\begin{align}\label{eq:thm5-3}
    &~\big\langle l^\prime(\Theta_t^r)Y_t^r, Y_t^r - Y_t^\star Q_t^r\big\rangle = \big\langle l^\prime(\Theta_t^r), \big(Y_t^r - Y_t^\star Q_t^r\big) Y_t^{r \mytrans}\big\rangle \notag\\
    = &~ \big\langle l^\prime(\Theta_t^r), Y_t^r Y_t^{r \mytrans} - Y_t^\star Y_t^{\star \mytrans} \big\rangle / 2 + \big\langle l^\prime(\Theta_t^r), Y_t^r Y_t^{r \mytrans} + Y_t^\star Y_t^{\star \mytrans} - 2Y_t^\star Q_t^r Y_t^{r \mytrans}\big\rangle / 2 \notag\\
    = &~ \big\langle l^\prime(\Theta_t^r), \Theta_t^r - \Theta_t^\star \big\rangle / 2 + \big\langle l^\prime(\Theta_t^r), \big(Y_t^r - Y_t^\star Q_t^r\big) \big(Y_t^r - Y_t^\star Q_t^r\big)^\mytrans \big\rangle / 2.
\end{align}
Combining \eqref{eq:thm5-2} with \eqref{eq:thm5-3} and decomposing $$ l^\prime(\Theta_t^r) = -\big(l^\prime(\Theta_t^\star) - l^\prime(\Theta_t^r)\big) + l^\prime(\Theta_t^\star)$$
shows 
\begin{align}\label{eq:thm5-4}
    \epsilon_t^{r+1} &\leqslant \epsilon_t^r - \eta_Y( D_1 -  D_2 -  D_3 - \eta_Y D_4) \\[5pt]
    \text{with\ }\quad\quad D_1 &= \big\langle l^\prime(\Theta_t^\star) - l^\prime(\Theta_t^r), \Theta_t^r - \Theta_t^\star \big\rangle, \notag\\
    D_2 &= \big| \big\langle l^\prime(\Theta_t^\star), \Theta_t^r - \Theta_t^\star \big\rangle \big|, \notag\\
    D_3 &= \big| \big\langle l^\prime(\Theta_t^r), \big(Y_t^r - Y_t^\star Q_t^r\big) \big(Y_t^r - Y_t^\star Q_t^r\big)^\mytrans \big\rangle \big|, \notag\\
    D_4 &= \big\|l^\prime(\Theta_t^r)Y_t^r\big\|_{\mathrm{F}}^2. \notag
\end{align}

We next derive a lower bound for $D_1$ and upper bounds for $D_2$, $D_3$, and $D_4$, respectively. Since $-l(\theta)$ is $\kappa_1$-strongly convex and $\kappa_2$-smooth in $[-2M_1^2, 2M_1^2]$ by Condition \ref{cond:parfunction} (i), Lemma \ref{adlem:nesterov2003} gives that
\begin{align} \label{eq:thm5-5}
   D_1 = \big\langle l^\prime(\Theta_t^\star) - l^\prime(\Theta_t^r), \Theta_t^r - \Theta_t^\star \big\rangle \geqslant \frac{\kappa_1 }{2} \big\|\Theta_t^r - \Theta_t^\star\big\|_{\mathrm{F}}^2 + \frac{1}{2\kappa_2}\big\|l^\prime(\Theta_t^r) - l^\prime(\Theta_t^\star)\big\|_{\mathrm{F}}^2.
\end{align}
 By the Schatten norm inequality, we have
\begin{align}\label{eq:thm5-6}
    D_2 &= \big| \big\langle l^\prime(\Theta_t^\star), \Theta_t^r - \Theta_t^\star \big\rangle \big| \leqslant \big\|l^\prime(\Theta_t^\star)\big\|_{\operatorname{op}} \big\|\Theta_t^r - \Theta_t^\star\big\|_* \notag\\
    & \leqslant  \sqrt{2 d_t} \, \big\|l^\prime(\Theta_t^\star)\big\|_{\operatorname{op}} \big\|\Theta_t^r - \Theta_t^\star\big\|_{\mathrm{F}} \leqslant    \frac{d_t}{2 c_2} \big\|l^\prime(\Theta_t^\star)\big\|_{\operatorname{op}}^2+c_2  \big\|\Theta_t^r - \Theta_t^\star\big\|_{\mathrm{F}}^2
\end{align}
for any constant $c_2 > 0$. Combining \eqref{eq:thm5-5} and \eqref{eq:thm5-6} with
\begin{align}\label{eq:thm5-7.0}
     \big\|\Theta_t^r - \Theta_t^\star\big\|_{\mathrm{F}}^2 \geqslant \sigma_{\operatorname{min}}^2(Y_t^\star Q_t^r) \big\|Y_t^r - Y_t^\star Q_t^r\big\|_{\mathrm{F}}^2/2 = \sigma_{\operatorname{min}}^2(Y_t^\star)\epsilon_t^r/2 \geqslant M_2 n \epsilon_t^r/2
\end{align}
by Lemma \ref{adlem:tu2016}, we obtain 
\begin{align} \label{eq:thm5-7}
    D_1 - D_2 \geqslant  \frac{(\kappa_1 - 2c_2)M_2 n }{4}\, \epsilon_t^r +  \frac{1}{2\kappa_2}\big\|l^\prime(\Theta_t^r) - l^\prime(\Theta_t^\star)\big\|_{\mathrm{F}}^2 - \frac{d_t}{2 c_2} \big\|l^\prime(\Theta_t^\star)\big\|_{\operatorname{op}}^2.
\end{align}
Similarly, for any constant $c_3 > 0$, we have
\begin{align} \label{eq:thm5-8}
    D_3 &= \big| \big\langle l^\prime(\Theta_t^r), \big(Y_t^r - Y_t^\star Q_t^r\big) \big(Y_t^r - Y_t^\star Q_t^r\big)^\mytrans \big\rangle \big| \notag\\
    &\leqslant \big\|l^\prime(\Theta_t^r)\big\|_{\operatorname{op}} \big\|\big(Y_t^r - Y_t^\star Q_t^r\big) \big(Y_t^r - Y_t^\star Q_t^r\big)^\mytrans\big\|_* \notag\\
    &\leqslant \left\{\big\|l^\prime(\Theta_t^\star)\big\|_{\operatorname{op}} + \big\|l^\prime(\Theta_t^r) - l^\prime(\Theta_t^\star)\big\|_{\operatorname{op}}\right\}\, \big\|Y_t^r - Y_t^\star Q_t^r \big\|_{\mathrm{F}}^2 \notag\\
    &\leqslant \frac{c_3}{2} \left\{\big\|l^\prime(\Theta_t^\star)\big\|_{\operatorname{op}} +  \big\|l^\prime(\Theta_t^r) - l^\prime(\Theta_t^\star)\big\|_{\mathrm{F}}\right\}^2 + \frac{1}{2c_3} \big\|Y_t^r - Y_t^\star Q_t^r \big\|_{\mathrm{F}}^4 \notag\\
    &\leqslant c_3 \big\|l^\prime(\Theta_t^\star)\big\|_{\operatorname{op}}^2 + c_3 \big\|l^\prime(\Theta_t^r) - l^\prime(\Theta_t^\star)\big\|_{\mathrm{F}}^2 + \frac{\epsilon_t^r}{2c_3}\, \epsilon_t^r.
\end{align}
Moreover, by Condition \ref{cond:truevalue} (i),  
\begin{align} \label{eq:thm5-9.0}
    \big\|Y_t^r\big\|_{\mathrm{F}}^2 \leqslant 2\big\|Y_t^\star\big\|_{\mathrm{F}}^2 + 2\big\|Y_t^r - Y_t^\star Q_t^r\big\|_{\mathrm{F}}^2 \leqslant 4M_1^2n +2\epsilon_t^r, 
\end{align}
then
\begin{align} \label{eq:thm5-9}
    D_4 &= \big\|l^\prime(\Theta_t^r) Y_t^r\big\|_{\mathrm{F}}^2 \leqslant \big\|l^\prime(\Theta_t^r)\big\|_{\operatorname{op}}^2 \big\|Y_t^r\big\|_{\mathrm{F}}^2 \notag \\
    &\leqslant 2 \big\|l^\prime(\Theta_t^\star)\big\|_{\operatorname{op}}^2 \big\|Y_t^r\big\|_{\mathrm{F}}^2 + 2 \big\|l^\prime(\Theta_t^r) - l^\prime(\Theta_t^\star)\big\|_{\mathrm{F}}^2 \big\|Y_t^r\big\|_{\mathrm{F}}^2 \notag\\
    &\leqslant (8M_1^2n+4\epsilon_t^r)  \big\|l^\prime(\Theta_t^\star)\big\|_{\operatorname{op}}^2  + (8M_1^2n+4\epsilon_t^r) \big\|l^\prime(\Theta_t^r) - l^\prime(\Theta_t^\star)\big\|_{\mathrm{F}}^2.
\end{align}

Combining \eqref{eq:thm5-4} with \eqref{eq:thm5-7}, \eqref{eq:thm5-8}, \eqref{eq:thm5-9}, and the choice $\eta_Y = \eta/n$ for a sufficiently small constant $\eta>0$ in Condition \ref{cond:parestY}, we obtain
\begin{align} \label{eq:thm5-10}
    \epsilon_t^{r+1} &\leqslant \epsilon_t^r - \eta \left( \frac{(\kappa_1 - 2c_2)M_2}{4} - \frac{\epsilon_t^r}{2c_3 n}\right)\epsilon_t^r \notag \\
    & \quad \quad - \eta \left(\frac{1}{2 \kappa_2} - c_3 - 8M_1^2 \eta - \frac{4\eta\, \epsilon_t^r}{n} \right) \frac{1}{n} \big\|l^\prime(\Theta_t^r) - l^\prime(\Theta_t^\star)\big\|_{\mathrm{F}}^2 \notag\\
    &\quad \quad + \eta \left(\frac{d_t}{2c_2} + c_3 + 8M_1^2\eta + \frac{4\eta\, \epsilon_t^r}{n}\right) \frac{1}{n}\big\|l^\prime(\Theta_t^\star)\big\|_{\operatorname{op}}^2.
\end{align}
 Since \eqref{eq:thm5-10} holds for any positive constants $c_2$ and $c_3$, we can choose sufficiently small values;
for clarity, we next 
set $c_2=\kappa_1/4$ and $c_3={1}/({4\kappa_2})$ as an example. Substituting these specific choices and \eqref{eq:pfb2-3} into \eqref{eq:thm5-10} yields
\begin{align*}
  \epsilon_t^{r+1}   &   \leqslant (1-c )\epsilon_t^r - \eta\left(\frac{1}{4\kappa_2}-8M_1^2\eta - 4\eta c_0  \right)\frac{1}{n}\big\|l^\prime(\Theta_t^r) - l^\prime(\Theta_t^\star)\big\|_{\mathrm{F}}^2+  \frac{C}{n}\big\|l^\prime(\Theta_t^\star)\big\|_{\operatorname{op}}^2
\end{align*}
with probability $1 - O((nT)^{-\varepsilon})$, where 
\begin{align} \label{eq:thm5-c}
    c =\frac{\eta \kappa_1M_2}{16}\quad \text{ and }\quad 
    C = \eta\left(\frac{2d_t}{\kappa_1}+ \frac{1}{4\kappa_2} + 8M_1^2\eta + 4\eta c_0\right).
\end{align}
When choosing $\eta<\min\{\frac{16}{\kappa_1M_2}, \frac{1}{16\kappa_2(2M_1^2+c_0)}\}$,  we have $c\in (0,1)$ and 
\begin{align*}
      \epsilon_t^{r+1}   &   \leqslant (1-c )\epsilon_t^r +  \frac{C}{n}\big\|l^\prime(\Theta_t^\star)\big\|_{\operatorname{op}}^2
\end{align*}
with probability $1 - O((nT)^{-\varepsilon})$. Meanwhile, we point out that similar conclusion can also be obtained with other choices of $c_0,c_2, c_3,$ and $\eta$, as long as they are sufficiently small. 



\begin{remark}
If the $T$ networks do not contain any shared factor ($k=0$), Theorem \ref{algor:estY} already yields oracle error rates for estimating latent embeddings within each network, thus a joint analysis is unnecessary. 
The subsequent proofs of Theorems  \ref{thm:initial} and \ref{thm:onestep} are relevant only  if  $k>0$, though this is not a restrictive assumption, as the arguments remain valid by simply omitting the terms relevant to $Z$. 
For conciseness, we do not explicitly distinguish between the cases $k=0$ or $k>0$ below.  Similarly, if any $k_t=0$, the terms relevant to $W_t$ can be omitted without affecting the validity of the proofs.  
\end{remark}

\subsection{Proof of Theorem \ref{thm:initial}} \label{sec:pf:thm:initial}

We first show that $\hat{\mathcal{T}} \neq \varnothing $ happens with a high probability, i.e., there exist $(t, s)$ such that 
\begin{align} \label{eq:thm1-1}
    \Pr\left\{ \mathcal{R}_{1,k+k_t+k_s}(\mathring Y_{t,s}) 
    >  \tau_1 \right\} = O((nT)^{-\varepsilon}).
\end{align}
Denote $\mathring Q_t = \arg \min_{Q \in \mathcal{O}(d_t)} \|\mathring Y_t - Y_t^\star Q\|_{\mathrm{F}}^2$ and $\mathring Q_{t,s} = \operatorname{diag}(\mathring Q_t, \mathring Q_s)$. Similar to \eqref{eq:prop2-2}, we define
\begin{align*} 
    V_{t,s}^\star= \frac{1}{\sqrt{2}}\begin{bmatrix}
    \mathring Q_t^\mytrans & \mathbf{0}\\
    \mathbf{0} & \mathring Q_s^\mytrans
\end{bmatrix} \begin{bmatrix}
     L_{t,z}\\
     -L_{s,z}
 \end{bmatrix}  \quad\text{ and }\quad  V_{t,s}^{\star \perp} =  \begin{bmatrix}
   \mathring Q_t^\mytrans & \mathbf{0}\\
    \mathbf{0} & \mathring Q_s^\mytrans
\end{bmatrix} \begin{bmatrix}
      L_{t,z}/\sqrt{2} &  L_{t,w} &  \mathbf{0}\\
      L_{s,z}/\sqrt{2}  &  \mathbf{0} &  L_{s,w}
    \end{bmatrix}
\end{align*}
for any $1 \leqslant t < s \leqslant T$. We still have 
\begin{align*}
    Y_{t,s}^{\star} \mathring Q_{t,s} V_{t,s}^\star = \frac{1}{\sqrt{2}}\, Y_t^\star L_{t,z} - \frac{1}{\sqrt{2}}\, Y_s^\star L_{s,z} = \mathbf 0 \quad \text{ and }\quad Y_{t,s}^{\star} \mathring Q_{t,s} V_{t,s}^{\star \perp} = \big[\sqrt{2}Z^\star, W_t^\star, W_s^\star\big],
\end{align*}
then for any $(t,s)$ pair satisfying Condition \ref{cond:truevalue} (iii),
\begin{align} \label{eq:thm1-2}
\sigma_{k+k_{t}+k_{s}}^2\big(Y_{t,s}^\star \mathring Q_{t,s}\big)&=\sigma_{\min }^2\big(Y_{t,s}^\star \mathring Q_{t,s} V_{t,s}^{\star \bot}\big)
=\sigma_{\min }^2\big(\big[\sqrt{2} Z^{\star},  W_{t}^{\star}, W_{s}^{\star}\big]\big) \geqslant \sigma_{\min}(nG_{t,s}^\star) \geqslant M_3 n.
\end{align}
Furthermore, Theorem \ref{thm:estY} indicates that for any $(t,s)$
\begin{align} \label{eq:thm1-3}
    \Pr\left\{\big\|\mathring Y_{t,s} - Y_{t,s}^\star \mathring Q_{t,s}\big\|_{\operatorname{op}} > C_\varepsilon \log^{1/2}(nT)\right\} = O((nT)^{-\varepsilon}).
\end{align}
for a constant $C_\varepsilon > 0$. Combining \eqref{eq:thm1-3} with Condition \ref{cond:truevalue} (i) and \eqref{eq:thm1-2}, we obtain $\sigma^2_1(\mathring Y_{t,s}) \asymp \sigma^2_{k + k_t + k_s}(\mathring Y_{t,s}) \asymp n$ with  probability greater than $1 - O((nT)^{-\varepsilon})$, which implies \eqref{eq:thm1-1} for any $\tau_1 \gg 1$.

\bigskip

We next derive the error bound of $\|\mathring F - Z^\star Z^{\star \mytrans}\|_{\mathrm{F}}$. Denote $\mathring F_{t,s} = \mathring Y_{t,-s} \mathring V_{t,s} \mathring  V_{t,s}^\mytrans \mathring Y_{t,-s}^{\mytrans}/2$ for any $(t,s) \in \hat{\mathcal{T}}$, then
\begin{align} \label{eq:thm1-4}
    \big\|\mathring F_{t,s} - Z^\star Z^{\star \mytrans}\big\|_{\mathrm{F}} 
    &=  \big\|\mathring Y_{t,-s} \mathring V_{t,s} \mathring  V_{t,s}^\mytrans \mathring Y_{t,-s}^{\mytrans} - 2Z^\star Z^{\star \mytrans}\big\|_{\mathrm{F}} /2 \notag\\
    &\leqslant \big\|\mathring Y_{t,-s} \big(\mathring V_{t,s} \mathring  V_{t,s}^\mytrans - V_{t,s}^\star   V_{t,s}^{\star\mytrans}\big) \mathring Y_{t,-s}^{\mytrans}\big\|_{\mathrm{F}} \notag\\
    &\quad+ \big\|\mathring Y_{t,-s} V_{t,s}^\star   V_{t,s}^{\star \mytrans} \mathring Y_{t,-s}^\mytrans 
    - (Y_{t,-s}^\star \mathring Q_{t,s}) V_{t,s}^\star   V_{t,s}^{\star \mytrans} (Y_{t,-s}^\star \mathring Q_{t,s})\|_{\mathrm{F}} \notag\\
    &\leqslant \big\|\mathring Y_{t,-s}\big\|_{\mathrm{F}}^2 \big\|\mathring V_{t,s} \mathring  V_{t,s}^\mytrans - V_{t,s}^\star   V_{t,s}^{\star\mytrans}\big\|_{\operatorname{op}} \notag\\
    &\quad+ \big\|\mathring Y_{t,-s} - Y_{t,-s}^\star \mathring Q_{t,s}\big\|_{\mathrm{F}} \big\|\mathring Y_{t,-s}\big\|_{\operatorname{op}} + \big\| Y_{t,-s}^\star\big\|_{\operatorname{op}} \big\|\mathring Y_{t,-s} - Y_{t,-s}^\star \mathring Q_{t,s}\big\|_{\mathrm{F}},
\end{align}
where the first inequality follows from $\sqrt{2}Z^\star = Y_{t,-s}^\star \mathring Q_{t,s} V_{t,s}^\star$. To examine the similarity between $\mathring V_{t,s}$ and $V_{t,s}^\star$,
 we take $X = Y_{t,s}^\star \mathring Q_{t,s}$ and $\hat{X}= \mathring Y_{t,s}$ in Lemma \ref{adlem:cai2018}, which gives 
\begin{align} \label{eq:thm1-5}
    \big\|\mathring V_{t,s} \mathring  V_{t,s}^\mytrans - V_{t,s}^\star V_{t,s}^{\star \mytrans}\big\|_{\operatorname{op}}  &=  \big\|\mathring V_{t,s}^\perp \mathring  V_{t,s}^{\perp \mytrans} - V_{t,s}^{\star \perp} V_{t,s}^{\star \perp \mytrans}\big\|_{\operatorname{op}} \notag\\
    &\leqslant C\sigma^{-1}_{k+ k_{t} +{k_{s}}}(X) \big\|X- \hat{X}\big\|_{\operatorname{op}} \notag\\
    &\leqslant C_\varepsilon \tau_1 n^{-1/2} \log^{1/2}(nT)
\end{align}
with probability $1-O((nT)^{-\varepsilon-2})$ by \eqref{eq:thm1-3} and $\sigma_{k + k_t + k_s}(Y_{t,s}^\star) \gtrsim \tau_1^{-1} n^{1/2}$. Combining \eqref{eq:thm1-4} with \eqref{eq:thm1-3}, \eqref{eq:thm1-5}, and Condition \ref{cond:truevalue} (i) shows that
\begin{align*} 
    \big\|\mathring F_{t,s} - Z^\star Z^{\star \mytrans}\big\|_{\mathrm{F}}  \leqslant C_\varepsilon \tau_1 n^{1/2} \log^{1/2}(nT) 
\end{align*}
with probability $1-O((nT)^{-\varepsilon-2})$. Since $\| \mathring F - Z^\star Z^{\star \mytrans}\|_{\mathrm{F}} \leqslant {\max_{(t,s) \in \hat{\mathcal{T}}}} \|\mathring F_{t,s} - Z^\star Z^{\star \mytrans}\|_{\mathrm{F}}$ we obtain
\begin{align} \label{eq:thm1-6}
    &\Pr\left\{\big\|\mathring F - Z^\star Z^{\star \mytrans}\big\|_{\mathrm{F}}  > C_{\varepsilon}\, \tau_1 n^{1/2} \log^{1/2}(nT)\right\} \notag\\
    \leqslant ~&\sum_{(t,s) \in \hat{\mathcal{T}}} \Pr\left\{\big\|\mathring F_{t,s} - Z^\star Z^{\star \mytrans}\big\|_{\mathrm{F}}  > C_{\varepsilon}\, \tau_1 n^{1/2} \log^{1/2}(nT)\right\} \notag \\
    = ~&O((nT)^{-\varepsilon-2} \,T^2) = O((nT)^{-\varepsilon}).
\end{align}



Finally we present the error bounds of $\mathring Z$ and $\mathring W_t$. Since $\operatorname{rank}(Z^\star Z^{\star \mytrans}) = k$, we apply the Eckart–Young–Mirsky theorem to obtain 
\begin{align} \label{eq:thm1-7}
    \big\|\mathring Z \mathring Z^\mytrans - Z^\star Z^{\star \mytrans}\big\|_{\mathrm{F}} 
    \leqslant \big\|\mathring Z \mathring Z^\mytrans - \mathring F\big\|_{\mathrm{F}} + \big\| \mathring F - Z^\star Z^{\star \mytrans}\big\|_{\mathrm{F}}
    \leqslant 2 \big\| \mathring F - Z^\star Z^{\star \mytrans}\big\|_{\mathrm{F}}.
\end{align}
Combining \eqref{eq:thm1-6}, \eqref{eq:thm1-7}, Lemma \ref{adlem:tu2016}, and Condition \ref{cond:truevalue} 
 (ii), there exists a constant $C_\varepsilon > 0$ such that 
\begin{align}\label{eq:thm1-8}
\Pr\left\{\operatorname{dist}^2(\mathring Z, Z^\star) > C_\varepsilon \tau_1^2 \log(nT)\right\} = O((nT)^{-\varepsilon}).
\end{align}
The argument for $\mathring W_t$ can be obtained by the similar technique and 
\begin{align}
    \big\|\mathring Y_t  \mathring Y_t^{\mytrans} - \mathring F - W_t^\star W_t^{\star \mytrans}\big\|_{\mathrm{F}} &\leqslant \big\| \mathring Y_t \mathring Y_t^{\mytrans} - Y_t^\star Y_t^{\star \mytrans}\big\|_{\mathrm{F}} + \big\|Z^\star Z^{\star \mytrans} - \mathring F\big\|_{\mathrm{F}}\notag \\
    &\leqslant\big\|\mathring Y_t - Y_t^\star \mathring Q_t\big\|_{\mathrm{F}} \big(\big\|\mathring Y_t\big\|_{\operatorname{op}} + \big\|Y_t^\star\big\|_{\operatorname{op}}\big) + \big\|Z^\star Z^{\star \mytrans} - \mathring F \big\|_{\mathrm{F}}\notag \\
    &\leqslant C_\varepsilon \tau_1 n^{1/2} \log^{1/2}(nT)\label{eq:thm1-9}
\end{align}
with probability $1 - O((nT)^{-\varepsilon})$. In summary, the conclusion of Theorem \ref{thm:initial} holds for any  $1 \ll \tau_1 \lesssim \log^{1/2}(n)$.

\subsection{Proof of Theorem \ref{thm:onestep}} \label{sec:pf:thm:onestep}
Recall that  Theorem \ref{thm:onestep} presents \eqref{eq:zhaterr} and \eqref{eq:wthaterr}, estimation error rates of the refined estimators $\hat{Z}$ and $\hat{W}_t$'s from  Algorithm \ref{algor:refine}, respectively.  
Note that $\hat{Z}$ and $\hat{W}_t$'s are derived from \eqref{eq:newtononev} with $\check{Z}$ and $\check{W}_t$'s in Algorithm \ref{algor:refine} as initialization.
We organize the proof of  Theorem \ref{thm:onestep} as follows. 

\smallskip 




Section \ref{sec:revisealgor} provides details on the adjustment in Remark \ref{rmk:pseudolik}. 

 Section \ref{sec:pdfthm2-1} establishes  two-to-infinity error bounds of the initializations $\check{Z}$ and $\check{W_t}$'s.  
Specifically, for any constant $\varepsilon > 0$, there exists a constant $C_\varepsilon > 0$ such that 
\begin{align} \label{eq:thm2-1}
    \Pr\left[ \max_{1\leqslant t \leqslant T} \left\{ \big\| \check Z - Z^{\star} \check Q\big\|_{2 \to \infty}^2 + \big\|\check W_t - W_t^{\star} \check Q_t\big\|_{2 \to \infty}^2 \right\} > \frac{C_\varepsilon\log^4(nT)}{n} \right] = O((nT)^{-\varepsilon}),
\end{align}
where $\check{Q} = \arg\min_{Q \in \mathcal{O}(k)} \|\check{Z} - Z^\star Q\|_{\mathrm{F}}^2$ and $\check{Q}_t = \arg\min_{Q \in \mathcal{O}(k_t)} \|\check{W}_t - W_t^\star Q\|_{\mathrm{F}}^2$. 

 Section \ref{sec:newtonform} derives analytical formula of \eqref{eq:newtononev} through expressing $\dot{\ell}(v)$ and $I(v)$. 

Section \ref{sec:proplemmaiv}  establishes useful properties on $I(v)$ that will be used in subsequent proofs. 

Section \ref{sec:pdfzhaterr} 
proves \eqref{eq:zhaterr} in Theorem \ref{thm:onestep} for $\hat{Z}$. 



Section \ref{sec:pdfwthaterr} 
proves \eqref{eq:wthaterr} in Theorem \ref{thm:onestep} for $\hat{W}_t$ with $ t=1,\ldots, T$. 

{\subsubsection{Revised algorithm in  Remark \ref{rmk:pseudolik}} \label{sec:revisealgor} }
As mentioned in Remark \ref{rmk:pseudolik}, the objective function $\ell(Z,W)$ of the projected gradient descent in Algorithm \ref{algor:refine} {(lines 2--5)} is replaced with $p\ell(Z,W)$ to facilitate the theoretical derivation. 
{Specifically,  $\tilde{Z}^{r+1}=Z^{r} + {\eta_Z}\,  \partial {\ell}(Z^r,W^r) / \partial Z $ and $\tilde{W}_t^{r+1}=W_t^{r} + \eta_W\,  \partial{\ell}(Z^r,W^r)/ \partial W_t$  in lines 3 and 4 of Algorithm \ref{algor:refine}  are replaced by 
\begin{align}\label{eq:pseudogd}
    \tilde{Z}^{r+1}=Z^{r} + {\eta_Z}\, \frac{\partial p\ell(Z^r,W^r)}{\partial Z} \quad \text{ and } \quad  \tilde{W}_t^{r+1}=W_t^{r} + \eta_W\,  \frac{\partial p\ell(Z^r,W^r)}{\partial W_t},
\end{align} respectively, where ${\partial p\ell(Z,W)}/{\partial Z}$ and ${\partial p\ell(Z,W)}/{\partial W_t}$ denote the partial derivatives of $p\ell(Z,W)$ with respect to $Z$ and $W_t$, respectively. By the definition of $p\ell(Z,W)$, we have 
\begin{align*}
    \frac{\partial p\ell(Z,W)}{\partial Z} =  \sum_{t = 1}^T l^\prime(Z\mathring Z^{\top}+W_t \mathring{W}_t^{\top}) \mathring Z \quad \text{and} \quad  \frac{\partial p\ell(Z,W)}{\partial W_t} =   l^\prime(Z\mathring Z^{\top}+W_t \mathring{W}_t^{\top}) \mathring W_t,
\end{align*}
where $(\mathring Z, \mathring W_t)$ represents the initializations from Algorithm \ref{algor:initial}. 
Plugging $Z=Z^r$ and $W_t=W_t^r$ for each iteration of the projected gradient descent,  \eqref{eq:pseudogd} becomes 
\begin{align} \label{eq:gdupdate}
    \tilde Z^{r+1} =&~ Z^r + \eta_Z \sum_{t = 1}^T l^\prime(\mathring \Theta_t^{r})\mathring Z \quad \quad \text{and} \quad \quad  \tilde W_t^{r+1} = W_t^r + \eta_W \,l^\prime(\mathring \Theta_t^{r}) \mathring W_t, 
\end{align}where we let 
\begin{align*}
    \mathring \Theta_t^r =  Z^r \mathring Z^\mytrans + W_t^r \mathring W_t^\mytrans. 
\end{align*}
Theorem \ref{thm:onestep} is established for Algorithm \ref{algor:refine} when $\tilde Z^{r+1}$ and $ \tilde W_t^{r+1} $ in lines 3--4 are computed by \eqref{eq:gdupdate} above. 
Moreover, as a clarification, the one-step update, line 6 of  Algorithm \ref{algor:refine},  does not require the use of the pseudo log-likelihood and is still computed by \eqref{eq:newtononev} with the log-likelihood $\ell(Z,W)$ defined in \eqref{eq:zwlikelihood}; see details in  Section \ref{sec:newtonform}. 
}

\bigskip 

\subsubsection{Proof of \eqref{eq:thm2-1}} \label{sec:pdfthm2-1}
We next outline the key steps to prove \eqref{eq:thm2-1} and then provide details afterwards. 

\paragraph{Outline.} 
To prove \eqref{eq:thm2-1}, we will first show that there exists a constant $C>0$ such that 
\begin{equation} \label{eq:rotationfix}
    \begin{aligned} 
      \big\|\check{Z} - Z^\star \check{Q}\big\|_{2 \to \infty }   &\leqslant C \big\|\check Z - Z^\star \mathring{Q}\big\|_{2\to \infty} \\ \text{and}\quad \quad    \big\|\check{W}_t - W_t^\star \check{Q}_t\big\|_{2 \to \infty }   &\leqslant C \big\|\check W_t - W_t^\star \mathring{Q}_t\big\|_{2\to \infty},
\end{aligned}
\end{equation}
where $ \mathring Q = \arg \min_{Q \in \mathcal{O}(k)} \|\mathring Z - Z^\star Q\|_{\mathrm{F}}^2$  and $\mathring Q_t = \arg \min_{Q \in \mathcal{O}(k_t)} \|\mathring W_t - W_t^\star Q\|_{\mathrm{F}}^2.$ 
Moreover, by our construction $\check{Z}=Z^R$ and $\check{W}_t = W_t^R$ , to show \eqref{eq:thm2-1}, it suffices to prove
\begin{align} 
    \Pr\left[   \big\| Z^R - Z^{\star} \mathring Q\big\|_{2 \to \infty}^2  > C_\varepsilon n^{-1}\log^4(nT) \right] &= O((nT)^{-\varepsilon})\label{eq:thm2-14}\\
    \text{and} \quad\quad \Pr\Big[ \max_{1 \leqslant t \leqslant T} \Big\{ \big\| W_t^R - W_t^{\star} \mathring Q_t\big\|_{2 \to \infty}^2 \Big\}  > C_\varepsilon n^{-1}\log^4(nT) \Big] &= O((nT)^{-\varepsilon}). \label{eq:thm2-14-2}
\end{align}
To prove \eqref{eq:thm2-14} and \eqref{eq:thm2-14-2}, note that $Z^R$ and $W_t^R$'s are outputs from the  projected gradient descent procedure.   
We will analyze the errors with respect to iterations in the procedure similarly to   Section \ref{sec:pf_thm5}. 

We first examine $Z^R$. For each iteration $r = 0, \ldots, R$ and $i=1, \ldots,  n$, we define an $i$-th row  error as 
\begin{align} \label{eq:thm2-2}
    \epsilon_i^r &= T\,\big\|e_i^\mytrans \big(Z^r - Z^\star \mathring Q\big)\big\|_{2}^2 + \sum_{t = 1}^T \big\|e_i^\mytrans \big(W_t^r - W_t^\star \mathring Q_t\big)\big\|_{2}^2, 
\end{align}
where rotations $\mathring Q$ and $\mathring Q_t$'s are defined same as in \eqref{eq:rotationfix}, and $e_i \in \mathbb R^n$ denotes an indicator vector with a $1$ in the $i$-th position and $0$ in all other entries. 
As $Z^0$ and $W_t^0$ are initializations from Algorithm \ref{algor:initial}, by Theorem \ref{thm:initial}, we have
\begin{align} \label{eq:thm2-initerr}
   \max_{1 \leqslant i \leqslant n} \epsilon_i^0 \leqslant T\,\big\|Z^r - Z^\star \mathring Q\big\|_{\mathrm{F}}^2 + \sum_{t = 1}^T \big\|W_t^r - W_t^\star \mathring Q_t\big\|_{\mathrm{F}}^2 \leqslant C_\varepsilon T \log^2(nT)
\end{align}
with probability $1 - O((nT)^{-\varepsilon})$. We will then prove 
that there exist positive constants $C$ and $c$ such that
\begin{align} \label{eq:iterateerr}
    \epsilon_i^{r+1} \leqslant (1-c) \epsilon_i^r + \frac{CT}{n} \left[ B_{\nu} + \frac{1}{n}  (B_{z}+B_{w})\right]
\end{align}
holds with probability $1 - O((nT)^{-\varepsilon})$, where we define $\mathring \Theta_t^\star = Z^\star \mathring Q \mathring Z^\mytrans + W_t^\star \mathring Q_t \mathring W_t^\mytrans$ and  
\begin{align}
B_{\nu} = &~  \max_{1 \leqslant t \leqslant T}  \big\|l^\prime(\Theta_t^\star) - l^\prime(\mathring \Theta_t^\star)\big\|_{2 \to \infty}^2,  \label{eq:boundbvzw}\\    
    B_{z}=&~  \max_{1 \leqslant t \leqslant T} \big\|l^\prime(\Theta_t^\star)  \mathring Z \big\|_{2 \to \infty}^2,  \notag\\  
    B_{w} =  &~ \max_{1 \leqslant t \leqslant T}  \big\|l^\prime(\Theta_t^\star) \mathring W_t \big\|_{2 \to \infty}^2. \notag
\end{align} 
As the above inequality holds for all $1 \leqslant r \leqslant R-1$, we can repeatedly substitute to obtain
\begin{align}\label{eq:iterateerr3}
    \epsilon_i^r \leqslant (1-c)^r \epsilon_i^0 +\frac{CT}{cn} \left[  B_{\nu} + \frac{1}{n}  (B_{z}+B_{w})\right].
\end{align}
Given \eqref{eq:thm2-initerr}, to show  \eqref{eq:thm2-14} when $R\gg \log(nT)$,  
it remains to show that
\begin{equation}\label{eq:thm2-bbounds}
    \begin{aligned}
   \Pr\left[B_{\nu} > C_\varepsilon  \log^2(nT) \right] = O((nT)^{-\varepsilon}), \\   
     \Pr\left[B_{z}+B_{w} > C_\varepsilon n \log^4(nT) \right] = O((nT)^{-\varepsilon}). 
\end{aligned}
\end{equation}

We next analyze $W_t^R$ similarly. In particular, for $r = 1,\ldots, R$, $i = 1,\ldots, n$, and $t = 1, \ldots, T$, we define the error term 
\begin{align}\label{eq:heterogeneous_error_it}
    \epsilon_{it}^r = \big\|e_i^\mytrans \big(W_t^r - W_t^\star \mathring Q_t\big)\big\|_{2}^2.
\end{align}
Similarly to \eqref{eq:iterateerr}, we will show
that there exist positive constants $C$ and $c$ such that
\begin{align} \label{eq:iterateerr2}
    \epsilon_{it}^{r+1} \leqslant (1-c)\epsilon_{it}^r + \frac{C}{n} \Big(B_{\nu} + \frac{1}{n} B_{w} + n\big\|e_i^\mytrans \big(Z^r - Z^\star \mathring Q\big)\big\|_2^2 \Big)
\end{align}
holds with probability $1 - O((nT)^{-\varepsilon})$. Combining \eqref{eq:thm2-initerr}, \eqref{eq:iterateerr3}, and \eqref{eq:thm2-bbounds} above gives that for any $r \geqslant R/2 \gg \log(nT)$, $\mathrm{Pr}[\max_{1 \leqslant i \leqslant n}\big\|e_i^\mytrans \big(Z^r - Z^\star \mathring Q\big)\big\|_2^2 > C_\varepsilon n^{-1} \log^{4}(nT)]=O((nT)^{-\varepsilon})$. 
Consequently, \eqref{eq:thm2-14-2} follows from repeated substitution of \eqref{eq:iterateerr2} and \eqref{eq:thm2-bbounds}.

In summary, to prove \eqref{eq:thm2-1}, it remains to establish  \eqref{eq:rotationfix}, \eqref{eq:iterateerr}, \eqref{eq:thm2-bbounds}, and \eqref{eq:iterateerr2}, whose proofs are given on Pages \pageref{sec:pfrotationfix}, \pageref{sec:pfiterateerr}, \pageref{sec:pfthm2-bbounds}, and \pageref{sec:pfiterateerr2} below, respectively. In the analysis below, we use the following facts on the initial estimators: for any $\varepsilon>0$, {
\begin{align}  
   \mathrm{Pr}\left[ \max_{1\leqslant t\leqslant T} \big\|\big[\mathring Z,\mathring W_t\big]\big\|_{\mathrm{F}}^2 > 3M_1^2n \right]  &= O((nT)^{-\varepsilon} ),
   \label{eq:thm2-0} \\
  \mathrm{Pr}\left[ \min_{1\leqslant t\leqslant T} \sigma_{\min}^2\big(\big[\mathring Z,\mathring W_t\big]\big) < M_2n/2 \right]  &= O((nT)^{-\varepsilon} ),
  \label{eq:thm2-00}
\end{align}}which follow from 
\begin{align*}
   \max_{1\leqslant t \leqslant T} \big\|\big[Z^\star \mathring Q, W_t^\star \mathring Q_t\big]\big\|_{\mathrm{F}}^2 \leqslant 2M_1^2 n, \quad \min_{1 \leqslant t \leqslant T} \sigma_{\operatorname{min}}^2\big(\big[  Z^\star \mathring Q, W_t^\star \mathring Q_t\big]\big) \geqslant M_2 n \quad &\text{by Condition \ref{cond:truevalue}}, \\ \text{and} \quad
 \Pr\Big( \max_{1 \leqslant t \leqslant T} \big\| 
 \big[\mathring Z - Z^\star \mathring Q,\, \mathring W_t - W_t^\star \mathring Q_t \big] \big\|_{\mathrm{F}}^2 > C_{\varepsilon}\log^2(nT) \Big) = O((nT)^{-\varepsilon}) \quad &\text{by Theorem  \ref{thm:initial}}.
\end{align*}



\paragraph{Proof of \eqref{eq:rotationfix}.} \label{sec:pfrotationfix}

Note that
\begin{align*}
   \big\|\check{Q} - \mathring Q  \big\|_{\operatorname{op}} & \leqslant \sigma_k^{-1} (Z^\star) \big\|Z^\star \big(\check{Q} - \mathring Q\big)\big\|_{\mathrm{F}} 
   \leqslant (M_2n)^{-1/2} \big(\big\|\check Z - Z^\star \check{Q}\big\|_{\mathrm{F}} + \big\|\check Z - Z^\star \mathring{Q}\big\|_{\mathrm{F}}\big) \\&\leqslant 2 (M_2n)^{-1/2} \big\|\check Z - Z^\star \mathring{Q}\big\|_{\mathrm{F}} \leqslant 2 M_2^{-1/2} \big\|\check Z - Z^\star \mathring{Q}\big\|_{2\to \infty},
\end{align*}
where the third inequality follows from the definition of $\check{Q}$. As a result, 
\begin{align*}
    \big\|\check{Z} - Z^\star \check{Q}\big\|_{2 \to \infty }\leqslant  \big\|\check{Z} - Z^\star \mathring{Q}\big\|_{2 \to \infty } + \big\|Z^\star \big\|_{2 \to \infty} \big\|\check{Q} - \mathring Q\big\|_{\operatorname{op}} \leqslant C \big\|\check Z - Z^\star \mathring{Q}\big\|_{2\to \infty}.
\end{align*}
Then conclusion for $W_t$ can be similarly obtained. 
\paragraph{Proof of \eqref{eq:iterateerr}.} \label{sec:pfiterateerr}

By \eqref{eq:gdupdate}, we can write
\begin{align} \label{eq:thm2-3}
    \big\|e_i^\mytrans \big(Z^{r+1} - Z^\star \mathring Q\big)\big\|_{2}^2 &\leqslant  \big\|e_i^\mytrans \big(\tilde Z^{r+1} - Z^\star \mathring Q\big)\big\|_{2}^2 = \big\|e_i^\mytrans \big(Z^r - Z^\star \mathring Q\big)\big\|_{2}^2 \notag \\ 
    &+ 2 \eta_Z \sum_{t=1}^T e_i^\mytrans l^\prime(\mathring \Theta_t^r) \mathring Z \big(Z^r - Z^\star \mathring Q \big)^\mytrans e_i + \eta_Z^2\, \Big\|\sum_{t=1}^T e_i^\mytrans l^\prime(\mathring \Theta_t^r) \mathring Z\Big\|_2^2.
\end{align}
 Similarly, 
\begin{align} \label{eq:thm2-4}
    \big\|e_i^\mytrans \big(W_t^{r+1} - W_t^\star \mathring Q_t\big)\big\|_{2}^2 &\leqslant  \big\|e_i^\mytrans \big(\tilde W_t^{r+1} - W_t^\star \mathring Q_t\big)\big\|_{2}^2 = \big\|e_i^\mytrans \big(W_t^r - W_t^\star \mathring Q_t\big)\big\|_{2}^2 \notag\\[3pt]
    &+ 2 \eta_W\,  e_i^\mytrans l^\prime(\mathring \Theta_t^r) \mathring W_t \big(W_{t}^r - W_t^\star \mathring Q_t\big)^\mytrans e_i + \eta_W^2 \big\| e_i^\mytrans l^\prime(\mathring \Theta_t^r) \mathring W_t\big\|_2^2. 
\end{align}
Combining \eqref{eq:thm2-3} and \eqref{eq:thm2-4} with \eqref{eq:thm2-2} and plugging in $\eta_Z = \eta / (nT)$  and $\eta_W = \eta/n$ shows 
\begin{align} \label{eq:thm2-5}
    \epsilon_i^{r+1} &\leqslant \epsilon_i^r + \frac{2 \eta}{n}\, \sum_{t=1}^T e_i^\mytrans l^\prime(\mathring \Theta_t^r) \Big[\mathring Z \big(Z^r - Z^\star \mathring Q\big)^\mytrans + \mathring W_t \big(W_{t}^r - W_t^\star \mathring Q_t\big)^\mytrans \Big] e_i\notag\\
    &\quad + \frac{\eta^2}{n^2T}\, \Big\|\sum_{t=1}^T e_i^\mytrans l^\prime(\mathring \Theta_t^r) \mathring Z\Big\|_2^2 + \frac{\eta^2}{n^2}\,\sum_{t=1}^T \big\| e_i^\mytrans l^\prime(\mathring \Theta_t^r) \mathring W_t\big\|_2^2.
\end{align}
Applying the decomposition 
\begin{align} \label{eq:thm2-6}
    l^\prime(\mathring \Theta_t^r) = - \big(l^\prime(\mathring \Theta_t^\star) - l^\prime(\mathring \Theta_t^r)\big) + \big(l^\prime(\mathring \Theta_t^\star) - l^\prime(\Theta_t^\star)\big) + l^\prime(\Theta_t^\star)
\end{align}
with $\mathring \Theta_t^\star = Z^\star \mathring Q \mathring Z^\mytrans + W_t^\star \mathring Q_t \mathring W_t^\mytrans$ into \eqref{eq:thm2-5} yields 
\begin{align} \label{eq:thm2-7}
    \epsilon_i^{r+1} &\leqslant \epsilon_i^r - \eta D_1 + \eta D_2 + \eta D_3 + \eta^2 D_4 \\[5pt]
    \text{with\ }\quad D_1 &= \frac{2}{n} \sum_{t=1}^T e_i^\mytrans \big( l^\prime(\mathring \Theta_t^\star) - l^\prime(\mathring \Theta_t^r)\big) \Big[\mathring Z \big(Z^r - Z^\star \mathring Q\big)^\mytrans + \mathring W_t \big(W_{t}^r - W_t^\star \mathring Q_t\big)^\mytrans \Big] e_i \notag\\
    D_2 &= \frac{2}{n} \sum_{t=1}^T \Big| e_i^\mytrans \big(l^\prime( \Theta_t^\star) - l^\prime(\mathring \Theta_t^\star)\big)\Big[\mathring Z \big(Z^r - Z^\star \mathring Q\big)^\mytrans + \mathring W_t \big(W_{t}^r - W_t^\star \mathring Q_t\big)^\mytrans \Big] e_i\Big| \notag\\
    D_3 &= \frac{2}{n} \sum_{t=1}^T \Big| e_i^\mytrans l^\prime(\Theta_t^\star)\Big[\mathring Z \big(Z^r - Z^\star \mathring Q\big)^\mytrans + \mathring W_t \big(W_{t}^r - W_t^\star \mathring Q_t\big)^\mytrans \Big] e_i\Big| \notag\\
    D_4 &= \frac{1}{n^2T}\, \Big\|\sum_{t=1}^T e_i^\mytrans l^\prime(\mathring \Theta_t^r) \mathring Z\Big\|_2^2 + \frac{1}{n^2} \sum_{t=1}^T\big\| e_i^\mytrans l^\prime(\mathring \Theta_t^r) \mathring W_t\big\|_2^2. \notag
\end{align}

We next establish a lower bound for $D_1$ and upper bounds for $D_2$, $D_3$, and $D_4$. Since $-l(\theta)$ is $\kappa_1$-strongly convex and $\kappa_2$-smooth in $[-2M_1^2, 2M_1^2]$ by Condition \ref{cond:parfunction}, Lemma \ref{adlem:nesterov2003} indicates that
\begin{align} \label{eq:thm2-8}
    D_1 &= \frac{2}{n}\sum_{t=1}^T \Big\langle e_i^\mytrans \big( l^\prime(\mathring \Theta_t^\star) - l^\prime(\mathring \Theta_t^r)\big), e_i^\mytrans \big(\mathring \Theta_t^r -  \mathring \Theta_t^\star\big)\Big\rangle \notag\\
    &\geqslant \frac{\kappa_1 }{n}\sum_{t=1}^T\big\|e_i^\mytrans \big(\mathring \Theta_t^r -  \mathring \Theta_t^\star\big)\big\|_2^2 + \frac{1}{\kappa_2 n} \sum_{t=1}^T\big\|e_i^\mytrans \big( l^\prime(\mathring \Theta_t^r) -  l^\prime(\mathring \Theta_t^\star)\big)\big\|_2^2.
\end{align}
Then for any sufficiently small constant $c_2 > 0$, 
\begin{align} \label{eq:thm2-9}
    D_2 &= \frac{2}{n} \sum_{t=1}^T \Big|\Big\langle e_i^\mytrans \big( l^\prime( \Theta_t^\star) -  l^\prime(\mathring \Theta_t^\star)\big), e_i^\mytrans \big(\mathring \Theta_t^r -  \mathring \Theta_t^\star\big)\Big\rangle\Big| \notag\\
    &\leqslant \frac{2}{n}\sum_{t=1}^T \big\|e_i^\mytrans \big( l^\prime( \Theta_t^\star) -  l^\prime(\mathring \Theta_t^\star)\big)\big\|_2 \big\|e_i^\mytrans \big(\mathring \Theta_t^r -  \mathring \Theta_t^\star\big)\big\|_2 \notag \\
    &\leqslant \frac{c_2}{n}  \sum_{t=1}^T\big\|e_i^\mytrans \big(\mathring \Theta_t^r -  \mathring \Theta_t^\star\big)\big\|_2^2 + \frac{1}{c_2 n } \sum_{t=1}^T\big\|e_i^\mytrans \big( l^\prime( \Theta_t^\star) -  l^\prime(\mathring \Theta_t^\star)\big)\big\|_2^2,
\end{align}
where the last inequality follows by H\"older's inequality. 
{Note that 
\begin{align}
       &  \sum_{t=1}^T \big\| {e}_i^\top (\mathring{{\Theta}}_t^r - \mathring{{\Theta}}_t^\star) \big\|_2^2  \notag\\
  =  &~  \sum_{t=1}^T \big\| \big[\mathring{Z}, \mathring{{W}}_t\big] \big[{Z}^r - {Z}^\star \mathring{{Q}}, {W}_t^r - {W}_t^\star \mathring{{Q}}_t \big]^\mytrans {e}_i  \big\|_2^2 \notag\\
    \geqslant  &~ \sum_{t=1}^T \sigma_{\min}^2\big(\big[\mathring Z,\mathring W_t\big]\big) \Big\{ \big\|\big({Z}^r - {Z}^\star \mathring{{Q}}\big)^\mytrans {e}_i \big\|_2^2+ \big\|\big({W_t}^r - {W_t}^\star \mathring{{Q_t}}\big)^\mytrans {e}_i \big\|_2^2\Big\}\notag \\
    \geqslant &~  \min_{1\leqslant t\leqslant T}{\sigma_{\min}^2\big(\big[\mathring Z,\mathring W_t\big]\big) } \epsilon_i^r \geqslant \frac{M_2 n}{2}   \epsilon_i^r
    \label{eq:thm2-R2C10}
\end{align}
 with probability $1-O((nT)^{-\varepsilon})$, where the last inequality follows by \eqref{eq:thm2-00}. } 
Combining \eqref{eq:thm2-8}, \eqref{eq:thm2-9}, and \eqref{eq:thm2-R2C10}, we have 
\begin{align} \label{eq:thm2-10}
    D_1 - D_2 &\geqslant {\frac{M_2 (\kappa_1 - c_2) }{2}}\, \epsilon_i^r + \frac{1}{\kappa_2 n} \sum_{t=1}^T\big\|e_i^\mytrans \big( l^\prime(\mathring \Theta_t^r) -  l^\prime(\mathring \Theta_t^\star)\big)\big\|_2^2 \notag\\
    &\quad -  \frac{1}{c_2 n } \sum_{t=1}^T\big\|e_i^\mytrans \big( l^\prime( \Theta_t^\star) -  l^\prime(\mathring \Theta_t^\star)\big)\big\|_2^2.
\end{align}

Similarly, for any sufficiently small constant $c_3 >0$, we have
\begin{align} \label{eq:thm2-11}
    D_3 &= \frac{2}{n} \sum_{t=1}^T \Big| e_i^\mytrans l^\prime(\Theta_t^\star)\Big[\mathring Z \big(Z^r - Z^\star \mathring Q\big)^\mytrans + \mathring W_t \big(W_{t}^r - W_t^\star \mathring Q_t\big)^\mytrans \Big] e_i\Big| \notag\\
    &\leqslant \frac{2}{n} \sum_{t=1}^T \big\| e_i^\mytrans l^\prime(\Theta_t^\star)\mathring Z \big\|_{2} \big\|e_i^\mytrans\big(Z^r - Z^\star \mathring Q\big)\big\|_2 + \frac{2}{n} \sum_{t=1}^T \big\| e_i^\mytrans l^\prime(\Theta_t^\star) \mathring W_t \big\|_{2} \big\|e_i^\mytrans\big(W_t^r - W_t^\star \mathring Q_t\big)\big\|_2 \notag\\
    &\leqslant {c_3}\, \epsilon_i^r + \frac{1}{c_3 n^2} \sum_{t=1}^T\big\|e_i^\mytrans l^\prime(\Theta_t^\star) \big[\mathring Z, \mathring W_t\big]\big\|_2^2.
\end{align}
Finally, applying the same decomposition as \eqref{eq:thm2-6},
\begin{align}  \label{eq:thm2-12}
    D_4 &= \frac{1}{n^2T}\, \Big\|\sum_{t=1}^T e_i^\mytrans l^\prime(\mathring \Theta_t^r) \mathring Z\Big\|_2^2 + \frac{1}{n^2} \sum_{t=1}^T\big\| e_i^\mytrans l^\prime(\mathring \Theta_t^r) \mathring W_t\big\|_2^2 \notag\\
    &\leqslant \frac{1}{n^2} \sum_{t=1}^T\big\| e_i^\mytrans l^\prime(\mathring \Theta_t^r) \big[\mathring Z,\mathring W_t\big]\big\|_2^2 \leqslant \frac{3}{n^2} \sum_{t=1}^T \big\|e_i^\mytrans\big(l^\prime(\mathring \Theta_t^r) - l^\prime(\mathring \Theta_t^\star)\big) \big[\mathring Z,\mathring W_t\big]\big\|_2^2 \notag\\&\quad+ \frac{3}{n^2} \sum_{t=1}^T \big\|e_i^\mytrans\big(l^\prime(\Theta_t^\star) - l^\prime(\mathring \Theta_t^\star)\big) \big[\mathring Z,\mathring W_t\big]\big\|_2^2 + \frac{3}{n^2} \sum_{t=1}^T \big\|e_i^\mytrans l^\prime(\Theta_t^\star) \big[\mathring Z,\mathring W_t\big]\big\|_2^2 \notag\\
    &\leqslant \frac{9M_1^2}{n}  \sum_{t=1}^T \big\|e_i^\mytrans\big(l^\prime(\mathring \Theta_t^r) - l^\prime(\mathring \Theta_t^\star)\big)\big\|_2^2 + \frac{9M_1^2}{n}  \sum_{t=1}^T \big\|e_i^\mytrans\big(l^\prime(\Theta_t^\star) - l^\prime(\mathring \Theta_t^\star)\big)\big\|_2^2 \notag\\
    &\quad+ \frac{3}{n^2} \sum_{t=1}^T \big\|e_i^\mytrans l^\prime(\Theta_t^\star) \big[\mathring Z,\mathring W_t\big]\big\|_2^2,
\end{align}
where the last inequality follows from \eqref{eq:thm2-0}.

Combining \eqref{eq:thm2-7} with \eqref{eq:thm2-10}, \eqref{eq:thm2-11}, and \eqref{eq:thm2-12} gives
\begin{align} \label{eq:thm2-13}
    \epsilon_i^{r+1} &\leqslant \epsilon_i^r - \eta\Big({\frac{M_2 (\kappa_1 - c_2)}{2}} - c_3 \Big)\epsilon_i^r - \eta \Big(\frac{1}{\kappa_2 n} - \frac{9M_1^2 \eta}{n}\Big) \sum_{t=1}^T \big\|e_i^\mytrans\big(l^\prime(\mathring \Theta_t^r) - l^\prime(\mathring \Theta_t^\star)\big)\big\|_2^2 \notag\\
    &\quad + \eta \Big(\frac{1}{c_2n} + \frac{9M_1^2 \eta}{n}\Big) \sum_{t=1}^T \big\|e_i^\mytrans\big(l^\prime(\Theta_t^\star) - l^\prime(\mathring \Theta_t^\star)\big)\big\|_2^2 \notag\\
    &\quad + \eta \Big(\frac{1}{c_3 n^2} + \frac{3\eta}{n^2}\Big) \sum_{t=1}^T \big\|e_i^\mytrans l^\prime(\Theta_t^\star) \big[\mathring Z,\mathring W_t\big]\big\|_2^2.
\end{align}
Since $c_2, c_3$, and $\eta$ are sufficiently small constants, \eqref{eq:thm2-13} indicates that there exist positive constants $C$ and $c$ such that
\begin{align*}
    \epsilon_i^{r+1} \leqslant (1-c) \epsilon_i^r + C \Big( \frac{1}{n} \sum_{t=1}^T \big\|e_i^\mytrans\big(l^\prime(\Theta_t^\star) - l^\prime(\mathring \Theta_t^\star)\big)\big\|_2^2 + \frac{1}{n^2} \sum_{t=1}^T \big\|e_i^\mytrans l^\prime(\Theta_t^\star) \big[\mathring Z,\mathring W_t\big]\big\|_2^2  \Big).
\end{align*}

\paragraph{Proof of \eqref{eq:thm2-bbounds}.}\label{sec:pfthm2-bbounds} 


By Condition \ref{cond:parfunction} (i), we have
\begin{align*}
\big\|e_i^\mytrans\big(l^\prime(\Theta_t^\star) - l^\prime(\mathring \Theta_t^\star)\big)\big\|_2^2 \leqslant&~ \kappa_2^2 \big\|e_i^\mytrans\big(\Theta_t^\star - \mathring \Theta_t^\star\big)\big\|_2^2\\
    = &~ \kappa_2^2 \big\| \big( \mathring Z \mathring Q^\mytrans - Z^\star\big) z_i^{\star} +  \big( \mathring W_t \mathring Q_t^\mytrans - W_t^\star\big) w_{t,i}^{\star}\big\|_2^2 \\
    \leqslant&~ 2\kappa_2^2 \big\|\mathring Z - Z^\star \mathring Q\big\|_{\mathrm{F}}^2 \big\|z_i^\star\big\|_2^2 + 2\kappa_2^2\big\|\mathring W_t - W_t^\star \mathring Q_t\big\|_{\mathrm{F}}^2 \big\|w_{t,i}^\star\big\|_2^2 \\
    \leqslant&~  C_\varepsilon  \log^2(nT), 
\end{align*}
with probability $1 - O((nT)^{-\varepsilon})$, where the last inequality follows by Theorem \ref{thm:initial} and Condition \ref{cond:truevalue}. By Lemma \ref{lem:concentration2}, we have
\begin{align*}
    \big\|e_i^\mytrans l^\prime(\Theta_t^\star) \big[\mathring Z,\mathring W_t\big]\big\|_2^2 
    \leqslant &~ 2\, \big\|e_i^\mytrans l^\prime(\Theta_t^\star) \big[ Z^\star, W_t^\star\big]\big\|_2^2 + 2\, \big\|e_i^\mytrans l^\prime(\Theta_t^\star) \big\|_2^2\,  \big\| \big[\mathring Z,\mathring W_t\big] - \big[ Z^\star \mathring Q, W_t^\star \mathring Q_t\big]\big\|_{\mathrm{F}}^2 \\
   \leqslant  &~ C_\varepsilon n \log^4(nT)
\end{align*}
with probability $1 - O((nT)^{-\varepsilon})$. The above probabilistic arguments still hold for $B_\nu$ and $B_z + B_w$ due to the uniform control over $1 \leqslant i \leqslant n$ and $1 \leqslant t \leqslant T$ in Theorem \ref{thm:initial} and Lemma \ref{lem:concentration2}.

\paragraph{Proof of \eqref{eq:iterateerr2}.}\label{sec:pfiterateerr2} 

Applying the decomposition 
\begin{align} \label{eq:pfd36-1}
    l^\prime(\mathring \Theta_t^r) = - \big( l^\prime(\mathring \Theta_t^{r,\star}) - l^\prime(\mathring \Theta_t^r)\big) + \big(l^\prime(\mathring \Theta_t^{r,\star}) - l^\prime(\Theta_t^{\star})\big) + l^\prime(\Theta_t^\star)
\end{align}
with $\mathring \Theta_t^{r,\star} = Z^r  \mathring Z^\mytrans + W_t^\star \mathring Q_t \mathring W_t^\mytrans$ into \eqref{eq:thm2-4} yields 
\begin{align} \label{eq:pfd36-2}
    \epsilon_{it}^{r+1} &\leqslant \epsilon_{it}^r - \eta D_1 + \eta D_2 + \eta D_3 + \eta^2 D_4 \\[5pt]
    \text{with\ }\quad D_1 &= \frac{2}{n} \, e_i^\mytrans \big(  l^\prime(\mathring \Theta_t^{r,\star}) - l^\prime(\mathring \Theta_t^r)\big) \mathring W_t \big(W_{t}^r - W_t^\star \mathring Q_t\big)^\mytrans  e_i \notag\\
    D_2 &= \frac{2}{n}  \big| e_i^\mytrans \big(l^\prime(\Theta_t^\star) - l^\prime(\mathring \Theta_t^{r,\star})\big)\mathring W_t \big(W_{t}^r - W_t^\star \mathring Q_t\big)^\mytrans e_i\big| \notag\\
    D_3 &= \frac{2}{n}  \big| e_i^\mytrans l^\prime(\Theta_t^\star) \mathring W_t \big(W_{t}^r - W_t^\star \mathring Q_t\big)^\mytrans  e_i\big| \notag\\
    D_4 &= \frac{1}{n^2} \big\| e_i^\mytrans l^\prime(\mathring \Theta_t^r) \mathring W_t\big\|_2^2. \notag
\end{align}
Note that $ \big(W_{t}^r - W_t^\star \mathring Q_t\big) \mathring W_t^\mytrans =  \mathring \Theta_t^r -  \mathring \Theta_t^{r,\star}$. Then by Lemma \ref{adlem:nesterov2003},
\begin{align}  \label{eq:pfd36-3}
    D_1 &= \frac{2}{n} \Big\langle e_i^\mytrans \big( l^\prime(\mathring \Theta_t^{r,\star}) - l^\prime(\mathring \Theta_t^r)\big), e_i^\mytrans \big(\mathring \Theta_t^r -  \mathring \Theta_t^{r,\star}\big)\Big\rangle \notag\\
    &\geqslant \frac{\kappa_1 }{n} \big\|e_i^\mytrans \big(\mathring \Theta_t^r -  \mathring \Theta_t^{r,\star}\big)\big\|_2^2 + \frac{1}{\kappa_2 n} \big\|e_i^\mytrans \big( l^\prime(\mathring \Theta_t^r) -  l^\prime(\mathring \Theta_t^{r,\star})\big)\big\|_2^2.
\end{align}
For any sufficiently small constant $c_2 > 0$, 
\begin{align} \label{eq:pfd36-4}
    D_2 &= \frac{2}{n} \Big|\Big\langle e_i^\mytrans \big( l^\prime( \Theta_t^\star) -  l^\prime(\mathring \Theta_t^{r,\star})\big), e_i^\mytrans \big(\mathring \Theta_t^r -  \mathring \Theta_t^{r,\star}\big)\Big\rangle\Big| \notag\\
    &\leqslant \frac{2}{n} \big\|e_i^\mytrans \big( l^\prime( \Theta_t^\star) -  l^\prime(\mathring \Theta_t^{r,\star})\big)\big\|_2 \big\|e_i^\mytrans \big(\mathring \Theta_t^r -  \mathring \Theta_t^{r,\star}\big)\big\|_2 \notag \\
    &\leqslant \frac{c_2}{n}  \big\|e_i^\mytrans \big(\mathring \Theta_t^r -  \mathring \Theta_t^{r,\star}\big)\big\|_2^2 + \frac{1}{c_2 n } \big\|e_i^\mytrans \big( l^\prime( \Theta_t^\star) -  l^\prime(\mathring \Theta_t^{r,\star})\big)\big\|_2^2.
\end{align}
Combining \eqref{eq:pfd36-3} and \eqref{eq:pfd36-4} with \eqref{eq:thm2-R2C10} gives that, with probability $1-O((nT)^{-\varepsilon}) $,
\begin{align} \label{eq:pfd36-5}
    D_1 - D_2 &\geqslant \frac{M_2(\kappa_1 - c_2)}{2}\, \epsilon_{it}^r + \frac{1}{\kappa_2 n} \big\|e_i^\mytrans \big( l^\prime(\mathring \Theta_t^r) -  l^\prime(\mathring \Theta_t^{r,\star})\big)\big\|_2^2  -  \frac{1}{c_2 n } \big\|e_i^\mytrans \big( l^\prime( \Theta_t^\star) -  l^\prime(\mathring \Theta_t^{r,\star})\big)\big\|_2^2.
\end{align}
Moreover, for any sufficiently small constant $c_3 >0$, we have
\begin{align} \label{eq:pfd36-6}
    D_3 
    \leqslant \frac{2}{n}  \big\| e_i^\mytrans l^\prime(\Theta_t^\star) \mathring W_t \big\|_{2} \big\|e_i^\mytrans\big(W_t^r - W_t^\star \mathring Q_t\big)\big\|_2 
    \leqslant {c_3}\, \epsilon_{it}^r + \frac{1}{c_3 n^2} \big\|e_i^\mytrans l^\prime(\Theta_t^\star)   \mathring W_t\big\|_2^2.
\end{align}
Finally, applying the decomposition \eqref{eq:pfd36-1} gives
\begin{align}\label{eq:pfd36-7}
    D_4 
    &\leqslant \frac{3}{n^2} \big\|e_i^\mytrans\big(l^\prime(\mathring \Theta_t^r) - l^\prime(\mathring \Theta_t^{r,\star})\big) \mathring W_t \big\|_2^2 \notag + \frac{3}{n^2}\big\|e_i^\mytrans\big(l^\prime(\Theta_t^\star) - l^\prime(\mathring \Theta_t^{r,\star})\big) \mathring W_t\big\|_2^2  + \frac{3}{n^2}  \big\|e_i^\mytrans l^\prime(\Theta_t^\star)\mathring W_t \big\|_2^2 \notag\\
    &\leqslant \frac{9M_1^2}{n}   \big\|e_i^\mytrans\big(l^\prime(\mathring \Theta_t^r) - l^\prime(\mathring \Theta_t^{r,\star})\big)\big\|_2^2 + \frac{9M_1^2}{n} \big\|e_i^\mytrans\big(l^\prime(\Theta_t^\star) - l^\prime(\mathring \Theta_t^{r,\star})\big)\big\|_2^2 + \frac{3}{n^2} \big\|e_i^\mytrans l^\prime(\Theta_t^\star) \mathring W_t\big\|_2^2,
\end{align}
where the last inequality follows from \eqref{eq:thm2-0}. Combining \eqref{eq:pfd36-2} with \eqref{eq:pfd36-5}, \eqref{eq:pfd36-6}, and \eqref{eq:pfd36-7} shows that, with probability $1-O((nT)^{-\varepsilon}) $,
\begin{align} \label{eq:pfd36-8}
    \epsilon_{it}^{r+1} &\leqslant \epsilon_{it}^r - \eta\Big(\frac{M_2 (\kappa_1 - c_2)}{2} - c_3 \Big)\epsilon_{it}^r - \eta \Big(\frac{1}{\kappa_2 n} - \frac{9M_1^2 \eta}{n}\Big) \big\|e_i^\mytrans\big(l^\prime(\mathring \Theta_t^r) - l^\prime(\mathring \Theta_t^{r,\star})\big)\big\|_2^2 \notag\\
    & + \eta \Big(\frac{1}{c_2n} + \frac{9M_1^2 \eta}{n}\Big) \big\|e_i^\mytrans\big(l^\prime(\Theta_t^\star) - l^\prime(\mathring \Theta_t^{r,\star})\big)\big\|_2^2 
     + \eta \Big(\frac{1}{c_3 n^2} + \frac{3\eta}{n^2}\Big)  \big\|e_i^\mytrans l^\prime(\Theta_t^\star) \mathring W_t\big\|_2^2.
\end{align}
Since $c_2, c_3$, and $\eta$ are sufficiently small constants, \eqref{eq:pfd36-8} indicates that there exist positive constants $C$ and $c$ such that, with probability $1-O((nT)^{-\varepsilon}) $,
\begin{align} \label{eq:pfd36-9}
    \epsilon_{it}^{r+1} \leqslant (1-c) \epsilon_{it}^r + C \Big( \frac{1}{n} \big\|e_i^\mytrans\big(l^\prime(\Theta_t^\star) - l^\prime(\mathring \Theta_t^{r,\star})\big)\big\|_2^2 + \frac{1}{n^2}  \big\|e_i^\mytrans l^\prime(\Theta_t^\star) \mathring W_t\big\|_2^2  \Big).
\end{align}
Combining \eqref{eq:pfd36-9} with 
\begin{equation} \label{eq:pfd36-10}
    \begin{aligned}
    \big\|e_i^\mytrans\big(l^\prime(\Theta_t^\star) - l^\prime(\mathring \Theta_t^{r,\star})\big)\big\|_2^2 &\leqslant 2\big\|e_i^\mytrans\big(l^\prime(\Theta_t^\star) - l^\prime(\mathring \Theta_t^{\star})\big)\big\|_2^2 + 2\big\|e_i^\mytrans\big(l^\prime(\mathring \Theta_t^\star) - l^\prime(\mathring \Theta_t^{r,\star})\big)\big\|_2^2 \\
    &\leqslant 2\big\|e_i^\mytrans\big(l^\prime(\Theta_t^\star) - l^\prime(\mathring \Theta_t^{\star})\big)\big\|_2^2 + 6 M_1^2 \kappa_2^2 n\big\|e_i^\mytrans\big(Z^r - Z^\star \mathring Q\big)\big\|_2^2
\end{aligned}
\end{equation}
gives \eqref{eq:iterateerr2}.

\subsubsection{Formula of \eqref{eq:newtononev}} \label{sec:newtonform}



To obtain \eqref{eq:newtononev}, it suffices to derive analytical formulae of $\dot \ell(v)$ and $I(v)$, given in \eqref{eq:ldotform} and \eqref{eq:ivform} below, respectively. 
Meanwhile, we point out that the Hessian matrix of $\ell(v)$ is closely related to $I(v)$ and will also be used in the following proofs, so we also derive its analytical formula on Page \pageref{sec:hvform}. 


\paragraph{Formula of $\dot \ell(v)$.}\label{sec:formldot}
Let
\begin{align*}
    v_z = [z_1^\mytrans, z_2^\mytrans, \ldots,  z_n^\mytrans]^\mytrans \in \mathbb R^{nk \times 1} \quad \text{and} \quad v_{wt} = [w_{t,1}^\mytrans, w_{t,2}^\mytrans, \ldots, w_{t,n}^\mytrans]^\mytrans \in \mathbb R^{nk_t \times 1} 
\end{align*}
denote the vectorizations of $Z$ and $W_t$, respectively. Then $\dot \ell(v)$ can be written as  
\begin{align} \label{eq:pre1}
    \dot \ell(v) = \frac{\partial \ell(v)}{\partial v} = \begin{pmatrix}
        \frac{\partial \ell(v)}{\partial v_z}^\mytrans & \frac{\partial \ell(v)}{\partial v_{w1}}^\mytrans & \ldots & \frac{\partial \ell(v)}{\partial v_{wT}}^\mytrans
    \end{pmatrix}^\mytrans \in \mathbb R^{n(k+\ksum) \times 1}.
\end{align}
By the chain rule of the first-order derivatives, we have
\begin{align} \label{eq:pre2}
     \frac{\partial \ell (v)}{ \partial v_z} = \sum_{t=1}^T\frac{\partial \Theta_{t,v}^\mytrans}{\partial v_z} \,\frac{\partial \ell (v)}{\partial \Theta_{t,v}} \in \mathbb R^{n k \times 1} \quad \text{and} \quad \frac{\partial \ell (v)}{ \partial v_{wt}} = \frac{\partial \Theta_{t,v}^\mytrans}{\partial v_{wt}} \,\frac{\partial \ell (v)}{\partial \Theta_{t,v}} \in \mathbb R^{n k_t \times 1},
\end{align}
where $\Theta_{t,v} = (\Theta_{t,11}/\sqrt 2, \ldots , \Theta_{t,nn}/\sqrt 2, \Theta_{t,12}, \Theta_{t,13}, \ldots, \Theta_{t,n-1,n}) \in \mathbb R^{n(n+1)/2 \times 1}$ denote the vectorization of $\Theta_t$ {(with diagonal entries scaled to facilitate the proof)}.  For the simplicity of notation, we define 
\begin{align} \label{eq:pre3}
    \dot \ell_{\Theta_t}(v) = \begin{pmatrix}
        \dot \ell_{\Theta_{t,11}}(v) \\
        \vdots \\
        \dot \ell_{\Theta_{t,nn}}(v) \\
        \dot \ell_{\Theta_{t,12}}(v) \\
        \vdots \\
        \dot \ell_{\Theta_{t,n-1,n}}(v)
    \end{pmatrix} := \frac{\partial \ell (v)}{\partial \Theta_{t,v}} = \begin{pmatrix}
        \frac{\partial \ell (v)}{\partial (\Theta_{t,11}/\sqrt{2})} \\
        \vdots \\
        \frac{\partial \ell (v)}{\partial (\Theta_{t,nn}/\sqrt{2})} \\[6pt]
        \frac{\partial \ell (v)}{\partial \Theta_{t,12}} \\
        \vdots \\
        \frac{\partial \ell (v)}{\partial \Theta_{t,n-1,n}} 
    \end{pmatrix} = \begin{pmatrix}
        \sqrt{2} l^{\prime}(\Theta_{t,11}) \\
        \vdots \\
        \sqrt{2} l^{\prime}(\Theta_{t,nn}) \\[6pt]
        l^{\prime}(\Theta_{t,12}) \\
        \vdots\\
        l^{\prime}(\Theta_{t,n-1,n}) \\
    \end{pmatrix} \in \mathbb R^{\frac{n(n+1)}{2} \times 1}, 
\end{align}
\begin{align} \label{eq:pre4}
    \DZ (v) &= \begin{pmatrix}
        \mathcal{D}_{Z \Theta_{11}}(v) & \ldots & \mathcal{D}_{Z \Theta_{nn}}(v) & \mathcal{D}_{Z \Theta_{12}}(v) & \ldots &
        \mathcal{D}_{Z \Theta_{n-1,n}}(v)\end{pmatrix} := \frac{\partial \Theta_{t,v}^\mytrans}{\partial v_z }\\
        &= \begin{pmatrix}
        \frac{\partial (\Theta_{t,11}/\sqrt{2})}{ \partial z_{1}}& \ldots &   \frac{\partial (\Theta_{t,nn}/\sqrt{2})}{ \partial z_{1}} &  \frac{\partial \Theta_{t,12}}{ \partial z_{1}} & \ldots & \frac{\partial \Theta_{t,n-1,n}}{ \partial z_{1}} \\
        \vdots & \ldots & \vdots & \vdots & \ldots & \vdots \\
        \frac{\partial (\Theta_{t,11}/\sqrt{2})}{ \partial z_{n}}& \ldots &   \frac{\partial (\Theta_{t,nn}/\sqrt{2})}{ \partial z_{n}} &  \frac{\partial \Theta_{t,12}}{ \partial z_{n}} & \ldots & \frac{\partial \Theta_{t,n-1,n}}{ \partial z_{n}}
    \end{pmatrix} \notag\\
    &= \begin{pmatrix}
        \sqrt{2} z_{1} & \ldots & 0 & z_{2} & z_{3} & \ldots & z_{n} & 0& \ldots &0 \\
        0 & \ldots & 0 & z_{1} & 0 & \ldots & 0 & z_{3}& \ldots &0 \\
        0 & \ldots & 0 & 0 & z_{1} & \ldots & 0 & z_{2}& \ldots &0 \\
        \vdots & \ddots & \vdots & \vdots & \vdots & \ddots & \vdots & \vdots& \ddots &\vdots\\
        0 & \ldots & \sqrt{2} z_{n} & 0 & 0 & \ldots & z_{1} & 0& \ldots & z_{n-1} \\
    \end{pmatrix} \in \mathbb R^{nk \times \frac{n(n+1)}{2}}, \notag
\end{align}
where we point out that ${\partial \Theta_{t,v}^\mytrans}/{\partial v_z } $ has the same expression  across $t$, and thus the notation $ \DZ (v)$ omits the explicit dependence on $t$ for simplicity.  
Similarly, we define
\begin{align} \label{eq:pre5}
    \Dwt (v) &= \begin{pmatrix}
        \mathcal{D}_{W_t \Theta_{11}}(v) & \ldots & \mathcal{D}_{W_t \Theta_{nn}}(v) & \mathcal{D}_{W_t \Theta_{12}}(v) & \ldots &
        \mathcal{D}_{W_t \Theta_{n-1,n}}(v)\end{pmatrix} := \frac{\partial \Theta_{t,v}^\mytrans}{\partial v_{wt} } \\
        &= \begin{pmatrix}
        \frac{\partial (\Theta_{t,11}/\sqrt{2})}{ \partial w_{t,1}}& \ldots &   \frac{\partial (\Theta_{t,nn}/\sqrt{2})}{ \partial w_{t,1}} &  \frac{\partial \Theta_{t,12}}{ \partial w_{t,1}} & \ldots & \frac{\partial \Theta_{t,n-1,n}}{ \partial w_{t,1}} \\
        \vdots & \ldots & \vdots & \vdots & \ldots & \vdots \\
        \frac{\partial (\Theta_{t,11}/\sqrt{2})}{ \partial w_{t,n}}& \ldots &   \frac{\partial (\Theta_{t,nn}/\sqrt{2})}{ \partial w_{t,n}} &  \frac{\partial \Theta_{t,12}}{ \partial w_{t,n}} & \ldots & \frac{\partial \Theta_{t,n-1,n}}{ \partial w_{t,n}}
    \end{pmatrix} \notag\\
    &= \begin{pmatrix}
        \sqrt{2} w_{t,1} & \ldots & 0 & w_{t,2} & w_{t,3} & \ldots & w_{t,n} & 0& \ldots &0 \\
        0 & \ldots & 0 & w_{t,1} & 0 & \ldots & 0 & w_{t,3}& \ldots &0 \\
        0 & \ldots & 0 & 0 & w_{t,1} & \ldots & 0 & w_{t,2}& \ldots &0 \\
        \vdots & \ddots & \vdots & \vdots & \vdots & \ddots & \vdots & \vdots& \ddots &\vdots\\
        0 & \ldots & \sqrt{2} w_{t,n} & 0 & 0 & \ldots & w_{t,1} & 0& \ldots & w_{t,n-1} \\
    \end{pmatrix} \in \mathbb R^{nk_t \times \frac{n(n+1)}{2}}. \notag
\end{align}
Combining \eqref{eq:pre1}--\eqref{eq:pre5}, we can write
\begin{align}\label{eq:ldotform}
    \dot \ell(v) = \begin{pmatrix}
        \sum_{t=1}^T \DZ(v) \dot \ell_{\Theta_t}(v) \\
        \mathcal{D}_{W_1 \Theta}(v) \dot \ell_{\Theta_1}(v) \\
        \vdots \\
        \mathcal{D}_{W_T \Theta}(v) \dot \ell_{\Theta_T}(v)
    \end{pmatrix} \in \mathbb R^{n(k + \ksum) \times 1}.
\end{align}

\paragraph{Formula of $I(v)$.}\label{sec:formIv}

To facilitate the proof below, we further define 
\begin{align}
\DZt(v) := &~ \DZ^\mytrans(v) \in \mathbb R^{n(n+1)/2 \times nk}, \notag\\
\Dwtt(v) := &~  \Dwt^\mytrans(v) \in \mathbb R^{n(n+1)/2 \times nk_t},\notag\\
    \Dmut(v) := &~  \EXPT\big\{\dot \ell_{\Theta_t}(v) \dot \ell_{\Theta_t}(v)^\mytrans\big\} \in \mathbb R^{n(n+1)/2 \times n(n+1)/2 } \notag\\
    =&~ - \operatorname{diag}\big(2 l^{\prime \prime}(\Theta_{t,11}), \ldots, 2 l^{\prime \prime}(\Theta_{t,nn}),  l^{\prime \prime}(\Theta_{t,12}), \ldots, l^{\prime \prime}(\Theta_{t,n-1,n})\big),  \label{eq:dmutl}
\end{align}
where $\Dmut(v)$ is independent with $\mathbf{A}_t$ given fixed $v$ by Remark \ref{rm:lprimenotation}.  
By \eqref{eq:ldotform},  $I(v) = \EXPT\big\{\dot \ell(v) \dot \ell(v)^\mytrans\big\}=[I_{qm}(v)]_{q,m\in\{Z,W_1,\ldots,W_T\}}$ is a $(1+T)\times (1+T)$ block matrix  
with each block defined as
\begin{align*}
  I_{ZZ}(v) &:=  \EXPT\Big[ \big(\textstyle \sum_{t=1}^T \DZ(v) \dot \ell_{\Theta_t}(v)\big) \big(\textstyle \sum_{t=1}^T \DZ(v) \dot \ell_{\Theta_t}(v)\big)^\mytrans\Big] = \textstyle \sum_{t=1}^T \DZ(v)  \Dmut(v)  \DZt(v), \\
 I_{ZW_s}(v) &:=   \EXPT\Big[ \big(\textstyle \sum_{t=1}^T \DZ(v) \dot \ell_{\Theta_t}(v)\big) \big(\textstyle  \mathcal{D}_{W_s \Theta}(v) \dot \ell_{\Theta_s}(v)\big)^\mytrans\Big] =  \DZ(v)  \mathcal{D}_{\mu_s}(v)  \mathcal{D}_{\Theta W_s}(v), \\
 I_{W_sZ}(v) &:= \EXPT\Big[ \big(\textstyle  \mathcal{D}_{W_s \Theta}(v) \dot \ell_{\Theta_s}(v)\big)\big(\textstyle \sum_{t=1}^T \DZ(v) \dot \ell_{\Theta_t}(v)\big)^\mytrans\Big] =  \mathcal{D}_{ W_s \Theta}(v)  \mathcal{D}_{\mu_s}(v)  \DZt(v),\\
  I_{W_tW_s}(v) &:=   \EXPT\Big[ \big( \Dwt(v) \dot \ell_{\Theta_t}(v)\big) \big(\textstyle  \mathcal{D}_{W_s \Theta}(v) \dot \ell_{\Theta_s}(v)\big)^\mytrans\Big] =  \Dwt(v)  \Dmut(v)  \Dwtt(v) \mathbb I(t = s), 
\end{align*}
which are computed by $\EXPT\big\{\dot \ell_{\Theta_t}(v) \dot \ell_{\Theta_s}(v)^\mytrans\big\} = 0$ for any $1\leqslant t \neq s \leqslant T$. Specifically, we obtain  
\begin{small}
\begin{align} \label{eq:ivform}
 I(v)=  
  \begin{pmatrix}
        \sum_{t=1}^T \DZ(v) \Dmut(v) \DZt(v) &  \DZ(v) \mathcal{D}_{\mu_1}(v) \mathcal{D}_{\Theta W_1}(v) & \ldots & \DZ(v) \mathcal{D}_{\mu_T}(v)\mathcal{D}_{\Theta W_T}(v) \\
        \mathcal{D}_{W_1 \Theta}(v) \mathcal{D}_{\mu_1}(v)  \DZt(v) & \mathcal{D}_{W_1 \Theta}(v) \mathcal{D}_{\mu_1}(v)  \mathcal{D}_{\Theta W_1}(v) & \ldots & 0 \\
        \vdots & \vdots & \ddots & \vdots \\
        \mathcal{D}_{W_T \Theta}(v) \mathcal{D}_{\mu_T}(v)  \DZt(v) & 0 & \ldots & \mathcal{D}_{W_T \Theta}(v) \mathcal{D}_{\mu_T}(v)  \mathcal{D}_{\Theta W_T}(v) \end{pmatrix}.
\end{align}
\end{small}


\paragraph{Formula of $H(v)$.} \label{sec:hvform}
In the following analysis, we also need the analytical expression of $H(v)$,  the Hessian matrix of $\ell(v)$. 
In particular,
\begin{align}\label{eq:hvform}
    H(v) := \frac{\partial^2 \ell(v)}{\partial v \partial v^\mytrans} = [H_{qm}(v)]_{ q,m\in \{Z,W_1,\ldots, W_T\}} 
    \in \mathbb R^{n(k + \ksum) \times n(k + \ksum)}
\end{align}
is a $(1+T)\times (1+T)$ block matrix with each block defined as
\begin{align}\label{eq:hvform_blocks}
    &H_{ZZ}(v)=\frac{\partial^2 \ell (v)}{\partial v_z \partial v_z^{\top}}, \quad \  H_{ZW_t}(v) = H_{W_tZ}(v)^{\top} =   \frac{\partial^2 \ell (v)}{\partial v_z \partial v_{wt}^{\top}}, \quad \  
    H_{W_tW_{s}}(v) =  \frac{\partial^2 \ell (v)}{\partial v_{wt} \partial v_{ws}^{\top}} 
\end{align}
for $1\leqslant t, s\leqslant T$. 
By the chain rule, we have
\begin{align*}
    \frac{\partial  \dot \ell_{\Theta_t}(v)}{\partial v_z^\mytrans} &= \frac{\partial  \dot \ell_{\Theta_{t}}(v)}{\partial \Theta_{t,v}^\mytrans} \frac{\partial \Theta_{t,v}}{\partial v_z^\mytrans} = -\Dmut(v) \DZt(v), \\
    \text{and }\quad \frac{\partial  \dot \ell_{\Theta_t}(v)}{\partial v_{wt}^\mytrans} &= \frac{\partial  \dot \ell_{\Theta_{t}}(v)}{\partial \Theta_{t,v}^\mytrans} \frac{\partial \Theta_{t,v}}{\partial v_{wt}^\mytrans} = -\Dmut(v) \Dwtt(v). 
\end{align*}
Combining the above results and  \eqref{eq:pre2}, we obtain 
\begin{equation}
    \label{eq:hvform_4blocks}
\begin{aligned}
   H_{ZZ}(v)&= \sum_{t=1}^T  \frac{\partial [\DZ(v) \dot \ell_{\Theta_t}(v)]}{\partial v_z^\mytrans} = - \sum_{t=1}^T  \DZ(v) \Dmut(v) \DZt (v) + \sum_{t=1}^T \sum_{1 \leqslant i \leqslant j \leqslant n}\frac{\partial \mathcal{D}_{Z\Theta_{ij}}(v) }{\partial v_z^\mytrans}  \dot \ell_{\Theta_{t,ij}}(v),\\
 H_{ZW_t}(v)  &=   \frac{\partial [ \DZ(v) \dot \ell_{\Theta_t}(v)]}{\partial v_{wt}^\mytrans} = - \DZ(v) \Dmut(v) \Dwtt (v), \\
  H_{W_tW_t}(v)  &=  \frac{\partial [\Dwt(v) \dot \ell_{\Theta_t}(v)]}{\partial v_{wt}^\mytrans} = - \Dwt(v) \Dmut(v) \Dwtt (v) + \sum_{1 \leqslant i \leqslant j \leqslant n}\frac{\partial \mathcal{D}_{W_t \Theta_{ij}}(v) }{\partial v_{wt}^\mytrans}  \dot \ell_{\Theta_{t,ij}}(v),\\
  H_{W_tW_s}(v) & =\frac{\partial [ \Dwt(v) \dot \ell_{\Theta_{t}}(v)]}{\partial v_{ws}^\mytrans}  = 0, 
\end{aligned}
\end{equation}
for $1\leqslant t\neq s\leqslant T$,  
where we use ${\partial \DZ(v)}/{\partial v_{wt}} =0$, $ {\partial \Dwt(v)}/{\partial v_{z}}  = 0$,  $ {\partial \Dwt(v)}/{\partial v_{ws}}  = 0$, and ${\partial \dot{\ell}_{\Theta_t}(v)}/{\partial v_{ws}} =0$, 
by the analytical forms in \eqref{eq:pre3}, \eqref{eq:pre4}, and \eqref{eq:pre5}.
Comparing with the form of $I(v)$ in \eqref{eq:ivform} and by the form of $\dot{\ell}_{\Theta_{t,ij}}(v)$, $\mathcal{D}_{Z\Theta_{ij}}(v)$, and $\mathcal{D}_{W_t \Theta_{ij}}(v)$ in \eqref{eq:pre3}, \eqref{eq:pre4}, and \eqref{eq:pre5}, we can write
\begin{align} \label{eq:hv}
    H(v) = - I(v) + N(v) 
\end{align}
with 
\begin{equation}
\label{eq:def_Nv_Ntv}
\begin{aligned}
N(v)  &= \operatorname{diag}\big(\textstyle \sum_{t=1}^T N_t(v) \otimes \mathrm{I}_k, N_1(v) \otimes \mathrm{I}_{k_1}, \ldots, N_T(v) \otimes \mathrm{I}_{k_T}\big), \\
    N_t(v) &= \big[\omega_{ij} \, l^{\prime}(\Theta_{t,ij}) \big]_{1\leqslant i,j\leqslant n}  \in \mathbb R^{n \times n},\quad \text{with}\quad \omega_{ij}=\begin{cases}
    2 & i=j\\
    1 & i\neq j
\end{cases}.
\end{aligned}
\end{equation}

\subsubsection{Properties of $I(v)$} \label{sec:proplemmaiv}

To facilitate the subsequent proofs, we next establish several useful properties on $I(v)$. 

\begin{lemma} \label{lm:ivproperties}
Assume true latent vectors  $\{Z^{\star}, W_1^{\star},\ldots, W_T^{\star}\}$ satisfy  Conditions \ref{cond:truevalue}--\ref{cond:truevalue2}, and let    $v^{\star}$ denote its vectorization similarly to  \eqref{eq:vectorization}.  
Consider another  set of fixed latent vectors $\{Z, W_1,\ldots, W_T\}$ with its vectorization denoted as $v$  following  \eqref{eq:vectorization}. 
Assume 
\begin{itemize}
\setlength{\itemsep}{0pt}
    \item[(a)]    $\|Z\|_{2 \to \infty} \leqslant M_1$, $\|W_t\|_{2 \to \infty} \leqslant M_1$ for $1 \leqslant t \leqslant T$;  
    \item[(b)] there exists a constant $\epsilon$ such that $\|v - v_q^\star\|_{\infty} = O(n^{-1/2} \log^{\epsilon}(nT))$, where $v_q^\star =  Q_{nv}^\mytrans v^\star $ with $Q_{nv} = \operatorname{diag}(\mathrm{I}_n \otimes Q_z, \mathrm{I}_n \otimes Q_{w1}, \ldots, \mathrm{I}_n \otimes Q_{wT})$,
\begin{align*}
    Q_z = \argmin_{Q \in \mathcal{O}(k)} \big\|Z - Z^\star Q\big\|_{\mathrm{F}}^2, \quad \text{and} \quad Q_{wt} = \argmin_{Q \in \mathcal{O}(k_t)} \big\|W_t - W_t^\star Q\big\|_{\mathrm{F}}^2.
\end{align*}
\end{itemize}
 Then for any $u \in [0,1]$, $v_u =  {v} + u(v_{q}^\star -  v)$ satisfies the following properties when $ \log^{2\epsilon}(nT)/n$ is sufficiently small.   
 \begin{enumerate}
        \item[(i)] $\operatorname{rank
        }(I(v)) = \mathrm{r}_{n,k} + \sum_{t= 1}^T\mathrm{r}_{n,k_t}$ with  $\mathrm{r}_{n,k} = nk - k(k-1)/2$, 
        \item[(ii)] $v_u \in \operatorname{col}(I(v))$,
        \item[(iii)] $\big\|\DnT^{1/2}\, I(v_u)^+ \DnT^{1/2}\big\|_{\operatorname{op}} = O(1)$ with $\DnT = \operatorname{diag}(nT \mathrm{I}_{nk}, n \mathrm{I}_{n\ksum})$,
        \item[(iv)] 
        $\big\| \DnT^{-1/2} \big( I({v}) - I(v_u) \big) \DnT^{-1/2}\big\|_{\operatorname{op}} = O(n^{-1/2} \log^\epsilon(nT))$,
        \item[(v)]$ \big\| \DnT^{-1/2} \big( N(v_q^\star) - N(v_u) \big) \DnT^{-1/2}\big\|_{\operatorname{op}} = O(n^{-1/2} \log^\epsilon(nT))$,
       \item[(vi)] $\operatorname{col}(I(v_u)) = \operatorname{col}(I_d(v_u))$ for $I_d(v) = \operatorname{diag}(I_{ZZ}(v), I_{W_1 W_1}(v), \ldots, I_{W_T W_T}(v))$. 
    \end{enumerate}

\end{lemma}

\begin{remark}  \label{lm:ivpropertiescheck}
Consider the initial estimates $\{\check{Z}, \check{W}_1, \ldots, \check{W}_T\}$ satisfying \eqref{eq:thm2-1}. 
Let $\check{v}$ denote its vectorization   similarly to  \eqref{eq:vectorization}.  The above properties (i)--(vi) hold for  $v = \check{v}$ and $\epsilon = 2$ with probability greater than   $1-O((nT)^{-\varepsilon})$. 
\end{remark}

We next prove Lemma \ref{lm:ivproperties} and Remark \ref{lm:ivpropertiescheck} on Pages \pageref{sec:lemmaiv1} and \pageref{sec:ivpropertiescheck}, respectively.

\paragraph{Proof of (i)--(iii).} \label{sec:lemmaiv1}

To establish properties on $I(v)$, we define intermediate matrices 
\begin{align} \label{eq:xiv} 
   \Xi(v) = \begin{pmatrix} 
         T \DZ(v) \DZt(v) &  \DZ(v)  \mathcal{D}_{\Theta W_1}(v) & \ldots & \DZ(v) \mathcal{D}_{\Theta W_T}(v) \\
        \mathcal{D}_{W_1 \Theta}(v)   \DZt(v) & \mathcal{D}_{W_1 \Theta}(v) \mathcal{D}_{\Theta W_1}(v) & \ldots & 0 \\
        \vdots & \vdots & \ddots & \vdots \\
        \mathcal{D}_{W_T \Theta}(v) \DZt(v) & 0 & \ldots & \mathcal{D}_{W_T \Theta}(v)  \mathcal{D}_{\Theta W_T}(v) \end{pmatrix},
\end{align} 
and 
\begin{align}\label{eq:xidv} 
   \Xi_d(v) = \begin{pmatrix}
         T \DZ(v) \DZt(v) &  0 & \ldots & 0 \\
        0 & \mathcal{D}_{W_1 \Theta}(v) \mathcal{D}_{\Theta W_1}(v) & \ldots & 0 \\
        \vdots & \vdots & \ddots & \vdots \\
        0 & 0 & \ldots & \mathcal{D}_{W_T \Theta}(v)  \mathcal{D}_{\Theta W_T}(v) \end{pmatrix},
\end{align}  
which contains the diagonal blocks of $\Xi(v)$. 
{To prove Lemma \ref{lm:ivproperties} (i)--(iii), we first study the properties of $\Xi_d(v)$ and then examine the difference $\Xi(v)-\Xi_d(v)$ to reach conclusions on $I(v)$ based on its  formula in \eqref{eq:ivform}. }

\bigskip 
\noindent {\textit{Step 1: Properties of $\Xi_d(v)$.}}
Lemma \ref{adlem:he2023} allows us to construct a matrix $\UZ(v) \in \mathbb R^{nk \times \mathrm{r}_{n,k}}$ with column vectors in $\mathcal{B}_{C,1} \cup \mathcal{B}_{C,2} \cup \mathcal{B}_{C,3}$, satisfying
\begin{align*}
     \operatorname{rank}(\UZ(v)) = \mathrm{r}_{n,k}, \quad \quad \quad v_z \in \operatorname{col}(\DZ(v)\DZt(v)) = \operatorname{col}(\mathcal{U}_Z(v))&,\\
     \text{ and }\quad\quad \quad\quad\quad\quad\sigma_{\operatorname{min}}^2(Z)\leqslant\sigma\big(\UZ(v)^\mytrans \DZ(v) \DZt(v) \UZ(v)\big) \leqslant 2 \|Z\|_{\operatorname{op}}^2&.
\end{align*} 
Combining with $\|Z\|_{\operatorname{op}}^2 \leqslant n\|Z\|_{2 \to \infty}^2 = M_1^2n$ and 
\begin{align*}
    \sigma_{\operatorname{min}}(Z) \geqslant \sigma_{\operatorname{min}}(Z^\star) - \|Z - Z^\star Q_z\|_{\operatorname{op}} \geqslant (M_2 n)^{1/2} - C \log^{\epsilon}(nT) \geqslant (0.99 M_2 n)^{1/2}
\end{align*}
by Condition \ref{cond:truevalue} (ii) and $\log^{2 \epsilon}(nT)/n$ is sufficiently small, we obtain
\begin{align} \label{eq:lem3-9}
0.99M_2n \leqslant\sigma\big(\UZ(v)^\mytrans \DZ(v) \DZt(v) \UZ(v)\big) \leqslant 2 M_1^2 n.
\end{align}
Furthermore, we demonstrate $(\mathrm{I}_n \otimes Q_z)^\mytrans v_z^\star \in \operatorname{col}(\mathcal{U}_Z(v))$ by checking $(\mathrm{I}_n \otimes Q_z)^\mytrans v_z^\star\  \bot\  \mathcal{B}_{N} $. Specifically, by the property of Kronecker product we have 
\begin{align*}
   v_z^{\star \mytrans} (\mathrm{I}_n \otimes Q_z) ( \lambda_i \beta_{ij} - \lambda_j \beta_{ji}) &= \lambda_i v_z^{\star \mytrans} [U_i \otimes (Q_z V_j)] - \lambda_j v_z^{\star \mytrans} [U_j \otimes (Q_z V_i)] \\
   &= \lambda_i U_i^\mytrans Z^\star Q_z V_j - \lambda_j U_j^\mytrans Z^\star Q_z V_i \\
   &= V_i^\mytrans Z^\mytrans Z^\star Q_z V_j - V_j^\mytrans Z^\mytrans Z^\star Q_z V_i = 0,
\end{align*}
where in the third equation, we used $\lambda_i U_i = Z V_i$, and in the last equation, we used the symmetricity of $Z^\mytrans Z^\star Q_z$, which follows from the property of the orthogonal Procrustes problem.

For each $W_t$,  
 we can obtain analogous results by noticing the similarity of the expressions of $\DZ(v)$ in \eqref{eq:pre4} and $\Dwt(v)$ in \eqref{eq:pre5}. 
 Then by the diagonality of $\Xi_d(v)$ in \eqref{eq:xidv}, we obtain
\begin{align}\label{eq:xidrank}
     \operatorname{rank}(\Xi_d(v)) = \mathrm{r}_{n,k} + \sum_{t= 1}^T\mathrm{r}_{n,k_t},\quad \quad
 v, v_q^\star\in  \operatorname{col}(\Xi_d(v)) = \operatorname{col}(\mathcal{U}(v)) 
\end{align}
with 
\begin{align} \label{eq:uvdef}
    \mathcal{U}(v) = \begin{pmatrix}
        \mathcal{U}_Z(v) & 0 & \ldots & 0\\
        0 & \mathcal{U}_{W_1}(v) & \ldots & 0\\
        \vdots & \vdots & \ddots & \vdots \\
        0 & 0 & \ldots & \mathcal{U}_{W_T}(v) 
    \end{pmatrix} \in \mathbb R^{n(k+\ksum) \times (\mathrm{r}_{n,k} + \sum_{t = 1}^T \mathrm{r}_{n,k_t})},
\end{align}
and 
\begin{align}\label{eq:xideigenvalbd}
   0.99M_2 \leqslant \sigma\big(\mathcal{U}(v)^\mytrans \DnT^{-1/2} \,\Xi_d(v) \DnT^{-1/2}\, \mathcal{U}(v) \big)\leqslant 2M_1^2.
\end{align}

\noindent {\textit{Step 2: Difference  $\Xi(v)-\Xi_d(v)$.}}
By \eqref{eq:xiv} and \eqref{eq:xidv} we have
\begin{align} \label{eq:lem3-1}
    \big\|\DnT^{-1/2}\,\big(\Xi(v) - \Xi_d(v)\big)\DnT^{-1/2} \big\|_{\operatorname{op}}^2 = \frac{1}{n^2T} \big\|\sum_{t = 1}^T \DZ(v) \Dwtt(v) \Dwt(v) \DZt(v)\big\|_{\operatorname{op}}.
\end{align}
For any $x \in \mathbb R^{nk}$,
\begin{align} \label{eq:lem3-2}
    &\sum_{t = 1}^T x^\mytrans \DZ(v) \Dwtt(v) \Dwt(v) \DZt(v) x = \sum_{t = 1}^T \big\|\Dwt(v) \DZt(v) x\big\|_2^2 \notag\\
    \leqslant~ &2 \sum_{t = 1}^T \Big(\big\|\Gamma_{t,1} x\big\|_2^2 + \big\|\Gamma_{t,2} x\big\|_2^2\Big) \leqslant 2  \Big(\big\|\sum_{t=1}^T\Gamma_{t,1}^\mytrans \Gamma_{t,1}\big\|_{\operatorname{op}} + \big\|\sum_{t=1}^T\Gamma_{t,2}^\mytrans\Gamma_{t,2}\big\|_{\operatorname{op}}\Big) \|x\|_2^2,
\end{align}
where we define 
    \begin{align*}
        \Gamma_{t, 1}=&~\left(\begin{array}{cccc}
\sum_{l=1}^n w_{t, l} z_{l}^{\top} & 0 & \cdots & 0 \\
0 & \sum_{l=1}^n w_{t, l} z_{l}^{\top} & \cdots & 0 \\
\vdots & \vdots & \ddots & \vdots \\
0 & 0 & \cdots & \sum_{l=1}^n w_{t,l} z_{l}^{\top}
\end{array}\right), \\
    \text{ and }\quad   \Gamma_{t, 2}= &~\left(\begin{array}{cccc}
w_{t, 1} z_1^{\top} & w_{t, 2} z_1^{\top} & \cdots & w_{t, n} z_1^{\top} \\
w_{t, 1} z_2^{\top} & w_{t, 2} z_2^{\top} & \cdots & w_{t, n} z_2^{\top} \\
\vdots & \vdots & \ddots & \vdots \\
w_{t, 1} z_n^{\top} & w_{t, 2} z_n^{\top} & \cdots & w_{t, n} z_n^{\top}
\end{array}\right).
    \end{align*}
    Note that $\Gamma_{t, 1}^{\top} \Gamma_{t, 1}=\mathrm{I}_n \otimes\left(Z^{\top} W_t W_t^{\top} Z\right)$ and $\Gamma_{t, 2}^{\top} \Gamma_{t, 2}=\left(W_t W_t^{\top}\right) \otimes\left(Z^{\top} Z\right)$. Then by Conditions \ref{cond:truevalue} and \ref{cond:truevalue2} we have
    \begin{align} 
        & \big\|\sum_{t=1}^T \Gamma_{t, 1}^{\top} \Gamma_{t, 1}\big\|_{\mathrm{op}}=\big\|\sum_{t=1}^T Z^{\top} W_t W_t^{\top} Z\big\|_{\mathrm{op}} \leqslant n M_1^2\big\|\sum_{t=1}^T W_t W_t^{\top}\big\|_{\mathrm{op}} \leqslant 1.5 n^2T M_1^2 M_5, \label{eq:lem3-3}\\
        & \big\|\sum_{t=1}^T \Gamma_{t, 2}^{\top} \Gamma_{t, 2}\big\|_{\mathrm{op}}=\big\|\sum_{t=1}^T  W_t W_t^{\top}\big\|_{\mathrm{op}}\big\|Z^{\top} Z\big\|_{\mathrm{op}} \leqslant n M_1^2\big\|\sum_{t=1}^T W_t W_t^{\top}\big\|_{\mathrm{op}} \leqslant 1.5 n^2T M_1^2 M_5, \label{eq:lem3-4}
    \end{align}
where we used
\begin{align*}
    \big\|\sum_{t=1}^T  W_t W_t^{\top}\big\|_{\mathrm{op}} &\leqslant \big\|\sum_{t=1}^T  W_t^\star W_t^{\star \top}\big\|_{\mathrm{op}} + \sum_{t = 1}^T \big\| W_t W_t^{\top} - W_t^\star W_t^{\star \top}\big\|_{\mathrm{F}}\\
    &\leqslant nT \big\|\mathcal{G}(W^\star)/T\big\|_{\operatorname{op}} + \sum_{t = 1}^T 3 \big\|W_t\big\|_{\operatorname{op}} \big\|W_t - W_t^\star Q_{wt}\big\|_{\mathrm{F}} \\
    & \leqslant M_5 nT + C n^{1/2} T \log^{\epsilon}(nT) \leqslant 1.5 M_5 nT
\end{align*}
by Lemma \ref{adlem:tu2016B}, Condition \ref{cond:truevalue2}, and $\log^{2\epsilon}(nT)/n$ is sufficiently small. 
Combining \eqref{eq:lem3-1} with \eqref{eq:lem3-2}, \eqref{eq:lem3-3}, and \eqref{eq:lem3-4}, we obtain
\begin{align} \label{eq:lem3-5}
    \big\|\DnT^{-1/2}\,\Xi(v)\DnT^{-1/2} - \DnT^{-1/2}\,\Xi_d(v)\DnT^{-1/2}\big\|_{\operatorname{op}}^2 \leqslant 6 M_1^2 M_5 = 3M_2^2/4.
\end{align}

\noindent {\textit{Step 3: Properties on $I(v)$.}} 
By the expressions of $I(v)$ and $\Xi_d(v)$, we have $\operatorname{null}(\Xi_d(v)) \subset \operatorname{null}(I(v))$, which implies $\operatorname{col}(I(v)) \subset \operatorname{col}(\Xi_d(v)) = \operatorname{col}(\mathcal{U}(v))$. 
We further demonstrate that {$\operatorname{col}(I(v)) = \operatorname{col}(\Xi_d(v))$} by showing $x^\mytrans I(v) x > 0$ for any $x \in \operatorname{col}(\Xi_d(v))$, which is sufficient to examine the eigenvalues of $\mathcal{U}(v)^\mytrans I(v) \mathcal{U}(v)$.
Combining \eqref{eq:xideigenvalbd} and \eqref{eq:lem3-5} we obtain
\begin{align} \label{eq:lem3-6}
    0.99 M_2 - \sqrt{3}M_2/2 \leqslant \sigma\big(\mathcal{U}(v)^\mytrans \DnT^{-1/2} \,\Xi(v) \DnT^{-1/2}\, \mathcal{U}(v) \big)\leqslant 2M_1^2 + \sqrt{3}M_2/2.
\end{align}
Further note that 
\begin{align} \label{eq:lem3-6.1}
    I(v) = \sum_{t=1}^T I_t(v) := \sum_{t=1}^T \mathcal{D}_{Y_t \Theta}(v) \Dmut(v) \mathcal{D}_{\Theta Y_t}(v) \ \  \text{and}\ \  \Xi(v) = \sum_{t=1}^T \Xi_t(v) = \sum_{t=1}^T \mathcal{D}_{Y_t \Theta}(v)  \mathcal{D}_{\Theta Y_t}(v)
\end{align}
with
\begin{align} \label{eq:lem3-6.2}
    \mathcal{D}_{\Theta Y_t }(v) = \begin{pmatrix}
      \DZt(v) &  0 & \ldots & \mathcal{D}_{\Theta W_t}(v) & \ldots & 0 
\end{pmatrix} \in \mathbb R^{\frac{n(n+1)}{2} \times n(k + \ksum)}  , \quad \mathcal{D}_{Y_t 
 \Theta }(v) = \mathcal{D}_{ 
 \Theta Y_t}^\mytrans(v).
\end{align}
Since $\kappa_1 \leqslant -l^{\prime \prime}(\Theta_{t,ij}) \leqslant \kappa_2$ by $\|Z\|_{2 \to \infty} \leqslant M_1$, $\|W_t\|_{2 \to \infty} \leqslant M_1$, and Condition \ref{cond:parfunction}, for any $x \in \operatorname{col}(\mathcal{U}(v))$ we have
\begin{align*}
    x^\mytrans I(v) x =  \sum_{t = 1}^T x^\mytrans I_t(v) x \asymp \sum_{t = 1}^T x^\mytrans \Xi_t(v) x = x^\mytrans \Xi(v) x,
\end{align*}
which indicates
\begin{align}  \label{eq:lem3-7}
    \sigma\big(\mathcal{U}(v)^\mytrans \DnT^{-1/2} \,I(v) \DnT^{-1/2}\, \mathcal{U}(v) \big) \asymp \sigma\big(\mathcal{U}(v)^\mytrans \DnT^{-1/2} \,\Xi(v) \DnT^{-1/2}\, \mathcal{U}(v) \big).
\end{align}
Combining \eqref{eq:lem3-6} and \eqref{eq:lem3-7} shows $\operatorname{col}(\Xi_d(v)) \subset \operatorname{col}(I(v))$. In summary, we obtain 
\begin{align} 
    v, v_q^\star \in \operatorname{col}(I(v)) = &~ \operatorname{col}(\Xi_d(v)) \label{eq:lem3-8-1}\\
\text{ and }\quad\quad\quad \big\|\DnT^{1/2} I(v)^+ \DnT^{1/2}\big\|_{\operatorname{op}} = &~ O(1). \label{eq:lem3-8-2} 
\end{align} 
As \eqref{eq:lem3-8-1} implies $\operatorname{rank}(I(v)) = \operatorname{rank}(\Xi_d(v)) $,  combined with \eqref{eq:xidrank}, (i) is obtained. 
Moreover, since $v_u$ is a convex  combination of $v$ and $v_q^\star$,  \eqref{eq:lem3-8-1} gives (ii). Also, as $v_u$ exhibits similar properties to $v$,  (iii) can be derived following the same proof as  \eqref{eq:lem3-8-2}.  


\paragraph{Proof of (iv).} We first decompose the error as
\begin{align} \label{eq:lem3-10}
 &\quad\ \big\| \DnT^{-1/2} \big( I({v}) - I(v_u) \big) \DnT^{-1/2}\big\|_{\operatorname{op}}   \notag\\
 &\leqslant \frac{1}{nT} \big\|I_{ZZ}(v) - I_{ZZ}(v_u)\big\|_{\operatorname{op}} + \frac{1}{n} \big\|I_{WW}(v) - I_{WW}(v_u)\big\|_{\operatorname{op}} 
 +\frac{2}{n T^{1/2}} \big\|I_{ZW}(v) - I_{ZW}(v_u)\big\|_{\operatorname{op}} \notag\\
 &\leqslant \frac{2}{n} \Big( \frac{1}{T} \big\|I_{ZZ}(v) - I_{ZZ}(v_u)\big\|_{\operatorname{op}} +  \max_{1 \leqslant t \leqslant T} \big\|I_{W_tW_t}(v) - I_{W_tW_t}(v_u)\big\|_{\operatorname{op}} 
 +\max_{1 \leqslant  t \leqslant T }\big\|I_{ZW_t}(v) - I_{ZW_t} (v_u)\big\|_{\operatorname{op}} \Big),
\end{align}
where the last inequality follows from the diagonality of $I_{WW}(v)$. Furthermore, we have 
\begin{equation} \label{eq:lem3-10.5}
    \begin{aligned}
    &\quad\ \big\|\DZ(v)\Dmut(v)\DZt(v) - \DZ(v_u)\Dmut(v_u)\DZt(v_u)\big\|_{\operatorname{op}} \\
    &\leqslant \big\|\DZ(v) - \DZ(v_u)\big\|_{\operatorname{op}} \big\|\Dmut(v)\big\|_{\operatorname{op}} \big\|\DZt(v)\big\|_{\operatorname{op}} \\
    &\quad+ \big\|\DZ(v_u)\big\|_{\operatorname{op}} \big\|\Dmut(v) - \Dmut(v_u)\big\|_{\operatorname{op}} \big\|\DZt(v)\big\|_{\operatorname{op}} \\
    &\quad+ \big\|\DZ(v_u)\big\|_{\operatorname{op}} \big\|\Dmut(v_u)\big\|_{\operatorname{op}} \big\|\DZt(v) - \DZt(v_u)\big\|_{\operatorname{op}}.
    \end{aligned}
\end{equation}
Combining with $\|\DZ(v_u)\|_{\operatorname{op}} = \|\DZt(v_u)\|_{\operatorname{op}} = O(n^{1/2})$, $\|\Dmut(v_u)\|_{\operatorname{op}} = O(1)$,
\begin{align*}
    \big\|\DZ(v) - \DZ(v_u)\big\|_{\operatorname{op}} &= \big\|\DZ(v - v_u) \big\|_{\operatorname{op}} \lesssim \big\|Z - Z^u\big\|_{\operatorname{op}} \\& \leqslant \big\|Z - Z^\star Q_z\big\|_{\operatorname{op}} \lesssim n^{1/2} \big\|v - v_q^\star\big\|_{\infty} = O(\log^\epsilon(nT)), \\
    \big\|\Dmut(v) - \Dmut(v_u)\big\|_{\operatorname{op}} &\leqslant \max_{1 \leqslant i \leqslant j \leqslant n} |l^{\prime \prime}(\Theta_{t,ij}) - l^{\prime \prime}(\Theta_{t,ij}^u)| \lesssim \max_{1 \leqslant i \leqslant j \leqslant n} |\Theta_{t,ij}-\Theta_{t,ij}^u| \\
    &\lesssim \|v\|_{\infty} \|v - v_u\|_{\infty} \leqslant\|v\|_{\infty} \|v - v_q^\star\|_{\infty} = O(n^{-1/2} \log^\epsilon(nT)),
\end{align*}
where $Z^u = Z + u(Z^\star Q_z - Z)$, $W_t^u = W_t + u (W_t^\star Q_{wt} - W_t)$, and $\Theta_t^u = Z^uZ^{u\mytrans} + W_t^uW_t^{u\mytrans}$. We obtain $\|\DZ(v)\Dmut(v)\DZt(v) - \DZ(v_u)\Dmut(v_u)\DZt(v_u)\|_{\operatorname{op}} = O(n^{1/2}  \log^\epsilon(nT))$ and then
\begin{align} \label{eq:lem3-11}
    \big\|I_{ZZ}(v) - I_{ZZ}(v_u)\big\|_{\operatorname{op}} = O(n^{1/2} T \log^\epsilon(nT)).
\end{align}
Similarly, we can obtain
\begin{align}
    \big\|I_{ZW_t}(v) - I_{ZW_t} (v_u)\big\|_{\operatorname{op}} &= O(n^{1/2} \log^\epsilon(nT)) \label{eq:lem3-12}\\  \text{and} \quad \quad\big\|I_{W_tW_t}(v) - I_{W_tW_t} (v_u)\big\|_{\operatorname{op}} &= O(n^{1/2} \log^\epsilon(nT)) \label{eq:lem3-13}
\end{align}
for any $1 \leqslant t \leqslant T$. Combining \eqref{eq:lem3-10} with \eqref{eq:lem3-11}--\eqref{eq:lem3-13} gives (iv).

\paragraph{Proof of (v).} For any $1 \leqslant t \leqslant T$, we have 
\begin{align*}
    \big\|N_t(v_q^\star) - N_t(v_u) \big\|_{\operatorname{op}} & \leqslant 2\, \big\|l^\prime(\Theta_t^u) - l^\prime(\Theta_t^\star) \big\|_{\mathrm{F}} \lesssim \big\|\Theta_t^u - \Theta_t^\star\big\|_{\mathrm{F}} \\
    &\lesssim \big\|Z^\star\big\|_{\operatorname{op}} \big\|Z^u - Z^\star Q_z\big\|_{\mathrm{F}} \lesssim n^{1/2} \big\|Z^\star\big\|_{\operatorname{op}}  \big\|v - v_q^\star\big\|_{\infty} = O(n^{1/2}\log^\epsilon(nT)),
\end{align*}
where the second inequality follows from Condition \ref{cond:parfunction} (i) and the third inequality follows from Lemma \ref{adlem:tu2016B}. Combining with the diagonality of $N(v)$ we obtain (v).

\paragraph{Proof of (vi).} We have $\operatorname{col}(I_d(v)) \subset \operatorname{col}(I(v))$ due to $\operatorname{null}(I(v)) = \operatorname{null}(\Xi_d(v)) \subset \operatorname{null}(I_d(v))$ by \eqref{eq:lem3-8-1} and the expression of $I_d(v)$. We next show $\operatorname{col}(I_d(v)) = \operatorname{col}(I(v)) = \operatorname{col}(\mathcal{U}(v))$ by examining the eigenvalues of $\mathcal{U}(v)^\mytrans I_d(v) \mathcal{U}(v)$. For any $x \in  \operatorname{col}(\mathcal{U}_Z(v))$, we have 
\begin{align*}
    x^\mytrans I_{ZZ}(v) x = 
    T x^\mytrans \DZ(v) \Dmut(v) \DZt(v) x \asymp T x^\mytrans \DZ(v) \DZt(v) x \asymp nT \|x\|_2^2
\end{align*}
by $ \kappa_1 \leqslant -l^{\prime \prime}(\Theta_{t,ij}) \leqslant \kappa_2$ and \eqref{eq:lem3-9}. Combining with the similar arguments for $I_{W_tW_t}(v)$ and the diagonality of $I_d(v)$, we obtain 
\begin{align}\label{eq:Iduvsigma}
    \sigma\big(\mathcal{U}(v)^\mytrans \DnT^{-1/2} I_d(v) \DnT^{-1/2}\, \mathcal{U}(v)\big) \asymp 1.
\end{align}

\paragraph{Proof of Remark \ref{lm:ivpropertiescheck}.} \label{sec:ivpropertiescheck}
It suffices to show that assumption  (a) and (b) in Lemma \ref{lm:ivproperties} are satisfied by $\{\check{Z},\check{W}_1,\ldots, \check{W}_T\}$. (a) We know $\|\check Z \|_{2 \to \infty} \leqslant M_1$ and  $\|\check{W}_t\|_{2 \to \infty} \leqslant M_1$ by the projection step in Algorithm \ref{algor:refine}. 
(b) 
As $k$ and $k_t$'s are fixed,  $$  \big\|\check{v} - v_q^\star\big\|_{\infty}^2 \lesssim   \big\| \check Z - Z^{\star} \check Q\big\|_{2 \to \infty}^2 +  \max_{1\leqslant t \leqslant T} \big\|\check W_t - W_t^{\star} \check Q_t\big\|_{2 \to \infty}^2.$$ 
Therefore, by \eqref{eq:thm2-1},  $\Pr\big[\|\check{v} - v_q^\star\|_{\infty} > C_\varepsilon n^{-1/2} \log^2(nT)\big] = O((nT)^{-\varepsilon})$. 


\subsubsection{Proof of \eqref{eq:zhaterr}} \label{sec:pdfzhaterr}


By the definition of distance measure, $\operatorname{dist}^2(\hat Z,Z^\star ) \leqslant \big\|\hat Z - Z^\star \check Q\big\|_{\mathrm{F}}^2$. 
By our construction, 
$\hat{Z}$ is a matricization of $\mathcal{D}_{Zv} \hat{v} $ where 
\begin{align}\label{eq:dzvdef}
    \mathcal{D}_{Zv} = \frac{\partial v^\mytrans}{\partial v_z} =   \big[\mathrm I_{nk}, \mathbf 0_{nk \times n\ksum}\big] 
    \in \mathbb R^{nk \times n(k+\ksum)}.
\end{align} 
Similarly, 
letting 
$v_{q}^\star$ denote the vectorization of 
\begin{align}\label{eq:vstarmatrix}
\big[Z^\star,\, W_1^\star,\,  \ldots, \,  W_T^\star\big] \check{Q}_v \in \mathbb{R}^{n\times (k+\ksum)} \quad \ \text{with} \quad & \check{Q}_v = \mathrm{diag}( \check{Q},  \check{Q}_1,\ldots,  \check{Q}_T)
\end{align}
in the same manner as in  \eqref{eq:vectorization}, we have that $ Z^\star \check{Q}$ is a matricization of  $\mathcal{D}_{Zv}  v_{q}^\star  $. 
{Then 
\begin{align} \label{eq:thm2-15-1}
       \operatorname{dist}^2(\hat Z,Z^\star ) \leqslant  &~\big\|\hat Z - Z^\star \check Q\big\|_{\mathrm{F}}^2= \big\|  \mathcal{D}_{Zv} \big(\hat{v} - v_{q}^\star\big) \big\|_2^2. 
\end{align}
To prove \eqref{eq:zhaterr}, we note   
\begin{align}
    \eqref{eq:thm2-15-1} = &~ \big\| \mathcal{D}_{Zv} \big\{  \hat{v} - v_{q}^\star - I(v_q^{\star})^{+}\dot{\ell}(v_q^{\star}) + I(v_q^{\star})^{+}\dot{\ell}(v_q^{\star}) \big\} \big\|_2^2 \notag\\
    \leqslant &~    2\big\|\mathcal{D}_{Zv} I(v_q^{\star})^{+}\dot{\ell}(v_q^{\star}) \big\|_2^2 + 2 \big\|  \mathcal{D}_{Zv} \big\{\hat{v} - v_{q}^\star - I(v_q^{\star})^{+}\dot{\ell}(v_q^{\star}) \big\}   \big\|_2^2 \notag\\
    = &~ 2\big\|\mathcal{D}_{Zv} I(v_q^{\star})^{+}\dot{\ell}(v_q^{\star}) \big\|_2^2 + 2 \big\|  \mathcal{D}_{Zv} \big\{\check{v} - v_{q}^\star + I(\check{v})^{+}\dot{\ell}(\check{v})- I(v_q^{\star})^{+}\dot{\ell}(v_q^{\star}) \big\}   \big\|_2^2 \notag\\
    \leqslant &~ 2S_1+4S_2+4S_3,\label{eq:thm2-15}
\end{align} 
}
where the third equation follows from the construction $\hat{v} = \check{v}+ I(\check{v})^+ \dot \ell(\check v)$ and we define   
\begin{equation}
\label{eq:def_S1S2S3}
\begin{aligned}
    S_1 & =  \big\| \mathcal{D}_{Zv}\, I({v}_{q}^\star)^+ \dot \ell({v}_{q}^\star)\big\|_2^2, \\
    S_2 & =   \big\| \mathcal{D}_{Zv} \big\{\check{v} - v_{q}^\star +  I(\check{v})^+ \big(\dot \ell(\check{v}) - \dot \ell({v}_{q}^\star)\big) \big\}\big\|_2^2, \\
    S_3 & = \big\| \mathcal{D}_{Zv} \big\{I(\check{v})^+ - I({v}_{q}^\star)^+\big\}  \dot \ell({v}_{q}^\star)\big\|_2^2.
\end{aligned}    
\end{equation}
Then it suffices to show that for any $\varepsilon > 0$, there exists a constant $C_\varepsilon > 0 $ such that
\begin{align}
    \Pr\big[ S_1 > C_\varepsilon T^{-1} \log(nT)
    \big] &= O((nT)^{-\varepsilon}), \label{eq:thm2-17} \\
    \Pr\big[ S_2 > C_\varepsilon n^{-1} \log^8(nT)
    \big] &= O((nT)^{-\varepsilon}), \label{eq:thm2-18}  \\
    \Pr\big[ S_3 > C_\varepsilon n^{-1} \log^5(nT)
    \big] &= O((nT)^{-\varepsilon}) \label{eq:thm2-19},
\end{align}
on Pages  \pageref{sec:pfthm2-17}, \pageref{sec:pfthm2-18}, and \pageref{sec:pfthm2-19} below, respectively. 
Combining \eqref{eq:thm2-17}--\eqref{eq:thm2-19}  with \eqref{eq:thm2-15} gives
\begin{align*}
    \Pr\left[
\operatorname{dist}^2(\hat Z, Z^\star) >
   C_\varepsilon \max\left\{\frac{1}{T} ,\ \frac{1}{n}\right\} \log^{8}(nT)
\right] = O((nT)^{-\varepsilon}).
\end{align*}



\paragraph{Proof of \eqref{eq:thm2-17}.} \label{sec:pfthm2-17}
Let $\check{Q}_{nv} = \operatorname{diag}(\mathrm{I}_n \otimes \check{Q}, \mathrm{I}_n \otimes \check{Q}_1, \ldots, \mathrm{I}_n \otimes \check{Q}_T ) \in \mathbb R^{n(k+\ksum) \times n(k+\ksum)}$.  By the analytical forms of $\dot \ell(v)$ in \eqref{eq:ldotform} and $I(v)$ in \eqref{eq:ivform}, we have 
\begin{align*}
 \dot \ell({v}_{q}^\star) &= \check{Q}_{nv}^\mytrans\, \dot \ell(v^\star) 
    \quad\text{ and }\quad 
   I({v}_{q}^\star)^+  
    = \big\{ \check{Q}_{nv}^\mytrans\, I(v^\star) \check{Q}_{nv}\big\}^+ 
    = \check{Q}_{nv}^\mytrans\, I(v^\star)^+ \check{Q}_{nv}, 
\end{align*}
{where the last equation follows by $\check{Q}_{nv}^{-1}=\check{Q}_{nv}^{\top}$.}
As a result,  
\begin{align*}
    \mathcal{D}_{Zv}\, I({v}_{q}^\star)^+ \dot \ell({v}_{q}^\star) = \mathcal{D}_{Zv}\, \check Q_{nv}^\mytrans\, I({v}^\star)^+ \dot \ell({v}^\star) = (\mathrm{I}_n \otimes \check Q)^\mytrans \mathcal{D}_{Zv}\, I({v}^\star)^+ \dot \ell({v}^\star),
\end{align*}
where the second equation follows by the block form of $\check{Q}_{nv}$ and the definition of $  \mathcal{D}_{Zv}$ in \eqref{eq:dzvdef}. It follows that
\begin{align}\label{eq:s1_rotation_remove}
    S_1 = \big\|  \mathcal{D}_{Zv}\, I({v}_{q}^\star)^+ \dot \ell({v}_{q}^\star)\big\|_2^2 = \big\|\mathcal{D}_{Zv}\, I({v}^\star)^+ \dot \ell({v}^\star)\big\|_2^2. 
\end{align}
To apply a Bernstein inequality to bound $S_1$, we express each entry of $\mathcal{D}_{Zv}\, I({v}^\star)^+ \dot \ell({v}^\star)$ as a weighted sum of independent zero-mean sub-exponential random variables as follows:
\begin{align*}
    \mathcal{D}_{Zv}\, I({v}^\star)^+ \dot \ell({v}^\star) = \mathcal{D}_{Zv}\, I({v}^\star)^+ \begin{pmatrix}
        \DZ(v^\star)  & \ldots & \DZ(v^\star)\\
        \mathcal{D}_{W_1 \Theta}(v^\star) &\ldots & 0\\
        \vdots & \ddots & \vdots \\
        0 & \ldots & \mathcal{D}_{W_T \Theta}(v^\star)
    \end{pmatrix} \begin{pmatrix}
        \dot \ell_{\Theta_1}(v^\star) \\
        \vdots \\
        \dot \ell_{\Theta_T}(v^\star)
    \end{pmatrix},
\end{align*}
where the entries of
\begin{align} \label{eq::S_1:weightmatrix}
   \mathcal{D}_{Zv}\, I({v}^\star)^+ \begin{pmatrix}
        \DZ(v^\star)  & \ldots & \DZ(v^\star)\\
        \mathcal{D}_{W_1 \Theta}(v^\star) &\ldots & 0\\
        \vdots & \ddots & \vdots \\
        0 & \ldots & \mathcal{D}_{W_T \Theta}(v^\star)
    \end{pmatrix}  
\end{align}
    give the weights, and the entries of $(\dot \ell_{\Theta_1}(v^\star),\ldots, \dot \ell_{\Theta_T}(v^\star) )^\top $ are independent mean-zero sub-exponential random variables as shown in \eqref{eq:pre3}.
    Denote by $\big[\mathcal{D}_{Zv}\, I({v}^\star)^+ \dot \ell({v}^\star)\big]_i$ the $i$-th entry of $\mathcal{D}_{Zv}\, I({v}^\star)^+ \dot \ell({v}^\star)$ for $1\leqslant i\leqslant nk$. Then applying Lemma \ref{adlem:wellner2005} on the $i$-th entry shows 
\begin{align} \label{eq:thm2-20}
    \Pr\Big( \big[\mathcal{D}_{Zv}\, I({v}^\star)^+ \dot \ell({v}^\star)\big]_i^2 >  C_\varepsilon \big(V_i \log(nT) + B_i^2 \log^2(nT)\big)\Big)  = O((nT)^{-\varepsilon -1}) 
\end{align} by
Condition \ref{cond:parfunction} (ii),
where 
$ V_i = \VAR\big\{\big[\mathcal{D}_{Zv}\, I({v}^\star)^+ \dot \ell({v}^\star)\big]_i\big\}$ and $B_i$ denotes
    the infinity norm of the $i$-th row of the weight matrix given in \eqref{eq::S_1:weightmatrix}. 
By $S_1=\sum_{i=1}^{nk} [\mathcal{D}_{Zv}\, I({v}^\star)^+ \dot \ell({v}^\star)]_i^2$ and applying the union bound inequality, we have 
\begin{align} \label{eq:thm2-21}
     \Pr\Big( S_1 > \sum_{i = 1}^{nk}  C_\varepsilon  \big(V_i\log(nT) + B_i^2 \log^2(nT)\big)\Big) \leqslant &~ \sum_{i=1}^{nk} \eqref{eq:thm2-20}   \notag\\
= &~ 
 nk\,  O((nT)^{-\varepsilon-1}) = O((nT)^{-\varepsilon}).
\end{align}
The maximum weight term in \eqref{eq:thm2-21} satisfies
\begin{align} \label{eq:thm2-20-0}
    \max_{1 \leqslant i \leqslant nk} B_i &\leqslant  \big\|\mathcal{D}_{Zv}\, I({v}^\star)^+\big\|_{1 \to \infty} \big(\big\|\DZ(v^\star)\big\|_1 + \max_{1 \leqslant t \leqslant T }\big\|\Dwt(v^\star)\big\|_1\big) \notag\\
    &\leqslant\big\|\mathcal{D}_{Zv}\, I({v}^\star)^+\big\|_{\operatorname{op}} \big(\big\|\DZt(v^\star)\big\|_{\infty} + \max_{1 \leqslant t \leqslant T }\big\|\Dwtt(v^\star)\big\|_{\infty}\big) \notag\\
    &\leqslant n^{-1} T^{-1/2} \big\|\DnT^{1/2}\, I({v}^\star)^+ \DnT^{1/2} \big\|_{\operatorname{op}} \big(\big\|\DZt(v^\star)\big\|_{\infty} + \max_{1 \leqslant t \leqslant T }\big\|\Dwtt(v^\star)\big\|_{\infty}\big) \notag\\
    &= O(n^{-1} T^{-1/2})
\end{align}
The variance term in inequality \eqref{eq:thm2-21} can be bounded by
\begin{align} 
    \sum_{i = 1}^{nk} V_i &= \operatorname{tr}\big[\operatorname{cov}\big(\mathcal{D}_{Zv}\, I({v}^\star)^+ \dot \ell({v}^\star)\big)\big] \notag\\
    &= \operatorname{tr}\big[\mathcal{D}_{Zv} I(v^\star)^+ \mathcal{D}_{Zv}^\mytrans \big] \hspace{2em} (\text{by }I(v^{\star})=\mathrm{cov}\{\dot \ell({v}^\star)\} )\label{eq:variance_sum_bd} \\
    &= (nT)^{-1}\operatorname{tr}\big[\mathcal{D}_{Zv} \DnT^{1/2} I(v^\star)^+ \DnT^{1/2} \mathcal{D}_{Zv}^\mytrans \big] \notag \\
    &\leqslant k T^{-1} \, \big\|\mathcal{D}_{Zv} \DnT^{1/2} I(v^\star)^+ \DnT^{1/2} \mathcal{D}_{Zv}^\mytrans\big\|_{\operatorname{op}} \notag\\
    &\leqslant kT^{-1}\,  \big\|\DnT^{1/2}I({v}^\star)^+ \DnT^{1/2}\big\|_{\operatorname{op}} \leqslant C T^{-1}\label{eq:thm2-22}
\end{align}
for a constant $C > 0$, where the third equation follows by $\mathcal{D}_{Zv} \DnT^{1/2} = (nT)^{1/2} \mathrm{I}_{nk}$ and the last inequality follows by Lemma \ref{lm:ivproperties} (iii).

Combining \eqref{eq:thm2-21}, \eqref{eq:thm2-20-0}, and \eqref{eq:thm2-22}, we obtain \eqref{eq:thm2-17}.

\paragraph{Proof of \eqref{eq:thm2-18}.} \label{sec:pfthm2-18}
Lemma \ref{lm:ivproperties} (ii) indicates that $\check v - v_{q}^\star \in \operatorname{col}(I(\check{v}))$. Consequently, we have $\check v - v_{q}^\star = I(\check{v})^+  I(\check{v}) (\check{v} - v_{q}^\star)$ and then 
\begin{align} \label{eq:thm2-23}
     S_2 & =   \big\| \mathcal{D}_{Zv}\, I(\check{v})^+ \big\{ I(\check{v}) (\check{v} - v_{q}^\star) +   \dot \ell(\check{v}) - \dot \ell({v}_{q}^\star) \big\}\big\|_2^2 \notag \\
     & = \big\| \mathcal{D}_{Zv}\, I(\check{v})^+ \big( I(\check{v}) + \tilde H \big) (\check{v} - v_{q}^\star)  \big\|_2^2 \notag \\
     &\leqslant \big\| \mathcal{D}_{Zv}\, I(\check{v})^+ \DnT^{1/2} \big\|_{\operatorname{op}}^2 \big\| \DnT^{-1/2}\big( I(\check{v}) + \tilde H \big) \DnT^{-1/2} \big\|_{\operatorname{op}}^2 \big\|\DnT^{1/2} (\check{v} - v_{q}^\star)  \big\|_2^2
\end{align}
where the second equation follows from an integral form of the mean value theorem with
\begin{align}\label{eq:def_tildeH}
    \tilde H = \int_{0}^1 H\big(\check{v} + u(v_{q}^\star - \check v)\big) \ du
\end{align}
(the integration of a matrix means integrating each entry separately).
Similarly, we define 
\begin{align*}
    \tilde I = \int_{0}^1 I\big(\check{v} + u(v_{q}^\star - \check v)\big) \ du \quad \text{ and }\quad \tilde N = \int_{0}^1 N\big(\check{v} + u(v_{q}^\star - \check v)\big) \ du.
\end{align*}
Eq. \eqref{eq:hv} indicates that $\tilde H =- \tilde I + \tilde N$, then we have
\begin{align} \label{eq:thm2-24}
    &\ \quad\big\| \DnT^{-1/2} \big( I(\check{v}) + \tilde H \big) \DnT^{-1/2}\big\|_{\operatorname{op}}\notag\\
       &= \big\| \DnT^{-1/2} \big\{ \big( I(\check{v}) - \tilde I \big) - \big( N(v_q^\star) - \tilde N \big) +  N(v_q^\star) \big\}\DnT^{-1/2}\big\|_{\operatorname{op}} \notag \\
    &\leqslant \big\| \DnT^{-1/2} \big( I(\check{v}) - \tilde I \big) \DnT^{-1/2}\big\|_{\operatorname{op}}  + \big\| \DnT^{-1/2} \big( N(v_q^\star) - \tilde N \big) \DnT^{-1/2}\big\|_{\operatorname{op}} + \big\| \DnT^{-1/2} N(v_q^\star) \DnT^{-1/2}\big\|_{\operatorname{op}} \notag \\
    &\leqslant \max_{v = \check{v} + u(v_{q}^\star - \check v)}\Big\{\big\| \DnT^{-1/2} \big( I(\check{v}) -  I(v) \big) \DnT^{-1/2}\big\|_{\operatorname{op}}  + \big\| \DnT^{-1/2} \big( N(v_q^\star) - N(v) \big) \DnT^{-1/2}\big\|_{\operatorname{op}}\Big\} \notag\\
    &\quad+  \big\| \DnT^{-1/2} N(v_q^\star) \DnT^{-1/2}\big\|_{\operatorname{op}} \notag\\
    &\leqslant C_\varepsilon n^{-1/2} \log^2(nT)
\end{align}
with probability $1 - O((nT)^{-\varepsilon})$, where the last inequality follows from Lemma \ref{lm:ivproperties} 
 (iv)--(v) and Lemma \ref{lem:concentration}. 
%
Since $\mathcal{D}_{Zv} = (nT)^{-1/2} \mathcal{D}_{Zv} \DnT^{1/2}$, we have 
\begin{align} \label{eq:s2-1}
    \big\|\mathcal{D}_{Zv}\, I(\check{v})^+ \DnT^{1/2} \big\|_{\operatorname{op}}^2 &= (nT)^{-1} \big\|\mathcal{D}_{Zv}\DnT^{1/2}\, I(\check{v})^+ \DnT^{1/2} \big\|_{\operatorname{op}}^2 \notag\\
    &\leqslant (nT)^{-1} \big\|\DnT^{1/2}\, I(\check{v})^+ \DnT^{1/2} \big\|_{\operatorname{op}}^2 \leqslant C_\varepsilon (nT)^{-1}
\end{align}
with probability $1 - O((nT)^{-\varepsilon})$ by Lemma \ref{lm:ivproperties} (iii).
By \eqref{eq:thm2-1}, we have
\begin{align} \label{eq:s2-2}
    \big\|\DnT^{1/2} (\check{v} - v_{q}^\star)  \big\|_2^2 = nT \big\|\check Z - Z^\star \check{Q}\big\|_{\mathrm{F}}^2 + n \sum_{t = 1}^T \big\|\check W_t - W_t^\star \check{Q}_t \big\|_{\mathrm{F}}^2 \leqslant C_\varepsilon nT\log^4(nT)
\end{align}
with probability $1 - O((nT)^{-\varepsilon})$.
Combining \eqref{eq:thm2-23} with \eqref{eq:thm2-24}, \eqref{eq:s2-1}, and \eqref{eq:s2-2} gives \eqref{eq:thm2-18}.




\paragraph{Proof of \eqref{eq:thm2-19}.}\label{sec:pfthm2-19}

Note that 
\begin{align} \label{eq:thm2-25}
S_3&\leqslant { (nT)^{-1} \big\|\DnT^{1/2} \big\{I(\check{v})^+ - I(v_q^\star)^+ \big\} \dot \ell({v}_{q}^\star)\big\|_2^2}   \notag \\
&\leqslant (nT)^{-1} \big\|\DnT^{1/2} \big\{I(\check{v})^+ - I(v_q^\star)^+ \big\} \DnT^{1/2} \big\|_{\operatorname{op}}^2 \big\|  \DnT^{-1/2} \dot \ell({v}_{q}^\star)\big\|_2^2  \notag\\
    &\leqslant 2(nT)^{-1} \big\|\DnT^{1/2} I(\check{v})^+ \DnT^{1/2}  \big\|_{\operatorname{op}}^2 \  \big\|\DnT^{1/2}  I(v_q^\star)^+  \DnT^{1/2} \big\|_{\operatorname{op}}^2\notag  \\
    &\quad\quad \big\|\DnT^{-1/2} \big(I(\check{v}) - I(v_q^\star) \big) \DnT^{-1/2} \big\|_{\operatorname{op}}^2 \ \big\|  \DnT^{-1/2} \dot \ell({v}_{q}^\star)\big\|_2^2,
\end{align}
where the last inequality follows from applying Lemma \ref{adlem:wedin1973} to $\|\DnT^{1/2} \{I(\check{v})^+ - I(v_q^\star)^+ \} \DnT^{1/2} \|_{\operatorname{op}}$ and the invertibility of $\DnT$. 
Using a similar technique as in \eqref{eq:thm2-20-0}--\eqref{eq:thm2-22}, we obtain 
\begin{align} \label{eq:thm2-26}
    \big\|  \DnT^{-1/2} \dot \ell({v}_{q}^\star)\big\|_2^2  \leqslant C_\varepsilon \operatorname{tr}\big(\DnT^{-1/2}I(v^\star)\DnT^{-1/2}\big) \log(nT) \leqslant C_\varepsilon nT \log(nT)
\end{align}
with probability $1- O((nT)^{-\varepsilon})$. Consequently, \eqref{eq:thm2-19} follows from  \eqref{eq:thm2-25}, \eqref{eq:thm2-26}, and Lemma \ref{lm:ivproperties} (iii)--(iv).

\subsubsection{Proof of \eqref{eq:wthaterr}} \label{sec:pdfwthaterr}

To prove \eqref{eq:wthaterr}, we note that by the union bound inequality, 
\begin{align} \label{eq:wtsummationbound}
    \Pr\left[\max_{1 \leqslant t \leqslant T} \operatorname{dist}^2(\hat{W}_t,W_t^\star) > C_\varepsilon \log(nT)\right] \leqslant \sum_{t=1}^T \Pr\left[\operatorname{dist}^2(\hat{W}_t,W_t^\star) > C_\varepsilon \log(nT)\right]. 
\end{align}
Then it suffices to show that 
\begin{align} \label{eq:wt_individual_bound}
\Pr\big[\operatorname{dist}^2(\hat{W}_t,W_t^\star) > C_{\varepsilon} \log(nT)\big] =  O((nT)^{-\varepsilon-1} ), 
\end{align}
which gives $\eqref{eq:wtsummationbound} =  O((nT)^{- \varepsilon - 1} T) = O( (nT)^{-\varepsilon} )$.

{Similarly to  \eqref{eq:thm2-15-1}, we have
\begin{align}\label{eq:thm2-27-1}
      \operatorname{dist}^2(\hat{W}_t, W_t^\star) &\leqslant \big\|\hat{W}_t - W_t^\star \check{Q}_t\big\|_{\mathrm{F}}^2 = \big\|  \mathcal{D}_{W_t v} \big(\hat{v} - v_{q}^\star\big) \big\|_2^2,  
\end{align}
where $\mathcal{D}_{W_t v} = [\mathbf{0}_{nk_t \times nk}, \mathbf{0}_{nk_t \times nk_1}, \ldots, \mathrm{I}_{nk_t}, \ldots, \mathbf{0}_{nk_t \times nk_T}] \in \mathbb R^{nk_t \times n(k + \ksum)}$. 
To prove \eqref{eq:wthaterr}, similarly to \eqref{eq:thm2-15}, we note    
\begin{align}
    \eqref{eq:thm2-27-1}=&~  \big\| \mathcal{D}_{W_tv} \big\{  \hat{v} - v_{q}^\star - I(v_q^{\star})^{+}\dot{\ell}(v_q^{\star}) + I(v_q^{\star})^{+}\dot{\ell}(v_q^{\star}) \big\} \big\|_2^2 \notag\\
   \leqslant &~    2\big\|\mathcal{D}_{W_tv} I(v_q^{\star})^{+}\dot{\ell}(v_q^{\star}) \big\|_2^2 + 2 \big\|  \mathcal{D}_{W_tv} \big\{\hat{v} - v_{q}^\star - I(v_q^{\star})^{+}\dot{\ell}(v_q^{\star}) \big\}   \big\|_2^2 \notag\\
    = &~  2\big\|\mathcal{D}_{W_tv} I(v_q^{\star})^{+}\dot{\ell}(v_q^{\star}) \big\|_2^2 + 2 \big\|  \mathcal{D}_{W_tv} \big\{\check{v} - v_{q}^\star + I(\check{v})^{+}\dot{\ell}(\check{v})- I(v_q^{\star})^{+}\dot{\ell}(v_q^{\star}) \big\}   \big\|_2^2 \notag\\
    \leqslant &~ 2S_4+4S_5+4S_6,   \label{eq:thm2-27}
\end{align}
where 
} 
\begin{equation}
\label{eq:def_S4S5S6}
\begin{aligned}
    S_4 & =  \big\| \mathcal{D}_{W_t v}\, I({v}_{q}^\star)^+ \dot \ell({v}_{q}^\star)\big\|_2^2, \\
    S_5 & =   \big\| \mathcal{D}_{W_t v} \big[\check{v} - v_{q}^\star +  I(\check{v})^+ \big\{\dot \ell(\check{v}) - \dot \ell({v}_{q}^\star)\big\} \big]\big\|_2^2, \\
    S_6 & = \big\| \mathcal{D}_{W_t v} \big\{I(\check{v})^+ - I({v}_{q}^\star)^+\big\}  \dot \ell({v}_{q}^\star)\big\|_2^2.
\end{aligned}
\end{equation}
{Similarly to \eqref{eq:thm2-17}--\eqref{eq:thm2-19}, to prove \eqref{eq:wthaterr}, we next show that  
\begin{align} 
    \Pr\big[ S_4 > C_\varepsilon  \log(nT)
    \big] = &~  O((nT)^{-\varepsilon-1}), \label{eq:thm2-28-0} \\
    \Pr\big[ S_5 > C_\varepsilon n^{-1} \log^8(nT)
    \big] = &~  O((nT)^{-\varepsilon-1}), \label{eq:thm2-29} \\
       \Pr\big[ S_6 > C_\varepsilon n^{-1} \log^5(nT)
    \big]  = &~  O((nT)^{-\varepsilon-1}). \label{eq:thm2-30}
\end{align}}

Using a similar technique to that in \eqref{eq:thm2-20-0}--\eqref{eq:thm2-22}, we have 
\begin{align} \label{eq:thm2-28}
    S_4  = \big\|\mathcal{D}_{W_t v}\, I({v}^\star)^+ \dot \ell({v}^\star)\big\|_2^2 \leqslant C_\varepsilon \operatorname{tr} \big(\mathcal{D}_{W_t v}I(v^\star)^+ \mathcal{D}_{W_t v}^\mytrans\big) \log(nT) \leqslant C_\varepsilon \log(nT)
\end{align}
with probability $1- O((nT)^{-\varepsilon-1})$, giving \eqref{eq:thm2-28-0}. 

We next establish upper bounds for $S_5$ and $S_6$. 
By Lemma \ref{lm:ivproperties} (vi), we have 
\begin{align} \label{eq:thm2-28-1}
    I(v)^+ = {I_d(v)^{+} \,  I_d(v) \, I(v)^{+} }
    = I_d(v)^{+} - I_d(v)^{+} \big\{I(v) - I_d(v)\big\} I(v)^{+}.
\end{align}
 The block structures of $I(v)$ and $I_d(v)$ indicate that
 \begin{align*}
     \mathcal{D}_{W_t v}\, I_d(v)^{+} = I_{W_t W_t}(v)^+ \mathcal{D}_{W_t v}\quad \text{ and }\quad \mathcal{D}_{W_t v}\, \big\{I(v) - I_d(v)\big\} = I_{W_t Z}(v) \mathcal{D}_{Zv}, 
 \end{align*}  
then combining with \eqref{eq:thm2-28-1} we obtain
\begin{align} \label{eq:thm2-31}
    \mathcal{D}_{W_t v}\, I(v)^{+} = I_{W_t W_t}(v)^+ \mathcal{D}_{W_t v} + I_{W_t W_t}(v)^+ I_{W_t Z}(v) \mathcal{D}_{Zv}\, I(v)^+.
\end{align}
Applying similar arguments for $S_2$ and \eqref{eq:thm2-31}, we have
\begin{align} \label{eq:thm2-32}
    S_5 &= \big\| \mathcal{D}_{W_t v} \,I(\check{v})^+ \big( I(\check{v})(\check{v} - v_{q}^\star) +  \dot \ell(\check{v}) - \dot \ell({v}_{q}^\star)\big)\big\|_2^2 \notag \\
    &= \big\| \mathcal{D}_{W_t v} \,I(\check{v})^+ \big( I(\check{v}) + \tilde H\big)(\check{v} - v_{q}^\star) \big\|_2^2 \leqslant 2S_{51} + 2S_{52}
\end{align}
with 
\begin{align*}
    S_{51} &= \big\|I_{W_t W_t}(\check v)^+ \mathcal{D}_{W_t v} \big( I(\check{v}) + \tilde H\big)(\check{v} - v_{q}^\star)\big\|_2^2, \\
    S_{52} &= \big\|I_{W_t W_t}(\check v)^+ I_{W_t Z}(\check v) \mathcal{D}_{Zv}\, I(\check v)^+ \big( I(\check{v}) + \tilde H\big)(\check{v} - v_{q}^\star)\big\|_2^2.
\end{align*}
By the block diagonality of $I(\check{v})$ and $\tilde{H}$, we have $\mathcal{D}_{W_t v} \big( I(\check{v}) + \tilde H\big) \mathcal{D}_{W_s v}^\mytrans = 0$ for any $s \neq t$, then
\begin{align} \label{eq:s51}
    S_{51}&\leqslant 2\, \big\|I_{W_t W_t}(\check v)^+\big\|_{\operatorname{op}}^2 \big\|\mathcal{D}_{W_t v} \big( I(\check{v}) + \tilde H\big) \mathcal{D}_{Zv}^\mytrans \big\|_{\operatorname{op}}^2 \big\|\mathcal{D}_{Zv}(\check{v} - v^\star_q)\big\|_{2}^2 \notag\\
    &\quad+  2\, \big\|I_{W_t W_t}(\check v)^+\big\|_{\operatorname{op}}^2 \big\|\mathcal{D}_{W_t v} \big( I(\check{v}) + \tilde H\big) \mathcal{D}_{W_t v}^\mytrans \big\|_{\operatorname{op}}^2 \big\|\mathcal{D}_{W_t v}(\check{v} - v^\star_q)\big\|_{2}^2 \notag\\
    &\leqslant C_\varepsilon n^{-1} \log^8(nT)
\end{align}
with probability $1 - O((nT)^{-\varepsilon-1})$, where the last inequality follows from
\begin{align}
    &\big\|I_{W_t W_t}(\check v)^+\big\|_{\operatorname{op}}^2 \leqslant n^{-2}\big\|\DnT^{1/2}\,I(\check v)^+ \DnT^{1/2}\big\|_{\operatorname{op}}^2 \leqslant C_\varepsilon n^{-2}, \label{eq:s51-1}\\
    &\big\|\mathcal{D}_{W_t v} \big( I(\check{v}) + \tilde H\big) \mathcal{D}_{Zv}^\mytrans \big\|_{\operatorname{op}}^2 = \big\|\mathcal{D}_{W_t v} \big( I(\check{v}) - \tilde I\big) \mathcal{D}_{Zv}^\mytrans \big\|_{\operatorname{op}}^2 \leqslant C_\varepsilon n \log^4(nT), \notag\\
    &\big\|\mathcal{D}_{W_t v} \big( I(\check{v}) + \tilde H\big) \mathcal{D}_{W_t v}^\mytrans \big\|_{\operatorname{op}}^2 \leqslant n^{2} \big\|\DnT^{-1/2}\, \big( I(\check{v}) + \tilde H\big) \DnT^{-1/2}\big\|_{\operatorname{op}}^2  \leqslant C_\varepsilon n\log^4(nT)\notag
\end{align}
by Lemma \ref{lm:ivproperties} (iii)--(v). 
By \eqref{eq:thm2-18} and \eqref{eq:s51-1}, we have
\begin{align} \label{eq:s52}
    S_{52} &\leqslant  \big\|I_{W_t W_t}(\check v)^+\big\|_{\operatorname{op}}^2 \big\|I_{W_t Z}(\check v)\big\|_{\operatorname{op}}^2  \big\| \mathcal{D}_{Zv}\, I(\check v)^+ \big( I(\check{v}) + \tilde H\big)(\check{v} - v_{q}^\star)\big\|_2^2 \notag\\
    &= \big\|I_{W_t W_t}(\check v)^+\big\|_{\operatorname{op}}^2 \big\|I_{W_t Z}(\check v)\big\|_{\operatorname{op}}^2\, S_2 \leqslant C_\varepsilon n^{-1} \log^8(nT)
\end{align}
with probability $1 - O((nT)^{-\varepsilon-1})$.
Combining \eqref{eq:thm2-32} with \eqref{eq:s51} and \eqref{eq:s52}, we obtain \eqref{eq:thm2-29}.

Finally, \eqref{eq:thm2-31} indicates that
\begin{align*}
    \mathcal{D}_{W_t v} \big\{I(\check{v})^+ - I({v}_{q}^\star)^+\big\}  &=  \big\{I_{W_t W_t}(\check{v})^+ - I_{W_t W_t}({v}_{q}^\star)^+\big\}  \mathcal{D}_{W_t v} \\
    &+ \big\{I_{W_t W_t}(\check{v})^+ I_{W_t Z}(\check{v})\mathcal{D}_{Zv}\, I(\check{v})^+ - I_{W_t W_t}({v_q^\star})^+ I_{W_t Z}({v_q^\star})\mathcal{D}_{Zv}\, I(v_q^\star)^+\big\},
\end{align*}
then we have
\begin{align} \label{eq:thm2-33}
   S_6 &\leqslant 2 S_{61} + 2 S_{62}
\end{align}
with 
\begin{align*}
    S_{61} &= \big\| \big\{I_{W_t W_t}(\check{v})^+ - I_{W_t W_t}({v}_{q}^\star)^+\big\}  \mathcal{D}_{W_t v}\,\dot \ell({v}_{q}^\star)\big\|_2^2, \\
    S_{62} &=  \big\|\big\{I_{W_t W_t}(\check{v})^+ I_{W_t Z}(\check{v})\mathcal{D}_{Zv}\, I(\check{v})^+ - I_{W_t W_t}({v_q^\star})^+ I_{W_t Z}({v_q^\star})\mathcal{D}_{Zv}\, I(v_q^\star)^+\big\} \dot \ell({v}_{q}^\star)\big\|_2^2.
\end{align*}
Applying Lemma \ref{adlem:wedin1973} shows
\begin{align} \label{eq:thm2-34}
    S_{61} & \leqslant \big\| I_{W_t W_t}(\check{v})^+ - I_{W_t W_t}({v}_{q}^\star)^+ \big\|_{\operatorname{op}}^2 \big\|  \mathcal{D}_{W_t v}\,\dot \ell({v}_{q}^\star)\big\|_2^2 \notag\\
    &\leqslant  2\, \big\| I_{W_t W_t}(\check{v})^+\big\|_{\operatorname{op}}^2 \big\| I_{W_t W_t}(\check{v}) - I_{W_t W_t}({v}_{q}^\star)\big\|_{\operatorname{op}}^2 \big\|  I_{W_t W_t}({v}_{q}^\star)^+\big\|_{\operatorname{op}}^2\big\|  \mathcal{D}_{W_t v}\,\dot \ell({v}_{q}^\star)\big\|_2^2 \notag\\
    &\leqslant  C_\varepsilon n^{-1} \log^5(nT)
\end{align}
with probability $1 - O((nT)^{-\varepsilon-1})$, where the last inequality follows from \eqref{eq:s51-1}, 
\begin{align*}
    &\big\| I_{W_t W_t}(\check{v}) - I_{W_t W_t}({v}_{q}^\star)\big\|_{\operatorname{op}}^2 \leqslant n^{2} \big\|\DnT^{-1/2}\big( I(\check{v}) - I({v}_{q}^\star)\big)\DnT^{-1/2}\big\|_{\operatorname{op}}^2 \leqslant C_\varepsilon n \log^4(nT), \\
    &\big\|  \mathcal{D}_{W_t v}\,\dot \ell({v}_{q}^\star)\big\|_2^2 \leqslant C_\varepsilon   \operatorname{tr} \big(\mathcal{D}_{W_t v}I(v^\star) \mathcal{D}_{W_t v}^\mytrans\big) \log(nT)  \leqslant C_\varepsilon  n^2 \log(nT) .
\end{align*}
Further note that 
\begin{align} \label{eq:thm2-35}
    S_{62} &\leqslant 2\, \big\|I_{W_t W_t}(\check{v})^+   I_{W_t Z}(\check{v})  \mathcal{D}_{Zv} \big\{I(\check{v})^+ - I({v}_{q}^\star)^+\big\}  \dot \ell({v}_{q}^\star)\big\|_2^2 \notag\\
    &\quad+ 2\, \big\|\big\{I_{W_t W_t}(\check{v})^+ I_{W_t Z}(\check{v}) - I_{W_t W_t}({v_q^\star})^+ I_{W_t Z}({v_q^\star}) \big\} \mathcal{D}_{Zv}\, I(v_q^\star)^+\dot \ell({v}_{q}^\star)\big\|_2^2 \notag\\
    &\leqslant  2\, \big\| I_{W_t W_t}(\check{v})^+ \big\|_{\operatorname{op}}^2 \big\| I_{W_t Z}(\check{v}) \big\|_{\operatorname{op}}^2 \big\| \mathcal{D}_{Zv} \big(I(\check{v})^+ - I({v}_{q}^\star)^+\big)  \dot \ell({v}_{q}^\star)\big\|_2^2 \notag\\
   &\quad+ 2\, \big\|I_{W_t W_t}(\check{v})^+ I_{W_t Z}(\check{v}) - I_{W_t W_t}({v_q^\star})^+ I_{W_t Z}({v_q^\star}) \big\|_{\operatorname{op}}^2 \big\|\mathcal{D}_{Zv}\, I(v_q^\star)^+\dot \ell({v}_{q}^\star)\big\|_2^2 \notag\\
   &\leqslant C_\varepsilon n^{-1} \log^5(nT) 
\end{align}
with probability $1 - O((nT)^{-\varepsilon-1})$, where the last inequality follows from \eqref{eq:s51-1},  \eqref{eq:thm2-19}, \eqref{eq:thm2-17}, and 
\begin{align*}
    &\big\|I_{W_t W_t}(\check{v})^+ I_{W_t Z}(\check{v}) - I_{W_t W_t}({v_q^\star})^+ I_{W_t Z}({v_q^\star}) \big\|_{\operatorname{op}}^2 \\
    \leqslant& ~  2 \, \big\|\big\{I_{W_t W_t}(\check{v})^+  - I_{W_t W_t}({v_q^\star})^+\big\}  I_{W_t Z}(\check{v})\big\|_{\operatorname{op}}^2 +  2\, \big\|I_{W_t W_t}({v_q^\star})^+ \big\{I_{W_t Z}(\check{v}) - I_{W_t Z}({v_q^\star})\big\}\big\|_{\operatorname{op}}^2\\
\leqslant & ~ 2\,\big\|I_{W_t W_t}(\check{v})^+  - I_{W_t W_t}({v_q^\star})^+\big\|_{\operatorname{op}}^2 \big\| I_{W_t Z}(\check{v})  \big\|_{\operatorname{op}}^2  + 2\,\big\| I_{W_t W_t}({v_q^\star})^+ \big\|_{\operatorname{op}}^2 \big\|I_{W_t Z}(\check{v}) - I_{W_t Z}({v_q^\star}) \big\|_{\operatorname{op}}^2 \\
\leqslant&~ C_\varepsilon n^{-1} \log^4(nT)
\end{align*}
by Lemma \ref{lm:ivproperties} and Lemma \ref{adlem:wedin1973}.
Combining \eqref{eq:thm2-33} with \eqref{eq:thm2-34} and  \eqref{eq:thm2-35}
we obtain \eqref{eq:thm2-30}, which finishes the proof.

In summary,   combining  \eqref{eq:thm2-27-1}--\eqref{eq:thm2-30}, 
we obtain  \eqref{eq:wt_individual_bound} when  $n \gtrsim \log^8(nT)$. 




\begin{remark}[Connections with Cram\'{e}r-Rao lower bound]\label{rm:l2normvhat}
{Intuitively,  our derived results imply that up to the orthogonal transformation, $\hat{v}_q-v^\star$ is close to a random vector with covariance $ I(v^{\star})^{+} $, 
 generalizing classical Cram\'{e}r-Rao lower bound for high-dimensional embedding vectors.   
Specifically, we have $\mathrm{cov}\{I(v^{\star})^{+}\dot{\ell}(v^{\star})\}= I(v^{\star})^{+}$, and similar to the arguments in \eqref{eq:s1_rotation_remove}, 
\begin{align}
    &~   \big\|  \mathcal{D}_{Mv} \big\{\hat{v}_q - v^\star - I(v^{\star})^{+}\dot{\ell}(v^{\star}) \big\}   \big\|_2^2 \notag \\
=  &~ \big\|  \mathcal{D}_{Mv} \big\{\hat{v} - v_{q}^\star - I(v_q^{\star})^{+}\dot{\ell}(v_q^{\star}) \big\}   \big\|_2^2  \notag  \\
    \leqslant &~ 2(S_2+S_3+S_5+S_6) \quad\quad  (\text{by   \eqref{eq:thm2-15}  and  \eqref{eq:thm2-27}}) \notag \\
    \leqslant &~ C_{\varepsilon} n^{-1}\log^8(nT) \hspace{4.3em}  (\text{by \eqref{eq:thm2-18}, \eqref{eq:thm2-19}, \eqref{eq:thm2-29}, 
    and \eqref{eq:thm2-30}})  \label{eq:residual_bd_rm}
\end{align}
with probability $1-O((nT)^{-\varepsilon})$, 
where $M=Z$ and $M=W_t$ correspond to taking subcomponents of $\hat{v}_{q} - v^\star - I(v^{\star})^{+}\dot{\ell}(v^{\star})$ with respect to $Z$ and $W_t$, respectively. 
In summary, we establish closeness between $\hat{v}_q-v^\star$ and $ I(v^{\star})^{+}\dot{\ell}(v^{\star})$ for each group of  high-dimensional latent embeddings $Z^{\star}$ or $W_t^{\star}$ in terms of $\ell_2$ norm.   }
\end{remark}  

\begin{remark}[Comparison with \cite{xie2023efficient}]\label{rm:strongeff}
{As shown in \eqref{eq:residual_bd_rm}, our current results concern $\ell_2$ error for high-dimensional parameters. On the other hand,  \cite{xie2023efficient} proposed one-step update for single network, where  asymptotic normality  for low-dimensional  embeddings 
was established.  
Establishing comparable results under our current problem setting  corresponds to showing the $\infty$-norm discrepancy satisfies that up to logarithmic factors, 
\begin{align}\label{eq:infinitynorm_goal}
    \big\|  \hat{v}_q - v^\star - I(v^{\star})^{+}\dot{\ell}(v^{\star})   \big\|_{\infty}^2 = o_p((nT)^{-1}).
\end{align}}

{We emphasize that establishing \eqref{eq:infinitynorm_goal}  
in our problem is faced with unique challenges  that cannot be directly addressed by techniques in \cite{xie2023efficient}. 
Notably, as discussed in Section \ref{sec:challenges}, we consider multiple-network model with intricate factor dependence structures and general distribution family accommodating non-linear link functions, whereas \cite{xie2023efficient} considers random dot product graph for  a single network $T=1$. Due to our model complexity, many properties in \cite{xie2023efficient} cannot be directly established. 
For example,  \cite{xie2023efficient} requires that the initial estimator for one-step update satisfies an approximate linearization property (Definition 1). 
Such a property is shown to be satisfied by  Adjacency Spectral Embedding \citep{athreya2016limit}  under random dot product graph model in the existing literature  \citep{athreya2016limit,tang2018limit}. 
However, whether and how similar properties can be established in our general settings remain open problems, as the link function bridging expected adjacency matrix and embeddings can be non-linear, and naive spectral embeddings of individual networks cannot separate shared and individual embeddings for multiple networks. }  

{Nevertheless, our developments in this paper could pave the way for 
  future research on \eqref{eq:infinitynorm_goal}. 
  Most derivations can be directly generalized to  $\infty$ norm, such as 
   the decomposition idea used in \eqref{eq:thm2-15} and  \eqref{eq:thm2-27} and properties of the information and Hessian matrices. 
 But one fundamental challenge we expect  is that in \eqref{eq:thm2-24}, $\| \DnT^{-1/2} N(v_q^\star) \DnT^{-1/2} \|_{\operatorname{op}} \asymp n^{-1/2}\times \| \DnT^{-1/2} N(v_q^\star) \DnT^{-1/2} \|_{\infty}$ suggesting that examining $\infty$ norm of $ \DnT^{-1/2} N(v_q^\star) \DnT^{-1/2}$ is much larger than its operator norm. Intuitively, this is because $ N(v)$ is a dense matrix by its definition after \eqref{eq:hv}. Nevertheless, this could   be overcome if additional properties on the initial estimator are available. For example, if we make more assumptions on the initial estimator, such as a  linear approximation property similar to that assumed in \cite{xie2023efficient},   
 we could directly bound $\| \DnT^{-1/2} N(v_q^\star) (\check{v}-v_q^{\star}) \|_{\infty}$ by the closed-form expression of $N(v)$   and properties of initial $\check{v}$ together. Intuitively, this is possible because the entries in $N(v)$ are independent at fixed $v$, allowing more refined concentration inequalities to be applied.  
However,  a rigorous analysis would still require substantial amount of work and we leave it for future study.  }
\end{remark}


\medskip 
\begin{remark}[Usefulness of the second-order update] \label{rm:initialrelax}
{The above proof shows that the one-step update \eqref{eq:newtononev}  can attain the second-order efficiency with mild requirements on the input estimator $(\check{Z},\check{W})$. 
In particular, proofs in Sections \ref{sec:pdfzhaterr} and \ref{sec:pdfwthaterr} rely on  $(\check{Z},\check{W})$ (equivalently $\check{v}$) only through properties established in Lemma \ref{lm:ivproperties}.
Note that Lemma \ref{lm:ivproperties} generally holds for any estimators  satisfying 
 assumptions (a) and (b) in it, 
i.e., any estimators that are bounded and achieve entrywise errors of the order of $n^{-1/2}$, up to logarithmic factors. 
Following Section \ref{sec:challenges}, oracle estimation errors of $w_{t,i}$ and  $z_i$ are expected to be of orders of $n^{-1/2}$ and $(nT)^{-1/2}$, respectively. 
Therefore,  Lemma \ref{lm:ivproperties} requires   that  $\check{z}$ achieves  the same error rate as $\check{w}_{t,i}$ (up to logarithmic factors), but not its own oracle rate under $T$ networks, which is a mild requirement under our multiple-network setting.  }


{We emphasize that the required error rates for the initial estimators are upper bounds. 
When initial estimators achieve smaller or even oracle  error rates, the one-step update can still be proven to achieve the second-order efficiency following similar  arguments.   
The flexibility with respect to the input estimators shows that the proposed one-step update is a useful theoretical tool. 
For our constructed initial estimator $(\check{Z},\check{W})$, 
 directly establishing second-order efficiency could be challenging in theory.  
Numerically, we observe that $(\check{Z},\check{W})$  and $(\hat{Z},\hat{W})$ can be close, which is  demonstrated in Section \ref{sec:pseudolik}. 
In the future research, it would be of interest to theoretically examine the relationship  between $(\check{Z},\check{W})$ and $(\hat{Z},\hat{W})$. }

\end{remark}

\medskip 
\begin{remark}[Interpretation of $T\lesssim n$]\label{rm:interprtTn}
{As discussed in Section \ref{sec:challenges}, the oracle estimation error rates of the $Z$ and $W$ components can be interpreted from a semiparametric perspective, where $Z$ and $W$ are target and nuisance parameters, respectively.
We first elaborate on canonical results in classical  semiparametric analyses and explain the connections to our conclusions. 
Consider 
a canonical semiparametric problem with  
 target parameters $\theta$ and nuisance parameters $\eta$ 
 with their estimates denoted by $\hat{\theta}$ and $\hat{\eta}$, respectively. 
Under 
suitable regularity conditions, 
the contribution from the nuisance parameters to $\|\hat{\theta}-\theta\|_2$ can often  be bounded by $O(\|\hat{\eta}-\eta\|_2^2)$, a \emph{squared} $\ell_2$-error of nuisance parameters; 
see, e.g., Section 25.59 in \cite{van2000asymptotic}, Section 3 in \cite{murphy2000profile}, and Section 7.6 in \cite{bickel1993efficient}. Then  $\|\hat{\eta} -\eta\|_2^2\lesssim \|\hat{\theta}-\theta \|_2$ is a common requirement and is related to ``rate double robustness'' in the debiased machine learning literature  \citep{chernozhukiv.ectj.12097,bradic2019minimax}. }

 {Our constraint  $T\lesssim n$ aligns with the common double rate requirement on estimating nuisance parameters. 
Specifically, as discussed in Section \ref{sec:challenges} of the main text,
the oracle rates for estimating individual  target parameter $z_i$ and nuisance parameter  $w_{t,i}$ are expected to be 
$O_p((nT)^{-1/2})$ and $O_p(n^{-1/2})$, respectively. 
Following the ``squared contribution'' idea above,
 when 
each \emph{squared} oracle error rate of $w_{t,i}$
is no larger than the oracle error rate of $z_i$, i.e., $(n^{-1/2})^2 \lesssim (nT)^{-1/2}$, the constraint $T\lesssim n$ is needed. } 
\end{remark}

\subsection{Proof of Theorem \ref{thm:estk} }
We first demonstrate that the estimator $\hat{d}_t$ obtained through Step 1 is consistent. Note that $\hat{d}_t = d_t$ happens if and only if 
\begin{align} \label{eq:thm3-1}
    \mathcal{R}_{m+1,m}(\breve \Theta_t) > \tau_{2,m} \ \text{ for }\  1\leqslant m < d_t \quad\text{ and }\quad \mathcal{R}_{d_t+1,d_t}(\breve \Theta_t) \leqslant \tau_{2,d_t}.
\end{align}
Theorem \ref{thm:estTheta} indicates that there exists a constant $C_\varepsilon > 0$ such that for any $1 \leqslant t \leqslant T$,
\begin{align*}
    \sigma_{1}(\breve\Theta_t) &\leqslant \sigma_{1}(\Theta_t^\star) + \big\|\breve\Theta_t - \Theta_t^\star\big\|_{\mathrm{F}} \leqslant 2M_1^2n + C_\varepsilon n^{1-\frac{1}{2d_t+4}} \log^{\frac{1}{2}}(nT) \\
    \sigma_{d_t}(\breve\Theta_t) &\geqslant \sigma_{d_t}(\Theta_t^\star) - \big\|\breve\Theta_t - \Theta_t^\star\big\|_{\mathrm{F}} \geqslant M_2n - C_\varepsilon n^{1-\frac{1}{2d_t+4}} \log^{\frac{1}{2}}(nT)\\
    \sigma_{d_t+1}(\breve\Theta_t) &\leqslant \sigma_{d_t+1}(\Theta_t^\star) + \big\|\breve\Theta_t - \Theta_t^\star\big\|_{\mathrm{F}} \leqslant C_\varepsilon n^{1-\frac{1}{2d_t+4}} \log^{\frac{1}{2}}(nT)
\end{align*}
hold with probability $1 - O((nT)^{-\varepsilon})$. The above inequalities imply that  \eqref{eq:thm3-1} holds when $\log^{\dmax+2}(nT)/n$ is sufficiently small, because
\begin{align*}
     \mathcal{R}_{m+1,m}(\breve \Theta_t) &\geqslant \mathcal{R}_{d_t,1}(\breve \Theta_t) \geqslant M_2/(6M_1^2) > \tau_{2,m} \quad\text{ for }\quad  1\leqslant m < d_t \\
     \text{and }\quad \mathcal{R}_{d_t+1,d_t}(\breve \Theta_t) &\leqslant 2M_2^{-1} C_\varepsilon  n^{-\frac{1}{2d_t+4}} \log^{\frac{1}{2}}(nT) \leqslant \tau_{2,d_t}.
\end{align*}

We next show that the estimator $\hat k$ obtained through Algorithm \ref{algor:estk} is consistent when $\hat d_t = d_t$. By Theorem \ref{thm:estY}, with probability $1 - O((nT)^{-\varepsilon})$, we have
\begin{align*}
    \sigma_{1}(\mathring Y_{t,s}) &\leqslant \sigma_{1}( Y_{t,s}^\star) + \operatorname{dist}(\mathring Y_t, Y^\star_t) +  \operatorname{dist}(\mathring Y_s, Y^\star_s) \leqslant 2M_1n^{\frac{1}{2}} + 2C_\varepsilon \log^{\frac{1}{2}}(nT)\\
    \sigma_{1}(\mathring Y_{t,s}) &\geqslant \sigma_{1}( Y_{t,s}^\star) - \operatorname{dist}(\mathring Y_t, Y^\star_t) -  \operatorname{dist}(\mathring Y_s, Y^\star_s) \geqslant (M_2n)^{\frac{1}{2}} - 2C_\varepsilon \log^{\frac{1}{2}}(nT)\\
    \sigma_{K_{t,s}^r + 1}(\mathring Y_{t,s}) &\leqslant \sigma_{K_{t,s}^r + 1}( Y_{t,s}^\star) + \operatorname{dist}(\mathring Y_t, Y^\star_t) +  \operatorname{dist}(\mathring Y_s, Y^\star_s) \leqslant 2C_\varepsilon \log^{\frac{1}{2}}(nT)
\end{align*}
for all $1 \leqslant t < s \leqslant T$, and 
\begin{align*}
    \sigma_{K_{t,s}^r}(\mathring Y_{t,s}) &\geqslant \sigma_{K_{t,s}^r}( Y_{t,s}^\star) - \operatorname{dist}(\mathring Y_t, Y^\star_t) - \operatorname{dist}(\mathring Y_s, Y^\star_s) \geqslant (M_3n)^{\frac{1}{2}} - 2C_\varepsilon \log^{\frac{1}{2}}(nT)
\end{align*}
for any $(t,s)$ pair satisfying Condition \ref{cond:truevalue} (iii). The above inequalities indicate that when $\log(nT)/n$ is sufficiently small,
\begin{align*}
    \hat{K}_{t,s}^r \leqslant K_{t,s}^r \ \text{ for all } 1 \leqslant t < s \leqslant T \quad\text{ since }\ 
    \mathcal{R}_{K_{t,s}^r + 1,1} (\mathring Y_{t,s}) \leqslant 4 C_\varepsilon  (M_2n)^{-\frac{1}{2}}\log^{\frac{1}{2}}(nT) \leqslant \tau_3,
\end{align*}
and 
\begin{align*}
    \hat{K}_{t,s}^r \geqslant K_{t,s}^r \ \text{ for $(t,s)$ satisfying Condition \ref{cond:truevalue} (iii)} \quad\text{ since }\ \mathcal{R}_{K_{t,s}^r ,1} (\mathring Y_{t,s}) \geqslant M_3^{\frac{1}{2}} / (6 M_1) > \tau_3.
\end{align*}
Combining the above argument with $\hat{K}_{t,s} = \hat d_t + \hat d_s - \hat{K}_{t,s}^r$, ${K}_{t,s} = d_t + d_s - {K}_{t,s}^r$, $\hat d_t = d_t$, and Proposition \ref{prop:dimensionk}, we obtain
\begin{enumerate}
    \item[(i)] $k \leqslant  \hat{K}_{t,s}$ for all $1\leqslant t < s\leqslant T$,    
    \item[(ii)]  $k = \hat{K}_{t,s}$ holds for any $(t,s)$ pair satisfying Condition \ref{cond:truevalue} (iii)
\end{enumerate}
with probability $1 - O((nT)^{-\varepsilon})$, which shows $\Pr( \hat k \neq k ) = O((nT)^{-\varepsilon})$.

\newpage

\section{Additional Technical Lemmas}
In this section, we summarize technical lemmas from the existing literature that are used in the proofs. 

\begin{lemma}[\citealp{tu2016low}, Lemma 5.4]
\label{adlem:tu2016}
For any matrices $Z_1,Z_2\in \mathbb{R}^{n\times k}$, denote the distance between them as   $\operatorname{dist}(Z_1,Z_2)=\min_{Q \in \mathcal{O}(k)}\left\|Z_1-Z_2 Q\right\|_{\mathrm{F}}$, then 
\begin{align*}
\big\|Z_1Z_1^\top-Z_2Z_2^\top \big\|_{\mathrm{F}}^2\geqslant{2(\sqrt{2}-1)\sigma_k^2(Z_1)}\operatorname{dist}^2(Z_1,Z_2).
\end{align*}
\end{lemma}

\begin{lemma}[\citealp{tu2016low}, Lemma 5.3]
\label{adlem:tu2016B}
For any matrices $Z_1,Z_2\in \mathbb{R}^{n\times k}$, denote the distance between them as   $\operatorname{dist}(Z_1,Z_2)=\min_{Q \in \mathcal{O}(k)}\left\|Z_1-Z_2 Q\right\|_{\mathrm{F}}$. If $\operatorname{dist}(Z_1,Z_2) \leqslant \|Z_1\|_{\mathrm{op}}$, then
\begin{align*}
\big\|Z_1Z_1^\top-Z_2Z_2^\top\big\|_{\mathrm{F}}\leqslant3\|Z_1\|_{\mathrm{op}}\operatorname{dist}(Z_1,Z_2).
\end{align*}
\end{lemma}

\begin{lemma}[\citealp{tropp2012user}, Theorem 6.2]
\label{adlem:tropp2012}
Let $X_1,\ldots,X_m$ be independent, zero mean, $n\times n$ and symmetry
matrices, such that for $k = 1, \ldots, m$
$$\EXPT X_k^{l}\preceq \frac{l!}{2}\,B^{l-2}A_k^2 \quad\text{ for any integer } l \geqslant 2.$$
Then for all $x\geqslant0$, we have 
\begin{align*}
    \Pr\left[\Big\|\sum_{k=1}^m X_k\Big\|_{\mathrm{op}}\geqslant x\right]
    \leqslant 2n\,\exp{\left(\frac{-x^2/2}{\sigma^2+B x}\right)}, 
\end{align*}
where $\sigma^2=\|\sum_{k=1}^m A_k^2\|_{\mathrm{op}}$.
\end{lemma}

\begin{lemma}[\citealp{wellner2005empirical}, Pages 103--104]\label{adlem:wellner2005}
	Let $X_1,\ldots,X_n$ be independent zero mean random variables such that for $i = 1,\ldots, n$, $\mathbb{E}|X_i|^l\leqslant \mathbb{E}X_i^2L^{l-2}l!/2$ holds for any integer $l \geqslant 2$. Then for $M\geqslant\sum_{1 \leqslant i\leqslant n}\mathbb{E}X_i^2$ and $x\geqslant0$,
	$$\Pr\left(\sum_{i = 1}^{n} X_i\geqslant x\right)\leqslant\exp\left(\frac{-x^{2}/2}{M+Lx}\right).$$
\end{lemma}

\begin{lemma}[\citealp{chatterjee2015matrix}, Lemma 3.5]
\label{adlem:chatterjee2015}
Suppose $A$ and $B$ are two $n \times m$ matrices with $n \geqslant m$. Let $A=\sum_{i=1}^m\sigma_i u_i v_i^\top$ be the SVD of $A$. 
For any fixed $\delta> 2$, define
\begin{align*}
\widehat{B}=\sum_{i:\sigma_i> \delta \|A-B\|_{\mathrm{op}}}\sigma_i u_i v_i^\top. 
\end{align*}
Then there exists a constant $C > 0 $ such that
\begin{align*}
    \big\|\widehat{B}-B\big\|_{\mathrm{F}}^2
    \leqslant C \delta  \, \big\|A-B\big\|_{\mathrm{op}}  \big\|B\big\|_{\mathrm{*}},
\end{align*}
where $\|B\|_*$ represents the nuclear norm of $B$.
\end{lemma}

\begin{lemma}[\citealp{nesterov2003introductory}, Theorem 2.1.12] \label{adlem:nesterov2003}
For a continuously differentiable function $f$, if it is $\mu$-strongly convex and $L$-smooth on a convex domain $\mathcal{D}$, say for any $x,y\in \mathcal{D}$,
\begin{align*}
    \frac{\mu}{2}\,\big\|x-y\big\|_{\mathrm{2}}^2\leqslant
    f(y)-f(x)-\big\langle \nabla f(x),y-x\big\rangle
    \leqslant \frac{L}{2}\,\big\|x-y\big\|_{\mathrm{2}}^2,
\end{align*}
then
\begin{align*}
    \big\langle \nabla f(x)-\nabla f(y),x-y\big\rangle
    \geqslant \frac{\mu L}{\mu+L}\,\big\|x-y\big\|_{\mathrm{2}}^2
    +\frac{1}{\mu+L}\,\big\|\nabla f(x)-\nabla f(y)\big\|_{\mathrm{2}}^2.
\end{align*}
\end{lemma}

\begin{lemma} [\citealp{cai2018rate}, Theorem 1] \label{adlem:cai2018}
    Suppose $X \in \mathbb R^{n \times m}$ is a rank-$k$ matrix with the SVD $X = U\Sigma V^\mytrans$. Let $\hat{U} \hat{\Sigma} \hat{V}^{\mytrans}$ denote the top-$k$ singular components of a  perturbed matrix $\hat{X} \in \mathbb R^{n \times m}$. 
    If $\sigma_{k}(X) > 3\|X - \hat{X}\|_{\operatorname{op}}$, there exists a constant $C>0$ such that
    \begin{align*}
        \big\|\hat{V}\hat{V}^\mytrans - {V}{V}^\mytrans\big\|_{\operatorname{op}} \leqslant C\, \frac{\sigma_{k}(X) \big\|X - \hat{X}\big\|_{\operatorname{op}} }{\sigma^2_{k}(X) - 6\big\|X - \hat{X}\big\|_{\operatorname{op}}^2}.
    \end{align*}
\end{lemma}

\begin{lemma}[\citealp{wedin1973perturbation}, Theorem 4.1] \label{adlem:wedin1973}
    Suppose $A$ and $B$ are two $n \times m$ matrices. If $\operatorname{rank}(A) = \operatorname{rank}(B)$, then 
    \begin{align*}
        \big\|B^+ - A^+\big\|_{\operatorname{op}} \leqslant 2 \,\big\|B^+\big\|_{\operatorname{op}} \big\|B-A\big\|_{\operatorname{op}} \big\|A^+\big\|_{\operatorname{op}}.
    \end{align*}
\end{lemma}

\begin{lemma}[\citealp{he2023semiparametric}, Lemma B.2]\label{adlem:he2023}
    Given any $Z \in \mathbb R^{n \times k}$ satisfying $\operatorname{rank}(Z) = k$, consider its singular value decomposition $Z = \sum_{i=1}^k \lambda_i U_i V_i^\mytrans$.  Define $\beta_{ij} = U_i \otimes V_j \in \mathbb R^{nk}$ for $1 \leqslant i \leqslant n, 1 \leqslant j \leqslant k$, and
\begin{align*}
    \mathcal{B}_{C,1}=&~\{\beta_{ii} : i=1, \ldots,k\}, \hspace{6.5em} \mathcal{B}_{C,2}=\{\beta_{ji} : i=1, \ldots, k  \text{ and } j = k+1, \ldots, n\},\\
   \mathcal{B}_{C,3}=&\left\{\frac{\lambda_j \beta_{ij} + \lambda_i \beta_{ji}}{\sqrt{\lambda_i^2 + \lambda_j^2}} : 1\leqslant i < j \leqslant k \right\}, \ \   \mathcal{B}_{N} = \left\{ \frac{\lambda_i\beta_{ij}-\lambda_j\beta_{ji}}{\sqrt{\lambda_i^2+\lambda_j^2}}  : 1\leqslant i < j \leqslant k \right\}.  
\end{align*}
We have 
\begin{enumerate}
    \item Let $v_z = [z_1^\mytrans, z_2^\mytrans, \ldots,  z_n^\mytrans]^\mytrans \in \mathbb R^{nk \times 1}$ be the vectorization of $Z$, then $v_z \in \mathcal{B}_{C,1}$.
    \item Define $\DZ$ as in \eqref{eq:pre4} and $\Xi_Z = \DZ \DZt$, then vectors in $\cup_{j=1}^3\mathcal{B}_{C,j} \cup \mathcal{B}_{N}$ form a set of (normalized) eigenvectors  of $\Xi_Z$. Specifically, the vectors satisfy
    
    \quad $\Xi_Z \beta_{ii} = 2\lambda_i^2 \beta_{ii}$ for $i=1, \ldots,k$; 

    \quad $\Xi_Z \beta_{ji} = \lambda_i^2 \beta_{ji}$ for $j=k+1, \ldots,n$ and $i=1, \ldots,k$; 

    \quad $\Xi_Z(\lambda_j\beta_{ij} + \lambda_i\beta_{ji})/\sqrt{\lambda_i^2+\lambda_j^2} = (\lambda_i^2+\lambda_j^2)(\lambda_j\beta_{ij} + \lambda_i\beta_{ji})/\sqrt{\lambda_i^2+\lambda_j^2} $ for $1\leqslant i < j \leqslant k$;

    \quad $\Xi_Z(\lambda_i\beta_{ij} - \lambda_j\beta_{ji})/\sqrt{\lambda_i^2+\lambda_j^2} = 0 $ for $1\leqslant i < j \leqslant k$.
\end{enumerate}
\end{lemma}

\begin{proof}
    We note that while the original lemma considers only $Z$ satisfying $1_n^\mytrans Z = 0$, their proof can be directly generalized to all $Z$.
\end{proof}

\newpage

\section{Details on Literature Comparison} \label{sec:detailtbcompare}

For clarity, we provide detailed information on the specific results and conditions that Table \ref{table::summary_multilayer} refers to:
\begin{itemize}
    \item The error rate of \cite{arroyo2021inference} is cited from  ``Summary of Contributions'' in Section 1.1, which is a simplified version of Theorem 7 in that paper. 
    \item The error rate of \cite{macdonald2022latent} is cited from its Theorem 3, with the constraint $T=o(n^{{1}/{2}}) $ from Assumption 1 in that paper (obtained when $\tau=1$ and latent dimensions are fixed).
    \item The error rate for shared parameter and the constraint on $(n,T)$ of \cite{zhang2020flexible} are cited from its Theorem 2.
    \item The error rate and the constraint on $(n,T)$ of \cite{he2023semiparametric} are from its Theorem 1. 
\end{itemize}

\noindent As different papers may focus on distinct targets and challenges, we have made slight adjustments to make the results comparable in terms of estimation error rates with respect to   $n$ and $T$. 
For a more comprehensive understanding, 
we next give more details on the compared works. 

\begin{enumerate}
    \item 
    \cite{arroyo2021inference} focuses on estimating the left singular vectors of the shared latent factor that is normalized, whereas we target at the latent factors without normalization. As we consider settings where column-wise $\ell_2$ norms of latent factors are of the order of $n$, Table \ref{table::summary_multilayer} presents the rate from   \cite{arroyo2021inference} after amplifying by a factor of $n$, accounting for the normalization similarly to \eqref{eq:nfactornormalization} below.  Moreover, their Theorem 7 examines the expectation of estimation error, which may not align with  our target  of the error directly. 
    \item Among the compared works in Table \ref{table::summary_multilayer}, \cite{zhang2020flexible}, \cite{he2023semiparametric}, and this paper focus on dense networks in the sense that edge-wise mean values are of the order of constants, but \cite{arroyo2021inference} and \cite{macdonald2022latent} allow densities of networks to vary. 
For the purpose of understanding the essence of integrating multiple  networks,  we focus on dense-network scenarios so that 
   the intrinsic relationship between error rates and the size and number of networks would be clearer. 

In particular, although Theorem 7 in \cite{arroyo2021inference} gives more general results, Table \ref{table::summary_multilayer} presents the simplified conclusion in Section 1.1 of their paper. Moreover, 
Theorem 3 in \cite{macdonald2022latent} presents an error rate that  also depends on parameters on eigenvalues $(\tau)$ and eigengaps $(\xi)$ of latent factors.  Table \ref{table::summary_multilayer} presents Theorem 3 in \cite{macdonald2022latent} with $\tau=1$, corresponding to dense-network scenarios for a fair comparison, and $\xi=1$, corresponding to the largest eigengap to demonstrate the best possible rate. 

    \item  Among the compared works in Table \ref{table::summary_multilayer}, \cite{macdonald2022latent}, \cite{he2023semiparametric}, and this work consider  additive forms between shared and  network-specific parameters similar to \eqref{eq:model2}. It is thus meaningful to compare their estimation error rates of individual parameters, which are shown to be similar in Table \ref{table::summary_multilayer}.  

   On the other hand, \cite{arroyo2021inference} and \cite{zhang2020flexible} do not consider directly comparable estimands and thus are skipped in estimating individual parameters of Table \ref{table::summary_multilayer}. Specifically, \cite{arroyo2021inference} consider individual parameters as 
   network-specific kernel matrices with the total number being of the order of $O(kT)$, which is much smaller than $O(nT)$ in this work when $k$ is fixed. Moreover, \cite{zhang2020flexible} does not directly report theoretical estimation error rates of their network-specific parameters in each network individually but only their total summation.


\end{enumerate}

\bigskip 

\begin{remark} \label{rm:comparelei2023}
{Table \ref{table::summary_multilayer} focuses on studies with comparable model complexities and targets for fair comparisons. 
As mentioned in Section \ref{sec:relatedwork}, there are also other works on modeling multiple networks with shared and heterogeneous structures. Although distinct targets and challenges are considered, some of the studies also exhibit  fundamental connections with our established error rate $\max\{1/T,1/n\}$.}
 
{For example, as a reviewer points out, 
\cite{lei2023bias} studies clustering of $n$ nodes when their memberships are shared across $T$ networks. Their Theorem 1 and Remark 2  show that with a high probability, the proportion of misclustered nodes is of the order of 
\begin{align} 
    \frac{1}{n^2} + \frac{\log(T+n)}{Tn^2\rho^2}, \hspace{2em}    &\text{when } n\rho \lesssim 1, \label{eq:lei_regime1_res_supp}\\
 \frac{1}{n^2} + \frac{\log(T+n)}{Tn\rho},  \hspace{2em}     &\text{when } n\rho \gg 1, \label{eq:lei_regime2_res_supp}
\end{align} 
 respectively, where $\rho$ characterizes the sparsity level. 
 When $T\gtrsim \log(T+n)/\rho^2 \gtrsim n^2\log(T+n)$ in \eqref{eq:lei_regime1_res_supp} and $T\gtrsim n\log(T+n)/\rho$ in \eqref{eq:lei_regime2_res_supp}, both two error rates reduce to the order of $1/n^2$, which  does not decrease as $T$ increases. 
 Therefore in both two regimes, combining more network layers cannot improve error rates if  $T$ is above certain thresholds, which is similar to our conclusion.
Moreover, as a heuristic comparison, our paper considers a dense regime that corresponds to \eqref{eq:lei_regime2_res_supp} when $\rho$ is a constant. 
 In this comparable regime, $\eqref{eq:lei_regime2_res_supp}\asymp n^{-1}\times \max\{1/T, 1/n\} $ up to logarithmic factors. 
This aligns with our error rate as   \eqref{eq:lei_regime2_res_supp} intrinsically relies on  estimating  normalized latent embeddings, corresponding to estimating $Z^{\star}/\sqrt{n}$ in our problem, and we have 
\begin{align}\label{eq:nfactornormalization}
    \operatorname{dist}^2\left(\frac{\hat{Z}}{\sqrt{n}}, \frac{Z^\star}{\sqrt{n}}\right) = \frac{1}{n} \operatorname{dist}^2(\hat{Z}, Z^\star)  \lesssim  n^{-1}\times \max\{1/T, 1/n\}. 
\end{align} The similar error rate further supports our interpretations  under a semiparametric perspective  in Remark \ref{rm:interprtTn}.}

{Despite the  similarities, we emphasize that there exist unique challenges  under our model settings. For example, the number of nuisance parameters differs significantly from that in \cite{lei2023bias}. Under their model, 
the unshared  (heterogeneous)  parameters are $O(K^2)$ connecting probabilities across $T$ layers. Its total number is of the order of $O(T)$ when the latent dimension $K$ is fixed. But under our considered model, the total number of unshared parameter is of the order of $O(Tn)$. 
Also, we note that stochastic block models have intrinsic discreteness for latent communities, which could induce unique properties that may not be available for analyzing continuous latent vectors in general. Therefore, analyses of stochastic block models may not always be directly comparable to our considered settings. }   
\end{remark}

\newpage
\section{Supplementary Results for Simulation Studies} \label{sec:suppsim}

\subsection{Estimation Errors of $\hat{W}_1$} \label{sec:suppsim_hetero}
\begin{figure}[!htbp]
\centering
\begin{subfigure}{0.32\textwidth}
	\centering
	\includegraphics[width=1\linewidth]{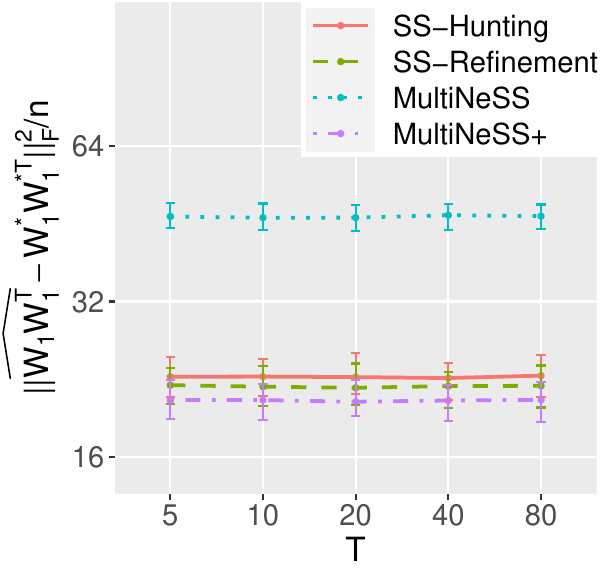}
    \caption{Bernoulli distribution}
\end{subfigure}
\begin{subfigure}{0.32\textwidth}
	\centering
	\includegraphics[width=1\linewidth]{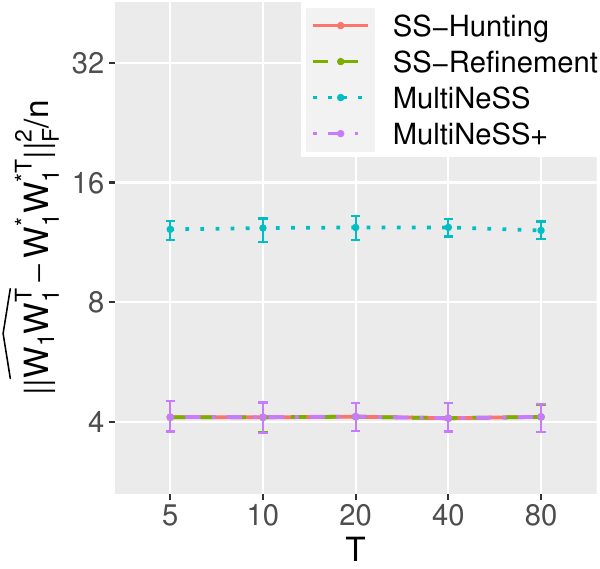}
    \caption{Gaussian distribution}
\end{subfigure}
\begin{subfigure}{0.32\textwidth}
	\centering
	\includegraphics[width=1\linewidth]{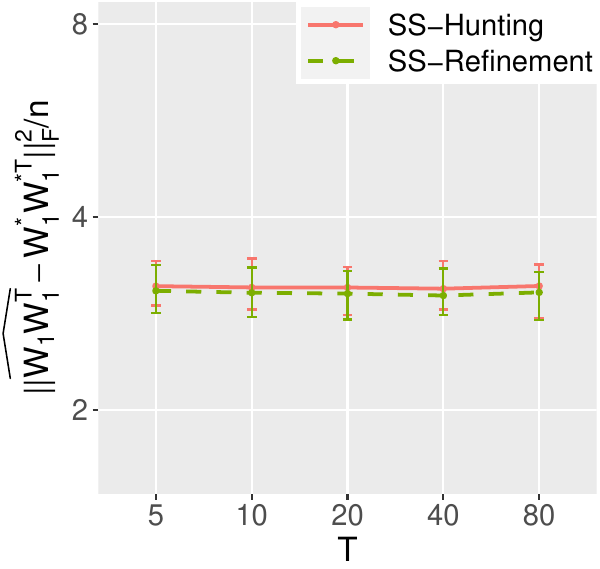}
    \caption{Poisson distribution}
\end{subfigure}
\caption{Empirical estimation errors of $\hat W_1$ versus $T$ under Case (A).  lines connect median estimation errors over 100 repetitions, and error bars are obtained from the 0.05 and 0.95 quantiles of those repetitions. Axes are in the log scale.}
\label{fig:simuWA}
\end{figure}

\begin{figure}[!htbp]
\centering
\begin{subfigure}{0.32\textwidth}
	\centering
	\includegraphics[width=1\linewidth]{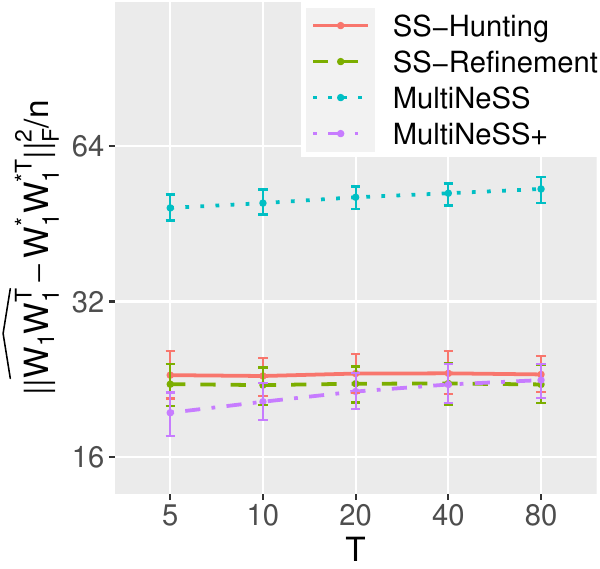}
    \caption{Bernoulli distribution}
\end{subfigure}
\begin{subfigure}{0.32\textwidth}
	\centering
	\includegraphics[width=1\linewidth]{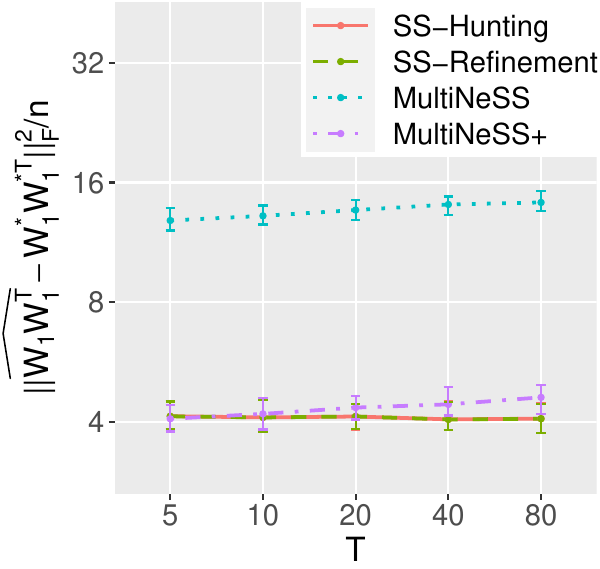}
    \caption{Gaussian distribution}
\end{subfigure}
\begin{subfigure}{0.32\textwidth}
	\centering
	\includegraphics[width=1\linewidth]{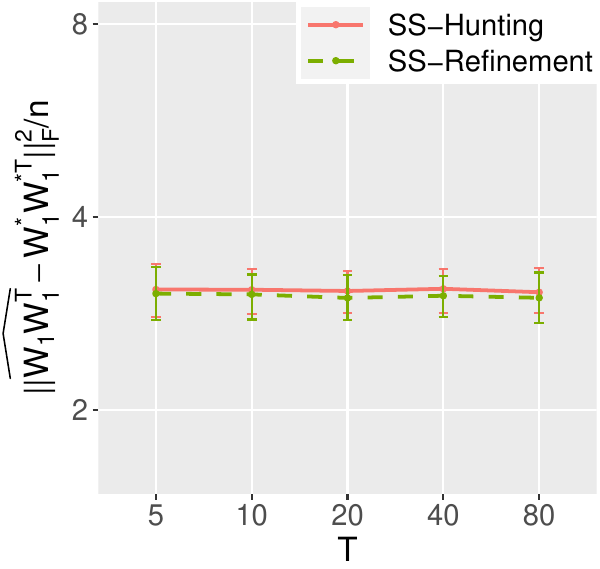}
    \caption{Poisson distribution}
\end{subfigure}
\caption{Empirical estimation errors  of $\hat W_1$ versus $T$ under Case (B). (Similar to Figure \ref{fig:simuWA}.)}
\label{fig:simuWB}
\end{figure}

\begin{figure}[!htbp]
\centering
\begin{subfigure}{0.32\textwidth}
	\centering
	\includegraphics[width=1\linewidth]{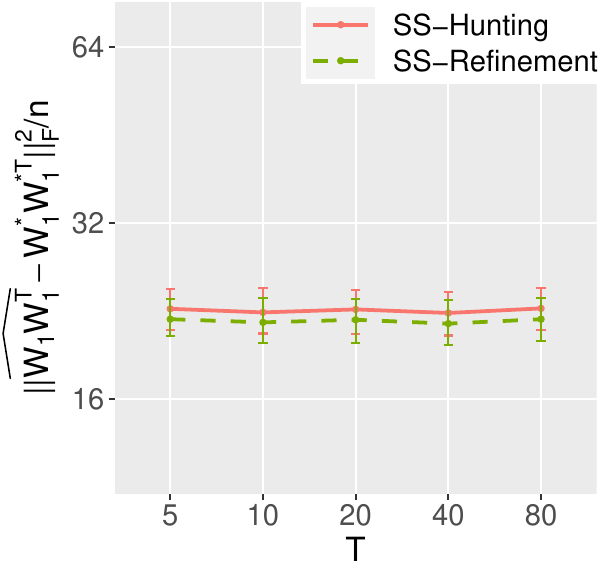}
    \caption{Bernoulli distribution}
\end{subfigure}
\begin{subfigure}{0.32\textwidth}
	\centering
	\includegraphics[width=1\linewidth]{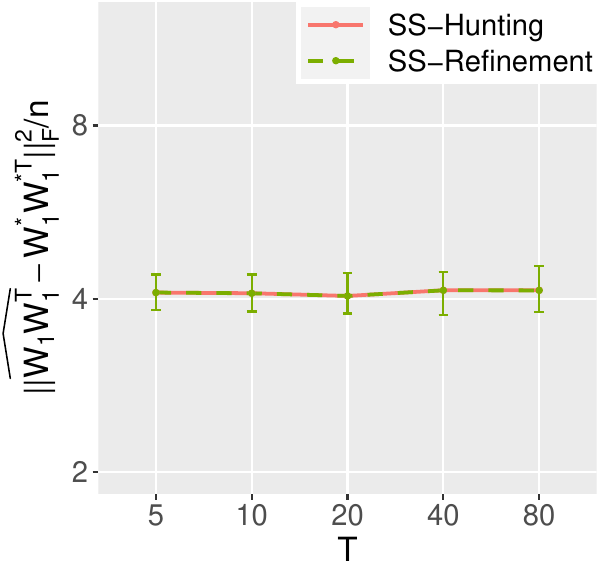}
    \caption{Gaussian distribution}
\end{subfigure}
\begin{subfigure}{0.32\textwidth}
	\centering
	\includegraphics[width=1\linewidth]{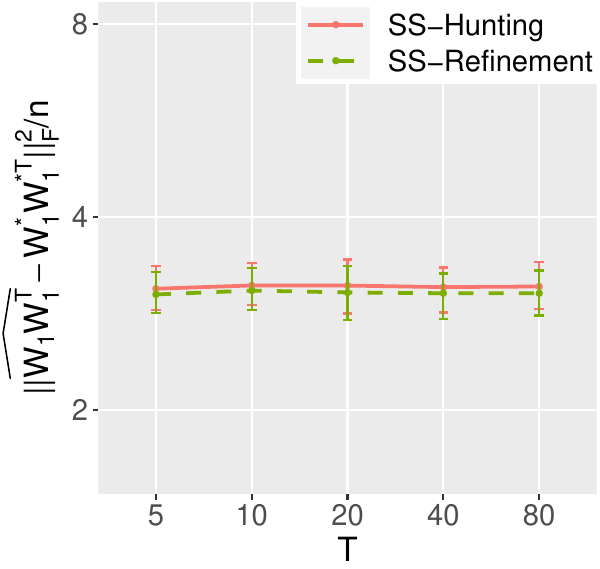}
    \caption{Poisson distribution}
\end{subfigure}
\caption{Empirical estimation errors of $\hat W_1$ versus $T$ under Case (C). (Similar to Figure \ref{fig:simuWA}.)}
\label{fig:simuWC}
\end{figure}

Recall Cases (A)--(C) and simulations on estimating latent vectors in Section \ref{sec:simulation}. 
We next present the empirical estimation errors of one individual latent factor $\hat{W}_1$ under Cases (A)--(C) in Figures \ref{fig:simuWA}--\ref{fig:simuWC}, respectively. 
Estimations of all other $W_t$'s give similar results by the symmetricity of our data generation over $1\leqslant t\leqslant T$.   
To ensure a fair comparison with \cite{macdonald2022latent}, we present the estimation error $\|\widehat{W_t W_t^\top} - W_t^\star W_t^{\star \top}\|_{\mathrm{F}}^2 / n$, which is equivalent to $\mathrm{dist}^2(\hat{W}_t, W_t^{\star})$ up to constants multiplied \citep{tu2016low}.  
Since the codes of MultiNeSS and MultiNeSS+ are not available under Poisson distribution, their errors are not presented under Poisson in the three figures. 

In Figure \ref{fig:simuWA}, all errors remain relatively stable with respect to $T$, which is consistent with the oracle error rate  discussed in Section \ref{sec:challenges}. 
SS-Hunting, SS-Refinement, and MultiNeSS+ exhibit comparable performance, all significantly outperforming MultiNeSS, with MultiNeSS+ achieving the smallest error.  
Nevertheless, Figure  \ref{fig:simuWB} shows that the performance of MultiNeSS+ may not be robust and could deteriorate as $T$ increases, whereas  
the proposed SS-Hunting and SS-Refinement yield consistent results.  
Under Case (C), errors of MultiNeSS and MultiNeSS+ escalate dramatically, and thus are not presented in Figure \ref{fig:simuWC}. 

\subsection{{Comparison of Variants of Algorithm \ref{algor:refine}}} \label{sec:pseudolik}

\paragraph{Variants of Algorithm \ref{algor:refine}} 

Recall that original Algorithm \ref{algor:refine} in the main text  consists of the projected gradient descent procedure  (lines 2--5) and the one-step update (line 6), and both are based on the joint  log-likelihood $\ell(Z,W)$ in \eqref{eq:zwlikelihood}.  This original version is first presented as using the joint log-likelihood is consistent with the existing literature and sets a foundational framework \citep{ma2020universal,zhang2020flexible,he2023semiparametric}.  

\quad Two variants of Algorithm \ref{algor:refine} are proposed for the convenience of obtaining theoretical and numerical results, respectively. 
The ``Theory'' variant  
is considered in Theorem \ref{thm:onestep} to facilitate the proof. 
As discussed in Remark \ref{rmk:pseudolik}, it  replaces the log likelihood $\ell(Z,W)$ used in the projected gradient descent (lines 2-5 of Algorithm \ref{algor:refine}) with the pseudo log-likelihood $p\ell(Z,W)$. 
Intuitively, this helps the proof because  $(Z,W)$ under $ p\ell(Z, W) $ can be identified without   ambiguity with respect to orthogonal transformations as half of $z_j$ and $w_{t,j}$ in the inner products are fixed. 
The ``Simulation'' variant skips  certain steps in the original  Algorithm \ref{algor:refine} to speed up the computation. 
As discussed in Section \ref{sec:algodetail}, the projection steps 
in lines 3--4 
and the second-order update in line 6 can be skipped, though  
we clarify that these steps can be implement in practice. (See Remark \ref{rm:initialrelax} for the usefulness of second-order update in theory.) 
For a clear comparison, we summarize  the two variants of  Algorithm \ref{algor:refine} in Algorithms \ref{algor:refine_theory}--\ref{algor:refine_simulation} below, where we highlight their differences compared to the original version in \blue{blue}. 

The original version and the two variants yield similar results in our numerical studies, which will be demonstrated as follows.  It would be an interesting future research direction to establish theoretical guarantee for the closeness of three variants observed in numerical studies.


\setcounter{algocf}{0}
\renewcommand{\thealgocf}{G.\arabic{algocf}}
\begin{algorithm}
	\caption{``Theory'' variant of Algorithm \ref{algor:refine}.}
	\label{algor:refine_theory}
	\KwIn{Data: $\mathbf A_1, \ldots, \mathbf A_T \in \mathbb{R}^{n\times n}$. Initial estimates: $\mathring Z$ and $ \mathring W_t$ for $1\leqslant t\leqslant T$.\\
    \hspace{3.3em} Parameters: $\eta_Z$,  $\eta_W$ (step sizes), $R$ (number of iterations), \\
    \hspace{9.3em}$\mathcal{C}_Z,  \mathcal{C}_{W_t}$ for $1\leqslant t\leqslant T$ (constraint sets for projection).}

\KwOut{$\hat{Z}$ and $\hat{W}_t$ for $1\leqslant t\leqslant T$.}


Let $Z^0 = \mathring{Z}$ and $W_t^0 = \mathring{W}_t$ for $1\leqslant t\leqslant T$.
	
\For{$r = 0, \ldots, R - 1$}{


Let \blue{$ \tilde Z^{r+1} =  Z^r + \eta_Z \sum_{t = 1}^T l^\prime(Z^r \mathring Z^\mytrans + W_t^r \mathring W_t^\mytrans)\mathring Z$}  and   $Z^{r+1}={\mathcal{P}_{\mathcal C_Z}}(\tilde{Z}^{r+1})$.  

   Let \blue{$\tilde W_t^{r+1} = W_t^r + \eta_W \,l^\prime(Z^r \mathring Z^\mytrans + W_t^r \mathring W_t^\mytrans) \mathring W_t $} and $W_t^{r+1}=\mathcal{P}_{\mathcal C_{W_t}}(\tilde{W}_t^{r+1})$ for $1\leqslant t\leqslant T$.

}

Construct $\hat{v}$ by \eqref{eq:newtononev} with $(\check{Z},\check{W})=(Z^R, W^R)$, and let $(\hat{Z},\hat{W})$ be its matrix version. 

\end{algorithm}

\begin{algorithm}
	\caption{``Simulation'' variant of Algorithm \ref{algor:refine}.}
	\label{algor:refine_simulation}
	\KwIn{Data: $\mathbf A_1, \ldots, \mathbf A_T \in \mathbb{R}^{n\times n}$. Initial estimates: $\mathring Z$ and $ \mathring W_t$ for $1\leqslant t\leqslant T$.\\
    \hspace{3.3em} Parameters: $\eta_Z$,  $\eta_W$ (step sizes), $R$ (number of iterations).}

\KwOut{$\hat{Z}$ and $\hat{W}_t$ for $1\leqslant t\leqslant T$.}


Let $Z^0 = \mathring{Z}$ and $W_t^0 = \mathring{W}_t$ for $1\leqslant t\leqslant T$.
	
\For{$r = 0, \ldots, R - 1$}{

Let $\tilde{Z}^{r+1}=Z^{r} + {\eta_Z}\, \dot{\ell}_{Z}(Z^r,W^r) $ and \blue{${Z}^{r+1}=\tilde{Z}^{r+1}$}.  

   Let $\tilde{W}_t^{r+1}=W_t^{r} + \eta_W\,  \dot{\ell}_{W_t}(Z^r,W^r)$ and \blue{${W}_t^{r+1}=\tilde{W}_t^{r+1}$ } for $1\leqslant t\leqslant T$.

}

Let \blue{$(\hat{Z}, \hat{W})= (Z^R, {W}^R)$}. 
\end{algorithm}

\paragraph{Numerical closeness}

\begin{figure}[!htbp] 
\centering 
\begin{subfigure}{0.32\textwidth}
	\centering
	\includegraphics[width=1\linewidth]{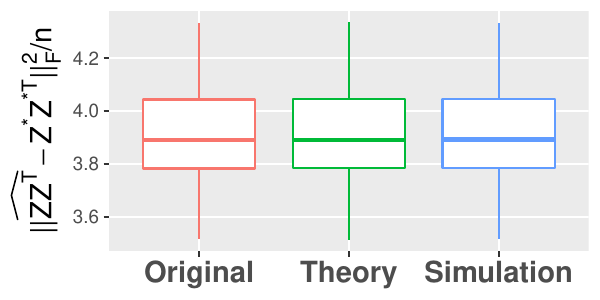}
    \caption{Bernoulli distribution}
\end{subfigure}
\begin{subfigure}{0.32\textwidth}
	\centering
	\includegraphics[width=1\linewidth]{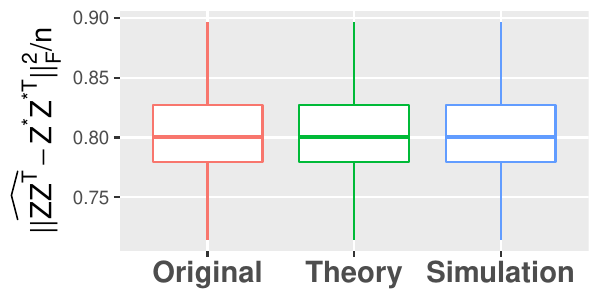}
    \caption{Gaussian distribution}
\end{subfigure}
\begin{subfigure}{0.32\textwidth}
	\centering
	\includegraphics[width=1\linewidth]{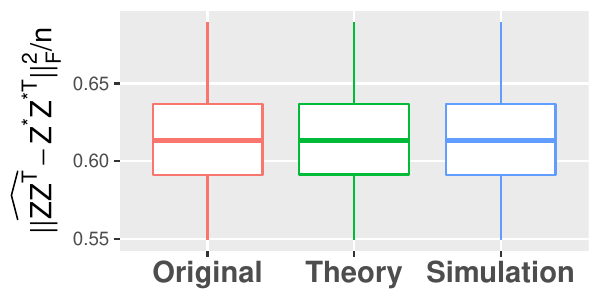}
    \caption{Poisson distribution}
\end{subfigure} 
\caption{Empirical estimation errors of the original Algorithm \ref{algor:refine} and ``Theory'' and ``Simulation'' variants under Case (A). Each box's upper and lower boundaries correspond to the 0.75 and 0.25
 quantiles over $100$ repetitions, while the central horizontal line marks the median of those repetitions.} 
\label{fig:rmk5A}
\end{figure}

\begin{figure}[!htbp]
\centering
\begin{subfigure}{0.32\textwidth}
	\centering
	\includegraphics[width=1\linewidth]{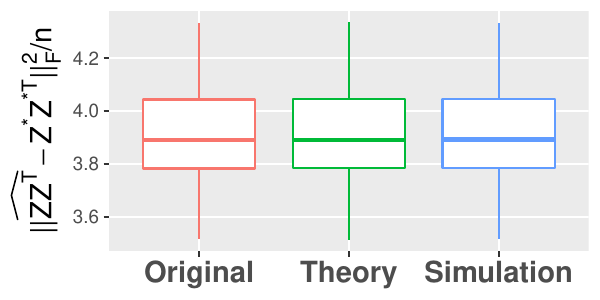}
    \caption{Bernoulli distribution}
\end{subfigure}
\begin{subfigure}{0.32\textwidth}
	\centering
	\includegraphics[width=1\linewidth]{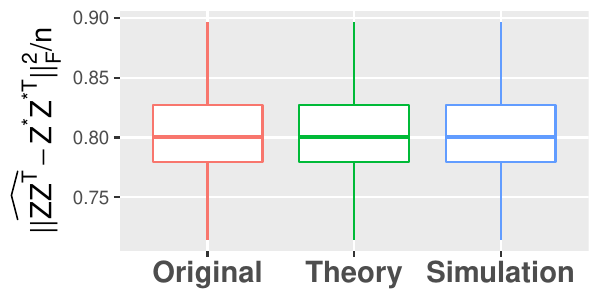}
    \caption{Gaussian distribution}
\end{subfigure}
\begin{subfigure}{0.32\textwidth}
	\centering
	\includegraphics[width=1\linewidth]{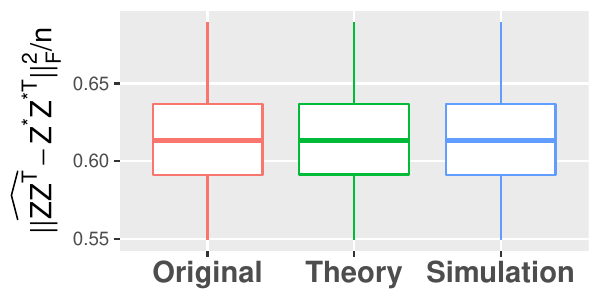}
    \caption{Poisson distribution}
\end{subfigure} 
\caption{Empirical estimation errors of the original Algorithm \ref{algor:refine} and ``Theory'' and ``Simulation'' variants under Case (B). (Similar to Figure \ref{fig:rmk5A}.) } 
\label{fig:rmk5B}
\end{figure}

\begin{figure}[!htbp]
\centering
\begin{subfigure}{0.32\textwidth}
	\centering
	\includegraphics[width=1\linewidth]{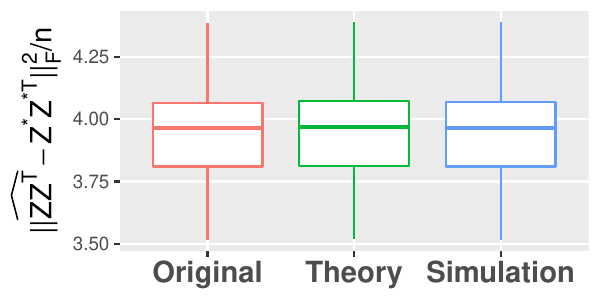}
    \caption{Bernoulli distribution}
\end{subfigure}
\begin{subfigure}{0.32\textwidth}
	\centering
	\includegraphics[width=1\linewidth]{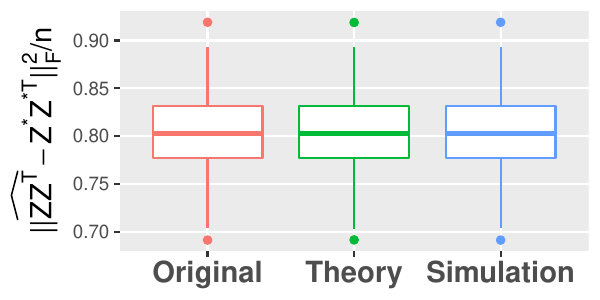}
    \caption{Gaussian distribution}
\end{subfigure}
\begin{subfigure}{0.32\textwidth}
	\centering
	\includegraphics[width=1\linewidth]{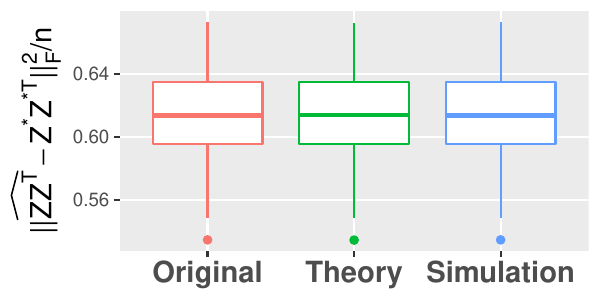}
    \caption{Poisson distribution}
\end{subfigure}  
\caption{Empirical estimation errors of the original Algorithm \ref{algor:refine} and ``Theory'' and ``Simulation'' variants under Case (C). (Similar to Figure \ref{fig:rmk5A}.)  } 
\label{fig:rmk5C}
\end{figure}

We next demonstrate that the original version and the two variants of  Algorithm \ref{algor:refine} perform similarly in simulations. 
We generate data under the three distributions and Cases (A)--(C) as in Section \ref{sec:simulation} with $T=5$. 
In Algorithms \ref{algor:refine} and \ref{algor:refine_theory}, we set the constraint sets $\mathcal{C}_Z$ and $\mathcal{C}_{W_t}$ as in \eqref{eq:constraintsets} with $M_1 = \sqrt{2.5}$.  We employ Barzilai-Borwein step sizes and set $R = 1000$ across all three versions.
Similarly to Section \ref{sec:simulation}, 
we compute the empirical estimation error $\|\widehat{ZZ^\mytrans} - Z^\star Z^{\star \mytrans}\|_{\mathrm{F}}^2/n$ for each version. Figures \ref{fig:rmk5A}-\ref{fig:rmk5C} present the boxplots of these empirical errors over 100 repetitions under Cases (A)--(C), respectively. The results show that the original  Algorithm \ref{algor:refine} and its two variants perform  similarly. 




\newpage 

\subsection{{Details for MultiNeSS and MultiNeSS+ under Case (C)}}\label{sec:suppsim_multiness}

           

In this section, we present the errors of MultiNeSS and MultiNeSS+ under Case (C) and then discuss our intuition for their performance in this case. 

\begin{figure}[!htbp]
    \centering
 \begin{subfigure}{0.29\textwidth}
	\centering
	\includegraphics[width=1\linewidth]{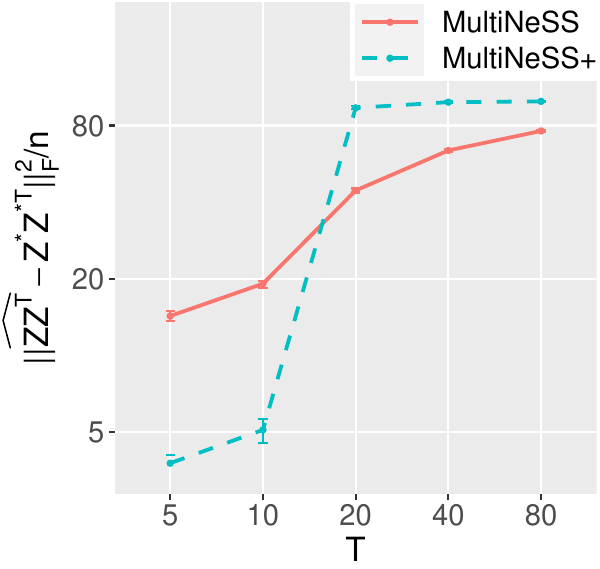}
    \caption{Bernoulli distribution}
\end{subfigure}
\begin{subfigure}{0.29\textwidth}
	\centering
	\includegraphics[width=1\linewidth]{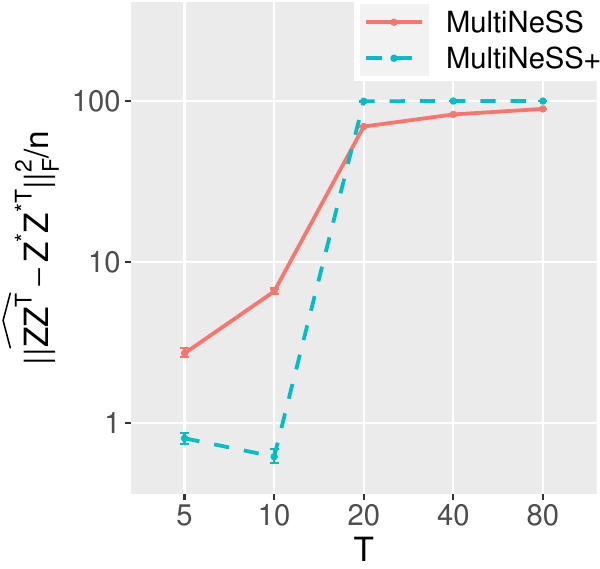}
    \caption{Gaussian distribution}
\end{subfigure} 
    \caption{Empirical $\|\widehat{ZZ^{\top}}-Z^{\star}Z^{\star\top}\|_{\mathrm{F}}^2/n$ versus $T$ under Case (C).  lines connect median estimation errors over 100 repetitions, and error bars are obtained from the 0.05 and 0.95 quantiles of those repetitions. Axes are in the log scale.} 
    \label{fig:ee_multine_zz_supp}
\end{figure}

\paragraph{Numerical results} Figure \ref{fig:ee_multine_zz_supp} shows that  the discrepancies between $Z^{\star}Z^{\star\top}$ and the shared factor estimate $\widehat{ZZ^{\top}}$ by \cite{macdonald2022latent} explode significantly as $T$ increases. 

\begin{figure}[!htbp]
    \centering
 \begin{subfigure}{0.29\textwidth}
	\centering
	\includegraphics[width=1\linewidth]{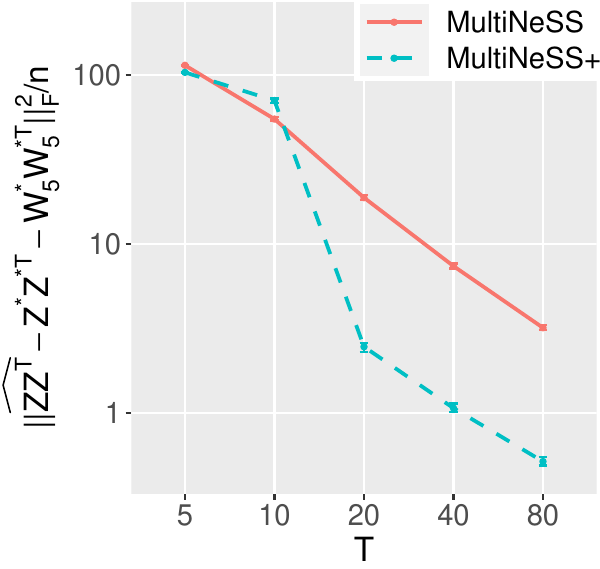}
    \caption{Bernoulli distribution}
\end{subfigure}
\begin{subfigure}{0.29\textwidth}
	\centering
	\includegraphics[width=1\linewidth]{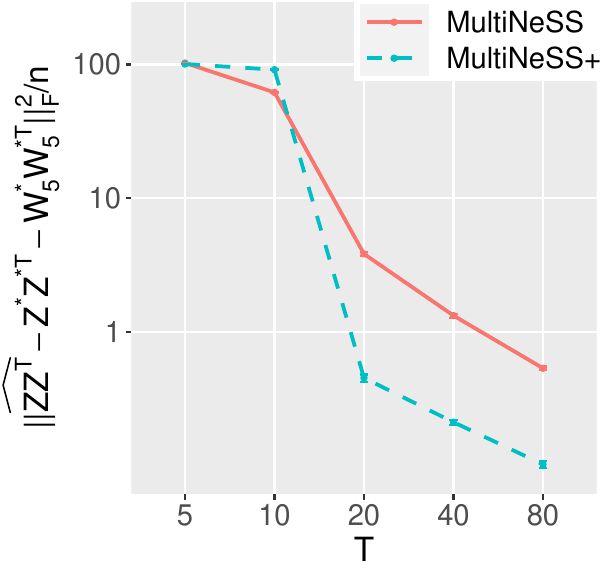}
    \caption{Gaussian distribution}
\end{subfigure} 
    \caption{Empirical $\|\widehat{ZZ^{\top}}-Z^{\star}Z^{\star\top}-W_5^{\star}W_5^{\star\top}\|_{\mathrm{F}}^2/n$ versus $T$ under Case (C). (Similar to Figure \ref{fig:ee_multine_zz_supp}.)} 
    \label{fig:ee_multine_zz_ww_supp}
\end{figure}

On the other hand, Figure \ref{fig:ee_multine_zz_ww_supp} presents the discrepancies between their estimates $\widehat{ZZ^{\top}}$ and  $Z^{\star}Z^{\star\top}+W_5^{\star}W_5^{\star\top}$. It shows that errors decrease with respect to $T$. 

Moreover,  Table \ref{tab:dimension_multiness_supp} summarizes the dimensions of latent spaces estimated by MultiNeSS and MultiNeSS+. It shows that the dimension of the shared factor is mistakenly estimated as $\hat{k}=4$ when $T\geqslant 10$, and the individual factors are mistakenly estimated as $\hat{W}_t=0$ for $5\leqslant t \leqslant T$ when $T\geqslant 20$. 

\begin{table}[!htbp]
 \renewcommand{\arraystretch}{1.3}
  \setlength{\tabcolsep}{15pt} 
    \centering
    \begin{tabular}{c | c c c} \hline 
     $T$     & $\hat{k}$ & $\hat{k}_t \  (1\leqslant t\leqslant 4)$ &   $\hat{k}_t\  (t\geqslant 5)$ \\ \hline 
     5    &  2 & 2  & 2\\ 
     10  &  \textbf{4} & 2  & 2\\
 $\geqslant 20$ & \textbf{4} & \textbf{4} & \textbf{0} \\ \hline 
    \end{tabular}
    \caption{Latent dimensions estimated by MultiNeSS and MultiNeSS+ under Case (C) and $T \in \{5, 10, 20, 40, 80\}$. Bold  numbers highlight mistakenly estimated dimensions.}  
    \label{tab:dimension_multiness_supp}
\end{table}

In summary, the above numerical results suggest MultiNeSS and MultiNeSS+ tend to mistakenly estimate the shared factor across all $T$ networks by including  the space of $W_5^{\star}$.

\paragraph{Intuition}

For the ease of understanding, we first provide a visualization of heterogeneous factors in Case (C) in Figure \ref{fig:casecfactor_supp}. In particular, each $W_t^{\star} \in \mathbb{R}^{n\times 2}$,  $W_1^{\star},W_2^{\star},\ldots, W_5^{\star}$ are orthogonal from each other, whereas $W_5^{\star}=\cdots =W_T^{\star}$. 
A notable feature  of this  case is that $T-4$ heterogeneous factors are the same out of $T$, so the proportion of them increases with respect to $T$. 
Intuitively, when $T$ is larger, it is harder to distinguish $W_5^{\star}=\cdots=W_T^{\star}$ from $Z^{\star}$ that is defined to be the shared factor across all $T$ networks.


\begin{figure}[!htbp]
    \centering
\begin{tikzpicture}[scale=0.8]
    
    \def\spacing{2}
    \def\wheightloc{3.3}
    \def\shiftleft{0.5}
    \def\rectwidth{1-\shiftleft}
    \def\rectheight{2.8}

    \node at (0, \wheightloc ) {$W_1^{\star}$};
    \node at (\spacing, \wheightloc) {$W_2^{\star}$};
    \node at (2*\spacing, \wheightloc) {$W_3^{\star}$};
    \node at (3*\spacing, \wheightloc) {$W_4^{\star}$};
    \node at (4*\spacing, \wheightloc) {$W_5^{\star}$};
    \node at (5*\spacing, \wheightloc) {$W_6^{\star}$};
    \node at (7*\spacing, \wheightloc) {$W_T^{\star}$}; 
    
    \draw[pattern=north east lines, pattern color=red] (0-\shiftleft,0) rectangle (\rectwidth,\rectheight);
    
    \draw[pattern=north west lines, pattern color=orange!80!yellow] (\spacing-\shiftleft,0) rectangle (\spacing+\rectwidth,\rectheight);
    
    \draw[pattern=horizontal lines, pattern color=green!70!black] (2*\spacing-\shiftleft,0) rectangle (2*\spacing+\rectwidth,\rectheight);
    
    \draw[pattern={Dots[distance=2pt]}, pattern color=red!80!black] (3*\spacing-\shiftleft,0) rectangle (3*\spacing+\rectwidth ,\rectheight);
    
    \draw[pattern=vertical lines, pattern color=blue!80!black] (4*\spacing-\shiftleft,0) rectangle (4*\spacing+\rectwidth ,\rectheight);
    
    \draw[pattern=vertical lines, pattern color=blue!80!black] (5*\spacing-\shiftleft,0) rectangle (5*\spacing+\rectwidth,\rectheight);
    
    \foreach \i in {0,1,2} {
        \fill (6*\spacing-\shiftleft + \i*0.4, 2) circle (0.1);
    }
    
    \draw[pattern=vertical lines, pattern color=blue!80!black] (7*\spacing-\shiftleft,0) rectangle (7*\spacing+\rectwidth,\rectheight);
    
\end{tikzpicture}
   \caption{Visualizations of true heterogeneous factors under Case (C). Each $W_t^{\star}\in \mathbb{R}^{n\times 2}$, $\{W_1^{\star},W_2^{\star},W_3^{\star},W_4^{\star},W_5^{\star}\}$   are orthogonal from each other, and $W_5^{\star}=\cdots=W_T^{\star}$.} 
    \label{fig:casecfactor_supp} 
\end{figure}

    Intuitively, we think MultiNeSS and MultiNeSS+ do not perform well  under Case (C)  because their shared space is identified based on the nuclear norm penalty that does not directly adapt to the relationship between heterogeneous factors. 
 Specifically,
let  $F=ZZ^{\top}$ and $G_t=W_t W_t^\top $.  
\cite{macdonald2022latent} solves $F$ and $G$ through a convex relaxation based on the nuclear norm penalty $\lambda \|F\|_* + \sum_{t=1}^{T} \lambda \alpha_t \|G_t\|_*, $
where $\lambda$ and $\alpha_t$ are tuning parameters of the orders of $\sqrt{nT}$ and $1/\sqrt{T}$, respectively. 
Their  estimation results depend on the trade-off between maximizing the likelihood and minimizing the penalty.  
Under our considered settings, true parameters $F^{\star}=Z^{\star}Z^{\star\top}$ and $G_t^{\star}=W_t^{\star}W_t^{\star\top}$ satisfy  $ \|F^{\star}\|_*\asymp n$, $ \|G_t^{\star}\|_* \asymp n$, and 
\begin{align*}
      \lambda \|F^{\star}\|_* + \sum_{t=1}^{T} \lambda \alpha_t \|G_t^{\star}\|_* \asymp \left( \sqrt{nT}  + T \sqrt{n} \right) \times n.
\end{align*} 
Under Case (C) with a large $T$, numerical results above suggest that for  $5\leqslant t\leqslant T$, $\|\hat{G}_t\|_*=\|\hat{W}_t\hat{W}_t^{\top}\|_*=0$,   $\hat{\Theta}_t=\hat{F}+0$ approximates $\Theta_t^{\star} = Z^{\star}Z^{\star\top}+W_t^{\star}W_t^{\star\top}$ well, and we expect  
\begin{align*}
     \lambda \|\hat{F}\|_* + \sum_{t=1}^{T} \lambda \alpha_t \|\hat{G}_t\|_* \asymp \left(\sqrt{nT} + 4\sqrt{n} +  (T-4)\times 0\right) \times n 
\end{align*}
with a high probability. 
Therefore,   
  mistakenly estimating $\|\hat{G}_t\|_*=0$ for  $5\leqslant t\leqslant T$  
  could greatly reduce the order of the penalty without introducing significant bias on estimating corresponding $\Theta_t^{\star}$. 
Although the estimation of $\{\Theta_1^{\star},\Theta_2^{\star}, \Theta_3^{\star},\Theta_4^{\star}\}$ still carries large errors when the shared space is mis-identified, 
networks $5\leqslant t\leqslant T$ takes a more dominant proportion as $T$ increases.

\newpage

\section{{Supplementary Results for Data Analysis}}\label{sec:dataanalysis_supp} 


\subsection{{Latent Dimensions Estimated by MultiNeSS+}}
Table \ref{tab:multiness_dimensions_data} presents dimensions of latent embeddings estimated by MultiNeSS+. 
 \begin{table}[!htbp]
  \renewcommand{\arraystretch}{1.2}
  \setlength{\tabcolsep}{7pt} 
      \centering 
      \begin{tabular}{c|c|c|c|c} \hline 
       Embeddings  &  $Z$ (Shared) &  $W_1$ (Coworker) &   $W_2$ (Friendship) &  $W_3$ (Advice)  \\ \hline 
        Dimensions   & 8 & 3 & 4 & 2 \\ \hline 
      \end{tabular}
      \caption{Estimated latent dimensions by MultiNeSS+  for Lawyers Data.}
      \label{tab:multiness_dimensions_data}
  \end{table}

\vspace{-1em}
\subsection{{Additional Information about  Latent Embeddings}}\label{sec:embeddingdetails}

 For the clarity of presentation, Section \ref{sec:data} in the main text only presents 2-dimensional scatterplots. In the following, we 
discuss more details of the analysis and visualization. 
 Note that latent embeddings are multivariate and  there are three nodewise features in the data. We acknowledge that there is no ground truth and there may also exist other ways of interpenetrating and visualizing latent embeddings.

\begin{table}[!htbp]
\centering
\begin{minipage}[t]{0.3\linewidth}
\centering
\small 
\caption*{(a) $Z$ vs Office}
\begin{tabular}{cc}
\hline 
Embeddings & Office \\
\hline
$\boldsymbol{Z_1}$     & \textbf{0.3} \\
$\boldsymbol{Z_2}$     & \textbf{0.7} \\
$Z_3$     & 0.1 \\
$Z_4$     & 0.1 \\
$Z_5$     & 0.1 \\
$Z_6$     & 0.0 \\
$Z_7$     & 0.1 \\
$Z_8$     & 0.0 \\
\hline 
\end{tabular}
\end{minipage}
\begin{minipage}[t]{0.3\linewidth}
\centering
\caption*{(b) $W_1$ vs Practice}
\begin{tabular}{cc}
\hline 
Embeddings & Practice \\
\hline
$\boldsymbol{W_{1,1}}$  & \textbf{0.1} \\
$W_{1,2}$  & 0.0 \\
$\boldsymbol{W_{1,3}}$  & \textbf{0.4} \\
\hline 
\end{tabular}
\end{minipage}
\begin{minipage}[t]{0.3\linewidth}
\centering
\caption*{(c) $W_2$  vs Status}
\begin{tabular}{cc}
\hline 
Embeddings & Status \\
\hline
$W_{2,1}$  & 0.0 \\
$\boldsymbol{W_{2,2}}$  & \textbf{0.8} \\
$\boldsymbol{W_{2,3}}$  & \textbf{0.1} \\
$W_{2,4}$  & 0.0 \\
\hline 
\end{tabular}
\end{minipage}
\caption{Correlations  (rounded to one digit) between embeddings and nodewise features. Bolded rows correspond to the largest two correlations.} 
\label{tab:multiness_corr}
\end{table}

\begin{figure}[!htbp]
    \centering
    \includegraphics[width=0.95\linewidth]{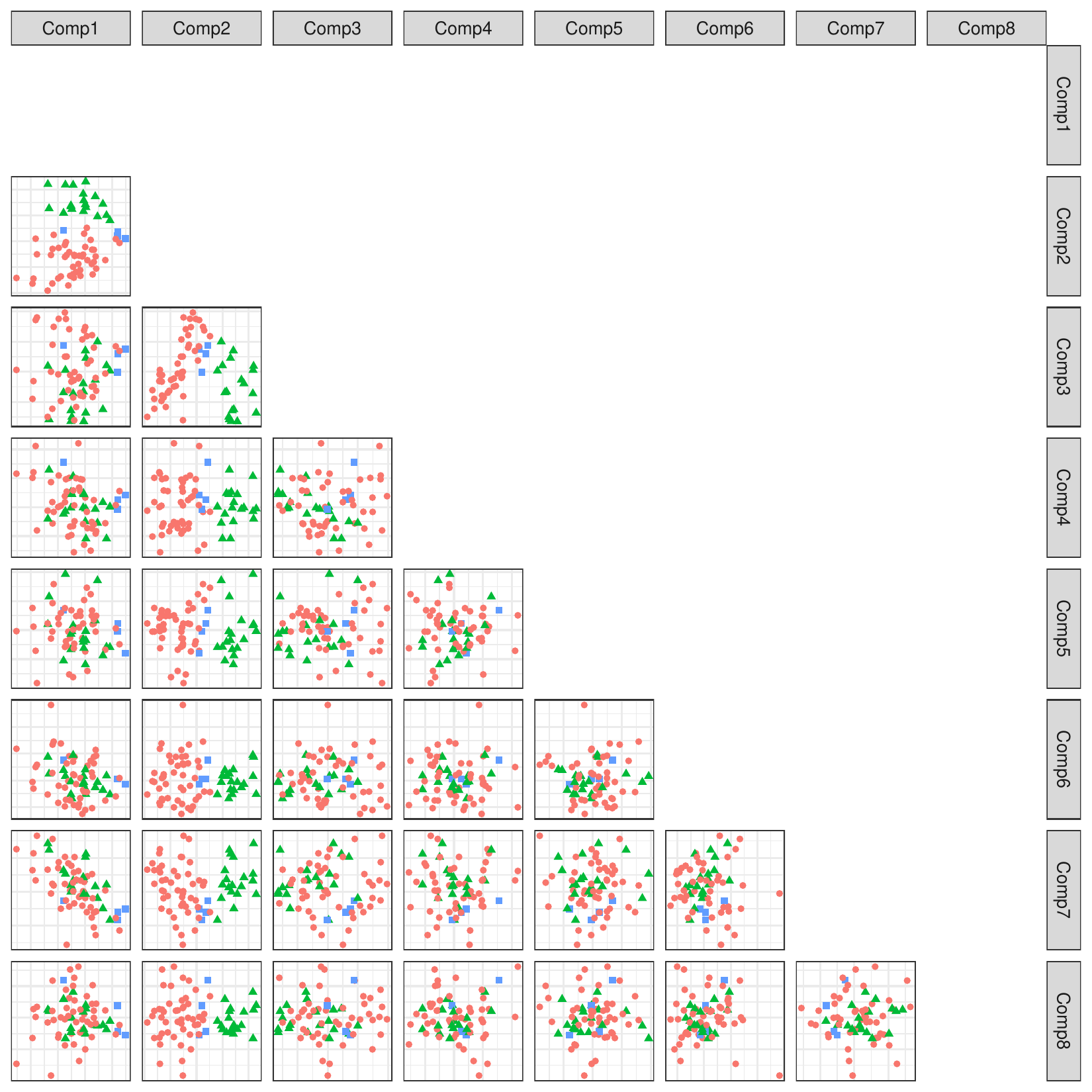}
    \caption{Pairwise scatterplots of eight dimensions of $Z$ estimated by MultiNeSS+. Points are colored based on lawyers' office:  Boston (round),     Hartford (triangle),  Providence (square).} 
    \label{fig:Zpairscatter}
\end{figure}

\begin{figure}[!htbp]
    \centering
    \includegraphics[width=0.5\linewidth]{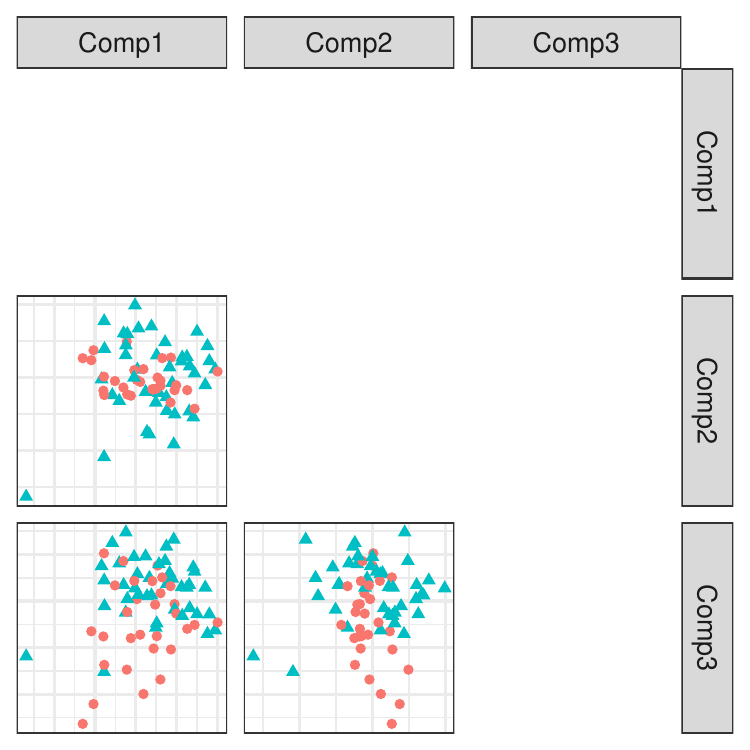}
    \caption{Pairwise scatterplots of three dimensions of $W_1$ estimated by MultiNeSS+. Points are colored based on lawyers' practice: corporate (round) and litigation (triangle).} 
    \label{fig:W1pairscatter}
\end{figure}

\begin{figure}[!htbp]
    \centering
    \includegraphics[width=0.65\linewidth]{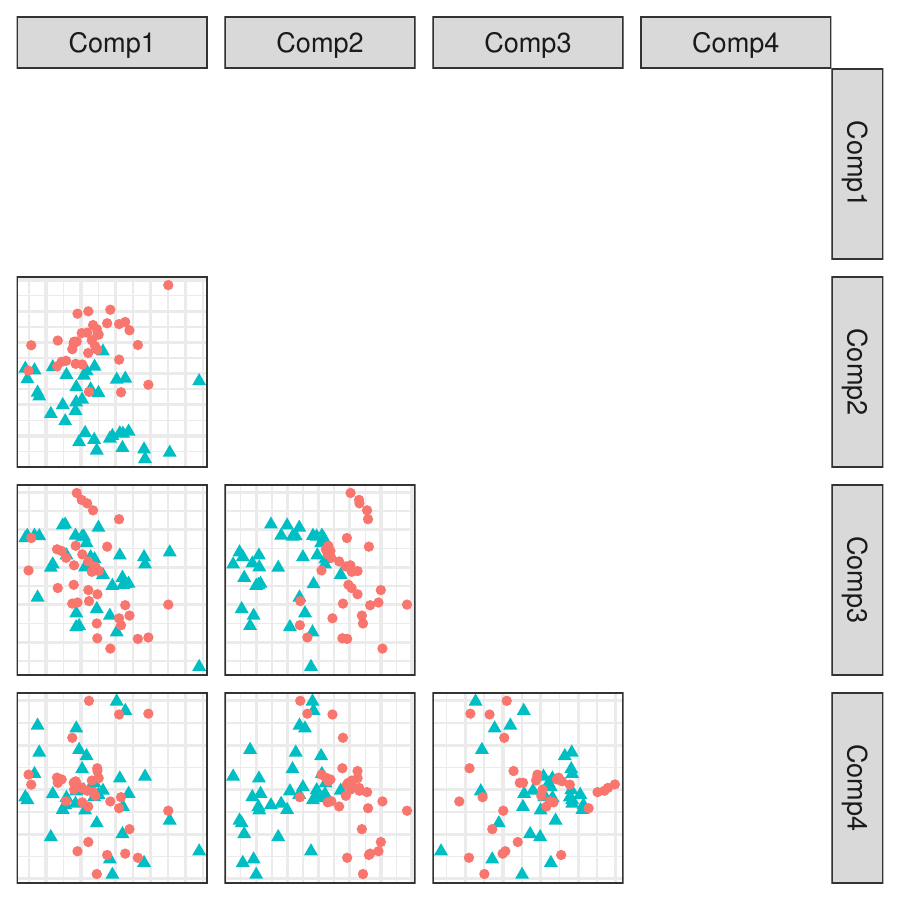}
    \caption{Pairwise scatterplots of four dimensions of $W_2$ estimated by MultiNeSS+. Points are colored based on lawyers' status: associate (round) and partner (triangle).} 
    \label{fig:W2pairscatter}
\end{figure}

For SS-Refinement, Components 1 and 2 in Figure \ref{fig:Z}--\ref{fig:estW2} are the two  principal components scores associated with the  two largest  variances, which follows from the convention in principal component analysis. It turns out to give interpretable separation based on three nodewise features shown in Section \ref{sec:data}. 
For MultiNeSS+, we find that similarly visualizing leading  principal components scores may not give clear interpretation.  
Instead, we compute Pearson's correlations between each dimension of latent embeddings and nodewise features, shown in Table \ref{tab:multiness_corr}. To facilitate interpretation and comparison, scatterplots of MultiNeSS+ in  Figure \ref{fig:Z}--\ref{fig:estW2} visualize the two dimensions that achieve the largest two correlations in Table \ref{tab:multiness_corr}, highlighted in bolded rows.  
To provide more comprehensive information about other embeddings estimated by MultiNeSS+, we 
 further provide scatterplots of pairwise columns in $Z, W_1,$ and $W_2$ in Figures  \ref{fig:Zpairscatter}--\ref{fig:W2pairscatter}, respectively.


\newpage 

\newpage


\section{{Extensions Beyond Natural Exponential Family}} \label{sec:expfam}

In the main text, Condition  \ref{cond:parfunction} assumes $l(\theta;x)$ belongs to natural exponential family distributions so that proofs above can be neater to read and understand. 
Nevertheless, we emphasize that the proposed methodology in Section \ref{sec:estall} is general and remains valid under conditions beyond natural exponential family distributions. 
We next discuss generalized conditions in Section \ref{sec:expfam_theory}, with theory and proofs given in Section \ref{sec:pfthmgeneraldist}. 


\subsection{{Model Conditions}}
\label{sec:expfam_theory}


Our theoretical results  remain valid when Condition \ref{cond:parfunction} on natural exponential family distribution is relaxed and replaced by the following regularity conditions on $l(\theta; x)$  in \eqref{eq:zwlikelihood}  and $\mu(\theta)$ in \eqref{eq:model2}. 



\begin{condition} \label{cond:parfunction1}
Let $\mathcal{X} = \{x \in \mathbb R : p(\,x\mid \theta\,) > 0\}$ denote the support of $p(\,\cdot \mid \theta\,)$.
Assume $l(\theta;x)$ in \eqref{eq:zwlikelihood}  satisfies the following conditions. 
\begin{itemize}\setlength{\itemsep}{0pt}
\item[(i)] For any fixed $x\in \mathcal{X}$, $l(\theta;x)$ is three times differentiable with respect to $\theta$. Moreover, there exist positive constants $\kappa_1$, $\kappa_2$, and $\kappa_3$ such that $\kappa_1 \leqslant -l^{\prime\prime}(\theta;x) \leqslant \kappa_2$ and $|l^{\prime\prime\prime}(\theta;x)| \leqslant \kappa_3$ for any  $x \in \mathcal{X}$ and $|\theta| \leqslant 2M_1^2$, where $M_1$ is given in Condition \ref{cond:truevalue}. 

\item[(ii)] There exists a constant $L>0$ such that $\EXPT |l^{\prime}(\Theta_{t,ij}^\star;A_{t,ij})|^{m} \leqslant \VAR\{l^{\prime}(\Theta_{t,ij}^\star;A_{t,ij})\} L^{m -2} m ! /2$ for any  $1 \leqslant i \leqslant j \leqslant n$, $1 \leqslant t \leqslant T$, and any integer $m \geqslant 2$, {where $m!=m(m-1)\cdots 1$ represents the factorial of $m$.} 
\end{itemize}
\end{condition}

\begin{condition} \label{cond:expfam}
Assume $\mu(\theta)$ in \eqref{eq:model2} satisfies the following conditions. 
\begin{itemize}\setlength{\itemsep}{0pt}

\item[(i)] The function $\mu( \theta)$ is continuously differentiable, and its derivative satisfies $\kappa_1 \leqslant |\mu^{\prime}(\theta)| \leqslant \kappa_2$ for any $|\theta| \leqslant  2M_1^2$, where $\kappa_1$ and $\kappa_2$ are given in Condition \ref{cond:parfunction1} (i).


\item[(ii)] For any  $1 \leqslant i \leqslant j \leqslant n$, $1 \leqslant t \leqslant T$, and any integer $m \geqslant 2$, $\EXPT|A_{t,ij} - \mu(\Theta_{t,ij}^\star) |^{m} \leqslant  \VAR(A_{t,ij}) L^{m-2} m ! /2$, where $L$ is given in Condition \ref{cond:parfunction1} (ii).
\end{itemize}
\end{condition}

Condition \ref{cond:parfunction1} differs from  Condition \ref{cond:parfunction} only by removing the requirement that $l(\theta; x)$ belongs to  natural exponential family distributions. 
Condition \ref{cond:expfam} (i) assumes the differentiability and bounded derivative of the mean function $\mu(\theta)$, which is satisfied by most standard models. Condition \ref{cond:expfam} (ii) assumes a  Bernstein moment condition for all edges $A_{t,ij}$, which provides convenience to use  concentration inequalities in theoretical proofs.
Notably, when Condition \ref{cond:parfunction} holds (i.e., under the natural exponential family assumption), Condition \ref{cond:expfam} is implied by Condition \ref{cond:parfunction1} given  
$ - l^{\prime \prime}(\theta;x) = \mu^{\prime}(\theta)  \text{ and } l^{\prime}(\Theta_{t,ij}^\star;A_{t,ij}) = A_{t,ij} - \mu(\Theta_{t,ij}^\star)$. 


\begin{remark} \label{rm:generaldistassumption}
Conditions \ref{cond:parfunction1}--\ref{cond:expfam} can be satisfied under models beyond natural exponential families. 
For example, for binary variables,
besides the canonical form of logistic link function, 
another prevalent modeling approach considers probit link function \citep{agresti2015foundations}, i.e., success probability takes the  cumulative distribution function of standard normal distribution $\Phi(\cdot)$ as the link. 
This model does not belong to natural exponential families,  but we argue that it can satisfy Conditions  \ref{cond:parfunction1}--\ref{cond:expfam}. 

Specifically, under the probit model, Condition \ref{cond:expfam} (i) follows by the property of $\Phi'(\theta)$, i.e., density of standard normal distribution, and Conditions \ref{cond:parfunction1} (ii) and \ref{cond:expfam} (ii) follow from the sub-Gaussian behavior of bounded random variables and 
$ l'(\theta;x)=\{x-\Phi(\theta)\}\Phi'(\theta)/[\Phi(\theta)\{1-\Phi(\theta)\}]$. 
In addition, $l(\theta, x)$ satisfies Condition \ref{cond:parfunction1}  (i)  for bounded $\theta$, by noticing that 
\begin{align*}
    l^{\prime \prime}(\theta;0)  = -\left\{1 - \theta   \frac{1 - \Phi(\theta)}{\Phi'(\theta)}\right\}\left\{ \frac{\Phi'(\theta)}{1-\Phi(\theta)} \right\}^2, \ \   l^{\prime \prime}(\theta;1)  = -\left\{ 1 + \theta  \frac{\Phi(\theta)}{\Phi'(\theta)}\right\} \left\{ \frac{\Phi'(\theta)}{\Phi(\theta)} \right\}^2, 
\end{align*}
 $\Phi'(\theta)>0$,  and  Mill's ratio inequality for standard normal distribution \citep[e.g.,][7.1.13]{abramowitz1965handbook}: 
\begin{align*}
    \frac{1}{\theta+\sqrt{\theta^2+4}}  < \frac{1-\Phi(\theta)}{\Phi'(\theta)} < \frac{1}{\theta+\sqrt{\theta^2+8/\pi}} \hspace{2em} \text{ when } \theta >0.
\end{align*}

 On the other hand, although our methods 
may be applied to  the random dot product graph (RDPG) model \citep{young2007random},
developing rigorous theoretical guarantee requires proper  modification to the analysis. 
In particular,  our current theoretical analysis requires that the log likelihood function $l(\theta;x)$ is well-behaved in 
a ball $\{\theta: |\theta| \leqslant \text{constant}\}$ for convenience. 
This is the natural domain for the corresponding success probability $\mu(\theta)$ to be bounded away from 0 and 1 under logistic and probit links. 
But the RDPG model takes $\mu(\theta)=\theta $ and thus requires $\theta\in (0,1)$ for proper modeling.  
Although we think that our technical idea could be  extended similarly, it would require 
a customized analysis 
for the specific parameter domain of the RDPG model. 
It would be an interesting future research direction to develop  detailed and rigorous analysis for the RDPG model.  
\end{remark}

\subsection{{Theory}} \label{sec:pfthmgeneraldist}

The following theorem shows that our theoretical conclusions still hold when replacing Condition \ref{cond:parfunction} by Conditions \ref{cond:parfunction1} and \ref{cond:expfam}. 

\begin{theorem} \label{thm:generaldist}
Under Conditions \ref{cond:tuning12}--\ref{cond:truevalue}, \ref{cond:truevalue2}, \ref{cond:parestTheta1}, and \ref{cond:parfunction1}--\ref{cond:expfam}, Theorems \ref{thm:estTheta}, \ref{thm:estY}, \ref{thm:initial}, and  \ref{thm:onestep} hold. 
\end{theorem}

Since the proofs follow a similar structure to those in Section \ref{sec:proof}, we only highlight arguments in the previous proofs that depend  on Condition \ref{cond:parfunction}  and explain how they may be established under Conditions \ref{cond:parfunction1} and \ref{cond:expfam} instead. 


\subsubsection{Proof of Theorem \ref{thm:estTheta}}

For the proof of Theorem \ref{thm:estTheta} in Section \ref{sec:pf_thmA1},
only derivations of \eqref{eq:thm4-3} and 
\eqref{eq:at_bernstein_a1} depend on Condition \ref{cond:parfunction}. 
First, \eqref{eq:thm4-3} still holds by the Lipschitz continuity of $\mu(\cdot)$ and $\mu^{-1}(\cdot)$  ensured by 
Condition \ref{cond:expfam} (i). 
Second, \eqref{eq:at_bernstein_a1} still holds by Bernstein moment conditions of $A_{t,ij} -\mu(\Theta_{t,ij}^{\star})$ in Condition \ref{cond:expfam} (ii) and the arguments in Lemma \ref{lem:concentration} when replacing $l'(\Theta_t^{\star})$ with $\mathbf{A}_t-\mu(\Theta_t^{\star})$.  



\subsubsection{Proof of Theorem \ref{thm:estY}}

For the proof of Theorem \ref{thm:estY} in 
Section \ref{sec:pf_thm5}, 
only the derivation of \eqref{eq:thm5-5} and the use of Lemma \ref{lem:concentration} depend on Condition \ref{cond:parfunction}. 
First, \eqref{eq:thm5-5} still holds by the smoothness and convexity assumptions on $-l(\theta;x)$ in Condition \ref{cond:parfunction1} (i).
Second, Lemma \ref{lem:concentration} still holds by the Bernstein moment conditions of $l'(\Theta_{t,ij}^{\star};A_{t,ij})$ in  Condition \ref{cond:parfunction1} (ii).


\subsubsection{Proof of Theorem \ref{thm:initial}}
Given Theorems \ref{thm:estTheta} and \ref{thm:estY} established, 
the proof of Theorem \ref{thm:initial} in Section \ref{sec:pf:thm:initial} does not depend on Condition \ref{cond:parfunction} and thus, all the arguments still hold under Conditions \ref{cond:parfunction1} and \ref{cond:expfam}. 
                                                           

\subsubsection{Proof of Theorem \ref{thm:onestep}}

{Recall that the proof of Theorem \ref{thm:onestep} in Section \ref{sec:pf:thm:onestep} consists of several components. We next discuss them one-by-one. }

\paragraph{Proof of \eqref{eq:thm2-1}}
For the proof of \eqref{eq:thm2-1} in Section \ref{sec:pdfthm2-1}, only the derivation of \eqref{eq:thm2-8} and \eqref{eq:pfd36-3}, as well as the use of Lemma \ref{lem:concentration2}, depend on Condition \ref{cond:parfunction}. First, \eqref{eq:thm2-8} and \eqref{eq:pfd36-3} still hold by the smoothness and convexity assumptions on $-l(\theta;x)$ in Condition \ref{cond:parfunction1} (i).
Second, Lemma \ref{lem:concentration2} still holds by the Bernstein moment conditions of $l'(\Theta_{t,ij}^{\star};A_{t,ij})$ in  Condition \ref{cond:parfunction1} (ii). 


\paragraph{Formula of $\dot{\ell}(v)$ and $I(v)$ and properties} 

Formula of $\dot{\ell}(v)$ remains the same as in \eqref{eq:ldotform}. 
Formula of $I(v)$ relies on Condition \ref{cond:parfunction} only in \eqref{eq:dmutl} where we use $  \mathbb{E}\{l'(\Theta_{t,ij}) l'(\Theta_{t,ij})\} = -l''(\Theta_{t,ij})$, 
which is deterministic 
as mentioned in Remark \ref{rm:lprimenotation}. 
To simplify practical computation and theoretical analysis, we 
 redefine 
 \begin{small}
     \begin{align*} 
 I(v)=  
  \begin{pmatrix}
        \sum_{t=1}^T \DZ(v) \mathcal{D}_{\Theta_t}(v) \DZt(v) &  \DZ(v) \mathcal{D}_{\Theta_1}(v) \mathcal{D}_{\Theta W_1}(v) & \ldots & \DZ(v) \mathcal{D}_{\Theta_T}(v)\mathcal{D}_{\Theta W_T}(v) \\
        \mathcal{D}_{W_1 \Theta}(v) \mathcal{D}_{\Theta_1}(v)  \DZt(v) & \mathcal{D}_{W_1 \Theta}(v) \mathcal{D}_{\Theta_1}(v) \mathcal{D}_{\Theta W_1}(v) & \ldots & 0 \\
        \vdots & \vdots & \ddots & \vdots \\
        \mathcal{D}_{W_T \Theta}(v) \mathcal{D}_{\Theta_T}(v) \DZt(v) & 0 & \ldots & \mathcal{D}_{W_T \Theta}(v) \mathcal{D}_{\Theta_T}(v) \mathcal{D}_{\Theta W_T}(v) \end{pmatrix}
\end{align*}\end{small}with $\mathcal{D}_{\Theta_t}(v) = - \operatorname{diag}\big(2 l^{\prime \prime}(\Theta_{t,11}), \ldots, 2 l^{\prime \prime}(\Theta_{t,nn}),  l^{\prime \prime}(\Theta_{t,12}), \ldots, l^{\prime \prime}(\Theta_{t,n-1,n})\big).$ This definition is the same as \eqref{eq:ivform} and thus is equivalent to $\EXPT\{\dot \ell(v) \dot \ell(v)^\mytrans\}$ under natural exponential family distributions. 
Under general distributions, $l^{\prime \prime}(\Theta_{t,ij})$ could depend on $A_{t,ij}$ and thus be random even with fixed  $\Theta_{t,ij}$, which makes $I(v)$ defined above differ from $\EXPT\{\dot \ell(v) \dot \ell(v)^\mytrans\}$ but we have
\begin{align} \label{eq:exp_iv}
  \mathbb{E}\{I(v)\}=\EXPT\{\dot \ell(v) \dot \ell(v)^\mytrans\}.
\end{align}
{Moreover, the definition of Hessian matrix $H(v)$ in \eqref{eq:hvform} and formula \eqref{eq:hv} still follow.}
The properties of $I(v)$ in Lemma \ref{lm:ivproperties} still hold, since the derivations only rely on the formula in  \eqref{eq:ivform} and assumptions of boundedness and Lipschitz continuity of $l^{\prime \prime}(\theta;x)$ in Condition \ref{cond:parfunction} (i) which are implied by redefined $I(v)$ and Condition \ref{cond:parfunction1} (i). 

{In addition, to facilitate the following proofs, we show that $\mathbb{E}\{I(v)\}$ have the following properties.}

\begin{lemma} \label{lm:exp_ivproperties}
Under the conditions of Lemma   \ref{lm:ivproperties}, we have 
\begin{itemize}
    \item[(i)] $\big\|\mathcal{D}_{nT}^{1/2} \, \{\mathbb{E}I(v)\}^{+} \, \mathcal{D}_{nT}^{1/2}\big\|_{\mathrm{op}}=O(1)$,
\item[(ii)] $\operatorname{col}(I({v})) = \operatorname{col}(\EXPT I({v}))$.
\end{itemize} 
\end{lemma}

\begin{proof}
    According to the proof of Lemma \ref{lm:ivproperties},  the column space and non-zero eigenvalue order of $I(v)$ are determined by the properties of $\DZ(v)$ and $\Dwt(v)$ when $\mathcal{D}_{\Theta_t}(v)$ is a diagonal matrix with bounded entries. Since $\EXPT \{ I(v)\}$ differs from $I(v)$ only through the additional expectation over $\mathcal{D}_{\Theta_t}(v)$ and $\EXPT \{\mathcal{D}_{\Theta_t}(v)\} $ remains a diagonal structure with bounded entries, the same spectral and column space properties carry over to $\EXPT \{I(v)\}$.
\end{proof}

 


\paragraph{Proof of  \eqref{eq:zhaterr}}
For the proof of \eqref{eq:zhaterr} in Section \ref{sec:pdfzhaterr}, it still suffices to prove 
\eqref{eq:thm2-17}--\eqref{eq:thm2-19} following the same arguments. 
For \eqref{eq:thm2-18} and \eqref{eq:thm2-19}, the same  proofs follow given $I(v)$ redefined and properties in  Lemma \ref{lm:ivproperties} argued above.

To prove \eqref{eq:thm2-17}, 
most arguments still follow except that  \eqref{eq:thm2-20} and \eqref{eq:variance_sum_bd}  require that  $I(v^{\star})$ is deterministic and $I(v^{\star})= \EXPT\{\dot \ell(v^{\star}) \dot \ell(v^{\star})^\mytrans\}$, which may not hold under general distributions. 
Nevertheless, motivated by \eqref{eq:exp_iv}, we utilize the decomposition $I(v)^{+}=\{\mathbb{E}I(v)\}^{+} + [ I(v)^{+}- \{\mathbb{E}I(v)\}^{+}] $ and then 
\begin{align*}
    S_1 = \big\|  \mathcal{D}_{Zv}\, I({v}_{q}^\star)^+ \dot \ell({v}_{q}^\star)\big\|_2^2 = \big\|\mathcal{D}_{Zv}\, I({v}^\star)^+ \dot \ell({v}^\star)\big\|_2^2 \leqslant 2S_{11}+2S_{12},
\end{align*}
where we define
\begin{align*}
    S_{11} = \big\| \mathcal{D}_{Zv} \{\EXPT I({v}^\star)\}^+ \dot \ell({v}^\star)\big\|_2^2 \quad \text{ and }\quad 
    S_{12} = \big\| \mathcal{D}_{Zv} \big(I({v}^\star)^+ - \{\EXPT I({v}^\star)\}^+ \big) \dot \ell({v}^\star)\big\|_2^2.
\end{align*}
Since $\EXPT I({v}^\star) $ is deterministic, by \eqref{eq:exp_iv} and Lemma \ref{lm:exp_ivproperties} (i),  $S_{11}$ can be bounded following the same arguments in  Section \ref{sec:pdfzhaterr}. 

We now show $\| \mathcal{D}_{Zv} (I({v}^\star)^+ - \{\EXPT I({v}^\star)\}^+ ) \dot \ell({v}^\star)\|_2^2$ exhibits the same upper bound as in \eqref{eq:thm2-17}.
Lemma \ref{lm:exp_ivproperties} (ii) gives $\operatorname{col}(I({v}^\star)) = \operatorname{col}(\EXPT I({v}^\star))$, then 
\begin{align*}
    &\quad \ \big\| \mathcal{D}_{Zv} \big(I({v}^\star)^+ - \{\EXPT I({v}^\star)\}^+ \big) \dot \ell({v}^\star)\big\|_2^2 = \big\| \mathcal{D}_{Zv}\, I({v}^\star)^+ \big(I({v}^\star)- \EXPT I({v}^\star) \big) \{\EXPT I({v}^\star)\}^+ \dot \ell({v}^\star)\big\|_2^2 \\
    &\leqslant \big\| \mathcal{D}_{Zv}\, I({v}^\star)^+ \DnT^{1/2} \big\|_{\operatorname{op}}^2 \big\| \DnT^{-1/2}\big(I({v}^\star)- \EXPT I({v}^\star) \big) \DnT^{-1/2} \big\|_{\operatorname{op}}^2 \big\| \DnT^{1/2} \{\EXPT I({v}^\star)\}^+ \dot \ell({v}^\star)\big\|_2^2 \\
    &\leqslant C_\varepsilon (nT)^{-1} \cdot n^{-1}\log(nT) \cdot nT\log(nT)  = C_\varepsilon n^{-1} \log^2(nT)  
\end{align*}
with probability $1 - O((nT)^{-\varepsilon})$, where \begin{align*}
    \big\| \DnT^{1/2} \{\EXPT I({v}^\star)\}^+ \dot \ell({v}^\star)\big\|_2^2 \leqslant C_\varepsilon \operatorname{tr}\big(\DnT^{1/2} \{\EXPT I({v}^\star)\}^+ \DnT^{1/2} \big) \log(nT) \leqslant C_\varepsilon nT \log(nT)
\end{align*} 
follows from  the
same arguments in  \eqref{eq:thm2-20-0}--\eqref{eq:thm2-22}, and  $$\big\| \DnT^{-1/2}(I({v}^\star)- \EXPT I({v}^\star) ) \DnT^{-1/2} \big\|_{\operatorname{op}}^2 \leqslant  C_\varepsilon n^{-1}\log(nT)$$ follows from a block-wise analysis. For example, considering the left-top block of $\DnT^{-1/2}(I({v}^\star)- \EXPT I({v}^\star) ) \DnT^{-1/2}$:
\begin{align*}
    &\,n^{-1} \,\big\|\DZ({v}^\star) \mathcal{D}_{\Theta_t}({v}^\star) \DZt({v}^\star) - \DZ({v}^\star) \EXPT[\mathcal{D}_{\Theta_t}({v}^\star)] \DZt({v}^\star)\big\|_{\operatorname{op}} \\
   \leqslant &\, n^{-1}  \left\| \sum_{j=1}^n \begin{pmatrix}
       (2 + \mathbb I(j=1)) m_{t,1j}^\star z_j^\star z_j^{\star \mytrans}  & \cdots & 0 \\
        \vdots & \ddots & \vdots \\
        0  & \cdots & (2 + \mathbb I(j=n)) m_{t,nj}^\star z_j^\star z_j^{\star \mytrans} 
    \end{pmatrix} \right\|_{\operatorname{op}} \\
    &+ n^{-1} \left\|\begin{pmatrix}
        m_{t,11}^\star z_1^\star z_1^{\star \mytrans}  & \cdots & m_{t,n1}^\star z_n^\star z_1^{\star \mytrans}  \\
        \vdots & \ddots & \vdots \\
         m_{t,1n}^\star z_1^\star z_n^{\star \mytrans}  & \cdots & m_{t,nn}^\star z_n^\star z_n^{\star \mytrans} 
\end{pmatrix}\right\|_{\operatorname{op}} \\
    \leqslant &\,C_\varepsilon n^{-1/2} \log^{1/2}(nT)
\end{align*}
with probability $1 - O((nT)^{-\varepsilon})$, where $m_{t,ij }^\star := \EXPT [l^{\prime \prime}(\Theta_{t,ij}^\star)] - l^{\prime \prime}(\Theta_{t,ij}^\star)$ are bounded random variables, and the bounds follow from Bernstein and matrix Bernstein inequalities.


\paragraph{Proof of \eqref{eq:wthaterr}}
For the proof of \eqref{eq:wthaterr} in Section \ref{sec:pdfwthaterr}, only the deterministic property of $I(v)$ and Lemma \ref{lm:ivproperties} rely on Condition \ref{cond:parfunction}. Similar to the proof of \eqref{eq:zhaterr}, the main task reduces to bounding the additional term $$\big\| \mathcal{D}_{W_t v} \big(I({v}^\star)^+ - \{\EXPT I({v}^\star)\}^+ \big) \dot \ell({v}^\star) \big\|_2^2 = \big\| \mathcal{D}_{W_t v}\, I({v}^\star)^+ \big(I({v}^\star)- \EXPT I({v}^\star) \big) \{\EXPT I({v}^\star)\}^+ \dot \ell({v}^\star) \big\|_2^2.$$ 
Applying \eqref{eq:thm2-28-1} shows 
\begin{align*}
    &\quad \ \big\| \mathcal{D}_{W_t v}\, I({v}^\star)^+ \big(I({v}^\star)- \EXPT I({v}^\star) \big) \{\EXPT I({v}^\star)\}^+ \dot \ell({v}^\star)\big\|_2^2 \\
    &\lesssim \big\| I_{W_tW_t}({v}^\star)^+ \mathcal{D}_{W_t v} \big(I({v}^\star)- \EXPT I({v}^\star) \big) \{\EXPT I({v}^\star)\}^+ \dot \ell({v}^\star)\big\|_2^2 \\
    &+ \big\| I_{W_tW_t}({v}^\star)^+ I_{ZW_t}({v}^\star) \mathcal{D}_{Z v} \, I({v}^\star)^+ \big(I({v}^\star)- \EXPT I({v}^\star) \big) \{\EXPT I({v}^\star)\}^+ \dot \ell({v}^\star)\big\|_2^2 \\
    &\lesssim \big\| I_{W_tW_t}({v}^\star)^+ \big\|_{\operatorname{op}}^2 \big\| \mathcal{D}_{W_t v} \big(I({v}^\star)- \EXPT I({v}^\star)  \big) \mathcal{D}_{Zv}^\mytrans \big\|_{\operatorname{op}}^2 \big\| \mathcal{D}_{Zv} \{\EXPT I({v}^\star)\}^+ \dot \ell({v}^\star)\big\|_2^2 \\
    &+ \big\| I_{W_tW_t}({v}^\star)^+ \big\|_{\operatorname{op}}^2 \big\| \mathcal{D}_{W_t v} \big(I({v}^\star)- \EXPT I({v}^\star)  \big) \mathcal{D}_{W_t v}^\mytrans \big\|_{\operatorname{op}}^2 \big\| \mathcal{D}_{W_tv} \{\EXPT I({v}^\star)\}^+ \dot \ell({v}^\star)\big\|_2^2 \\
    &+ \big\| I_{W_tW_t}({v}^\star)^+ \big\|_{\operatorname{op}}^2 \big\| I_{ZW_t}({v}^\star) \big\|_{\operatorname{op}}^2\big\|\mathcal{D}_{Z v} \, I({v}^\star)^+ \big(I({v}^\star)- \EXPT I({v}^\star) \big) \{\EXPT I({v}^\star)\}^+ \dot \ell({v}^\star)\big\|_2^2 \\
    & \leqslant C_\varepsilon n^{-1} \log^2(nT)
\end{align*}
with probability $1 - O((nT)^{-\varepsilon})$.
In summary, we can obtain the error rates consistent with those established under the natural exponential family framework.

\newpage

\section{{Extensions Beyond Dense Networks}} \label{sec:sparse_extend}


Our model in \eqref{eq:model} 
can also be extended to allow flexible overall edge density.  
In the existing literature, 
one prevalent approach to adjust network density 
is to introduce an overall edge density parameter  $\rho$ such that  
$\mathbb{E}(A_{t,ij}) \asymp \rho$ 
\citep{lei2023bias},
where  $\rho \to 0$ as $n\to \infty$ gives sparser networks when edges are  binary. 
The idea can be similarly adopted under our model. 
We next present a generalized model, and then discuss how our proposed procedure and theory can be extended.

\subsection{{Model}}\label{sec:sparse_model}

In the following, we consider a generalized log likelihood of natural exponential family distribution satisfying
\begin{align} \label{eq:like_sparse}
    l(\theta;x) \propto \theta x - \rho \nu(\theta), 
\end{align}
where $\nu(\theta)$ is a log-partition function in classical natural exponential families, and $\rho \in (0,1]$ is an overall edge density parameter. 
Under the model \eqref{eq:like_sparse}, we have $\EXPT(A_{t,ij}) = \rho\nu^{\prime}(\Theta_{t,ij})$, and $\rho=1$ reduces to the classical natural exponential families with $\mu(\theta)=\nu'(\theta)$. 
We consider  \eqref{eq:like_sparse}  for the  ease of presenting theory as the relationship between log likelihood and expected adjacency has explicit  characterization, though we expect results under general distributions can be similarly extended following the idea in Section \ref{sec:expfam} with more detailed technical arguments.



\subsection{{Estimation Procedure}} \label{sec:sparse_estimation}

When $\rho$ is known, 
the proposed methods in Section \ref{sec:estall} can be similarly applied with slight adjustment to the tuning parameters by replacing $n$ with $n\rho$. 
We formalize these adjustments 
in the  regularity conditions below and then establish theory in Section \ref{sec:sparse_theory}. 
When  $\rho$ is unknown, we expect that its order can be estimated through the observed overall edge density, which could be sufficient to adjust tuning parameters. In the future research, it would be of interest to formalize the corresponding procedure and establish rigorous theoretical guarantee. 


\begin{condition}[Tuning parameters for Algorithms \ref{algor:estTheta1}--\ref{algor:estY}] \label{cond:sparseA}
In Algorithm \ref{algor:estTheta1}, 
(i) $(n\rho)^{1/2}\\ \log^{1/2}(nT) \ll \tau_4 \lesssim (n\rho)^{1/2} \log(nT)$, and (ii) 
$\mathcal{C}_E = \{E \in \mathbb R^{n \times n}: 
E_{ij} \in \mu([-2M_1^2,2M_1^2])
\text{ for } \\1\leqslant i,j \leqslant n\}$, where $M_1$ is given in Condition \ref{cond:truevalue}. 
In Algorithm \ref{algor:estY},   (i) step size $\eta_Y = \eta/(n\rho)$  for a sufficiently small constant $\eta > 0$, (ii)  number of iterations $R \gg \log(nT)$, and (iii) constraint set $\mathcal{C}_{Y_t} = \{Y_t \in \mathbb R^{n\times d_t }: \|Y_t\|_{2 \to \infty} \leqslant 2M_1\} $, where $M_1$ is given in Condition \ref{cond:truevalue}. 
\end{condition} 

\begin{condition}[Tuning parameters for Algorithms \ref{algor:initial}--\ref{algor:refine}] \label{cond:sparsetun}
Assume the hyperparameters in Algorithms   \ref{algor:initial} and \ref{algor:refine}  satisfy:  
(i) $1 \ll \tau_1 \lesssim \sqrt{\log n}$.   (ii) Step sizes  $ \eta_Z = \eta/(n\rho T)$ and $\eta_W = \eta/(n \rho)$ for a sufficiently small constant $\eta > 0$. (iii) Constraint sets $\mathcal{C}_Z$ and $\mathcal{C}_{W_t}$ are chosen as in \eqref{eq:constraintsets}. (iv) Number of iterations $R \gg \log(nT)$. 
\end{condition}


\subsection{{Theory}} \label{sec:sparse_theory}

Generalizing the discussions in Section \ref{sec:challenges}, the effective sample sizes for estimating latent vectors $z_{i}$ and $w_{t,i}$ should be replaced by $ n\rho T$ and $n\rho$, respectively under the model \eqref{eq:like_sparse}. As a result, the aggregated oracle estimation errors of $Z$ and $W_t$ are  expected to be of the order of  $O_p(1/(\rho T))$ and $O_p(1/\rho)$, respectively. 
To establish such oracle rates 
under \eqref{eq:like_sparse}, we can extend our conclusions in Theorems \ref{thm:estTheta}, \ref{thm:estY}, \ref{thm:initial}, and  \ref{thm:onestep} to the following Theorems \ref{thm:sparseA1}--\ref{thm:sparse2} by incorporating proper adjustments to the effective sample sizes. 
Since the log-likelihood function  $l(\theta;x)$ in \eqref{eq:like_sparse} differs from classical natural exponential families by a factor $\rho$, we update Condition \ref{cond:parfunction} to the following form. 



\begin{condition} \label{cond:sparse}
Let $\mathcal{X} = \{x \in \mathbb R : p(\,x\mid \theta\,) > 0\}$ denote the support of $p(\,\cdot \mid \theta\,)$. Assume $l(\theta;x)$ in \eqref{eq:like_sparse} satisfies the following conditions. 
\begin{itemize}\setlength{\itemsep}{0pt}
\item[(i)] For any fixed $x \in \mathcal{X}$, $l(\theta;x)$ is three times differentiable with respect to  $\theta$. Moreover, there exist positive constants $\kappa_1$, $\kappa_2$, and $\kappa_3$ such that $\kappa_1 \rho \leqslant -l^{\prime\prime}(\theta;x) \leqslant \kappa_2 \rho$ and $|l^{\prime\prime\prime}(\theta;x)| \leqslant \kappa_3 \rho$ for any $x \in \mathcal{X}$ and $|\theta| \leqslant  2M_1^2$, where $M_1$ is given in Condition \ref{cond:truevalue};
 
\item[(ii)] There exists a constant $L>0$ such that $\EXPT |l^{\prime}(\Theta_{t,ij}^\star;A_{t,ij})|^{m} \leqslant \VAR\{l^{\prime}(\Theta_{t,ij}^\star;A_{t,ij})\} L^{m -2} m ! /2$ for any  $1 \leqslant i \leqslant j \leqslant n$, $1 \leqslant t \leqslant T$, and any integer $m \geqslant 2$, {where $m!=m(m-1)\cdots 1$ represents the factorial of $m$.}
\end{itemize}
\end{condition}

Condition \ref{cond:sparse} (i) differs from Condition \ref{cond:parfunction} (i) with an extra scaling factor $\rho$ 
to bound the derivatives of $l(\theta;x)$. This is a natural requirement under  \eqref{eq:like_sparse} as $ - l^{\prime \prime}(\theta;x) = \rho\nu^{\prime}(\theta)$. 
 Condition \ref{cond:sparse} (ii) implies that $l'(\Theta_{t,ij}^\star; A_{t,ij})=A_{t,ij} - \rho \nu'(\Theta_{t,ij}^\star)=A_{t,ij}-\mathbb{E}(A_{t,ij})$ satisfies the Bernstein moment condition. It essentially imposes a regularity condition on $\nu(\cdot)$ that is weaker than Condition \ref{cond:parfunction} (ii),  justified by Lemma  \ref{lm:bernmoment_rho} below. 

\begin{lemma} \label{lm:bernmoment_rho}
If Condition \ref{cond:sparse} (ii) holds for $l(\theta;x)$ in \eqref{eq:like_sparse} with $\rho=1$ (equivalently, Condition \ref{cond:parfunction} (ii) holds), it is also satisfied for $l(\theta;x)$ with $\rho\in (0,1]$.    
\end{lemma} 
\begin{proof}

Given $\rho\in (0,1]$, and the similarity between \eqref{eq:like_sparse} and natural exponential family distribution, 
 we can employ classical moment-computing  arguments to obtain $\mathrm{var}\{l'(\Theta_{t,ij}^\star; A_{t,ij})\} = \rho \nu''(\Theta_{t,ij}^\star)$ and 
\begin{align} \label{eq:spare_moment_center}
\mathbb{E}
\{l^{\prime}(\Theta_{t,ij}^\star;A_{t,ij})\}^{m} = m\text{-th Bell polynomial} (0,\kappa_2(\rho), \ldots, \kappa_m(\rho))  
\end{align}
with $m$-th cumulant $\kappa_m(\rho):=\rho \nu^{(m)}(\Theta_{t,ij}^\star)$, where  $\nu^{(m)}(\cdot)$ represents the $m$-th order derivative of $\nu(\cdot)$.  
 By \eqref{eq:spare_moment_center}, $\rho\in (0,1]$, and property of polynomials, 
\begin{align}
    \mathbb{E}\{l^{\prime}(\Theta_{t,ij}^\star;A_{t,ij})\}^{m} &\leqslant \rho \cdot m\text{-th Bell polynomial} (0,\kappa_2(1), \ldots, \kappa_m(1)) \notag\\
    &\leqslant \rho \cdot \nu''(\Theta_{t,ij}^\star) L^{m-2} m!/2 \notag\\
    &= \VAR\{l^{\prime}(\Theta_{t,ij}^\star;A_{t,ij})\} L^{m-2} m!/2, \label{eq:bernstein_moment_rho}
\end{align} 
where the second inequality follows by   
Condition \ref{cond:parfunction} (ii). 
When $m$ is even,
Condition \ref{cond:sparse} (ii) holds  by \eqref{eq:bernstein_moment_rho}. When $m$ is odd,  
by Cauchy's inequality, 
\begin{align*}
	 \mathbb{E}|l^{\prime}(\Theta_{t,ij}^\star;A_{t,ij})|^{m} &\leqslant    \big[ \mathbb{E}\{l^{\prime}(\Theta_{t,ij}^\star;A_{t,ij})\}^{2}\, \mathbb{E}\{l^{\prime}(\Theta_{t,ij}^\star;A_{t,ij})\}^{2m-2}\big]^{1/2}\\
	&\leqslant  \big[ \mathrm{var}\{l^{\prime}(\Theta_{t,ij}^\star;A_{t,ij})\} \, \mathrm{var}\{l^{\prime}(\Theta_{t,ij}^\star;A_{t,ij})\} L^{2m-4}(2m-2)!/2\big]^{1/2}\quad (\text{by }\eqref{eq:bernstein_moment_rho})\\
	 &\lesssim    \mathrm{var}\{l^{\prime}(\Theta_{t,ij}^\star;A_{t,ij})\} (2L)^{m-2}m!, 
\end{align*}
where the last inequality follows by 
$ \frac{m^m}{e^{m-1}} \leqslant m!\leqslant \frac{m^{m+1}}{e^{m-1}}$. Therefore,  Condition \ref{cond:sparse} (ii) still holds with an adjusted value of $L$ in the bound.
\end{proof}

We next present extensions of our main theoretical results (Theorems \ref{thm:estTheta}, \ref{thm:estY}, \ref{thm:initial}, and  \ref{thm:onestep}) along with proofs, with a focus on the influences of the factor $\rho$.  

\vspace{1em} 


\begin{theorem}[Extension of Theorem \ref{thm:estTheta}] \label{thm:sparseA1}
 Assume Conditions \ref{cond:truevalue}, \ref{cond:sparseA}, and \ref{cond:sparse}. Let $\breve \Theta_t$ be the individual estimator obtained through Algorithm \ref{algor:estTheta1}. For any constant $\varepsilon >0$, there exist positive constants $c_\varepsilon$ and $C_\varepsilon$ such that when 
 $ \rho\geqslant c_\varepsilon  \log(nT)/n$, 
    $$
    \Pr\left[ \bigcup_{1\leqslant t \leqslant T} \left\{\big\|\breve \Theta_t - \Theta_t^\star\big\|_{\mathrm{F}}/n > C_\varepsilon n^{-\frac{1}{2 d_t + 4}} \rho^{-\frac{1}{4}}\log^{\frac{1}{2}}(nT)\right\}\right] = O\big((nT)^{-\varepsilon}\big).
    $$
\end{theorem}

\smallskip 
\begin{proof}
We follow the proof of  Theorem \ref{thm:estTheta} in Section \ref{sec:pf_thmA1}. 
As Condition \ref{cond:sparse} implies that $\mu^{-1}(\cdot)$  is $(\kappa_1\rho)^{-1}$ Lipschitz, 
by the construction of $\breve \Theta_t$ in Algorithm \ref{algor:estTheta1}, we now  have
\begin{align*}
    \big\|\breve \Theta_t - \Theta_t^\star\big\|_{\mathrm{F}} \leqslant \big\|\tilde \Theta_t - \Theta_t^\star\big\|_{\mathrm{F}} \leqslant (\kappa_1 \rho)^{-1} \big\|\breve E_t - \mu(\Theta_t^\star)\big\|_{\mathrm{F}} \leqslant (\kappa_1 \rho)^{-1} \big\|\tilde E_t - \mu(\Theta_t^\star)\big\|_{\mathrm{F}}.
\end{align*}
Moreover, under the model \eqref{eq:like_sparse},  we have the identities
\begin{align*}
    \mu(\theta) = \rho \nu^{\prime}(\theta), \quad l^{\prime}(\theta;x) = x - \mu(\theta), \quad \text{and} \quad l^{\prime \prime}(\theta;x) = - \mu^{\prime}(\theta).
\end{align*} 
Also, under Condition \ref{cond:sparse}, $\VAR\{l^{\prime}(\Theta_{t,ij}^\star;A_{t,ij})\} = - \EXPT[l^{\prime \prime}(\Theta_{t,ij}^\star;A_{t,ij})] \leqslant \kappa_2\rho.$ Then similarly following the proof of Lemma \ref{lem:concentration} gives   when $\log(nT)/(n\rho)$ is sufficiently small,
$$\Pr\left[\max_{1 \leqslant t \leqslant T} \Big\{\big\|\mathbf A_t - \mu(\Theta_t^\star)\big\|_{\operatorname{op}}^2\Big\} \geqslant  C_\varepsilon n \rho \log(nT) \right] \leqslant (nT)^{-\varepsilon},$$ corresponding to \eqref{eq:at_bernstein_a1} with the  adjustment to the $\rho$ factor. 
Thus,    \begin{align*}
         \big\|\tilde E_t - \mu(\Theta_t^\star)\big\|_{\mathrm{F}}^2 \lesssim (n\rho)^{\frac{1}{2}} \log(nT) \big\|\mu(\Theta_t^\star)\big\|_*
     \end{align*}
     with probability $1 - (nT)^{-\varepsilon}$,  corresponding to \eqref{eq:thm4-1} after adjusting the $\rho$ factor. 
  Since $\mu^{\prime}(\theta) = - l^{\prime \prime}(\theta;x
    ) \leqslant \kappa_2 \rho$,  
     $\mu(\cdot)$ has $\kappa_2 \rho$-Lipschitz continuity on $[-2M_1^2,2M_1^2]$, 
     we have $\|\mu(\Theta_t^\star)\|_* = O(\rho n^{\frac{3}{2} - \frac{1}{d_t+2}})$, which is obtained  following the proof of \eqref{eq:thm4-5} similarly. 
     Combining the above results, we obtain 
    \begin{align*}
    \Pr\left[ \bigcup_{1\leqslant t \leqslant T} \left\{\big\|\tilde E_t - \mu(\Theta_t^\star)\big\|_{\mathrm{F}}/n > C_\varepsilon n^{-\frac{1}{2 d_t + 4}}\rho^{\frac{3}{4}} \log^{\frac{1}{2}}(nT)\right\}\right] = O((nT)^{-\varepsilon}).
\end{align*}

\end{proof}

\bigskip 
\begin{theorem}[Extension of Theorem  \ref{thm:estY}] \label{thm:sparseA2}
 Assume Conditions \ref{cond:truevalue}, \ref{cond:sparseA}, and \ref{cond:sparse}. Let $\mathring Y_t$ be the output of Algorithm \ref{algor:estY} when using $\breve \Theta_t$ in Theorem \ref{thm:sparseA1} as initialization. For any constant $\varepsilon >0$, there exist positive constants $c_\varepsilon$ and $C_\varepsilon$ such that when $ \rho\geqslant  c_\varepsilon  n^{-2/(\dmax+2)} \log^2(nT) $, 
    $$
    \Pr\left[ \max_{1\leqslant t \leqslant T} \left\{\operatorname{dist}^2(\mathring Y_t, Y_t^\star) \right\} > C_\varepsilon \rho^{-1} \log(nT)\right] = O\big( (nT)^{-\varepsilon}\big).
    $$
\end{theorem}

\begin{proof}
    Similar to \eqref{eq:pfb2-1}, Theorem \ref{thm:sparseA1} indicates for all $1 \leqslant t \leqslant T$,
    $$\epsilon_t^0 \leqslant C_\varepsilon n^{1-\frac{1}{d_t+2}} \rho^{-\frac{1}{2}} \log(nT) \quad \text{with probability} \quad 1 - O((nT)^{-\varepsilon}).$$ Combining with the condition that $\rho^{-1} n^{-2/(\dmax+2)} \log^2(nT)$ is sufficiently small we have $\Pr[\max_{1 \leqslant t \leqslant T}\epsilon_t^0 > c_0 n/2] = O((nT)^{-\varepsilon})$, which yields $\Pr[\max_{0 \leqslant r \leqslant r_0, 1 \leqslant t \leqslant T}\epsilon_t^r > c_0 n] = O((nT)^{-\varepsilon})$ through induction. In the proof of \eqref{eq:pfb2-2}, since $-l(\theta;x)$ is $\kappa_1 \rho$-strongly convex and $\kappa_2 \rho$-smooth for any fixed $x \in \mathcal{X}$ by Condition \ref{cond:sparse}, applying Lemma \ref{adlem:nesterov2003} shows
    \begin{align*}
   D_1 = \big\langle l^\prime(\Theta_t^\star) - l^\prime(\Theta_t^r), \Theta_t^r - \Theta_t^\star \big\rangle \geqslant \frac{\kappa_1 \rho }{2} \big\|\Theta_t^r - \Theta_t^\star\big\|_{\mathrm{F}}^2 + \frac{1}{2\kappa_2 \rho}\big\|l^\prime(\Theta_t^r) - l^\prime(\Theta_t^\star)\big\|_{\mathrm{F}}^2.
   \end{align*} 
    We further replace $c_2$ and $c_3$ with $c_2 \rho$ and $c_3 \rho^{-1}$ in \eqref{eq:thm5-6} and \eqref{eq:thm5-8}, respectively. Also note that $\eta_Y = \eta/(n\rho)$, then \eqref{eq:thm5-10} transforms into 
    \begin{align*} 
    \epsilon_t^{r+1} &\leqslant \epsilon_t^r - \frac{\eta}{\rho} \left( \frac{(\kappa_1 \rho - 2c_2 \rho)M_2}{4} - \frac{\epsilon_t^r \rho}{2c_3 n}\right)\epsilon_t^r \notag \\
    & \quad \quad - \frac{\eta}{\rho} \left(\frac{1}{2 \kappa_2 \rho} - \frac{c_3}{\rho} - \frac{8M_1^2 \eta}{\rho}- \frac{4\eta\, \epsilon_t^r}{n\rho}\right) \frac{1}{n} \big\|l^\prime(\Theta_t^r) - l^\prime(\Theta_t^\star)\big\|_{\mathrm{F}}^2 \notag\\
    &\quad \quad + \frac{\eta}{\rho} \left(\frac{d_t}{2c_2 \rho} + \frac{c_3}{\rho} + \frac{8M_1^2\eta}{\rho} +  \frac{4\eta\, \epsilon_t^r}{n\rho}\right)  \frac{1}{n}\big\|l^\prime(\Theta_t^\star)\big\|_{\operatorname{op}}^2.
\end{align*}
This simplifies to
\begin{align*}
    \epsilon_t^{r+1} \leqslant (1 - c )\, \epsilon_t^r + \frac{C}{n \rho^2}\, \big\|l^\prime(\Theta_t^\star)\big\|_{\operatorname{op}}^2
\end{align*}
where $C$ and $c$ are the same constants as in \eqref{eq:thm5-c}. Finally, through iterative application of the preceding bounds, we obtain $ \epsilon_t^R \lesssim \|l^\prime(\Theta_t^\star)\|_{\operatorname{op}}^2/(n\rho^2) \leqslant C_\varepsilon \rho^{-1} \log(nT)$ with probability $1 - O((nT)^{-\varepsilon})$.
\end{proof}

\bigskip 
\begin{theorem}[Extension of Theorem \ref{thm:initial}] \label{thm:sparse1}
    Assume Conditions \ref{cond:truevalue},  \ref{cond:sparsetun}, and \ref{cond:sparse}. Let $(\mathring Z,\mathring W)$ be the estimators obtained through Algorithm \ref{algor:initial} with  $\mathring Y_t$ in Theorem \ref{thm:sparseA2} as initialization. 
     For any constant $\varepsilon >0$, there exist positive constants $c_\varepsilon$ and $C_\varepsilon$ such that when 
     $\rho\geqslant c_\varepsilon n^{-2/(\dmax+2)} \log^2(nT)$,  
    $$
    \Pr\left[ \left\{\operatorname{dist}^2(\mathring Z, Z^\star) +  \max_{1\leqslant t \leqslant T} \operatorname{dist}^2(\mathring W_t, W_t^\star) \right\} > C_\varepsilon \rho^{-1} \log^2(nT)\right] = O((nT)^{-\varepsilon}).
    $$
\end{theorem}
\begin{proof}
We follow the proof in Section \ref{sec:pf:thm:initial}. 
By Theorem \ref{thm:sparseA2}, we have $\Pr[\|\mathring Y_{t,s} - Y_{t,s}^\star \mathring Q_{t,s}\|_{\operatorname{op}}^2 > C_\varepsilon \rho^{-1} \log(nT)] = O((nT)^{-\varepsilon})$. 
 With $\rho^{-1} n^{-2/(\dmax+2)} \log^2(nT)$ is sufficiently small, we have $\Pr[\|\mathring Y_{t,s} - Y_{t,s}^\star \mathring Q_{t,s}\|_{\operatorname{op}}^2  \gtrsim n ]=O((nT)^{-\varepsilon})$. 
Combining this with Condition \ref{cond:truevalue} (i) and \eqref{eq:thm1-2}, we know \eqref{eq:thm1-1} remains valid. 
With an extra $\rho^{-1}$ factor in the upper bound of $\|\mathring Y_{t,s} - Y_{t,s}^\star \mathring Q_{t,s}\|_{\operatorname{op}}^2$ above,  \eqref{eq:thm1-5} should be  updated to $\big\|\mathring V_{t,s} \mathring  V_{t,s}^\mytrans - V_{t,s}^\star V_{t,s}^{\star \mytrans} \big\|_{\operatorname{op}} \leqslant  C_\varepsilon \tau_1 n^{-1/2} \rho^{-1/2} \log^{1/2}(nT)$ with probability $1- O((nT)^{-\varepsilon})$. Consequently,  \eqref{eq:thm1-6} is transformed into
\begin{align*} 
    \Pr\left\{\big\|\mathring F - Z^\star Z^{\star \mytrans}\big\|_{\mathrm{F}}  > C_\varepsilon \tau_1 n^{1/2} \rho^{-1/2} \log^{1/2}(nT)\right\} = O((nT)^{-\varepsilon}). 
\end{align*}
Finally, applying the arguments in \eqref{eq:thm1-7}, \eqref{eq:thm1-8} and \eqref{eq:thm1-9}, we obtain
\begin{align*}
    \Pr\left[ \left\{\operatorname{dist}^2(\mathring Z, Z^\star) +  \max_{1\leqslant t \leqslant T} \operatorname{dist}^2(\mathring W_t, W_t^\star) \right\} > C_\varepsilon \tau_1^2 \rho^{-1} \log(nT)\right] = O((nT)^{-\varepsilon}).
\end{align*}
The proof is finished by our choice $\tau_1\lesssim \sqrt{\log n}$.

\end{proof}

\bigskip 
\begin{theorem}[Extension of Theorem \ref{thm:onestep}]  \label{thm:sparse2}
Assume Conditions \ref{cond:truevalue}, \ref{cond:truevalue2}, \ref{cond:sparsetun}, and \ref{cond:sparse}.
 Let $(\hat Z, \hat{W})$ be the refined estimators from Algorithm \ref{algor:refine} with $(\mathring{Z},\mathring{W})$ in Theorem \ref{thm:sparse1} as initialization and the adjustment in Remark \ref{rmk:pseudolik}.  For any constant $\varepsilon >0$, there exist positive constants $c_\varepsilon$ and $C_\varepsilon$ such that when  $ \rho \geqslant c_{\varepsilon} \max\{n^{-2/(\dmax+2)}, n^{-1/2}\} \log^2(nT)$, 
\begin{align}
\Pr\left[
\operatorname{dist}^2(\hat Z, Z^\star) >
   C_\varepsilon \max\left\{\frac{1}{T \rho} ,\ \frac{1}{n \rho^4}\right\} \log^{8}(nT)
\right] = &~O\big((nT)^{-\varepsilon}\big), \label{eq:z_onestep_rho}\\
\Pr\left[ \max_{1\leqslant t \leqslant T} 
\operatorname{dist}^2(\hat W_t, W_t^\star) > 
   C_\varepsilon  \max\left\{\frac{1}{ \rho} ,\ \frac{1}{n \rho^4}\right\} \log^{8}(nT)
\right] = &~O\big((nT)^{-\varepsilon}\big). \label{eq:w_onestep_rho}
\end{align}
\end{theorem}

\smallskip 

Theorem \ref{thm:sparse2} implies that when  we further assume $\rho \gtrsim (T/n)^{1/3}$,  the final estimators $\hat{Z}$ and $\hat{W}_t$ would achieve that up to logarithmic factors, 
\begin{align*}
     \operatorname{dist}^2\big(\hat{Z}, Z^{\star}\big) = O_p\left( \frac{1}{T\rho} \right)\hspace{1.2em}\text{ and }\hspace{1.2em}    \max_{1\leqslant t \leqslant T} \operatorname{dist}^2(\mathring W_t, W_t^\star)=O_p\left(\frac{1}{\rho}\right),
\end{align*}
which are the expected oracle error rates as discussed at the beginning of this section. 
The lower bounds of $\rho$ above facilitate the theoretical derivation. In the future research, it would be of interest to  develop theory when $\rho$ is below the density lower bound too.

\bigskip  

\begin{proof}
The proof outline is similar to that in Section \ref{sec:pf:thm:onestep}. 
First,  we will establish a $\rho$-adjusted version of \eqref{eq:thm2-1}: 
    \begin{align} \label{eq:sparse1}
    \Pr\left[ \max_{1\leqslant t \leqslant T} \left\{ \big\| \check Z - Z^{\star} \check Q\big\|_{2 \to \infty}^2 + \big\|\check W_t - W_t^{\star} \check Q_t\big\|_{2 \to \infty}^2 \right\} > \frac{C_\varepsilon\log^4(nT)}{n \rho^2} \right] = O((nT)^{-\varepsilon}).
    \end{align}
Second, the analytical formula of \eqref{eq:newtononev} derived in  Section \ref{sec:newtonform} remains the same. Third, similarly to Section \ref{sec:proplemmaiv}, we will establish properties on $I(v)$ in Lemma \ref{lm:ivproperties1}, which generalizes Lemma \ref{lm:ivproperties}. 
Finally, to prove \eqref{eq:z_onestep_rho} and \eqref{eq:w_onestep_rho}, 
we follow the arguments in Sections \ref{sec:pdfzhaterr}--\ref{sec:pdfwthaterr} similarly. 
As the upper bounds in \eqref{eq:thm2-15} and \eqref{eq:thm2-27} still hold, 
it remains to  generalize \eqref{eq:thm2-17}--\eqref{eq:thm2-19} and \eqref{eq:thm2-28-0}--\eqref{eq:thm2-30} under the model \eqref{eq:like_sparse}. In particular, we will show 
\begin{align*}
    \Pr\big[S_1 > C_\varepsilon  \rho^{-1} T^{-1} \log(nT)\big] &= O((nT)^{-\varepsilon}), \quad \quad \quad \Pr\big[S_4 > C_\varepsilon  \rho^{-1} \log(nT)\big] = O((nT)^{-\varepsilon-1}),\\
     \Pr\big[S_2 > C_\varepsilon  \rho^{-4} n^{-1} \log^8(nT)\big] &= O((nT)^{-\varepsilon}), \quad \Pr\big[S_5 > C_\varepsilon \rho^{-4} n^{-1} \log^8(nT)\big] = O((nT)^{-\varepsilon-1}), \\ 
     \Pr\big[S_3 > C_\varepsilon  \rho^{-3} n^{-1} \log^5(nT)\big] &= O((nT)^{-\varepsilon}), \quad \Pr\big[S_6 > C_\varepsilon \rho^{-3} n^{-1} \log^5(nT)\big] = O((nT)^{-\varepsilon-1}).
\end{align*}
Substituting the updated upper bounds of $S_1$--$S_6$ above into \eqref{eq:thm2-15} and \eqref{eq:thm2-27} yields \eqref{eq:z_onestep_rho} and \eqref{eq:w_onestep_rho}. We next provide detailed arguments following the outlined steps.


\vspace{1.5em}
\noindent \textit{(i) Proof of \eqref{eq:sparse1}.}    The proof follows the outline in Section \ref{sec:pdfthm2-1}, with key modifications to handle the factor $\rho$. 
Recall that  the proof outline in Section \ref{sec:pdfthm2-1} relies on showing \eqref{eq:rotationfix}, \eqref{eq:iterateerr}, \eqref{eq:thm2-bbounds}, and \eqref{eq:iterateerr2}. 
First, \eqref{eq:rotationfix} still holds following the same arguments on Page \pageref{sec:pfrotationfix}. 
Second, we show  \eqref{eq:iterateerr} is generalized to 
\begin{align}  \label{eq:sparse2}
    \epsilon_i^{r+1} \leqslant (1-c ) \epsilon_i^r + \frac{CT}{n\rho^2} \left[ B_{\nu} + \frac{1}{n}  (B_{z}+B_{w})\right],
\end{align}
where 
$B_{\nu} =   \max_{1 \leqslant t \leqslant T}  \|l^\prime(\Theta_t^\star) - l^\prime(\mathring \Theta_t^\star)\|_{2 \to \infty}^2$, $   B_{z}=  \max_{1 \leqslant t \leqslant T} \|l^\prime(\Theta_t^\star)  \mathring Z \|_{2 \to \infty}^2$, and $ B_{w} =   \max_{1 \leqslant t \leqslant T}  \|l^\prime(\Theta_t^\star) \mathring W_t \|_{2 \to \infty}^2$ are defined same as in \eqref{eq:boundbvzw}. In particular, following the arguments on Page \pageref{sec:pfiterateerr}, and by $ \eta_Z = \eta/(n\rho T)$ and $\eta_W = \eta/(n \rho)$, the decomposition in \eqref{eq:thm2-7} similarly holds with $\eta$ replaced by $\eta/\rho$, i.e.,  $\epsilon_i^{r+1} \leqslant \epsilon_i^r - \frac{\eta}{\rho} D_1 + \frac{\eta}{\rho}  D_2 + \frac{\eta}{\rho}  D_3 + \frac{\eta^2}{\rho^2} D_4$, where the definitions of $\{D_i: i=1,2,3,4\}$ remain the same. Since $-l(\theta;x)$ is $\kappa_1 \rho$-strongly convex and $\kappa_2 \rho$-smooth for any fixed $x \in \mathcal{X}$, Lemma \ref{adlem:nesterov2003} gives  
    \begin{align*}
    D_1 &= \frac{2}{n}\sum_{t=1}^T \Big\langle e_i^\mytrans \big( l^\prime(\mathring \Theta_t^\star) - l^\prime(\mathring \Theta_t^r)\big), e_i^\mytrans \big(\mathring \Theta_t^r -  \mathring \Theta_t^\star\big)\Big\rangle \notag\\
    &\geqslant \frac{\kappa_1 \rho}{n}\sum_{t=1}^T\big\|e_i^\mytrans \big(\mathring \Theta_t^r -  \mathring \Theta_t^\star\big)\big\|_2^2 + \frac{1}{\kappa_2  \rho n} \sum_{t=1}^T\big\|e_i^\mytrans \big( l^\prime(\mathring \Theta_t^r) -  l^\prime(\mathring \Theta_t^\star)\big)\big\|_2^2.
\end{align*} 
Under the model \eqref{eq:like_sparse}, 
using the arguments of \eqref{eq:thm2-9} and \eqref{eq:thm2-11} with  $c_2$ and $c_3$ replaced by $c_2 \rho$ and $c_3 \rho$, we have 
\begin{align*}
    D_2\leqslant &~\frac{c_2 \rho}{n}\sum_{t=1}^T\big\|e_i^\mytrans \big(\mathring \Theta_t^r -  \mathring \Theta_t^\star\big)\big\|_2^2 + \frac{1}{c_2\rho n} \sum_{t=1}^T\big\|e_i^\mytrans \big( l^\prime( \Theta_t^\star) -  l^\prime(\mathring \Theta_t^\star)\big)\big\|_2^2,\\
    D_3 \leqslant &~ c_3\rho \,\epsilon_i^r + \frac{1}{c_3\rho n^2 } \sum_{t=1}^T\big\|e_i^\mytrans l^\prime(\Theta_t^\star) \big[\mathring Z, \mathring W_t\big]\big\|_2^2.
\end{align*}
Since $\eta_Z = \eta/(n\rho T)$ and $\eta_W = \eta/(n \rho)$, \eqref{eq:thm2-13} transforms into 
\begin{align*}
     \epsilon_i^{r+1} &\leqslant \epsilon_i^r - \frac{\eta}{\rho}\Big({\frac{M_2 (\kappa_1 \rho - c_2 \rho )}{2}} - c_3 \rho \Big)\epsilon_i^r - \frac{\eta}{\rho} \Big(\frac{1}{\kappa_2 \rho n} - \frac{9M_1^2 \eta}{n\rho}\Big) \sum_{t=1}^T \big\|e_i^\mytrans\big(l^\prime(\mathring \Theta_t^r) - l^\prime(\mathring \Theta_t^\star)\big)\big\|_2^2 \notag\\
    &\quad + \frac{\eta}{\rho} \Big(\frac{1}{c_2\rho n} + \frac{9M_1^2 \eta}{n\rho}\Big) \sum_{t=1}^T \big\|e_i^\mytrans\big(l^\prime(\Theta_t^\star) - l^\prime(\mathring \Theta_t^\star)\big)\big\|_2^2 \notag\\
    &\quad + \frac{\eta}{\rho} \Big(\frac{1}{c_3 \rho n^2} + \frac{3\eta}{n^2 \rho}\Big) \sum_{t=1}^T \big\|e_i^\mytrans l^\prime(\Theta_t^\star) \big[\mathring Z,\mathring W_t\big]\big\|_2^2,
\end{align*}
which implies \eqref{eq:sparse2}. Third, following the arguments on Page \pageref{sec:pfthm2-bbounds}, we generalize \eqref{eq:thm2-bbounds} to
\begin{align} \label{eq:sparse_thm2-bbounds}
   \Pr\big[ B_{\nu} > C_\varepsilon \rho \log^2(nT)\big] = O((nT)^{-\varepsilon}) \ \ \text{and} \ \  \Pr\big[ B_{z}+ B_w > C_\varepsilon n \log^4(nT)\big] = O((nT)^{-\varepsilon}),
\end{align}
where an extra $\rho$ factor arises in the bound of $B_{\nu}$ due to the $\kappa_2\rho$-Lipschitz continuity of $l^{\prime}(\cdot)$ and the use of  Theorem \ref{thm:sparse1} that extends Theorem  \ref{thm:initial} under the model \eqref{eq:like_sparse}, 
and the upper bound of $B_z+B_w$ is obtained using Theorem \ref{thm:sparse1}  and $\| l^\prime(\Theta_t^\star)\|_{2 \to \infty}^2 \leqslant C_\varepsilon n \rho \log^2(nT)$ following the arguments in Lemma \ref{lem:concentration2}. 
Fourth, following similar arguments to \eqref{eq:sparse2}, we know \eqref{eq:iterateerr2} can be extended to 
\begin{align}  \label{eq:sparse3}
    \epsilon_{it}^{r+1} \leqslant (1-c)\epsilon_{it}^r + \frac{C}{n \rho^2} \Big(B_{\nu} + \frac{1}{n} B_{w} + n \rho^2 \big\|e_i^\mytrans \big(Z^r - Z^\star \mathring Q\big)\big\|_2^2 \Big),
\end{align}
where the $\rho^2$ factor in the third term  
arises by applying the $\kappa_2\rho$-Lipschitz continuity of $l^{\prime}(\cdot)$ in \eqref{eq:pfd36-10}. The proof of \eqref{eq:sparse1} is completed by iteratively applying the inequalities \eqref{eq:sparse2} and \eqref{eq:sparse3}.



\vspace{1.5em}

\noindent \textit{(ii) Properties of $I(v)$.} Lemma \ref{lm:ivproperties1} below states properties of $I(v)$ with proofs given afterwards. 
\begin{lemma} \label{lm:ivproperties1}
Assume true latent vectors  $\{Z^{\star}, W_1^{\star},\ldots, W_T^{\star}\}$ satisfy  Conditions \ref{cond:truevalue}, \ref{cond:truevalue2}, and \ref{cond:sparse}. Let    $v^{\star}$ denote its vectorization similarly to  \eqref{eq:vectorization}.  
Consider another set of fixed latent vectors $\{Z, W_1,\ldots, W_T\}$ with its vectorization denoted as $v$  following  \eqref{eq:vectorization}. 
Assume 
\begin{itemize}
\setlength{\itemsep}{0pt}
    \item[(a)]    $\|Z\|_{2 \to \infty} \leqslant M_1$, $\|W_t\|_{2 \to \infty} \leqslant M_1$ for $1 \leqslant t \leqslant T$;  
    
    \item[(b)] there exists a constant $\epsilon$ such that $\|v - v_q^\star\|_{\infty} = O(n^{-1/2} \rho^{-1} \log^{\epsilon}(nT))$, where $v_q^\star =  Q_{nv}^\mytrans v^\star $ with $Q_{nv} = \operatorname{diag}(\mathrm{I}_n \otimes Q_z, \mathrm{I}_n \otimes Q_{w1}, \ldots, \mathrm{I}_n \otimes Q_{wT})$,
\begin{align*}
    Q_z = \argmin_{Q \in \mathcal{O}(k)} \big\|Z - Z^\star Q\big\|_{\mathrm{F}}^2, \quad \text{and} \quad Q_{wt} = \argmin_{Q \in \mathcal{O}(k_t)} \big\|W_t - W_t^\star Q\big\|_{\mathrm{F}}^2.
\end{align*}
\end{itemize}
 Then for any $u \in [0,1]$, $v_u =  {v} + u(v_{q}^\star -  v)$ satisfies the following properties when $ \log^{2\epsilon}(nT)/(n\rho^2)$ is sufficiently small.   
 \begin{enumerate}
        \item[(i)] $\operatorname{rank
        }(I(v)) = \mathrm{r}_{n,k} + \sum_{t= 1}^T\mathrm{r}_{n,k_t}$ with  $\mathrm{r}_{n,k} = nk - k(k-1)/2$, 
        \item[(ii)] $v_u \in \operatorname{col}(I(v))$,
        \item[(iii)] $\big\|\DnT^{1/2}\, I(v_u)^+ \DnT^{1/2}\big\|_{\operatorname{op}} = O(\rho^{-1})$ with $\DnT = \operatorname{diag}(nT \mathrm{I}_{nk}, n \mathrm{I}_{n\ksum})$,
        \item[(iv)] 
        $\big\| \DnT^{-1/2} \big( I({v}) - I(v_u) \big) \DnT^{-1/2}\big\|_{\operatorname{op}} = O(n^{-1/2} \log^\epsilon(nT))$,
        \item[(v)]$ \big\| \DnT^{-1/2} \big( N(v_q^\star) - N(v_u) \big) \DnT^{-1/2}\big\|_{\operatorname{op}} = O(n^{-1/2} \log^\epsilon(nT))$,
       \item[(vi)] $\operatorname{col}(I(v_u)) = \operatorname{col}(I_d(v_u))$ for $I_d(v) = \operatorname{diag}(I_{ZZ}(v), I_{W_1 W_1}(v), \ldots, I_{W_T W_T}(v))$ defined same as in Lemma \ref{lm:ivproperties}. 
    \end{enumerate}
\end{lemma}

Note $I(v)$ under \eqref{eq:like_sparse} still takes the same form as in \eqref{eq:ivform} following the same derivations in Section \ref{sec:formIv}. 
Therefore, the proof of Lemma \ref{lm:ivproperties1} can follow the arguments in Section \ref{sec:proplemmaiv} similarly, except that  
 a distinct $l(\theta;x)$ is considered, with  $-l''(\Theta_{t,ij})\asymp 1$ in  Section \ref{sec:proplemmaiv}, whereas  $-l''(\Theta_{t,ij})\asymp \rho$ here.  This makes orders of the elements in $\Dmut(v)$ change from $O(1)$ to $O(\rho)$. We next explain the proofs of Lemma \ref{lm:ivproperties1}  with a focus on the impacts from the extra scaling $\rho$ in $\Dmut(v)$. 

For the proof of (i)--(iii) in Section \ref{sec:proplemmaiv}, most arguments remain the same, except that when comparing the properties of $I(v)$ to $\Xi(v)$ (defined in \eqref{eq:xiv}), an extra $\rho$ multiplicative factor would arise, due to the presence of $\Dmut(v)$ in $I(v)$ but not $\Xi(v)$. Specifically,  the original property \eqref{eq:lem3-7} is updated to $$\sigma(\mathcal{U}(v)^\mytrans \DnT^{-1/2} \,I(v) \DnT^{-1/2}\, \mathcal{U}(v) ) \asymp \rho \cdot \sigma(\mathcal{U}(v)^\mytrans \DnT^{-1/2} \,\Xi(v) \DnT^{-1/2}\, \mathcal{U}(v) ).$$ As a result, $O(1)$ in  \eqref{eq:lem3-8-2} is replaced by $O(\rho^{-1})$, giving (iii) in Lemma \ref{lm:0p_ivproperties}. Meanwhile, the statements of (i) and (ii)  remain exactly the same because  they  primarily rely on showing $\mathrm{col}(I(v))=\mathrm{col}(\mathcal{U}(v))$ (defined same as in \eqref{eq:uvdef}), which would not change with the extra multiplicative factor $\rho>0$ introduced by $\Dmut(v)$. 

Moreover, (iv) and (v) in Lemma \ref{lm:ivproperties1}  achieve $\rho$-independent bounds due to cancellation: our condition  $\|v - v_q^\star\|_{\infty} = O( \rho^{-1} n^{-1/2} \log^{\epsilon}(nT) )$  gives  
\begin{align*}
    \big\|\DZ(v) - \DZ(v_u)\big\|_{\operatorname{op}} &= O(\rho^{-1} \log^{\epsilon}(nT))\\ \text{ and }\quad \quad \quad \max_{1 \leqslant i \leqslant j \leqslant n}\big|\Theta_{t,ij} - \Theta_{t,ij}^u\big| &= O(\rho^{-1} n^{-1/2} \log^{\epsilon}(nT)),
\end{align*}
which balance with $\|\Dmut(v)\|_{\operatorname{op}} = O(\rho)$ and the $\rho$-Lipschitz continuity of both $l^{\prime}(\cdot)$ and $l^{\prime \prime}(\cdot)$. 

Finally, (vi)  in Lemma \ref{lm:ivproperties1} can be established following the corresponding  proofs in  Section \ref{sec:proplemmaiv} similarly, except that extra $\rho$ factors can arise, e.g., \eqref{eq:Iduvsigma} should be updated to $ \sigma(\mathcal{U}(v)^\mytrans \DnT^{-1/2} I_d(v) \DnT^{-1/2}\, \mathcal{U}(v)) \asymp \rho$ instead of 1. But as $\rho>0$, this still implies that $\mathrm{col}(I_d(v))\subseteq \mathrm{col}(\mathcal{U}(v))$, and therefore, the conclusion on the column spaces in (vi) remains unchanged.

\vspace{1em}

\noindent \textit{(iii) Error bounds for $S_1$--$S_6$.} 
Before the analysis, we note that Lemma \ref{lm:ivproperties1} can be applied to $I(v)$ at  $v = \check v$, because the assumption (a) holds by the projection step in line 4 of Algorithm \ref{algor:estY} and the assumption (b) in Lemma \ref{lm:ivproperties1} holds with $\epsilon = 2$ and probability $1 - O((nT)^{-\varepsilon})$, as ensured by \eqref{eq:sparse1}. 
Therefore, the arguments for bounding $S_1$--$S_6$ in Section  \ref{sec:pdfzhaterr} can similarly apply with extra $\rho$ factors explained below. 
Specifically, the extra $\rho^{-1}$  in the bounds of $S_1$ and $S_4$ arise due to 
$\|\DnT^{1/2} I(v^\star)^+ \DnT^{1/2}\|_{\operatorname{op}}=O(\rho^{-1})$ with probability $1 - O((nT)^{-\varepsilon})$ by Lemma \ref{lm:ivproperties1} (iii). Then similarly, the extra $\rho^{-4}$  in the bounds of $S_2$ and $S_5$ follow by 
\begin{align*}
\big\|\DnT^{1/2} I(\check v)^+ \DnT^{1/2}\big\|_{\operatorname{op}}^2=O(\rho^{-2}) \quad\text{and}\quad \big\|\DnT^{1/2}(\check{v} - v_q^\star)\big\|_2^2=O(\rho^{-2}nT\log^4(nT))
\end{align*}
with probability $1 - O((nT)^{-\varepsilon})$. And the extra $\rho^{-3}$  in the bounds of $S_3$ and $S_6$ come from 
\begin{align*}
    \big\|\DnT^{1/2} I(\check v)^+ \DnT^{1/2}\big\|_{\operatorname{op}}^2=O(\rho^{-2}), \quad \quad \big\|\DnT^{1/2} I(v_q^\star)^+ \DnT^{1/2}\big\|_{\operatorname{op}}^2&=O(\rho^{-2}),  \\
    \text{and} \quad \quad  \quad \quad\big\|\DnT^{-1/2} I(v^\star) \DnT^{-1/2}\big\|_{\operatorname{op}}&=O(\rho)
\end{align*}
with probability $1 - O((nT)^{-\varepsilon})$.

\end{proof}

\newpage

\section{{Extensions Beyond Fully Connected  Networks}} \label{sec:fullyconnect}


In the model  \eqref{eq:model}, when  $p(\cdot \mid \theta)$ is a continuous distribution,  all edge weights $A_{t,ij}$ are non-zero almost surely, implying fully connected networks. 
Nonetheless, \eqref{eq:model} can be extended to model weighted networks with a substantial proportion of disconnected node pairs. We introduce the generalized model in Section \ref{sec:0p_model} below and develop corresponding estimation procedure and theoretical guarantees in Sections \ref{sec:0p_method} and \ref{sec:0p_theory}, respectively. 


\subsection{{Model}}\label{sec:0p_model}

We generalize model \eqref{eq:model} to the following mixture distribution 
\begin{align} 
    A_{t,ij} = A_{t,ji}\ \sim \pi \, p(\, \cdot \mid \Theta_{t,ij}^{\star} \,) + (1-\pi)\, \delta_0 , \quad \quad 1 \leqslant i \leqslant j \leqslant n,\ 1 \leqslant t \leqslant T,
    \label{eq::0p_model}
\end{align} 
independently, where  $p(\cdot\mid \theta)$ and 
$\Theta_{t,ij}^{\star} $ are defined same as in  \eqref{eq:model},  $\delta_0$ denotes a probability mass at zero, and $\pi\in (0,1]$. 
Under \eqref{eq::0p_model}, we consider $p(\cdot \mid \theta)$ is a continuous distribution, and then 
the atom at zero corresponds to a disconnected edge.
The model \eqref{eq::0p_model} implies that in each  
network, each node pair is disconnected with probability $1-\pi$, and otherwise is connected with edge weight drawn from $\ p(\, \cdot \mid \Theta_{t,ij}^\star \,) $.  Consequently, under \eqref{eq::0p_model},
\begin{align} \label{eq:0p_model_mean}
    \mathbb E(\mathbf A_t) = \pi\mu(\Theta_t^\star), 
\end{align}
where $\mu(\cdot)$ denotes the unique invertible function linking $\theta$ and the variable expectation under $p(\cdot \mid \theta) $.

\subsection{{Estimation Procedure}}\label{sec:0p_method}
Under \eqref{eq::0p_model}, we modify
the proposed methods in Sections \ref{sec:estall} and \ref{sec:ytestappendx} to accommodate the presence of zero mass in the edge distributions and to obtain similar statistical guarantees. 
We first summarize high-level ideas of key updates and then describe them in detail. 

First, as \eqref{eq:0p_model_mean} differs  from \eqref{eq:model2} by a scalar $\pi$,  
original steps based on the singular value decomposition of adjacency matrices (Algorithm \ref{algor:estTheta1}) need to be rescaled to accommodate for $\pi$. 
Second, steps based on the likelihood of data  (Algorithms \ref{algor:estY} and \ref{algor:refine}) need to be updated to accommodate distributional properties under \eqref{eq::0p_model}. 
The detailed modifications to the algorithms are described below with $\pi$ given for notational  simplicity.  If $\pi$ is unknown, it can be replaced by observed proportions of connected edges to apply the algorithms.



\begin{enumerate}
\item[(1)] For Algorithm \ref{algor:estTheta1}, 
its line 2 is changed to ``Let $\tilde{E}_t = \sum_{i: d_{t,i}> \tau_4 {{\pi}^{1/2}}} d_{t,i} u_{t,i} v_{t,i}^\top $ and $\breve{E}_t =\mathcal{P}_{\mathcal{C}_E}(\tilde{E}_{t} { / \pi}) $.'' 
\item[(2)] For Algorithm \ref{algor:estY}, in its line 3, 
we redefine
\begin{align}\label{eq:new_algo_a2_0p_model}
{l^{\prime}(\langle y_{t,i}^r, y_{t,j}^r \rangle; A_{t,ij} ) = A_{t,ij} - \pi \mu(\langle y_{t,i}^r, y_{t,j}^r \rangle)}, 
\end{align}  
while the other formulae remain the same. 
\item[(3)] For Algorithm \ref{algor:initial}, no modification is needed.
\item[(4)] For Algorithm \ref{algor:refine}, the procedural steps remain applicable with slight adjustments to likelihood-related quantities.  
 First, $\ell(Z,W)$ takes the same  form as in \eqref{eq:zwlikelihood} with updated 
$   l (\theta;x) = \theta x - \pi  \nu(\theta)$ and  $\nu^\prime(\theta) = \mu(\theta)$. 
Second, we redefine $I(v)$ in \eqref{eq:newtononev} to take the same form as in \eqref{eq:ivform} with
\begin{align}
\label{eq:0p_Dumtform}
    \Dmut(v) = - \operatorname{diag}\big(2 l^{\prime \prime}(\Theta_{t,11}), \ldots, 2 l^{\prime \prime}(\Theta_{t,nn}),  l^{\prime \prime}(\Theta_{t,12}), \ldots, l^{\prime \prime}(\Theta_{t,n-1,n})\big) 
\end{align} using $l''(\theta)=-\pi \mu'(\theta)$ by \eqref{eq:new_algo_a2_0p_model}. 
\end{enumerate}


\begin{remark}\label{rm:pseudo_0pmodel}
The above estimation procedure, particularly the newly defined $l'(\cdot)$ in \eqref{eq:new_algo_a2_0p_model}, is motivated by solving the unbiased estimating equation $\mathbb E(\mathbf A_t) - \pi\mu(\Theta_t^\star) =0$ based on  \eqref{eq:0p_model_mean}. 
In this framework, $l'(\cdot)$ induces a pseudo log likelihood function equal to $l (\theta;x) = \theta x - \pi  \nu(\theta)$ (up to additive constants) as used in Algorithm \ref{algor:refine}. 
When $\pi=1$, the induced pseudo log likelihood coincides with the log likelihood of the  natural exponential family, and the corresponding estimation procedure remains the same as under the original model \eqref{eq:model}. 

Similarly to Remark \ref{rm:lprimenotation}, we use $l'(\Theta_{t,ij})$ and $l''(\Theta_{t,ij})$ as shorthands for $l'(\Theta_{t,ij};A_{t,ij})$ and $l''(\Theta_{t,ij};A_{t,ij})$ in the following when there is no ambiguity. 
\end{remark}

\bigskip  

To establish desired statistical guarantee, we require appropriate choices of hyperparameters,
which are summarized below with highlights on the differences compared to Conditions \ref{cond:parestY} and \ref{cond:tuning12}.


\begin{condition}[Tuning parameters for Algorithms \ref{algor:estTheta1}--\ref{algor:estY}] \label{cond:0p_A}
In  Algorithms \ref{algor:estTheta1} and \ref{algor:estY},  assume      the hyperparameters still satisfy Condition \ref{cond:parestY}, except that  the step size in Algorithm \ref{algor:estY} satisfies $\eta_Y = \eta/(n\pi)$  for a sufficiently small constant $\eta > 0$. 
\end{condition} 

\begin{condition}[Tuning parameters for Algorithms \ref{algor:initial}--\ref{algor:refine}] \label{cond:0p_tun}
In in Algorithms \ref{algor:initial} and \ref{algor:refine}, assume the hyperparameters still satisfy Condition \ref{cond:tuning12}, except that the step sizes in Algorithm \ref{algor:refine} satisfy $ \eta_Z = \eta/(n\pi T)$ and $\eta_W = \eta/(n \pi)$ for a sufficiently small constant $\eta > 0$.
\end{condition}

\subsection{{Theory}}\label{sec:0p_theory}

Under the model \eqref{eq::0p_model}, we  expect similar efficiency gain from pooling multiple heterogeneous networks, as discussed in Section \ref{sec:challenges}. 
But different from  the original model \eqref{eq:model}, the effective sample sizes for   $w_{t,i}$ and $z_i$ under \eqref{eq::0p_model} are expected to be of the orders of $O(n\pi)$  and  $O(n\pi T)$, respectively, due to the $1-\pi$ proportion of disconnections in each network. 
As a result, the aggregated oracle estimation errors of $Z$ and $W_t$ are expected to be of the orders of $O_p(1/(\pi T))$ and $O_p(1/\pi)$, respectively, generalizing the discussion in Section \ref{sec:challenges}.  
We next demonstrate that our theoretical results can be generalized to achieve the oracle errors rates under \eqref{eq::0p_model}. 

As the mixture distribution \eqref{eq::0p_model} differs from the original model \eqref{eq:model},
we first adjust our Condition \ref{cond:parfunction} on the edgewise distribution to the non-zero-mass component $p(\cdot\mid\theta)$ under \eqref{eq::0p_model}.

\begin{condition}[Adjusted version of Condition \ref{cond:parfunction}] \label{cond:0p_parfunction}
Assume $p(\cdot\mid \theta)$ in \eqref{eq::0p_model} belongs to the natural  exponential family and satisfies the following conditions.  
\begin{itemize}\setlength{\itemsep}{0pt}
\item[(i)] 
The link function $\mu(\theta) $ induced by $p(\cdot \mid \theta)$ 
is twice differentiable with respect to $\theta$, with its first and second order derivatives with respect to $\theta$ denoted by $\mu^{\prime}(\theta)$ and $\mu^{\prime \prime}(\theta)$, respectively. Moreover, there exist positive constants $\kappa_1$, $\kappa_2$, and $\kappa_3$ such that $\kappa_1 \leqslant \mu^{\prime}(\theta) \leqslant \kappa_2$ and $|\mu^{\prime\prime}(\theta)| \leqslant \kappa_3$ for any 
$|\theta| \leqslant 2M_1^2$, 
 where $M_1$ is given in Condition \ref{cond:truevalue}. 
\item[(ii)] There exists a constant $L>0$ such that $$\EXPT \big\{|A_{t,ij} - \mu(\Theta_{t,ij}^{\star}) |^{m}\mid A_{t,ij} \neq 0 \big\} \leqslant \VAR\big\{A_{t,ij} - \mu(\Theta_{t,ij}^{\star})\mid A_{t,ij} \neq 0\big\} L^{m -2} m ! /2$$ for any  $1 \leqslant i \leqslant j \leqslant n$, $1 \leqslant t \leqslant T$, and any integer $m \geqslant 2$.
\end{itemize}
\end{condition}


\smallskip 
Under the mixture model \eqref{eq::0p_model}, $p(\cdot \mid \Theta_{t,ij}^\star)$ represents  the distribution of $A_{t,ij}$  conditional on $A_{t,ij}\neq 0$. Thus, the moment property  in Condition~\ref{cond:0p_parfunction} (ii) essentially implies regularity constraints on the non-zero-mass component $p(\cdot\mid \theta)$. 
Under the natural exponential family, original Condition~\ref{cond:parfunction} is equivalent to  Condition~\ref{cond:expfam} (discussed in Section~\ref{sec:expfam}). Since Condition \ref{cond:0p_parfunction} implies similar requirements on $p(\cdot \mid \theta)$ to   Condition \ref{cond:expfam}, it is reasonably expected to hold and reduces to  Condition~\ref{cond:parfunction} when $\pi=1$.  
Nevertheless, the natural exponential family is assumed in Condition \ref{cond:0p_parfunction} for the ease of understanding the connections and technical details.  This assumption can be relaxed by more intricate arguments similar to those in Section~\ref{sec:expfam}, which could be an interesting future research direction. 



\medskip 

We next present extensions of our main theoretical results (Theorems \ref{thm:estTheta}, \ref{thm:estY}, \ref{thm:initial}, and  \ref{thm:onestep})  along with their proofs.
For the ease of understanding, our proofs will  focus on illustrating the influences of the connection probability $\pi$ when it is known. The arguments can be similarly generalized when $\pi$ is estimated by the observed proportion of connected edges, with more detailed arguments handling the error of estimating $\pi$.

\begin{theorem}[Counterpart of Theorem \ref{thm:estTheta}] \label{thm:0p_A1}
 Assume Conditions \ref{cond:truevalue}, \ref{cond:0p_A}, and   \ref{cond:0p_parfunction}. 
 Let $\breve \Theta_t$ be the individual estimator obtained through the adjusted version of Algorithm \ref{algor:estTheta1} (described in Section \ref{sec:0p_method}). 
 For any constant $\varepsilon >0$, there exist positive constants $c_\varepsilon$ and $C_\varepsilon$ such that when
 $\pi \geqslant c_\varepsilon \log(nT)/n$, 
    $$
    \Pr\left[ \bigcup_{1\leqslant t \leqslant T} \left\{\big\|\breve \Theta_t - \Theta_t^\star\big\|_{\mathrm{F}}/n > C_\varepsilon n^{-\frac{1}{2 d_t + 4}} \pi^{-\frac{1}{4}}\log^{\frac{1}{2}}(nT)\right\}\right] = O\big((nT)^{-\varepsilon}\big).
    $$
\end{theorem}


\begin{proof}
Please see  Section \ref{page:pfofthmk1}.  
\end{proof}


\smallskip 

\begin{theorem}[Counterpart of Theorem  \ref{thm:estY}] \label{thm:0p_A2} 
 Assume Conditions \ref{cond:truevalue}, \ref{cond:0p_A}, and \ref{cond:0p_parfunction}.
 Let $\mathring{Y}_t$ be the output of the adjusted version of Algorithm \ref{algor:estY} (described in Section \ref{sec:0p_method}) when initialized with $\breve{\Theta}_t$ from Theorem \ref{thm:0p_A1}. For any constant $\varepsilon >0$, there exist positive constants $c_\varepsilon$ and $C_\varepsilon$ such that when $ \pi \geqslant c_\varepsilon n^{-2/(\dmax+2)} \log^2(nT)  $,  
    $$
    \Pr\left[ \max_{1\leqslant t \leqslant T} \left\{\operatorname{dist}^2(\mathring Y_t, Y_t^\star) \right\} > C_\varepsilon \pi^{-1} \log(nT)\right] = O\big( (nT)^{-\varepsilon}\big).
    $$
\end{theorem}

\begin{proof}
Please see Section \ref{page:pfofthmk2}. 
\end{proof}

\smallskip

\begin{theorem}[Counterpart of Theorem \ref{thm:initial}] \label{thm:0p_1}
    Assume Conditions  \ref{cond:truevalue}, \ref{cond:0p_tun}, and \ref{cond:0p_parfunction}. 
    Let $(\mathring Z,\mathring W)$ be the estimators obtained through Algorithm \ref{algor:initial} with  $\mathring Y_t$ in Theorem \ref{thm:0p_A2} as initialization. 
For any constant $\varepsilon >0$, there exist positive constants $c_\varepsilon$ and $C_\varepsilon$ such that when 
$\pi\geqslant c_{\varepsilon} n^{-2/(\dmax+2)} \log^2(nT)$, 
    $$
    \Pr\left[ \left\{\operatorname{dist}^2(\mathring Z, Z^\star) +  \max_{1\leqslant t \leqslant T} \operatorname{dist}^2(\mathring W_t, W_t^\star) \right\} > C_\varepsilon \pi^{-1} \log^2(nT)\right] = O((nT)^{-\varepsilon}).
    $$
\end{theorem}
\begin{proof} 
Given $\mathring Y_t$ achieving the error rate in Theorem \ref{thm:0p_A2}, 
the proof of Theorem \ref{thm:0p_1} follows the same arguments as that of Theorem \ref{thm:sparse1}, with $\rho$ replaced by $\pi$. 
\end{proof}

\medskip   




\begin{remark} \label{rm:pseudolik_extend}
Similarly to Remark \ref{rmk:pseudolik}, in the theoretical analysis for the adjusted version of Algorithm \ref{algor:refine} under the model \eqref{eq::0p_model}, following Section \ref{sec:revisealgor}, we replace
$\tilde{Z}^{r+1}=Z^{r} + {\eta_Z}\,  \partial {\ell}(Z^r,W^r) / \partial Z $ and $\tilde{W}_t^{r+1}=W_t^{r} + \eta_W\,  \partial{\ell}(Z^r,W^r)/ \partial W_t$  in lines 3 and 4 of Algorithm \ref{algor:refine} by 
\begin{equation}
\label{eq:0p_pseudogd}
\begin{aligned}
    &~\tilde{Z}^{r+1}=Z^{r} + {\eta_Z}\, \sum_{t = 1}^T l^\prime(Z^r\mathring Z^{\top}+W_t^r \mathring{W}_t^{\top}) \mathring Z  \\ \text{ and } \quad  &~\tilde{W}_t^{r+1}=W_t^{r} + \eta_W\,  l^\prime(Z^r\mathring Z^{\top}+W_t^r \mathring{W}_t^{\top}) \mathring W_t,
\end{aligned}
\end{equation}
 respectively, where $l^\prime(\cdot)$ is defined same as in \eqref{eq:new_algo_a2_0p_model}.
\end{remark}

\medskip 

\begin{theorem}[Counterpart of Theorem \ref{thm:onestep}]  \label{thm:0p_2}
Assume Conditions \ref{cond:truevalue},  \ref{cond:truevalue2}, \ref{cond:0p_tun}, and \ref{cond:0p_parfunction}.
Let $(\hat Z, \hat{W})$ be the refined estimators from the adjusted version of Algorithm \ref{algor:refine} (described in Section \ref{sec:0p_method}), with $(\mathring{Z},\mathring{W})$ in Theorem \ref{thm:0p_1} as initialization and the adjustment in Remark \ref{rm:pseudolik_extend}. For any constant $\varepsilon>0$, there exist positive constants $c_\varepsilon$ and $C_\varepsilon$ such that when $\pi \geqslant  c_\varepsilon \max\{n^{-2/(\dmax+2)}, n^{-1/2}\} \log^2(nT) $,
    \begin{align}
\Pr\left[
\operatorname{dist}^2(\hat Z, Z^\star) >
   C_\varepsilon \max\left\{\frac{1}{T \pi} ,\ \frac{1}{n \pi^4}\right\} \log^{8}(nT)
\right] = &~O\big((nT)^{-\varepsilon}\big), \label{eq:ThmK4-1}\\
\Pr\left[ \max_{1\leqslant t \leqslant T} 
\operatorname{dist}^2(\hat W_t, W_t^\star) > 
   C_\varepsilon  \max\left\{\frac{1}{ \pi} ,\ \frac{1}{n \pi^4}\right\} \log^{8}(nT)
\right] = &~O\big((nT)^{-\varepsilon}\big). \label{eq:ThmK4-2}
\end{align}
\end{theorem}

Theorem \ref{thm:0p_2} implies that when we further assume $\pi\gtrsim (T/n)^{1/3}$, the final estimators $\hat{Z}$ and $\hat{W}_t$ would achieve that up to logarithmic factors, 
\begin{align*}
     \operatorname{dist}^2\big(\hat{Z}, Z^{\star}\big) = O_p\left( \frac{1}{T\pi} \right)\hspace{1.2em}\text{ and }\hspace{1.2em}    \max_{1\leqslant t \leqslant T} \operatorname{dist}^2(\mathring W_t, W_t^\star)=O_p\left(\frac{1}{\pi}\right),
\end{align*}
aligning with the expected oracle error rates as discussed at the beginning of this section.
\medskip 

\begin{proof}
Similarly to Section \ref{sec:pf:thm:onestep}, we first establish two-to-infinity error bounds for $\check{Z}$ and $\check{W}_t$, which are now  obtained through lines 1-5 of Algorithm \ref{algor:refine} with adjustments in Section \ref{sec:0p_method} and Remark \ref{rm:pseudolik_extend}. In particular,  Section \ref{page:pfofpropk1} shows that for any constant $\varepsilon>0$, there exist positive constants $c_\varepsilon$ and $C_\varepsilon$ such that when  $\pi\geqslant c_{\varepsilon} \max\{n^{-2/(\dmax+2)}, n^{-1/2}\} \log^2(nT)$, 
\begin{align} \label{eq:0p_1}
    \Pr\left[ \max_{1\leqslant t \leqslant T} \left\{ \big\| \check Z - Z^{\star} \check Q\big\|_{2 \to \infty}^2 + \big\|\check W_t - W_t^{\star} \check Q_t\big\|_{2 \to \infty}^2 \right\} > \frac{C_\varepsilon\log^4(nT)}{n \pi^2} \right] = O((nT)^{-\varepsilon}),
    \end{align}
    which generalizes \eqref{eq:thm2-1}. 
Second, Section \ref{page:pfoflemk1} establishes the useful properties of $I(v)$ under \eqref{eq::0p_model}, extending results in Sections \ref{sec:newtonform} and \ref{sec:lemmaiv1}. 
Third, Section \ref{sec::pfofThmK4} establishes error rates for $\hat{Z}$ and $\hat{W}_t$, obtained through line 6 of Algorithm \ref{algor:refine} with adjustments in Section \ref{sec:0p_method}. 
\end{proof}

\bigskip 



\subsubsection{Proof of Theorem \ref{thm:0p_A1}}
\label{page:pfofthmk1}


As line 3 in Algorithm \ref{algor:estTheta1} is not changed in the adjusted version described in Section \ref{sec:0p_method}, similarly to the proof of Theorem \ref{thm:estTheta} in Section \ref{sec:pf_thmA1},
we have
\begin{align}
\label{eq:0p_thmK1-0}
    \big\|\breve \Theta_t - \Theta_t^\star\big\|_{\mathrm{F}} \leqslant \big\|\tilde \Theta_t - \Theta_t^\star\big\|_{\mathrm{F}} \leqslant \kappa_1^{-1} \big\|\breve E_t - \mu(\Theta_t^\star)\big\|_{\mathrm{F}} \leqslant \kappa_1^{-1}  \big\|\tilde E_t /\pi 
    - \mu(\Theta_t^\star)\big\|_{\mathrm{F}}.
\end{align} 
Therefore, by a union bound, it suffices to prove  
\begin{align} \label{eq:0p_tildeet_bound}
    \Pr   \left\{\big\|\tilde E_t /
    \pi - \mu(\Theta_t^\star)\big\|_{\mathrm{F}}/n > C_\varepsilon n^{-\frac{1}{2 d_t + 4}}\pi^{-\frac{1}{4}}\log^{\frac{1}{2}}(nT)\right\} = O((nT)^{-\varepsilon-1})
\end{align} 
for each $t=1,\ldots,T$.

Similarly to Section \ref{sec:pf_thmA1}, we want to apply Lemma \ref{adlem:chatterjee2015} to prove \eqref{eq:0p_tildeet_bound}. 
To this end, 
we first establish the  following matrix concentration result: when $\log(nT)/(n\pi)$ is sufficiently small,
\begin{align} \label{eq:0p_at_bernstein_a1}
\max_{1\leqslant t\leqslant T} \big\|\mathbf A_t - \pi \mu(\Theta_t^\star)\big\|_{\operatorname{op}}^2 \leqslant C_{\varepsilon} n\pi\log(nT)    
\end{align} 
with probability $1- (nT)^{-\varepsilon} $.
Following  the proof of  Lemma \ref{lem:concentration}, to prove \eqref{eq:0p_at_bernstein_a1}, it suffices to show  entries in the random matrix $\mathbf A_t - \pi \mu(\Theta_t^\star)$ satisfy the Bernstein moment condition  
\begin{align}
    \EXPT |A_{t,ij} - \pi \mu(\Theta_{t,ij}^\star)  |^{m} \leqslant \VAR\{A_{t,ij} - \pi \mu(\Theta_{t,ij}^\star) \} \tilde{L}^{m -2} m ! /2 \label{eq:bern_0p_model}
\end{align}
for a constant $\tilde{L}$.  
As this holds trivially at $m=2$, we consider $m\geqslant 3$ below. 
We denote $\Omega_{t,ij} = \mathbb{I}(A_{t,ij}\neq 0) $ for $1\leqslant i,j\leqslant n, 1\leqslant t\leqslant T $.
By H\"older's inequality, 
\begin{align} 
   \EXPT|A_{t,ij} - \pi \mu(\Theta_{t,ij}^\star)  |^{m} 
    \leqslant &~ 2^{m-1} \, \EXPT|A_{t,ij} - \Omega_{t,ij} \mu(\Theta_{t,ij}^\star)  |^{m} + 2^{m-1}  |\mu(\Theta_{t,ij}^\star)|^m \, \EXPT|\Omega_{t,ij} - \pi |^{m} \notag\\
= &~ 2^{m-1} \pi \, \EXPT(|A_{t,ij} -\mu(\Theta_{t,ij}^\star)  |^{m} \mid \Omega_{t,ij}=1) \label{eq:two_exp_bd_1}\\
&~ + 2^{m-1}|\mu(\Theta_{t,ij}^\star) |^m \{\pi(1-\pi)^m + (1-\pi)\pi^m \} \label{eq:two_exp_bd_2}
\end{align}
where \eqref{eq:two_exp_bd_1} follows by the law of iterated expectations and  $A_{t,ij}=0$ if $\Omega_{t,ij}=0$.
As the distribution of  $A_{t,ij}$  conditioning on $\Omega_{t,ij}=1 $ satisfies Condition \ref{cond:0p_parfunction} (ii), we have 
\begin{align*}
  \eqref{eq:two_exp_bd_1}  \leqslant &~2^{m-1}\pi \,  \mathrm{var}( A_{t,ij} \mid \Omega_{t,ij}=1) L^{m-2} m!/2;
\end{align*}
second, as  $\mu(\Theta_{t,ij}^{\star})\leqslant 2\kappa_2 M_1^2$ by Conditions \ref{cond:truevalue} and \ref{cond:0p_parfunction} and $\pi\in (0,1]$, we have
\begin{align*}
    \eqref{eq:two_exp_bd_2}  \leqslant &~ 2^{m-1}|\mu (\Theta_{t,ij}^{\star})|^2  \{\pi(1-\pi)^2 + (1-\pi)\pi^2  \}(2\kappa_2 M_1^2)^{m-2} \\
    =&~2^{m-1}\mathrm{var}\{\Omega_{t,ij} \, \mu(\Theta_{t,ij}^{\star})\} (2\kappa_2 M_1^2)^{m-2}.
\end{align*}
By the law of total variance,  $\mathrm{var}(A_{t,ij}-\pi \mu(\Theta_{ij}^{\star}))=\pi \mathrm{var}(A_{t,ij}\mid \Omega_{t,ij}=1) +  \mathrm{var}\{\Omega_{t,ij} \, \mu(\Theta_{t,ij}^{\star})\}$. 
In summary \eqref{eq:bern_0p_model} holds for $\tilde{L}=4L+4\kappa_2M_1^2$. 

 Now we are ready to apply Lemma \ref{adlem:chatterjee2015}. We specify $A = \mathbf{A}_t$, $B = \pi \mu(\Theta_t^\star)$, and $\delta = \tau_4 
 \pi^{1/2}/\|A - B\|_{\operatorname{op}}$. 
Notably, when $n$ is sufficiently large,  the condition $\delta>2$ in Lemma  \ref{adlem:chatterjee2015}  is satisfied with probability $1 - (nT)^{-\varepsilon-1}$, because of \eqref{eq:0p_at_bernstein_a1} 
and the fact that we choose  $\tau_4 \gg n^{1/2} \log^{1/2}(nT)$ by Condition \ref{cond:0p_A}.
Consequently,  with probability  $1 - (nT)^{-\varepsilon-1}$, 
 Lemma \ref{adlem:chatterjee2015} can be applied to give
\begin{align*} 
    \big\|\tilde E_t - \pi\mu(\Theta_t^\star)\big\|_{\mathrm{F}}^2  &\lesssim \delta \,\big\|\mathbf A_t - \pi\mu(\Theta_t^\star)\big\|_{\operatorname{op}} \big\|\pi\mu(\Theta_t^\star)\big\|_* = \tau_4 (
    \pi)^{1/2}\pi \, \big\|\mu(\Theta_t^\star)\big\|_* \notag \\
    &\lesssim n^{1/2} \pi^{3/2} \log(nT) \big\|\mu(\Theta_t^\star)\big\|_*, 
\end{align*}
 where the last inequality follows by $\tau_4\lesssim \sqrt{n}\log(nT)$ by Condition \ref{cond:0p_A}.

Combining the above result with the fact that $\|\mu(\Theta_t^\star)\|_* = O( n^{\frac{3}{2} - \frac{1}{d_t+2}})$ for any $1\leqslant t\leqslant T$ by \eqref{eq:thm4-4}, 
we have with probability $1- O((nT)^{-\varepsilon-1})  $ that
\begin{align*}
\big\|\tilde E_t / 
\pi - \mu(\Theta_t^\star)\big\|_{\mathrm{F}}
    \leqslant&~ \big\|\tilde E_t  - \pi\mu(\Theta_t^\star)\big\|_{\mathrm{F}}/ 
    \pi
    \lesssim  n^{1-\frac{1}{2d_t+4}} \pi^{-\frac{1}{4}}\log^{\frac{1}{2}}(nT).
\end{align*}
Plugging the above result into \eqref{eq:0p_thmK1-0} and \eqref{eq:0p_tildeet_bound} gives the conclusion of Theorem \ref{thm:0p_A1}.

\smallskip

\subsubsection{Proof of Theorem \ref{thm:0p_A2}}
\label{page:pfofthmk2}
We follow the steps in the proof of Theorem \ref{thm:estY} in Section \ref{sec:pf_thm5} and highlight differences below. First, analogous to the proof of \eqref{eq:pfb2-1.0}, by  Theorem \ref{thm:0p_A1}, Lemma \ref{adlem:tu2016}, and the assumption that 
$\pi^{-1} n^{-2/(\dmax+2)} \log^2(nT)$ is sufficiently small, 
we have 
\begin{align} \label{eq:0p_A2_pfb2-1.0}
    \Pr\Big(\max_{1 \leqslant t \leqslant T} \epsilon_t^0 > c_0 n /2\Big) = O((nT)^{-\varepsilon})
\end{align} for a sufficiently small constant $c_0$.
Using the same induction argument as in the proof of Theorem \ref{thm:estY},  
at each step $0 \leqslant r_0\leqslant R-1$, we assume
 \begin{align} \label{eq:0p_A2_pfb2-3}
    \Pr\Big(\max_{0\leqslant r\leqslant r_0,\   1\leqslant t\leqslant T} \, \epsilon_t^{r}  > c_0 n  \Big) = O((nT)^{-\varepsilon}).
\end{align}

We next prove a counterpart of \eqref{eq:pfb2-2}: there exist constants $c\in (0,1)$ and $C>0$ such that
\begin{align}\label{eq:error_decay_0p_model}
      \epsilon_t^{r+1}   &   \leqslant (1-c)\epsilon_t^r +  \frac{C}{n\pi^2}\big\|l^\prime(\Theta_t^\star)\big\|_{\operatorname{op}}^2
\end{align} 
for $0 \leqslant r \leqslant r_0 $ and $1\leqslant t \leqslant T$ with probability $1 - O((nT)^{-\varepsilon})$,
where we denote
\begin{align}\label{eq:lprime_0pmodel}
    l'(\Theta_{t}^*)=\mathbf{A}_{t}-\pi \mu(\Theta_{t}^*),
\end{align}
consistent with \eqref{eq:new_algo_a2_0p_model} and Remark \ref{rm:pseudo_0pmodel}. 
To prove \eqref{eq:error_decay_0p_model}, 
we note the decomposition \eqref{eq:thm5-4} still holds with $l'(\cdot)$ defined as in \eqref{eq:lprime_0pmodel} and  $\eta_Y = \eta /( n\pi) $. 
We next analyze the terms $D_1$--$D_4$ in \eqref{eq:thm5-4} under the new definitions one by one. 
Since $-l(\theta; x)$ is $\kappa_1\pi$-strongly convex and $\kappa_2\pi$-smooth by Condition \ref{cond:0p_parfunction},  Lemma \ref{adlem:nesterov2003} can be applied, and we obtain
\begin{equation}\label{eq:0p_A2_1}
\begin{aligned} 
   D_1 
   =&~ \big\langle  l^\prime(\Theta_t^\star) - l^\prime(\Theta_t^r) , \Theta_t^r  - \Theta_t^\star \big\rangle \\
    \geqslant&~ \frac{\kappa_1 \pi}{2} \big\| \Theta_t^r - \Theta_t^\star\big\|_{\mathrm{F}}^2 + \frac{1}{2\kappa_2\pi}\big\|l^\prime(\Theta_t^r) - l^\prime(\Theta_t^\star)\big\|_{\mathrm{F}}^2.
\end{aligned}
\end{equation}
For $D_2$ and $D_3$, we apply the same analyses as in \eqref{eq:thm5-6} and \eqref{eq:thm5-8}, 
with $c_2$ replaced by $c_2\pi$ and $c_3$ by $c_3\pi^{-1}$; the substitutions are justified as these constants can be freely chosen when applying H\"older's inequality in \eqref{eq:thm5-6} and \eqref{eq:thm5-8}.
It follows that 
\begin{align*}
    D_2 \leqslant&~    \frac{d_t}{2 c_2\pi} \big\|l^\prime(\Theta_t^\star)\big\|_{\operatorname{op}}^2+c_2\pi  \big\|\Theta_t^r - \Theta_t^\star\big\|_{\mathrm{F}}^2, \\
    D_3 \leqslant&~ \frac{c_3}{\pi}\big\|l^\prime(\Theta_t^\star)\big\|_{\operatorname{op}}^2 + \frac{c_3}{\pi} \big\|l^\prime(\Theta_t^r) - l^\prime(\Theta_t^\star)\big\|_{\mathrm{F}}^2 + \frac{\pi\epsilon_t^r}{2c_3}\, \epsilon_t^r.
\end{align*}
For $D_4$, we still use the bound of \eqref{eq:thm5-9}, as \eqref{eq:thm5-9.0} still holds. 
By \eqref{eq:thm5-7.0} and the choice of $\eta_Y = \eta/(n\pi)$, \eqref{eq:thm5-10} now transforms into
    \begin{align} 
    \epsilon_t^{r+1} &\leqslant \epsilon_t^r - \frac{\eta}{\pi} \left( \frac{(\kappa_1 - 2c_2)M_2\pi}{4} - \frac{\pi\epsilon_t^r}{2c_3 n}\right)\epsilon_t^r \notag \\
    & \quad \quad - \frac{\eta}{\pi}  \left(\frac{1}{2 \kappa_2\pi} - \frac{c_3}{\pi} - \frac{8M_1^2 \eta}{\pi} - \frac{4\eta\, \epsilon_t^r}{n\pi} \right) \frac{1}{n} \big\|l^\prime(\Theta_t^r) - l^\prime(\Theta_t^\star)\big\|_{\mathrm{F}}^2 \notag\\
    &\quad \quad + \frac{\eta}{\pi}  \left(\frac{d_t}{2c_2\pi} + \frac{c_3}{\pi} + \frac{8M_1^2\eta}{\pi} + \frac{4\eta\, \epsilon_t^r}{n\pi}\right) \frac{1}{n}\big\|l^\prime(\Theta_t^\star)\big\|_{\operatorname{op}}^2 \notag\\
 &\leqslant \epsilon_t^r - \eta \left( \frac{(\kappa_1 - 2c_2)M_2}{4} - \frac{c_0}{2c_3 }\right)\epsilon_t^r \notag \\
    & \quad \quad - \eta \left(\frac{1}{2 \kappa_2} - {c_3} - 10M_1^2 \eta\right) \frac{1}{n\pi^2} \big\|l^\prime(\Theta_t^r) - l^\prime(\Theta_t^\star)\big\|_{\mathrm{F}}^2 \notag\\
    &\quad \quad + \eta \left(\frac{d_t}{2c_2 } + {c_3} + {10M_1^2\eta}\right) \frac{1}{n\pi^2}\big\|l^\prime(\Theta_t^\star)\big\|_{\operatorname{op}}^2\label{eq:0p_thm5-10}
\end{align} 
with probability $1-O((nT)^{-\varepsilon}) $,
where in the second inequality we used the induction assumption of \eqref{eq:0p_A2_pfb2-3}.
When taking $c_2,c_3,\eta$ and $c_0$ sufficiently small, by \eqref{eq:0p_thm5-10}, 
we obtain \eqref{eq:error_decay_0p_model}. 
Thus,
\begin{align}\label{eq:0p_A2_descent}
      \epsilon_t^{r+1}   &   \leqslant (1-c)^{r+1} \epsilon_t^0 +  \frac{C}{n\pi^2}\big\|l^\prime(\Theta_t^\star)\big\|_{\operatorname{op}}^2
\end{align} for $0 \leqslant r \leqslant r_0 $ and $1\leqslant t \leqslant T$ with probability $1 - O((nT)^{-\varepsilon})$. 
Plugging in the bounds \eqref{eq:0p_at_bernstein_a1} and \eqref{eq:0p_A2_pfb2-1.0},
    we have 
     \begin{align*} 
    \Pr\Big(\max_{0\leqslant r\leqslant r_0+1,\   1\leqslant t\leqslant T} \, \epsilon_t^{r}  > c_0 n  \Big) = O((nT)^{-\varepsilon}).
\end{align*}
    with probability $1 - O((nT)^{-\varepsilon})$, i.e., \eqref{eq:0p_A2_pfb2-3} also holds for $r_0+1$. 
    By iteratively showing \eqref{eq:0p_A2_descent} and \eqref{eq:0p_A2_pfb2-3}, we have
    $$ \epsilon_{t}^R \leqslant \frac{C_{\varepsilon} \log(nT) }{{\pi}} $$ with probability $1 - O((nT)^{-\varepsilon})$. 

\subsubsection{Proof of \eqref{eq:0p_1}}\label{page:pfofpropk1}
    Since \eqref{eq:rotationfix} still holds following the same arguments on Page \pageref{sec:pfrotationfix}, it suffices to show
    \begin{align} 
    \Pr\left[   \big\| Z^R - Z^{\star} \mathring Q\big\|_{2 \to \infty}^2  > C_\varepsilon n^{-1}\log^4(nT)\pi^{-2} \right] &= O((nT)^{-\varepsilon})\label{eq:0p_thm2-14}\\
    \text{and} \quad \Pr\Big[ \max_{1 \leqslant t \leqslant T} \Big\{ \big\| W_t^R - W_t^{\star} \mathring Q_t\big\|_{2 \to \infty}^2 \Big\}  > C_\varepsilon n^{-1}\log^4(nT)\pi^{-2}  \Big] &= O((nT)^{-\varepsilon}). \label{eq:0p_thm2-14-2}
\end{align}

We first show \eqref{eq:0p_thm2-14}.
Similarly to Section \ref{sec:pdfthm2-1}, define
\begin{align*}
B_{\nu} = &~  \max_{1 \leqslant t \leqslant T}  \big\|l^\prime(\Theta_t^\star) - l^\prime(\mathring \Theta_t^\star)\big\|_{2 \to \infty}^2, \\    
    B_{z}=&~  \max_{1 \leqslant t \leqslant T} \big\|l^\prime(\Theta_t^\star) \mathring Z \big\|_{2 \to \infty}^2,  \\  
 \text{and} \qquad
 B_{w} =  &~ \max_{1 \leqslant t \leqslant T}  \big\|l^\prime(\Theta_t^\star) \mathring W_t \big\|_{2 \to \infty}^2
\end{align*}
taking the same form  as \eqref{eq:boundbvzw}. 
Since   $l(\theta;x)=\theta x - \pi \nu(\theta) $ is constructed to be similar to    \eqref{eq:like_sparse} with $\rho$ replaced by $\pi$, and the same regularity conditions are imposed on $\nu'(\theta)$ implied by Conditions \ref{cond:sparse} and \ref{cond:0p_parfunction}, we expect that similar properties on $l'(\cdot)$ hold and then the analyses of $B_{v}, B_z,$ and $B_w$ in Section \ref{sec:sparse_extend} can be similarly applied. In particular, following the proof of  \eqref{eq:sparse_thm2-bbounds} with $\rho$ replaced by $\pi$, we obtain  
\begin{align}
   \Pr\big[ B_{\nu} > C_\varepsilon \pi \log^2(nT)\big] =&~ O((nT)^{-\varepsilon}),\label{eq:0p_thm2-bbounds}\\
   \Pr\big[ B_{z}+ B_w > C_\varepsilon n \log^4(nT)\big] =&~  O((nT)^{-\varepsilon}),\label{eq:0p_thm2-bbounds2}
\end{align} 
where we verify that $l^{\prime}(\cdot)$ is  $\kappa_2\pi$-Lipschitz continuous, that the bound for $[\mathring Z, \mathring W]$ in Theorem \ref{thm:0p_1} is the same as that in Theorem \ref{thm:sparse1} (with $\rho$ replaced by $\pi$), and that $l^\prime(\Theta_{t,ij}^\star) $ satisfies the Bernstein condition by \eqref{eq:bern_0p_model},  giving $\| l^\prime(\Theta_t^\star)\|_{2 \to \infty}^2 \leqslant C_\varepsilon n \pi \log^2(nT)$ and $\| l^\prime(\Theta_t^\star)Y_t^\star\|_{2 \to \infty}^2 \leqslant C_\varepsilon n \pi \log^2(nT)$ with probability $1-O((nT)^{-\varepsilon})$ following the arguments in Lemma \ref{lem:concentration2}. 

On Page \pageref{page:pfofk21}, we will 
prove that there exist positive constants $C$ and $c$ such that
\begin{align} \label{eq:0p_iterateerr}
    \epsilon_i^{r+1} \leqslant (1-c) \epsilon_i^r + \frac{CT}{n\pi^2} \left[ B_{\nu} + \frac{1}{n}  (B_{z}+B_{w})\right]
\end{align}
holds with probability $1 - O((nT)^{-\varepsilon})$. 
Then repeating \eqref{eq:0p_iterateerr} and plugging in \eqref{eq:0p_thm2-bbounds} and \eqref{eq:0p_thm2-bbounds2} 
gives that, when $n$ is sufficiently large, for any $r \geqslant R/2 \gg \log(nT)$,
\begin{align} \label{eq:0p_thm2-14-R/2}
    \Pr\left[   \big\| Z^r - Z^{\star} \mathring Q\big\|_{2 \to \infty}^2  > C_\varepsilon n^{-1}\log^4(nT)\pi^{-2} \right] &= O((nT)^{-\varepsilon}),
\end{align} which includes the result of \eqref{eq:0p_thm2-14}.

Next we show \eqref{eq:0p_thm2-14-2}. Similarly to \eqref{eq:0p_iterateerr} and \eqref{eq:iterateerr2},  we have that there exist positive constants $C$ and $c$ such that
\begin{align}  \label{eq:0p_3}
    \epsilon_{it}^{r+1} \leqslant (1-c)\epsilon_{it}^r + \frac{C}{n \pi^2} \Big(B_{\nu} + \frac{1}{n} B_{w} + n\pi^2\big\|e_i^\mytrans \big(Z^r - Z^\star \mathring Q\big)\big\|_2^2 \Big)
\end{align} holds with probability $1 - O((nT)^{-\varepsilon})$.
Repeating \eqref{eq:0p_3},  plugging in \eqref{eq:0p_thm2-bbounds}, \eqref{eq:0p_thm2-bbounds2} and \eqref{eq:0p_thm2-14-R/2}, and taking maximum over $1\leqslant t\leqslant T$ 
gives the result of \eqref{eq:0p_thm2-14-2}.

\paragraph{Proof of \eqref{eq:0p_iterateerr}.} \label{page:pfofk21} Under the choice of $\eta_Z = \eta/(nT\pi), \eta_W = \eta/(n\pi)$, the decomposition \eqref{eq:thm2-7} still holds, with all the $\eta$ replaced by $\eta/\pi$, and $l'(\Theta_{t,ij}) $ now defined by \eqref{eq:new_algo_a2_0p_model} and \eqref{eq:lprime_0pmodel}. We next analyze the terms $D_1$--$D_4$ in \eqref{eq:thm2-7} one by one.

In the lower bound of $D_1$, similarly to the arguments for \eqref{eq:0p_A2_1}, 
we have
\begin{align*}
    D_1 &= \frac{2}{n}\sum_{t=1}^T \Big\langle e_i^\mytrans \big( l^\prime(\mathring \Theta_t^\star) - l^\prime(\mathring \Theta_t^r)\big) , e_i^\mytrans \big(\mathring \Theta_t^r -  \mathring \Theta_t^\star\big)\Big\rangle \notag\\
    &\geqslant \frac{\kappa_1\pi }{n}\sum_{t=1}^T\big\|e_i^\mytrans  \big(\mathring \Theta_t^r -  \mathring \Theta_t^\star\big) \big\|_2^2 + \frac{1}{\kappa_2 n\pi} \sum_{t=1}^T\big\|e_i^\mytrans \big( l^\prime(\mathring \Theta_t^\star) - l^\prime(\mathring \Theta_t^r)\big) \big\|_2^2.
\end{align*}
For  $D_2$ and $D_3$, we replace $c_2$ and $c_3$ with $c_2 \pi$ and $c_3 \pi$ in \eqref{eq:thm2-9} and \eqref{eq:thm2-11}, respectively, to obtain
\begin{align*}
    D_2\leqslant &~\frac{c_2 \pi}{n}\sum_{t=1}^T\big\|e_i^\mytrans \big(\mathring \Theta_t^r -  \mathring \Theta_t^\star\big)\big\|_2^2 + \frac{1}{c_2 n\pi} \sum_{t=1}^T\big\|e_i^\mytrans \big( l^\prime( \Theta_t^\star) -  l^\prime(\mathring \Theta_t^\star)\big)\big\|_2^2,\\
    D_3 \leqslant &~ c_3\pi \,\epsilon_i^r + \frac{1}{c_3 n^2\pi } \sum_{t=1}^T\big\|e_i^\mytrans l^\prime(\Theta_t^\star) \big[\mathring Z, \mathring W_t\big]\big\|_2^2.
\end{align*}
By the result of Theorem \ref{thm:0p_1}, \eqref{eq:thm2-0} and \eqref{eq:thm2-00} still hold. Therefore, the bound of \eqref{eq:thm2-12} still holds for $D_4$, and \eqref{eq:thm2-R2C10} also still holds true. By the choice of $\eta_Z = \eta/(nT\pi), \eta_W = \eta/(n\pi)$, \eqref{eq:pfd36-8} transforms into
\begin{align} \label{eq:0p_thm2-13}
    \epsilon_i^{r+1} &\leqslant \epsilon_i^r - \frac{\eta}{\pi}\Big({\frac{M_2\pi (\kappa_1 - c_2)}{2}} - c_3\pi \Big)\epsilon_i^r - \frac{\eta}{\pi} \Big(\frac{1}{\kappa_2 n\pi} - \frac{9M_1^2 \eta}{n\pi}\Big) \sum_{t=1}^T \big\|e_i^\mytrans\big(l^\prime(\mathring \Theta_t^r) - l^\prime(\mathring \Theta_t^\star)\big)\big\|_2^2 \notag\\
    &\quad + \frac{\eta}{\pi} \Big(\frac{1}{c_2n\pi} + \frac{9M_1^2 \eta}{n\pi}\Big) \sum_{t=1}^T \big\|e_i^\mytrans\big(l^\prime(\Theta_t^\star) - l^\prime(\mathring \Theta_t^\star)\big)\big\|_2^2 \notag\\
    &\quad + \frac{\eta}{\pi} \Big(\frac{1}{c_3\pi n^2} + \frac{3\eta}{n^2\pi}\Big) \sum_{t=1}^T \big\|e_i^\mytrans l^\prime(\Theta_t^\star) \big[\mathring Z,\mathring W_t\big]\big\|_2^2
\end{align}  with probability $1 - O((nT)^{-\varepsilon})$.
Since $c_2, c_3$, and $\eta$ are sufficiently small constants, \eqref{eq:0p_thm2-13} indicates that there exist positive constants $C$ and $c$ such that, with probability $1 - O((nT)^{-\varepsilon})$,
\begin{align*}
    \epsilon_i^{r+1} \leqslant (1-c) \epsilon_i^r + C \Big\{ \frac{1}{n\pi^2} \sum_{t=1}^T \big\|e_i^\mytrans \big(l^\prime(\Theta_t^\star) - l^\prime(\mathring \Theta_t^\star)\big)\big\|_2^2 + \frac{1}{ n^2\pi^2} \sum_{t=1}^T \big\|e_i^\mytrans l^\prime(\Theta_t^\star)\big[\mathring Z,\mathring W_t\big]\big\|_2^2  \Big\}.
\end{align*}

\subsubsection{Properties of $I(v)$ }
\label{page:pfoflemk1}

Recall that we define $I(v)$  to take the form of \eqref{eq:ivform}  with $\Dmut$ defined as in \eqref{eq:0p_Dumtform}. Lemma \ref{lm:0p_ivproperties} below establishes the properties of the constructed $I(v)$, which is similar to Lemma \ref{lm:ivproperties}.

\begin{lemma}[Counterpart of Lemma \ref{lm:ivproperties}] \label{lm:0p_ivproperties}
Assume true latent vectors  $\{Z^{\star}, W_1^{\star},\ldots, W_T^{\star}\}$ satisfy  Conditions \ref{cond:truevalue}, \ref{cond:0p_parfunction}, and \ref{cond:truevalue2}. Let    $v^{\star}$ denote its vectorization similarly to  \eqref{eq:vectorization}.  
Consider another set of fixed latent vectors $\{Z, W_1,\ldots, W_T\}$ with its vectorization denoted as $v$  following  \eqref{eq:vectorization}. 
Assume 
\begin{itemize}
\setlength{\itemsep}{0pt}
    \item[(a)]    $\|Z\|_{2 \to \infty} \leqslant M_1$, $\|W_t\|_{2 \to \infty} \leqslant M_1$ for $1 \leqslant t \leqslant T$;  
    
    \item[(b)] there exists a constant $\epsilon$ such that $\|v - v_q^\star\|_{\infty} = O(n^{-1/2} \pi^{-1} \log^{\epsilon}(nT))$, where $v_q^\star =  Q_{nv}^\mytrans v^\star $ with $Q_{nv} = \operatorname{diag}(\mathrm{I}_n \otimes Q_z, \mathrm{I}_n \otimes Q_{w1}, \ldots, \mathrm{I}_n \otimes Q_{wT})$,
\begin{align*}
    Q_z = \argmin_{Q \in \mathcal{O}(k)} \big\|Z - Z^\star Q\big\|_{\mathrm{F}}^2, \quad \text{and} \quad Q_{wt} = \argmin_{Q \in \mathcal{O}(k_t)} \big\|W_t - W_t^\star Q\big\|_{\mathrm{F}}^2.
\end{align*}
\end{itemize}
 Then for any $u \in [0,1]$, $v_u =  {v} + u(v_{q}^\star -  v)$ satisfies the following properties when $ \log^{2\epsilon}(nT)/(n\pi^2)$ is sufficiently small.   
  \begin{enumerate}
        \item[(i)] $\operatorname{rank
        }(I(v)) = \mathrm{r}_{n,k} + \sum_{t= 1}^T\mathrm{r}_{n,k_t}$ with  $\mathrm{r}_{n,k} = nk - k(k-1)/2$, 
        \item[(ii)] $v_u \in \operatorname{col}(I(v))$,
        \item[(iii)] $\big\|\DnT^{1/2}\, I(v_u)^+ \DnT^{1/2}\big\|_{\operatorname{op}} = O(\pi^{-1})$, \quad $\big\|\DnT^{-1/2}\, I(v_u) \DnT^{-1/2}\big\|_{\operatorname{op}} = O(\pi)$ \\ with $\DnT = \operatorname{diag}(nT \mathrm{I}_{nk}, n \mathrm{I}_{n\ksum})$,
        \item[(iv)] 
        $\big\| \DnT^{-1/2} \big( I({v}) - I(v_u) \big) \DnT^{-1/2}\big\|_{\operatorname{op}} = O(n^{-1/2} \log^\epsilon(nT))$,
        \item[(v)]$ \big\| \DnT^{-1/2} \big( N(v_q^\star) - N(v_u) \big) \DnT^{-1/2}\big\|_{\operatorname{op}} = O(n^{-1/2} \log^\epsilon(nT))$,
       \item[(vi)] $\operatorname{col}(I(v_u)) = \operatorname{col}(I_{d}(v_u))$ for $I_{d}(v) = \operatorname{diag}(I_{ZZ}(v), I_{W_1 W_1}(v), \ldots, I_{W_T W_T}(v))$, where 
       \begin{align*}
           I_{ZZ}(v)  = \textstyle \sum_{t=1}^T \DZ(v)  \Dmut(v)  \DZt(v) \ \ \text{ and } \ \   I_{W_tW_t}(v)  =  \Dwt(v)  \Dmut(v)  \Dwtt(v) 
       \end{align*}
       represent diagonal blocks in $I(v)$. 
    \end{enumerate}
\end{lemma}


\begin{proof}
The proof follows the arguments in Section \ref{sec:proplemmaiv} similarly,  as it also considers $I(v)$ in the form of  \eqref{eq:ivform}.  The only difference is that a distinct $l(\theta;x)$ is considered, with  $-l''(\Theta_{t,ij})\asymp 1$ in  Section \ref{sec:proplemmaiv}, whereas  $-l''(\Theta_{t,ij})\asymp \pi$ here.  This makes orders of the elements in $\Dmut(v)$ change from $O(1)$ to $O(\pi)$. Notably, this property is shared by $I(v)$ analyzed in Lemma \ref{lm:ivproperties1} with $\pi$ replaced by $\rho$.  
Since the forms of $l(\cdot)$ considered  in Lemmas \ref{lm:ivproperties1} and \ref{lm:0p_ivproperties} are comparable and similar assumptions are made,  the proof of Lemma \ref{lm:ivproperties1} can be directly applied here with $\rho$ replaced by $\pi$. 
\end{proof}

Lemma \ref{lm:0p_ivproperties} (i)--(vi) are the same as those in Lemma \ref{lm:ivproperties} except $O(\pi^{-1})$ and $O(\pi)$  in (iii). This difference leads to extra factors polynomial in $\pi$ in our conclusions when Lemma \ref{lm:0p_ivproperties} (iii) is applied in derivations, which will be elaborated in Section \ref{sec::pfofThmK4} below.

\smallskip
\begin{remark}[Relationship between $H(v)$ and $I(v)$]\label{rm::0p_HvIv}
Similarly to \eqref{eq:hv}, we still have  $H(v)=-I(v)+N(v)$, where $H(v)=\frac{\partial^2 \ell(v)}{\partial v \partial v^\mytrans}$,  and $N(v)$ is defined same as in \eqref{eq:def_Nv_Ntv}.  
In particular, for $l(\cdot)$ and $\ell(\cdot)$ defined in Section \ref{sec:0p_method}, the first-order derivatives $\dot \ell(v)$  and  $\dot \ell_{\Theta_t}(v)$ still take the same form as in \eqref{eq:ldotform} and \eqref{eq:pre3},  respectively. 
Then following the same calculations as in \eqref{eq:hvform_blocks}--\eqref{eq:hvform_4blocks}, we obtain that $H(v) = [H_{qm}(v)]_{ q,m\in \{Z,W_1,\ldots, W_T\}} $ is still the $(1+T)\times (1+T)$ block matrix with each block given by \eqref{eq:hvform_4blocks}, with $\Dmut$ defined as in \eqref{eq:0p_Dumtform}. Then, as we define $I(v)$  to take the form of \eqref{eq:ivform}, \eqref{eq:hv} still holds.
\end{remark}



\subsubsection{Proof of \eqref{eq:ThmK4-1} and \eqref{eq:ThmK4-2}} \label{sec::pfofThmK4}


To prove \eqref{eq:ThmK4-1}, we follow the proof of \eqref{eq:zhaterr} in Section \ref{sec:pdfzhaterr}. As the decomposition \eqref{eq:thm2-15} still holds, 
it remains to upper bound the terms $S_1$--$S_3$, whose definitions are the same as in \eqref{eq:def_S1S2S3}. 

To prove \eqref{eq:ThmK4-2}, we follow the proof of \eqref{eq:wthaterr} in Section \ref{sec:pdfwthaterr}. As the decomposition \eqref{eq:thm2-27} still holds, it remains to upper bound the terms $S_4$--$S_6$, whose definitions are the same as in \eqref{eq:def_S4S5S6}. 

To finish the proof, 
we will establish the following probabilistic bounds for the error terms $S_1$--$S_6$:
\begin{align*}
    \Pr\big[S_1 > C_\varepsilon  \pi^{-1} T^{-1} \log(nT)\big] &= O((nT)^{-\varepsilon}), \quad \quad \quad \Pr\big[S_4 > C_\varepsilon  \pi^{-1} \log(nT)\big] = O((nT)^{-\varepsilon-1}),\\
     \Pr\big[S_2 > C_\varepsilon  \pi^{-4} n^{-1} \log^8(nT)\big] &= O((nT)^{-\varepsilon}), \quad \Pr\big[S_5 > C_\varepsilon \pi^{-4} n^{-1} \log^8(nT)\big] = O((nT)^{-\varepsilon-1}), \\ 
     \Pr\big[S_3 > C_\varepsilon  \pi^{-3} n^{-1} \log^5(nT)\big] &= O((nT)^{-\varepsilon}), \quad \Pr\big[S_6 > C_\varepsilon \pi^{-3} n^{-1} \log^5(nT)\big] = O((nT)^{-\varepsilon-1}).
\end{align*}
Firstly, we note that each element of $\dot\ell_{\Theta_t}(v^\star) $, i.e., $A_{t,ij} - \pi\mu(\Theta_{t,ij}) $, satisfies the Bernstein condition verified in \eqref{eq:bern_0p_model}. Thus, by applying Lemma \ref{adlem:wellner2005}, we have that
the upper bound of $S_1$ follows the same form as in \eqref{eq:thm2-21}, 
where $
    V_i = \VAR\big\{\big[\mathcal{D}_{Zv}\, I({v}^\star)^+ \dot \ell({v}^\star)\big]_i\big\}$, and $B_i$ is the infinity norm of the $i$-th row of the weight matrix given in \eqref{eq::S_1:weightmatrix}. 
Following  the same argument as in \eqref{eq:thm2-20-0}, we have $\max_{1\leqslant i \leqslant nk} B_i =  O(n^{-1} T^{-1/2} \pi^{-1} )$, 
where the additional $\pi^{-1}$ factor arises from $\big\|\DnT^{1/2}\, I({v}^\star)^+ \DnT^{1/2} \big\|_{\operatorname{op}}=O(\pi^{-1})$ by Lemma \ref{lm:0p_ivproperties} (iii).
In addition, $ \sum_{i = 1}^{nk} V_i$ in the inequality \eqref{eq:thm2-21} satisfies 
\begin{align} 
    \sum_{i = 1}^{nk} V_i &= \operatorname{tr}\big[\operatorname{cov}\big(\mathcal{D}_{Zv}\, I({v}^\star)^+ \dot \ell({v}^\star)\big)\big] \notag\\
    &= \operatorname{tr}\big[\mathcal{D}_{Zv} I(v^\star)^+ I_1(v^{\star}) I(v^\star)^+ \mathcal{D}_{Zv}^\mytrans \big], \quad (I_1(v) := \mathrm{cov}\{\dot \ell({v})\} = \mathbb{E}\{\dot \ell({v}) \dot \ell({v})^\top\})  \notag\\
    &= (nT)^{-1}\operatorname{tr}\big[\mathcal{D}_{Zv} \DnT^{1/2} I(v^\star)^+ I_1(v^{\star}) I(v^\star)^+  \DnT^{1/2} \mathcal{D}_{Zv}^\mytrans \big]  \notag\\
    &\leqslant k T^{-1} \, \big\|\mathcal{D}_{Zv} \DnT^{1/2} I(v^\star)^+ I_1(v^{\star}) I(v^\star)^+  \DnT^{1/2} \mathcal{D}_{Zv}^\mytrans\big\|_{\operatorname{op}}  \notag\\
    &\leqslant kT^{-1}\,  \big\|\DnT^{1/2}I(v^\star)^+ I_1(v^{\star}) I(v^\star)^+ \DnT^{1/2}\big\|_{\operatorname{op}}, \label{eq:vsum_bound_0p_model}
\end{align}
which follows the same arguments as \eqref{eq:thm2-22} with $I(v^{\star})^+$ replaced by $I(v^\star)^+ I_1(v^{\star}) I(v^\star)^+$. 
 Lemma \ref{lm:0p_ivproperties_I1} below  will show that $I_1(v^{\star})$ exhibits properties similar to $I(v^{\star})$ in Lemma \ref{lm:0p_ivproperties}. Specifically, we have $\big\|\DnT^{-1/2}I_1(v^{\star})\DnT^{-1/2} \big\|_{\operatorname{op}}=O(\pi)$,  which, combined with Lemma \ref{lm:0p_ivproperties} (iii), gives
\begin{align}
    \eqref{eq:vsum_bound_0p_model} & \leqslant  kT^{-1}\,  
    \big\|\DnT^{1/2}I(v^\star)^+ \DnT^{1/2}\big\|_{\operatorname{op}}
    \big\|\DnT^{-1/2}I_1(v^{\star})\DnT^{-1/2} \big\|_{\operatorname{op}}
    \big\|\DnT^{1/2}I(v^\star)^+ \DnT^{1/2}\big\|_{\operatorname{op}}
\notag\\
    &\leqslant C T^{-1}\pi^{-1} \pi \pi^{-1} =   C T^{-1}\pi^{-1}. \label{eq:vsum_bound_0p_model_2} 
\end{align}
Plugging the bounds of $\sum_{i=1}^{nk} V_i$ and $\max_{1\leqslant i\leqslant nk}  B_i$ back into \eqref{eq:thm2-21}, we obtain the bound for $S_1$.

For the bound of $S_2$, using the same argument for \eqref{eq:thm2-23},
we have
\begin{align}\label{eq:0p_thm2-23}
    S_2 \leqslant  \big\| \mathcal{D}_{Zv}\, I(\check{v})^+ \DnT^{1/2} \big\|_{\operatorname{op}}^2 \big\| \DnT^{-1/2}\big( I(\check{v}) + \tilde H \big) \DnT^{-1/2} \big\|_{\operatorname{op}}^2 \big\|\DnT^{1/2} (\check{v} - v_{q}^\star)  \big\|_2^2,
\end{align}
where $\tilde H$ is defined same as in \eqref{eq:def_tildeH}. 
To analyze \eqref{eq:0p_thm2-23}, we note 
Lemma \ref{lm:0p_ivproperties} can be applied to $I(v)$ at $v=\check{v}$ (equivalently vectorization of  $[\check Z, \check W_1, \ldots, \check W_T]$), because the assumption  (a) holds by the projection step  in line 4 of Algorithm \ref{algor:estY} and the assumption (b) is ensured by \eqref{eq:0p_1} for $\epsilon=2$.  
Following the arguments in \eqref{eq:s2-1} and \eqref{eq:s2-2}, we now obtain that with probability $1-O((nT)^{-\varepsilon})$,
\begin{align*}
    \big\| \mathcal{D}_{Zv}\, I(\check{v})^+ \DnT^{1/2} \big\|_{\operatorname{op}}^2=O(\pi^{-2}n^{-1}T^{-1})\quad \text{ and }\quad \big\|\DnT^{1/2} (\check{v} - v_{q}^\star)  \big\|_2^2 = O(\pi^{-2}nT\log^4(nT)),
\end{align*} 
where  additional $\pi^{-2}$ factors  arise due to applying Lemma \ref{lm:0p_ivproperties} (iii) and \eqref{eq:0p_1}, respectively.  
Following the arguments in \eqref{eq:thm2-24}, as $\tilde{H}=-\tilde{I}+\tilde{N}$ still holds by Remark \ref{rm::0p_HvIv},  we similarly have
\begin{align*}
  &~  \big\| \DnT^{-1/2}\big( I(\check{v}) + \tilde H \big) \DnT^{-1/2} \big\|_{\operatorname{op}}  \\
\lesssim &~\max_{v = \check{v} + u(v_{q}^\star - \check v)}\Big\{\big\| \DnT^{-1/2} \big( I(\check{v}) -  I(v) \big) \DnT^{-1/2}\big\|_{\operatorname{op}}  
 + \big\| \DnT^{-1/2} \big( N(v_q^\star) - N(v) \big) \DnT^{-1/2}\big\|_{\operatorname{op}}\Big\} \\
&~\quad+  \big\| \DnT^{-1/2} N(v_q^\star) \DnT^{-1/2}\big\|_{\operatorname{op}}  =  O(n^{-1/2} \log^2(nT))
\end{align*}
with probability $1-O((nT)^{-\varepsilon})$,
where  the last equation follows by Lemma \ref{lm:0p_ivproperties} (iv)--(v) and applying the proof of  Lemma \ref{lem:concentration}. Plugging the above upper bounds into \eqref{eq:0p_thm2-23}, we obtain the bound for $S_2$.


For the bound for $S_3$, using the same argument as in \eqref{eq:thm2-25}, we have
\begin{align} \label{eq:0p_thm2-25}
S_3
     &\leqslant 2(nT)^{-1} \big\|\DnT^{1/2} I(\check{v})^+ \DnT^{1/2}  \big\|_{\operatorname{op}}^2 \  \big\|\DnT^{1/2}  I(v_q^\star)^+  \DnT^{1/2} \big\|_{\operatorname{op}}^2\notag  \\
    &\quad\quad \big\|\DnT^{-1/2} \big(I(\check{v}) - I(v_q^\star) \big) \DnT^{-1/2} \big\|_{\operatorname{op}}^2 \ \big\|  \DnT^{-1/2} \dot \ell({v}_{q}^\star)\big\|_2^2.
\end{align} 
By Lemma \ref{lm:0p_ivproperties}, we have
\begin{align*}
    \big\|\DnT^{1/2} I(\check v)^+ \DnT^{1/2}\big\|_{\operatorname{op}}=O(\pi^{-1}), \quad \big\|\DnT^{1/2} I(v_q^\star)^+ \DnT^{1/2}\big\|_{\operatorname{op}}&=O(\pi^{-1}), \\ \text{and} \quad \quad  \quad \quad
    \big\|\DnT^{-1/2} \big(I(\check{v}) - I(v_q^\star) \big) \DnT^{-1/2} \big\|_{\operatorname{op}} &= O(n^{-1/2} \log^2(nT))
\end{align*}
with probability $1-O((nT)^{-\varepsilon})$. 
Combining the argument in \eqref{eq:thm2-26} with Lemma \ref{lm:0p_ivproperties_I1}, we further have 
\begin{align} 
\label{eq:dotl_2bound_0p_model_2}
    \big\|  \DnT^{-1/2} \dot \ell({v}_{q}^\star)\big\|_2^2  \leqslant C_\varepsilon \operatorname{tr}\big(\DnT^{-1/2}I_1(v^\star)\DnT^{-1/2}\big) \log(nT) \leqslant C_\varepsilon nT\pi \log(nT)
\end{align} with probability $1-O((nT)^{-\varepsilon}) $.
Plugging these bounds into \eqref{eq:0p_thm2-25}, we obtain the bound for $S_3$.

   Similar arguments to those used in bounding $S_1$–$S_3$ under model
  \eqref{eq::0p_model} apply to $S_4$--$S_6$.
Finally, substituting these bounds into \eqref{eq:thm2-15} and \eqref{eq:thm2-27} yields the conclusions of Theorem \ref{thm:0p_2}.


\paragraph{Formula and Properties of $I_1(v) = \mathbb{E}\{ \dot \ell(v) \dot \ell(v)^\top \}$.}
\label{paragraph::0p_I1v} 
To complete the above proof, particularly \eqref{eq:vsum_bound_0p_model_2} and \eqref{eq:dotl_2bound_0p_model_2},  
we next examine $I_1(v) = \mathbb{E}\{ \dot \ell(v) \dot \ell(v)^\top \}$,  where with slight abuse use of notation, the  expectation is taken over the data  generated by true parameters  $v$ in the following. First, we derive the formula of  $I_1(v)$. As $\dot\ell(v)$ still follows the form of \eqref{eq:ldotform} with $\DZ (v) $ and $ \Dwt (v)$ defined as in \eqref{eq:pre4} and \eqref{eq:pre5}, we have that $I_1(v) = \EXPT\big\{\dot \ell(v) \dot \ell(v)^\mytrans\big\}=[I_{1,qm}(v)]_{q,m\in\{Z,W_1,\ldots,W_T\}}$ is a $(1+T)\times (1+T)$ block matrix  
with each block given by
\begin{align*}
  I_{1,ZZ}(v) &:=  \EXPT\Big[ \big(\textstyle \sum_{t=1}^T \DZ(v) \dot \ell_{\Theta_t}(v)\big) \big(\textstyle \sum_{t=1}^T \DZ(v) \dot \ell_{\Theta_t}(v)\big)^\mytrans\Big] = \textstyle \sum_{t=1}^T \DZ(v) \mathcal{D}_{1,\mu_t}(v) \DZt(v), \\
 I_{1,ZW_s}(v) &:=   \EXPT\Big[ \big(\textstyle \sum_{t=1}^T \DZ(v) \dot \ell_{\Theta_t}(v)\big) \big(\textstyle  \mathcal{D}_{W_s \Theta}(v) \dot \ell_{\Theta_s}(v)\big)^\mytrans\Big] =  \DZ(v)  \mathcal{D}_{1,\mu_s}(v) \mathcal{D}_{\Theta W_s}(v), \\
 I_{1,W_sZ}(v) &:= \EXPT\Big[ \big(\textstyle  \mathcal{D}_{W_s \Theta}(v) \dot \ell_{\Theta_s}(v)\big)\big(\textstyle \sum_{t=1}^T \DZ(v) \dot \ell_{\Theta_t}(v)\big)^\mytrans\Big] =  \mathcal{D}_{ W_s \Theta}(v)  \mathcal{D}_{1,\mu_s}(v)  \DZt(v),\\
  I_{1,W_tW_s}(v) &:=   \EXPT\Big[ \big( \Dwt(v) \dot \ell_{\Theta_t}(v)\big) \big(\textstyle  \mathcal{D}_{W_s \Theta}(v) \dot \ell_{\Theta_s}(v)\big)^\mytrans\Big] =  \Dwt(v) \mathcal{D}_{1,\mu_t}(v)  \Dwtt(v) \mathbb{I}(t = s), 
\end{align*}
which are computed by $\EXPT\big\{\dot \ell_{\Theta_t}(v) \dot \ell_{\Theta_s}(v)^\mytrans\big\} = 0$  for any $1\leqslant t \neq s \leqslant T$, and  
\begin{align}
\label{eq:0p_D1mut}
&  \mathcal{D}_{1,\mu_t}(v) :=  \EXPT\big\{\dot \ell_{\Theta_t}(v) \dot \ell_{\Theta_t}(v)^\mytrans\big\} \in \mathbb R^{\frac{n(n+1)}{2} \times \frac{n(n+1)}{2}} \notag\\
    =&  \operatorname{diag}\big(2 \EXPT\big[\big\{l^\prime(\Theta_{t,11})\big\}^2\big] , \ldots, 2 \EXPT\big[\big\{l^\prime(\Theta_{t,nn})\big\}^2\big] ,  \EXPT\big[\big\{l^\prime(\Theta_{t,12})\big\}^2\big] , \ldots, \EXPT\big[\big\{l^\prime(\Theta_{t,n-1,n})\big\}^2\big]  \big).
\end{align}
In \eqref{eq:0p_D1mut}, for any $1\leqslant i, j\leqslant n $ and $1\leqslant t\leqslant T$, 
\begin{align}
\EXPT\big[\big\{l^\prime(\Theta_{t,ij})\big\}^2\big]   =&~ \EXPT\big[\big\{A_{t,ij} - \pi \mu(\Theta_{t,ij})\big\}^2\big] = \VAR\big\{A_{t,ij} \big\} \notag\\
=&~ \EXPT\big[\VAR\big\{A_{t,ij}\mid\Omega_{t,ij}\big\}\big] + \VAR\big[\EXPT\big\{A_{t,ij} \mid\Omega_{t,ij}\big\}\big] \notag\\
=&~ \EXPT\big\{ \Omega_{t,ij} \mu^\prime(\Theta_{t,ij}) \big\}  +   \VAR\big\{\Omega_{t,ij}\mu(\Theta_{t,ij})\big\} \notag\\
=&~ \pi \mu^\prime(\Theta_{t,ij}) + \pi(1-\pi)\mu^2(\Theta_{t,ij}). \label{eq:0p_1mut2mut} 
\end{align}

Lemma \ref{lm:0p_ivproperties_I1}  below shows that the properties in Lemma  \ref{lm:0p_ivproperties} for $I(v)$ also hold for $I_1(v)$.

\begin{lemma}\label{lm:0p_ivproperties_I1}
Under the same assumptions as in Lemma  \ref{lm:0p_ivproperties}, (i)--(vi) in Lemma  \ref{lm:0p_ivproperties}  still hold with $I(\cdot)$ replaced by $I_1(\cdot)$. 
\end{lemma}
\begin{proof}
The formula of $I_1(v)$ is similar to $I(v)$ (defined in the form of \eqref{eq:ivform}) except that $\Dmut(v)$ in \eqref{eq:0p_Dumtform} is replaced by $ \mathcal{D}_{1,\mu_t}(v) $ in \eqref{eq:0p_D1mut}. 
Nevertheless, by \eqref{eq:0p_1mut2mut},  $\EXPT\big[\big\{l^\prime(\Theta_{t,ij})\big\}^2\big] \asymp \pi$ exhibits the rate similar to $-l''(\Theta_{t,ij})\asymp \pi$ used in the proof of Lemma  \ref{lm:0p_ivproperties}.
As a result, despite the differences between the formulae of $\Dmut(v)$ and $ \mathcal{D}_{1,\mu_t}(v) $, the proof of Lemma \ref{lm:0p_ivproperties} can be similarly applied to $I_1(v)$ to reach the same conclusions.     
\end{proof}

\bigskip
\begin{remark}[Relationship between $I(v)$ and $I_1(v)$]\label{rmk:0p_I12} 
When $\pi=1$ and 
$l(\cdot)$ is in the natural exponential family, we have 
$\EXPT[\{l^\prime(\Theta_{t,ij})\}^2] = -l''(\Theta_{t,ij})$, implying
$\Dmut(v)$ defined in \eqref{eq:0p_Dumtform} and $\mathcal{D}_{1,\mu}(t)$ defined in \eqref{eq:0p_D1mut} are equal, and thus  $I(v) = I_1(v)$. 
This equivalence, however, may not hold generally for $I(v)$ and $I_1(v)$ constructed based on the pseudo likelihood $\ell(\cdot)$ with $l(\theta; x) = \theta x - \pi \nu(\theta)$ under~\eqref{eq::0p_model}. This  necessitates the examination of $I_1(v)$ in addition to $I(v)$, which differs from the proof under the original model \eqref{eq:model}. 
Nevertheless, as Lemma \ref{lm:0p_ivproperties_I1} shows that  $I_1(v)$ and $I(v)$ enjoy similar properties, 
our main theoretical arguments carry through with slight modifications,
as demonstrated above.  
\end{remark}

\newpage
\section{{Evaluation by Edge Cross-Validation}} \label{sec:extend_full}

To evaluate the fit of network models, one practical assessment is applying the estimated models to link  prediction, which relies on estimating the underlying expected adjacency matrices \citep{zhang2017estimating}. 
To this end, we adopt the edge cross-validation framework in \cite{li2020network}. 
Specifically, we randomly mask a subset of edges in each network and fit the model using only the unmasked entries. The fitted model is then evaluated by how well it recovers the masked edges. 
In the following, we describe the evaluation procedure in detail, 
followed by numerical results on both simulated data and the Lazega lawyer dataset. 


\subsection{{Procedure}} \label{sec:extend_full_model}

\paragraph{Edge Subsampling} 
Let $\mathbf{A}_t^o=(A_{t,ij}^o)_{1\leqslant i,j\leqslant n}$ denote the originally observed network adjacency matrix. 
For each node pair $(i,j)$ in the $t$-th network, we independently sample $\Omega_{t,ij}$ from a Bernoulli distribution with the success probability $\pi\in (0,1]$, where $A_{t,ij}^o$ is masked if $\Omega_{t,ij}=0$ and kept if $\Omega_{t,ij}=1$. Then we estimate the network model based on the partially observed networks represented as  matrices $\mathbf{A}_t:=[A_{t,ij}^o\Omega_{t,ij}]_{1\leqslant i,j\leqslant n}$. 
The estimates of original expectation $\mathbb{E}(A_{t,ij}^o)$ for masked edges ($\Omega_{t,ij}=0$) are compared to the masked true values $A_{t,ij}^o$ to assess the model fit.




\paragraph{Model Estimation for Partially Masked Networks} 
When $\mathbb{E}(\mathbf{A}_t^o)=\mu(\Theta_t^\star)$,
 the subsampled data $\mathbf{A}_t$ satisfies $\mathbb{E}( \mathbf A_t)= \pi \mu(\Theta_t^\star)$, 
which is similar to the moment property \eqref{eq:0p_model_mean}  of $\mathbf{A}_t$ under the mixture model \eqref{eq::0p_model}.  
As a result, estimation methods and theoretical results remain the same as those in Sections \ref{sec:0p_method} and \ref{sec:0p_theory} and are therefore omitted for conciseness. 

\subsection{{Numerical Results}}
\paragraph{Simulations}
We evaluate the link prediction performance of our proposed method 
using simulated data with binary edges. 
To generate the data, we first simulate complete binary networks following the setup in Section \ref{sec:simulation} of the main text, considering the three parameter settings Cases (A)–(C). 
Each node pair is then independently masked with probability 0.1. 
This procedure is repeated 100 times. 
For each realization, we fit the model using the partially observed network data, obtain the estimated edge probabilities $\{\hat p_{t,ij}\}$, and assess how well they recover the masked binary edges.

How well the estimated $\{\hat p_{t,ij}, \ (t,i,j): \Omega_{t,ij} = 0 \}$ predicts the held-out binary links $\{A_{t,ij}^o, \ (t,i,j): \Omega_{t,ij} = 0 \}$
is evaluated via receiver operating characteristic (ROC) curves following \cite{zhang2017estimating}.  
In particular, for each $c\in (0,1)$, we define the false-positive and true-positive rates as
\begin{align}
 \mathrm{FPR}(c) =  &~ \sum_{(t,i,j):\Omega_{t,ij} =0} \mathbb I( \hat{p}_{t,ij} > c , A_{t,ij}^o=0)\, \biggr/\sum_{(t,i,j):\Omega_{t,ij}=0} \mathbb I( A_{t,ij}^o=0), \notag \\
\mathrm{TPR}(c) = &~ \sum_{(t,i,j):\Omega_{t,ij}=0} \mathbb I( \hat{p}_{t,ij} > c , A_{t,ij}^o=1) \, \biggr/\sum_{(t,i,j):\Omega_{t,ij}=0} \mathbb I( A_{t,ij}^o=1). \notag
\end{align}
By varying $c\in (0,1)$, we obtain the ROC curve. 
As a baseline, we compare with  MultiNeSS+ and  an oracle method that uses true expected adjacency matrices to predict removed edges, i.e., using true $\mathbb{E}(A_{t,ij}^o)$ in the place of $\hat{p}_{t,ij}$ in $\mathrm{FPR}(c)$ and $\mathrm{TPR}(c)$ above.

\begin{figure}[!htbp]
\centering
\begin{subfigure}{0.27\textwidth}
  \centering
      \caption{Case (A)}
  \includegraphics[width=1\linewidth]{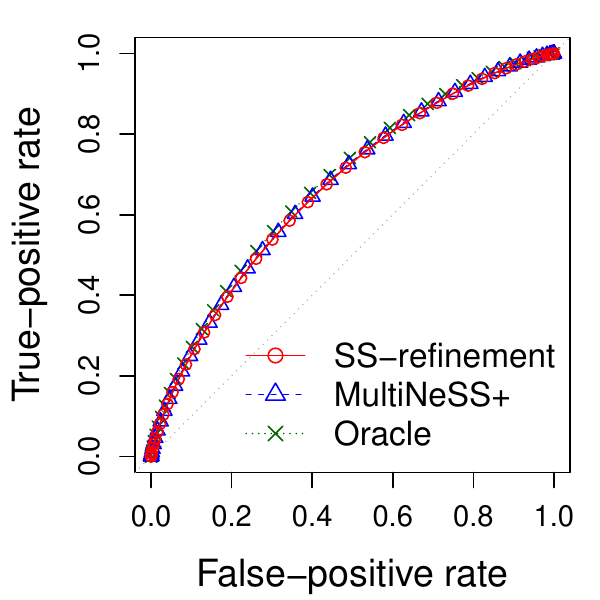}
\end{subfigure} \hfill 
\begin{subfigure}{0.27\textwidth}
  \centering
      \caption{Case (B)}
  \includegraphics[width=1\linewidth]{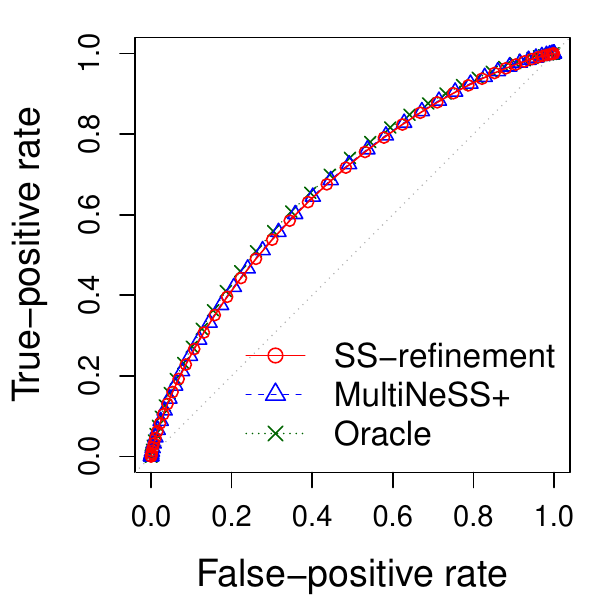}
\end{subfigure}\hfill
\begin{subfigure}{0.27\textwidth}
  \centering
      \caption{Case (C)}
  \includegraphics[width=1\linewidth]{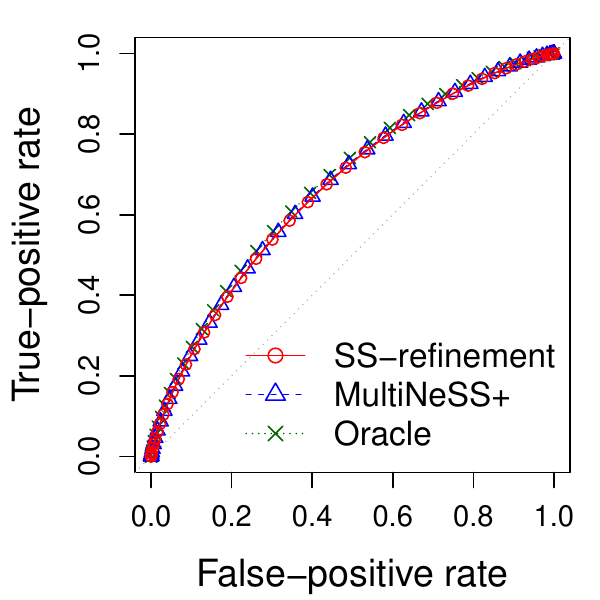}
\end{subfigure}
\caption{Average ROC curves of the proposed SS-Refinement, MultiNeSS+, and the oracle method under Cases (A)–(C) and $T=5$.} 
\label{fig:pred_simu_ber_rev_5}
\end{figure}

\begin{figure}
\centering
\begin{subfigure}{0.27\textwidth}
  \centering
      \caption{Case (A)}
  \includegraphics[width=1\linewidth]{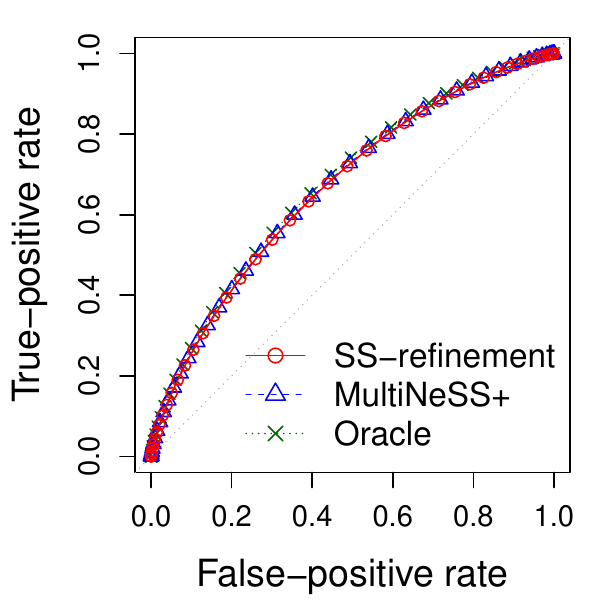}
\end{subfigure} \hfill 
\begin{subfigure}{0.27\textwidth}
  \centering
      \caption{Case (B)}
  \includegraphics[width=1\linewidth]{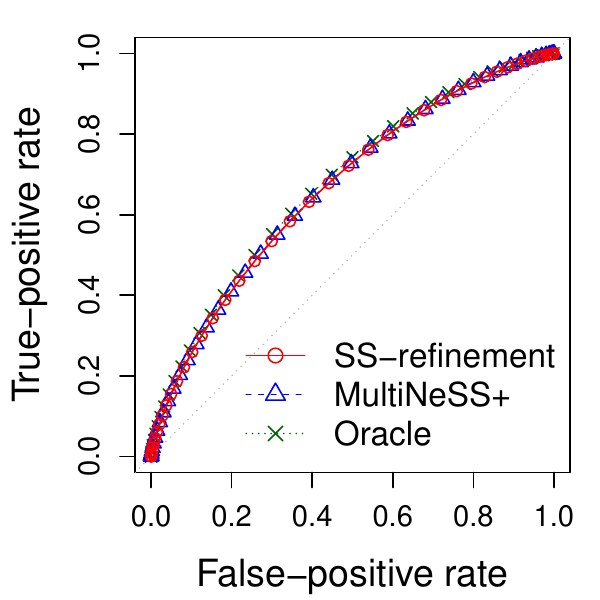}
\end{subfigure}\hfill
\begin{subfigure}{0.27\textwidth}
  \centering
      \caption{Case (C)}
  \includegraphics[width=1\linewidth]{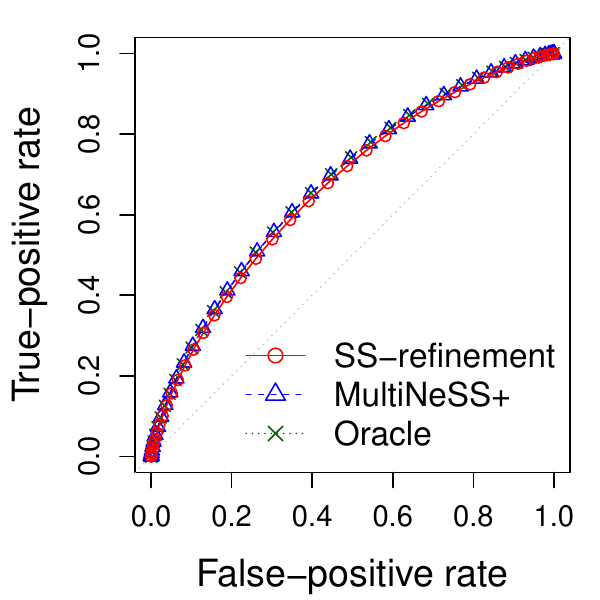}
\end{subfigure}
\caption{Average ROC curves of the proposed SS-Refinement, MultiNeSS+, and the oracle method under Cases (A)–(C) and $T=80$.} 
\label{fig:pred_simu_ber_rev_80}
\end{figure}

{We report the average ROC curves across the 100 realizations of the proposed SS-Refinement, MultiNeSS+, and the oracle method in Figures~\ref{fig:pred_simu_ber_rev_5} and \ref{fig:pred_simu_ber_rev_80}, under $T\in \{5, 80\}$, respectively. 
The ROC curves of the three methods remain close across all settings, suggesting that both our estimation procedure and MultiNeSS+ effectively recover the underlying connecting probabilities for link prediction.}





\paragraph{Real Data Analysis}

\begin{figure}[!htbp]
    \centering
    \includegraphics[width=0.26\linewidth]{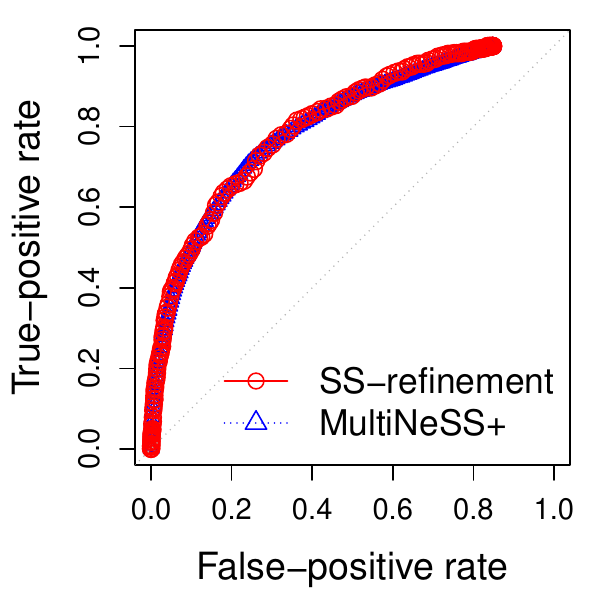}
    \caption{Average ROC curves of the proposed SS-Refinement and MultiNeSS+ under the same model structure.} 
    \label{fig:ROC_Lazega_rev_both}
\end{figure}

We further evaluate the proposed estimation procedure on the Lazega lawyer dataset. We similarly create partially observed networks with each node pair removed with probability 0.1 independently, and repeat this procedure 100 times. 
As the true model is unknown, we skip the oracle method, and fit SS-Refinement and MultiNeSS+ under the same model structure for a fair comparison. 
 Figure \ref{fig:ROC_Lazega_rev_both} presents the average ROC curves of SS-Refinement and MultiNeSS+, respectively.  Similarly to the simulations, the two types of fitting methods yield similar results. 
 
\end{document}